\documentclass[12pt,a4paper,twoside]{report}
\usepackage{fullpage}
\usepackage{amssymb}
\usepackage{amsmath}
\usepackage{commath}
\usepackage{abstract}
\usepackage{graphicx,graphics}
\usepackage{color}
\usepackage{float}
\usepackage{url}
\usepackage{xcolor}
\usepackage{hyperref}
\usepackage[symbol]{footmisc}
\usepackage[font={small,it}]{caption}
\usepackage{tikz}
\usetikzlibrary{calc}
\usepackage{filecontents,notoccite}
\usepackage{fancyhdr,layout}
\usepackage[titletoc]{appendix}

\usepackage{graphicx,graphics}
\usepackage{placeins}
\usepackage{amsmath}
\usepackage{amssymb}
\usepackage{mathtools}
\usepackage{onimage}
\usepackage{caption}
\usepackage{array}
\usepackage{subcaption}
\usepackage{lineno,hyperref}
\usepackage{color}
\usepackage{ulem}
\usepackage{dcolumn}
\usepackage{bm}

\usepackage[utf8]{inputenc}
\usepackage[LGR, T1]{fontenc}
\usepackage[english]{babel}
\usepackage{libertine}
\usepackage{afterpage}

\usepackage{pdfpages}

\newcommand\blankpage{%
    \null
    \thispagestyle{empty}%
    \addtocounter{page}{-1}%
    \newpage}

\newcommand{\p}{^}

\newcommand{\vR}{\mbox{{\bm{$R$}}}}

\newcommand{\vphi}{\mbox{{\bm{$\phi$}}}}

\newcommand{\vv}[1]{\boldsymbol{#1}}
\newcommand{\n}{\vv{\nabla}}
\renewcommand{\l}{\mathopen{}\mathclose\bgroup\left}
\renewcommand{\r}{\aftergroup\egroup\right}
\newcommand{\veps}{\mbox{{\bm{$\epsilon$}}}}
\newcommand{\diff} [1]{\mathrm{d}{#1}} 
\newcommand{\phia}{\phi_{\alpha}}
\newcommand{\phib}{\phi_{\theta}}
\newcommand{\gab}{\gamma_{\alpha \theta}}

\newcommand{\C}{\vv{\mathcal{C}}}

\let\originaleps=\epsilon
\let\epsilon=\varepsilon
\let\varepsilon=\originaleps
\usepackage{stmaryrd}
\newcommand{\ljump}{\llbracket}
\newcommand{\rjump}{\rrbracket}

\newcommand{\T}{\mathsf{\!T}}

\newcommand{\TT}{\vv{\mathcal{T}}}
\renewcommand{\S}{\vv{\mathcal{S}}}

\usepackage[refpage,intoc]{nomencl}
\usepackage{ifthen}
\usepackage{siunitx}

\makenomenclature

\renewcommand{\nomgroup}[1]{
  \ifthenelse{%
    \equal{#1}{A}%
  }{%
  \item[\textbf{Symbols}]%

  }{%
    \ifthenelse{\equal{#1}{B}}{%
    \item[\textbf{Acronyms}]%
    }{}%
  }%
}

\begin{document}

\begin{titlepage}

\begin{tikzpicture}[overlay,remember picture]
\draw [line width=1.0pt,rounded corners=10pt,]
    ($ (current page.north west) + (1cm,-1cm) $)
    rectangle
    ($ (current page.south east) + (-1cm,1cm) $);       
\end{tikzpicture}
\begin{figure}
\includegraphics[scale=0.08]{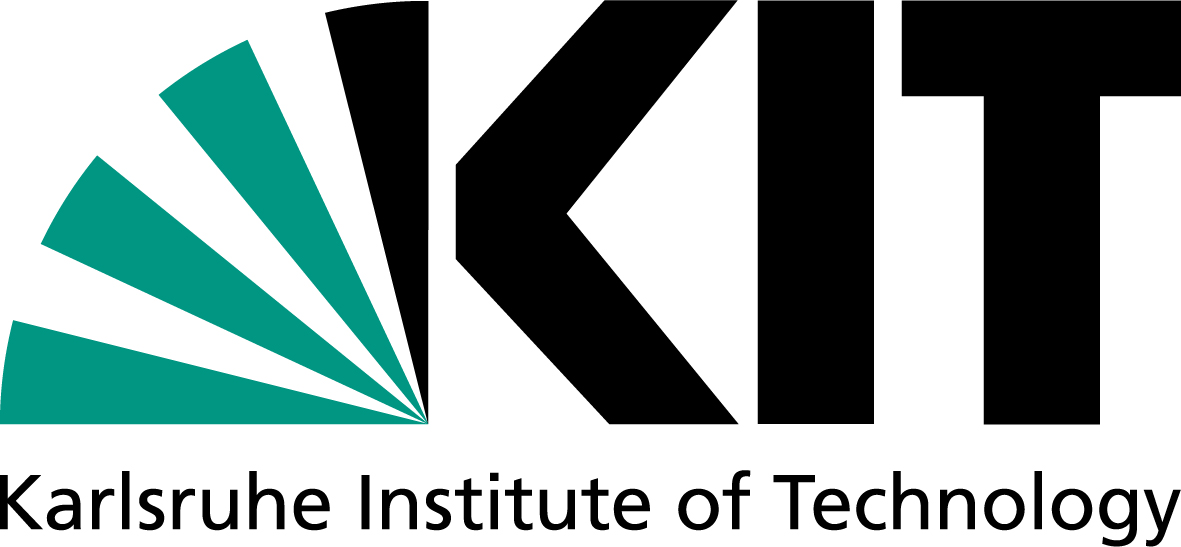}
\end{figure}\\
\Huge\centering\textbf{Understanding the volume-diffusion governed shape-instabilities in metallic systems}

\vspace*{1cm}
\Large\centering{Zur Erlangung des akademischen Grades \\ \textbf{Doktor der Ingenieurwissenschaften} \\ von der Fakult{\"a}t f{\"u}r Maschinenbau des \\ Karlsruher Institut f{\"u}r Technologie (KIT)}

\vspace*{0.5cm}
\Large\centering{genehmigte \\ \textbf{Dissertation} \\ von}

\vspace*{0.5cm}
\Large\centering\textbf{M.Tech. Prince Gideon Kubendran Amos}

\vspace*{1cm}
\begin{center}
 \Large{Tag der mundlichen Prufung : 23, May, 2019}\\
 \Large{Hauptreferent : Prof. Dr. rer. nat. Britta Nestler}\\
 \Large{Korreferent 1 : Prof. Dr. rer. nat. Wolfgang Wenzel}
\end{center}

\end{titlepage}


\includepdf[fitpaper= true, pages=-]{dedication.pdf}
 
\afterpage{\blankpage}

\setcounter{page}{1}
\setcounter{secnumdepth}{4} 
\setcounter{tocdepth}{4}
\pagenumbering{roman}
\cleardoublepage
\phantomsection

\pagestyle{fancy}
\fancyhf{}
\lhead[\thepage]{Abstract}      
\rhead[\thesection]{\thepage}
\chapter*{Abstract}

The reliability of any day-to-day material is critically dictated by its properties.
One factor which governs the behaviour of a material, under a given condition, is the microstructure.
Despite the absence of any phase transformation, a change in the microstructure would significantly alter the properties.
Therefore, a substantial understanding on the stability of the microstructure is vital to avert any unexpected catastrophic change in the material properties.

Employing conventional techniques, particularly, experimental investigations to explicate the evolution of the microstructure is an arduous task, mainly, because the spatial distribution of the phases extends beyond the regular two-dimensional representation.
Consequently, theoretical treatments are adopted to complement the experimental observations and enhance the understanding of the shape-instability in a microstructure.
With the increasing availability of the computational resources, the contribution of the numerical analyses in delineating the intricacies of complex phenomena has been progressively expanding.
In the present work, one such numerical approach called phase-field modelling in employed to analyse the stability of two- and three-dimensional finite structures, which dictate the curvature-driven evolution of the microstructure.  
A characteristic feature of this numerical approach is the introduction of a scalar variable, called the phase field, in addition to the other thermodynamic variables.
While the inclusion of the phase field obviates the need for the interface tracking, which is a strenuous aspect of the other conventional techniques, it replaces the sharp interface with a finite diffuse region.
Therefore, before adopting and extending the phase-field technique, it is shown that the model recovers the governing law, $i.e,$ Gibbs-Thomson relation, despite the introduction of the diffuse interface.
Subsequently, the numerical treatment is employed to investigate the volume-diffusion governed curvature-induced transformation.

The present study begins with the analysis of the morphological evolution of a discontinuous precipitate in a representative lamellar arrangement, in order to realise the influence of the neighbouring structures on the transformation of the individual precipitate.
Consistent with the existing reports, this analysis unravels that the role of the adjacent precipitates in guiding the evolution of an isolated precipitate decreases significantly with increase in the distance separating them.
Correspondingly, the transformation induced by the curvature difference associated with the inherent shape of the individual precipitate is extensively investigated.
As opposed to the existing studies, the present technique renders a cumulative, and an exhaustive, analysis of the mechanism and the kinetics of the volume-diffusion governed transformations that are prevalent in the metallic systems.
Consequently, critical aspects of the shape-instability which have hitherto been unknown, or conveniently assumed, is unraveled in this study.
\chapter*{Kurzfassung}

Die Verlässlichkeit jedes täglich vorkommenden Materials wird entscheidend durch seine Eigenschaften bestimmt.
Ein Faktor, durch den das Verhalten eines Materials unter einem gegebenen Zustand geregelt wird, ist die Mikrostruktur.
Obwohl keine Phasenumwandlung vorhanden ist, würde eine Veränderung in der Mikrostruktur die Eigenschaften erheblich verändern.
Deshalb ist ein umfassendes Verständnis der mikrostrukturellen Stabilität notwendig, um unerwartete katastropische Veränderungen in den Materialeigenschaften zu verhindern.

Es ist nicht einfach, die Mikrostrukturentwicklung durch die Verwendung von konventionellen Techniken, besonders von experimentellen Untersuchungen, zu erklären, da die räumliche Verteilung hauptsächlich über die reguläre zweidimensionale Darstellung hinausgeht.
Deshalb werden theoretische Behandlungen übernommen, um die experimentellen Beobachtungen zu vervollständigen und das Verständnis der Forminstabilität in einer Mikrostruktur zu verbessern.
Durch die zunehmende Verfügbarkeit leistungsfähiger Rechner hat der Beitrag der numerischen Analysen zu der Beschreibung der Komplikationen komplexer Phänomene schrittweise zugenommen.
In der vorliegenden Arbeit wird ein numerischer Ansatz verwendet, der als Phasenfeldmodellierung bezeichnet wird, um die Stabilität von zwei- und dreidimensionalen finiten Strukturen zu analysieren, durch die die krümmungsgetriebene Entwicklung der Mikrostruktur bestimmt wird.  
Mit diesem numerischen Ansatz wird das sogenannte Phasenfeld  zusätzlich zu den anderen thermodynamischen Variablen als eine skalare Variable eingeführt, was eine charakteristische Eigenschaft dieses numerischen Ansatzes ist.
Während die Einbindung des Phasenfelds keine Grenzflächenverfolgung benötigt, was ein schwieriger Aspekt der anderen konventionellen Techniken ist, ersetzt sie die scharfe Grenzfläche mit einem finit diffusen Bereich.
Bevor das Phasenfeldverfahren übernommen und erweitert wird, wird deshalb gezeigt, dass durch dieses Modell das geltende Gesetz, z. B. die Gibbs-Thomson-Beziehung, trotz der Einführung der diffusen Grenzfläche wiederhergestellt wird.
Anschließend wird die numerische Behandlung für die Untersuchung der krümmungsinduzierten Umwandlung verwendet, die durch die Volumendiffusion bestimmt wird.

Die vorliegende Studie beginnt mit der Analyse der morphologischen Entwicklung einer diskontinuierlichen Ablagerung in einer repräsentativen lamellaren Anordnung, um den Einfluss der benachbarten Strukturen auf die Umwandlung der einzelnen Ablagerungen verstehen zu können.
Im Einklang mit den bereits existierenden Berichten zeigt diese Analyse, dass die Funktion der benachbarten Ablagerungen, die daraus besteht, die Entwicklung einer isolierten Ablagerung zu steuern, deutlich abnimmt, wenn der Abstand, durch den die Ablagerungen voneinander getrennt werden, zunimmt.
Entsprechend wird die Umwandlung, die durch den Krümmungsunterschied induziert wird, der mit der inhärenten Form der einzelnen Ablagerungen verbunden ist, umfangreich untersucht.
Im Gegensatz zu den existierenden Untersuchungen liefert das vorliegende Verfahren eine kumulative und vollständige Analyse des Mechanismus und der Kinetik der Umwandlungen, die durch die Volumendiffusion geregelt werden und in Metallsystemen weit verbreitet sind.
Somit werden die kritischen Aspekte der Forminstabilität, die bis jetzt noch unbekannt waren oder nur vermutet wurden, in dieser Untersuchung erklärt.

\afterpage{\blankpage}
\afterpage{\blankpage}

\chapter*{Acknowledgement}

\section*{\lq nanos gigantum humeris insidentes\rq \thinspace}

Sir Isaac Newton humbly claimed \lq If I have seen further it is by standing on the shoulders of Giants.\rq \thinspace,
I dare not to say the same.
However, I attempt to mention few \lq Giants\rq \thinspace who helped me \lq see\rq \thinspace clearly in the realm of research.

I begin by thanking Prof. Britta Nestler who allowed me to join her group and helped me grow in a way that suited me.
The initial guidance of Dr. Arnab Mukherjee, Prof. Kumar Ankit, Dr. Avisor Bhattacharya and Dr. Sebastin Schulz in my early days, which opened doors to phase-field modelling, is remembered in gratitude.
I am greatly indebted to my group leader Dr. Daniel Schneider and my brothers in \lq keys\rq \thinspace,  Ephraim Schoof, Ramanathan Perumal, Tobias Mittnacht, Jay Santoki and Nikhil Kulkarni, who patiently nurtured my works by being an active part of it, despite my never ending questions.  
The support of Prof. Leslie T Mushongera who encouraged me sufficiently in writing manuscripts, is acknowledged with immense gratitude.
I thank Christoph Hermann, Christof Ratz and Lukas Schoeller for their aid, and for creating a wonderful environment to work in.
The buoyant ambiance maintained by my fellow scholars, who are always ready to help, is thankfully appreciated.
My works owe a lot, a very lot, to the genius minds, both past and present, behind the adroit package PACE-3D.

At this moment, I look upon my \lq AMMA\rq \thinspace, who always supported me through her prayers, and my \lq APPA\rq \thinspace, who never gives up on me.
The cheery support of Papa and Robin is remembered with great fondness.
I thank my wife, Dr. Med. Deepti Madurai Muthu.
Her sacrifices and efforts are often left unsung, but for the home she devised, my works would not have been possible.
By the way, she is responsible for the my odd working hours.

Ultimately, with all my heart, I thank the \textbf{triune God}, whose will I intended to fulfill everyday but failed miserably often.

%
%


\afterpage{\blankpage}

\pagestyle{fancy}
\fancyhf{}
\lhead[\thepage]{Contents}      
\rhead[\thesection]{\thepage}
\tableofcontents
\clearpage

\afterpage{\blankpage}


\pagestyle{fancy}
\fancyhf{}
\lhead[\thepage]{List of figures and tables}      
\rhead[\thesection]{\thepage}
\listoffigures
\rhead[\thesection]{\thepage}
\clearpage




\newpage\null\thispagestyle{empty}\newpage

\newpage
\pagenumbering{arabic}
\setcounter{page}{1}
\pagestyle{fancy}
\fancyhf{}
\lhead[\thepage]{Chapter \thechapter .}      
\rhead[\thesection]{\thepage}


\newpage
\thispagestyle{empty}
\vspace*{8cm}
\begin{center}
 \Huge \textbf{Part I} \\
 \Huge \textbf{Motivation and Introduction}
\end{center}

\newpage\null\thispagestyle{empty}\newpage

\chapter{Motivation}

The effect of capillarity in fluids is evident and macroscopically observed.
For instance, a steady flow of a liquid in the form of a cylindrical jet breaks-up into individual globules under a definite condition.
Owing to the macroscopic influence of capillarity, the stability of the fluid flow has been extensively analysed by both physicist and engineers.
In a system of two immiscible fluids, a definite region called the interface separates the constituent phases.
When the morphology of a fluid evolves in response to the capillarity-induced instability, generally, the mass enclosed by the interface offers negligible resistance to the transformation.
Therefore, the shape-change is predominantly governed by the mass transfer along the interface.
The mass transfer along the region distinguishing the fluids, which subsequently results in the migration of the interface, is referred to surface (or interface) diffusion.
In other words, the transformation of a three-dimensional structure is dictated by the mass transfer along two-dimensional interface.
Theoretical studies investigating the curvature-induced instability in fluids, appropriately assume that the consequent morphological evolution is exclusively governed by the surface diffusion~\cite{rayleigh1878instability,rayleigh1892xvi,rayleigh1892xix}.
This consideration averts the need for analysing the entire domain, and simplifies the approach by restricting the governing-physics to the interface region.

Analogous to fluids, although not macroscopically, the curvature-induced transformations are observed in solid-state systems on a microscopic scale.
The microstructure, which depicts the distribution of the phases in solid-state materials, include structures that inherently exhibit difference in the curvature.
Under an appropriate thermodynamic condition, governed by the curvature difference, the phases transform morphologically.
Since the driving force for this evolution implicitly emerges from the shape of the precipitate, the resulting phenomena is referred to as the shape-instability.

In metallic systems, the properties are considerably influenced by the morphology of the phases in the microstructure.
Thus, a comprehensive understanding of the shape-instability serves two critical purposes.
One, it aides in achieving the desired properties without any phase transformation.
Two, it predicts the stability of a microstructure, in turn a material, for a given condition.
Unlike fluids, since the shape-instability in metallic systems is observed only on a microscopic level, experimental studies demand both microscopic and in-situ analysis.
Therefore, theoretical treatments are employed to delineate the curvature-induced transformations in the microstructure.

Owing to the similarity in the driving force, often the approach formulated to explicate the effect of capillarity in fluid flow is adopted to analyse the shape-instability~\cite{nichols1965morphological,nichols1976spheroidization,qian1994thermodynamic}.
However, in metallic systems, the mass transfer is not predominantly confined to the interface.
In other words, the transformations induced by the shape-instability are not exclusively governed by surface diffusion.
Therefore, a quantitative analysis demands an approach which includes the role of the volume diffusion by considering the entire domain.
Since treating a entire domain is inherently more complex than the simplified technique that focuses on the interface dynamics, theoretical investigations encompassing volume diffusion are limited and largely confined to two-dimension~\cite{tian2014phase}.
Moreover, due to the lack of in-situ information on the volume-diffusion governed transformation, analytical studies assume the mechanism of evolution to predict the kinetics~\cite{courtney1989shape,semiatin2005prediction,park2012mechanisms}.
In order to render a comprehensive understanding of the shape-instability in metallic systems, in the present work, a well-established numerical approach, which includes volume diffusion and implicitly considers the temporal change in the entire domain, is adopted to analyse the curvature-driven transformations. 

\section{Outline}

Before delineating the numerical approach that is adopted to investigate the curvature-induced transformation, various forms of shape-instability observed in the metallic system is introduced in Chapter~\ref{chap:shape}.
Subsequently, a review of the theoretical advancements leading up to the model adopted in the present study is rendered in Chapter~\ref{chap:pfi}.
In Chapter~\ref{sec:grand_potentialM}, the model employed to analyse the shape-instability is introduced.
The thermodynamic consistency of the model in recovering the physical laws, and the framework to incorporate quantitative data is elucidated in Chapter~\ref{chap:quant}.
In Chapter~\ref{chap:chm_elast}, a chemo-elastic model which include elastic driving-force and distinguishes interstitial diffusion from the substitutional is presented.
In the subsequent chapters, the kinetics and the mechanism of the shape-instability induced transformations is extensively analysed.

To understand the influence of the neighbouring structures on the evolution of an isolated precipitate, the morphological evolution of a representative lamellar microstructure is studied in Chapter~\ref{chap:fault}.
In Chapters~\ref{chap:ribbon} and ~\ref{chap:rods}, the stability of the ribbon-like structures and three-dimensional rods are respectively investigated.
The morphological evolution of the pancake, elliptical and faceted plate are correspondingly elucidated in Chapters~\ref{chap:pancake},~\ref{chap:epplate} and ~\ref{chap:pearlite}

\chapter{Shape-instabilities in metallic systems}\label{chap:shape}

One prominent aspect of the vast field of material science involves understanding the influence of the microstructure on the properties of the materials~\cite{torquato2013random,bhadeshia2017steels}.
Since the applicability of any material is influenced by its behaviour under an imposed circumstance, the evolution of the microstructure which dictates the material behaviour is investigated comprehensively.
Microstructure, particularly in metallic (alloy) systems, comprises of polycrystalline arrangement of different phases, wherein each phases are characteristically distinguished by chemical composition and/or crystallographic arrangement of atoms~\cite{callister2011materials}.
Therefore, the microstructure influences the properties by governing the chemical composition and crystal structure of the material through the volume fraction of the constituent phases.
Consequently, a phase transformation which alters the volume fraction of the phases, introduces a appropriate change in the material properties.
Different manufacturing process like heat treatment techniques are meticulously devised to establish precise combination of the phases to render desired properties.

One feature of the microstructure, in addition to the phase fraction and crystal structure, which equally influences the material properties is the morphology of the constituent phases.
Besides governing the mechanical behaviour of the materials, the shape adopted by the phases in a microstructure influences other properties including magnetic and electrical conductivities~\cite{perrin1986microstructure,wu2009microstructure,jiles1988influence}.
Therefore, the process chain of a material is not confined to the phase transformation inducing treatments but combines techniques which establish a change in the morphology of the phases~\cite{qian2015titanium,bhadeshia2012steels}.
The shape assumed by the phases in a microstructure reflects the transformation mechanism which yields the corresponding phases.
For instance, the lamellar arrangement of the ferrite and cementite in pearlitic steel indicates the co-operative growth of the phases~\cite{de2006modern}.
However, to expand the applicability of the material, this characteristic morphology of the phases are disrupted through suitable techniques.

\begin{figure}
    \centering
      \begin{tabular}{@{}c@{}}
      \includegraphics[width=0.7\textwidth]{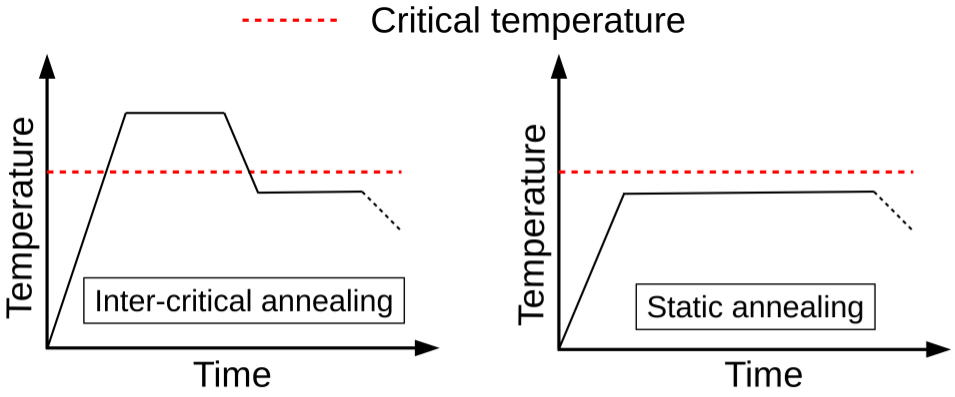}
    \end{tabular}
    \caption{ A schematic representation of the thermal cycle involved in inter-critical and static annealing treatment.
    \label{fig:thermal_cycle}}
\end{figure}

Heat treatment is the most common and well-established processing technique which is predominantly adopted to induce microstructural transformation in metallic systems~\cite{brooks1982heat,thelning2013steel}.
Every heat treatment is characterised by a specific thermal cycle which renders the microstructure corresponding to the desired properties.
A schematic illustration of the thermal cycle is rendered in Fig.~\ref{fig:thermal_cycle}.
Depending on the role of the phase transformation, the heat treatment techniques employed to institute a change in the morphological arrangement of the phases can be broadly categorised.
In techniques like inter-critical annealing, the shape-change is achieved with the aid of the appropriate phase transformation~\cite{hernandez1992spheroidization,ericsir2014effect}.
Whereas, certain other heat treatment processes, referred to as static annealing, establish the morphological transformation in the complete absence of any phase transformation~\cite{chen2011static,kaspar1988hot}. 
While the former heat treatment is dictated, and often restricted, by the chemical composition of the material, the latter provides a more generalized approach to obtain the essential microstructure.
Therefore, despite the apparently enhanced transformation-rate, the processing technique which accomplish the morphological change in the absence of the phase transformation is preferred, owing to its versatility.

As shown in Fig.~\ref{fig:thermal_cycle}, static annealing involves holding the material at high temperature, marginally below the critical temperature to avoid phase transformation, for a definite period of time~\cite{rajan2011heat}.
Accordingly, these treatments are also referred to as sub-critical annealing~\cite{guo2012subcritical,gang2009effect}.
The thermodynamical principle undergirding the microstructural evolution which ensues the static annealing is unambiguous and straightforward.
The morphological arrangement in a microstructure, which results from a phase transformation, often inherently exhibit a difference in the curvature.
Therefore, governed by the inherent curvature-difference, a gradient in the chemical potential is induced in accordance with the Gibbs-Thomson relation~\cite{gibbs1928collected,gibbs1957collected}.
The induced potential-gradient, eventuates mass transfer as the components migrate from the region of the high potential to the low potential region.
The migration of the components ultimately establishes the shape-change.
The secluded role of the curvature difference in the propelling the shape-change can be unraveled by considering the interfacial energy. 
With the interfacial energy density ($\gamma$) being a constant, the extensive variable, overall interfacial energy, increases proportionately with interfacial area.
In \textit{as-received} microstructures, the interfacial area are rarely at its minimal, and the morphological changes accompanying the annealing treatment reduces the overall interfacial energy.
Therefore, the microstructural evolution pertaining to the sub-critical annealing can  be viewed as the thermodynamical ability of the system to reduces it interfacial energy by suitable transformations.
The high temperature, at which the material is held during the sub-critical annealing, shown in Fig.~\ref{fig:thermal_cycle}, increases the diffusivity of the migrating component thereby enhancing the rate of transformation.
Since these transformations are primarily governed by the inherent curvature-difference in a shape, and ultimately disrupts the shape, the morphological evolution are called as the \textit{shape instabilities}~\cite{mullins1958shape}.
From this brief elucidation it is evident that the factors which govern the shape evolution, besides the curvature difference, are temperature, interfacial energy density and material parameters like diffusivity. 

In the addition to the morphology which emerges from the static annealing, the distribution of the phases plays a pivotal role in influencing the properties of the material~\cite{hafiz2001mechanical}.
The distribution of the phase during the morphological evolution is dictated by the transformation mechanism.
Therefore, a comprehensive understanding of the mechanism and kinetics of the morphological evolution is necessary to optimize the thermal cycle and perceive the distribution of the phases.
Despite the seemingly straightforward principle, the microstructural evolution pertaining to the shape-change have long been identified to be intricate~\cite{cline1971shape}.
Consequently, several investigations have been reported to unravel the complexity of the curvature-driven transformation~\cite{srolovitz1986capillary,livingston1974discontinuous,kampe1989shape,sharma2000instability}.
Since the shapes which temporally evolve during the transformation are convoluted three-dimensional structures, widening the current insight on the shape instabilities exclusively through the experimental observations is an arduous task.
Such an analysis would require three-dimensional projection of the microstructure and periodic observation of its transformation. 
Therefore, owing to the well-defined thermodynamical principle which undergird these evolutions, theoretical studies have complemented the experimental reports in delineating the complexities~\cite{martin1997stability}.

\begin{figure}
    \centering
      \begin{tabular}{@{}c@{}}
      \includegraphics[width=0.5\textwidth]{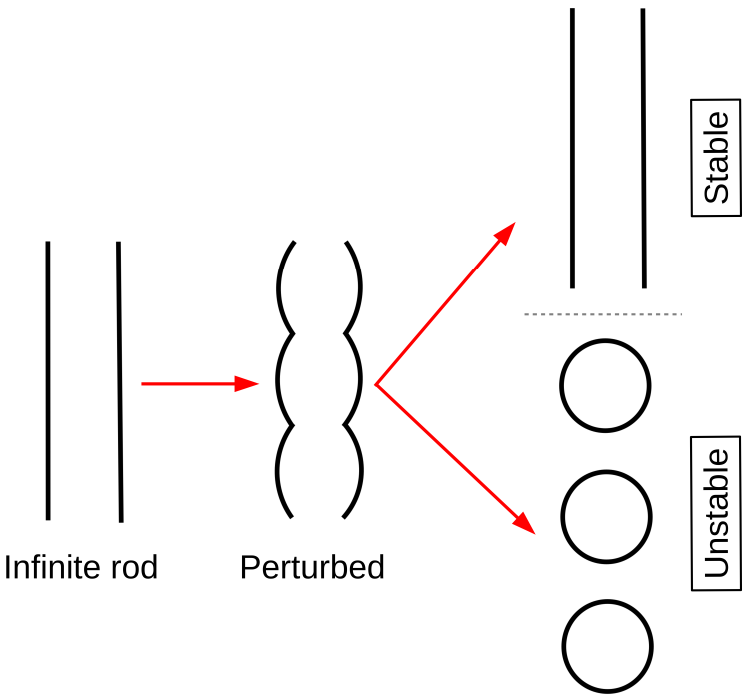}
    \end{tabular}
    \caption{ A schematic illustration of the \textit{caused} approach to investigate the stability of the infinitely-long cylindrical rod.
    \label{fig:induced_instability}}
\end{figure}

Considering the theoretical treatment of the curvature-driven transformations, the approach adopted to investigate shape instabilities can largely be categorised as \textit{caused} and \textit{inherent}.
Although these distinctions are entirely based on the theoretical approach and possess no relevance to the physical observations, whatsoever, such distinctions highlight the proximity of the theoretical considerations in relation to the physical conditions.
The \textit{caused}-approach was originally introduced around 1880s to analyse the stability of the cylindrical flow of fluid jets~\cite{rayleigh1878instability}.
Since an ideal infinitely-long cylindrical rod, owing to the lack of any curvature difference, remains perpetually stable, an external perturbation is introduced to \textit{cause} the instability.
Rayleigh extensively adopted this approach in seminal works on fluid jets wherein a numerically well-defined perturbation is introduced to the, otherwise, homogeneous rod and the criterion for the instability is derived based on the features of the perturbation, including amplitude and wavelength~\cite{rayleigh1878instability,rayleigh1892xvi,rayleigh1892xix}.
Consequently, these instabilities are referred to as Rayleigh instabilities.
Fig.~\ref{fig:induced_instability} broadly illustrates the caused-approach of understanding the instability.
To this day, this approach is the primary numerical tool for analysing the stability of the fluid flow~\cite{papageorgiou1995breakup,yarin1995impact,le1998progress,eggers2008physics}.

Later, Nichols and Mullins extended the caused-approach to understand the stability of the rods in solid systems~\cite{mullins1959flattening,nichols1965surface}. 
Typically, their semi-analytical treatment involved analysing the responsive behaviour of the rod to an imposed external disturbance, well-defined perturbation.
Similar to the Rayleigh instabilities, criticality is defined based on the nature of the imposed perturbation. 
Claiming that the external perturbation resembles the inhomogeneity in the cross-section of the rod, this approach has been employed to the metallic systems~\cite{brailsford1975influence,qian1994thermodynamic,qian1995breakup,qian1998non,mclean1978microstructural}.
Despite the thermodynamical consistency of this caused-approach, based on the experimental observations, it is argued that the inhomogeneities pertaining to the rods in the microstructure are rarely periodic and well-defined.
Furthermore, it is identified that the assumed continuity of the rods is often disturbed by the other forms of instabilities, thus nullifying the approach~\cite{kampe1989shape,sharma2000instability}.

\begin{figure}
    \centering
      \begin{tabular}{@{}c@{}}
      \includegraphics[width=1.0\textwidth]{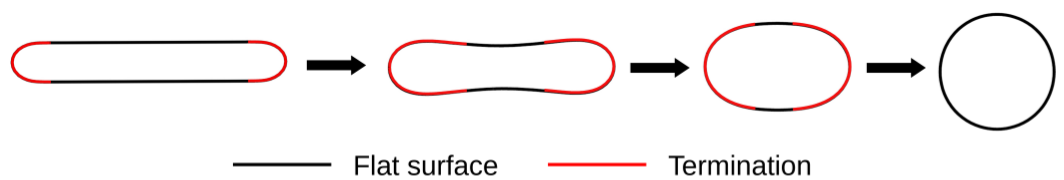}
    \end{tabular}
    \caption{ Morphological evolution governed by the \textit{inherent} curvature difference between the termination and the flat surface of the finite structure.
    \label{fig:2d}}
\end{figure}

The \textit{inherent} approach of investigating the stability of a structure can be elucidated by considering the evolution of the finite rods.
The fragmentation of the infinite rods into small finite structures obviates the need for any external disturbances.
As shown in Fig.~\ref{fig:2d}, the finite rod, innately, owing to the presence of the termination, sets up a difference in curvature between the boundaries and adjacent flat surfaces.
This curvature difference \textit{inherently} facilitates the morphological evolution of the rods.
In other words, the inherent theoretical-treatment involves analysing the morphological evolution of the structure governed by curvature-difference of its shape, devoid of any external perturbation.
Interestingly, one of the seminal work reporting on shape-change of the finite rod through inherent-approach adopts the framework of the caused technique~\cite{nichols1976spheroidization}.
Unlike the caused-approach which is largely confined to the rods, the inherent technique can be extended to various configuration. 
Since the microstructure is substantially complex in comparison to the regular arrangement of rods, and exhibits wide range of instabilities, the inherent-approach appears more pertinent than the caused method.

\begin{figure}
    \centering
      \begin{tabular}{@{}c@{}}
      \includegraphics[width=1.0\textwidth]{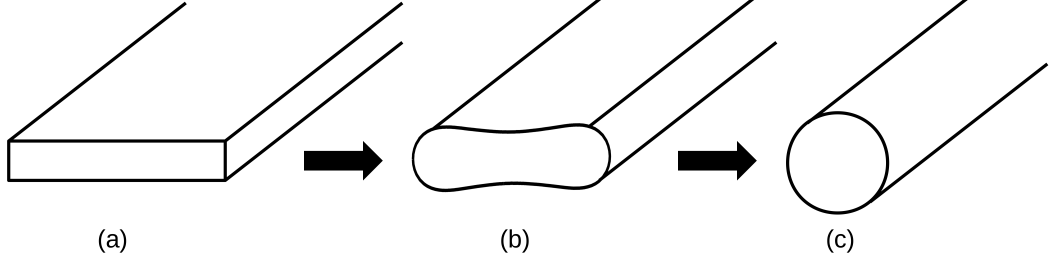}
    \end{tabular}
    \caption{ Plate morphology assumed by the precipitate in a microstructure and its evolution to cylindrical rod.
    \label{fig:cylinderization}}
\end{figure}

Owing to the intricacies in a microstructure, different forms of instabilities are observed in the metallic systems~\cite{kampe1989shape,sharma2000instability}.
One common form, shown in Fig.~\ref{fig:2d}, corresponds to the \textit{spheroidization} of the finite rod.
Apart from rod morphology, the phases assume a plate-like structure as shown in Fig.~\ref{fig:cylinderization}a.
If these plates are infinitely long, the caused-approach cannot be directly extended, since plates are resistant to the external perturbation~\cite{mullins1959flattening}.
However, it has been experimentally observed that such infinite plates transform to rods through the process of \textit{cylinderization}.
A schematic representation of the morphological transformation accompanying cylinderization is presented in Fig.~\ref{fig:cylinderization}.
Although the stability of the structure following the cylinderization can be analysed by adopting caused-approach, the inherent is apparently only viable option for understanding shape-change leading cylinderization.
Similar to the finite rod, the evolution of the infinite plate is governed by the disparity in the geometrical form of the sharp edges and face surfaces of plate, which translates to the curvature difference.

\begin{figure}
    \centering
      \begin{tabular}{@{}c@{}}
      \includegraphics[width=0.75\textwidth]{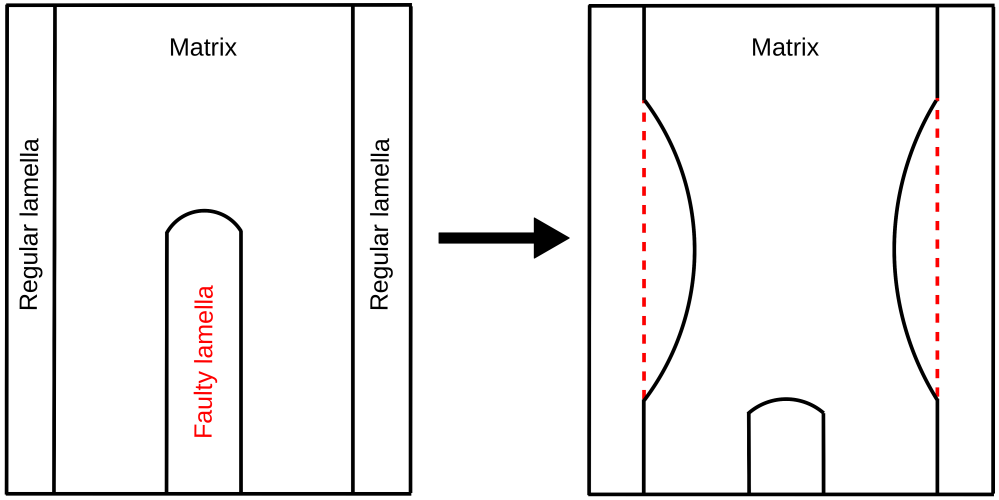}
    \end{tabular}
    \caption{ Schematic representation of fault migration wherein the discontinuous structure evolves by losing its mass to the surrounding regular lamellae.
    \label{fig:fault_migration}}
\end{figure}

The rod and plate morphology of the phases (precipitates) often are a component of the lamellar microstructure~\cite{cline1971structures,jackson1988lamellar,hillert1971diffusion,kirkaldy1980stability}.
Lamellar microstructure, which is characteristically recognised by the alternating arrangement of the phases, is one of well-known microstructure that spans across different materials.
The stability of the lamellar arrangement of the phases are extensively studied for two reasons.
One, as mentioned earlier, to optimize the properties and expand the applicability of the material~\cite{ogris2002silicon,shin2003spheroidization,sjolander2010heat}.
And the other is to prevent the failure of the material, owing to the behavioural change associated with the morphological evolution, in high temperature applications~\cite{ramanujan1996thermal,bartholomeusz1994effect,hong1998microstructural}.
The lamellar microstructure resulting the processing techniques are not necessarily ideal exhibiting an alternating arrangement of the continuous structures extending across the grain. 
During the growth of the phases, often discontinuous \lq faulty\rq \thinspace structure are formed which are sandwiched between the regular lamellae, as shown in Fig.~\ref{fig:fault_migration}.
The termination of this faulty precipitate introduces a curvature difference in relation to the flat surface of its own and neighbouring precipitates.
Consequently, the discontinuous structure evolves dictated by the mass transfer which is induced by the gradient in the chemical potential.
This morphological transformation wherein the discontinuous precipitate recedes is referred to as the fault migration~\cite{graham1966coarsening,marich1969solidification}.
A schematic representation of the fault migration is illustrated in Fig.~\ref{fig:fault_migration}.
The fault, when viewed as the migration of the discontinuous termination, is also referred to as termination migration~\cite{zherebtsov2011spheroidization,wang1999microstructural}.
Here, it is important to realize that the shape-change exhibited by any finite structure can be considered as the termination-migration assisted, since it is the recession of the edges, under the influence of the disparity in the curvature between the boundary (termination) and flat surface, which governs the transformation~\cite{fan2012mechanism,xu2014static}.
However, one distinguishing feature of the fault migration is the role of the neighbouring structures in augmenting the mass transferred from the discontinuous precipitate~\cite{cline1971shape,nakagawa1972stability}.

\begin{figure}
    \centering
      \begin{tabular}{@{}c@{}}
      \includegraphics[width=0.65\textwidth]{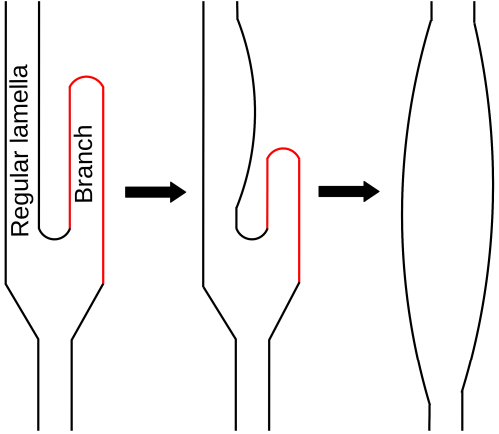}
    \end{tabular}
    \caption{ Faulty structure forming a branch of the regular lamella and its evolution leading to the inhomogeneity in the cross-section of the regular structure.
    \label{fig:branch_elimination}}
\end{figure}

Besides the faulty structure, the discontinuity is also introduced within a precipitate lamella leading to the formation of the \lq branch\rq \thinspace.
Fig.~\ref{fig:branch_elimination}a shows the schematic representation of a lamellar branch observed in a microstructure~\cite{gigliotti1970temperature,ardell1972isotropic}.
Similar to the fault structure the branches introduce curvature-difference which consequently leads to the morphological evolution.
Owing to the size of the branch and its proximity to the regular structure, the mass predominantly transfers from the branch which ultimately leads to its disappearance.
This morphological transition exhibited by the branch is referred to as branch elimination~\cite{semiatin2008branch,mason1995microstructure}.
The shape-change associated with the branch elimination is presented in Fig.~\ref{fig:branch_elimination}.
It is evident from the illustration that the branch elimination introduces a visible inhomogeneity in the regular structure which might disrupt the stability.

\begin{figure}
    \centering
      \begin{tabular}{@{}c@{}}
      \includegraphics[width=1.0\textwidth]{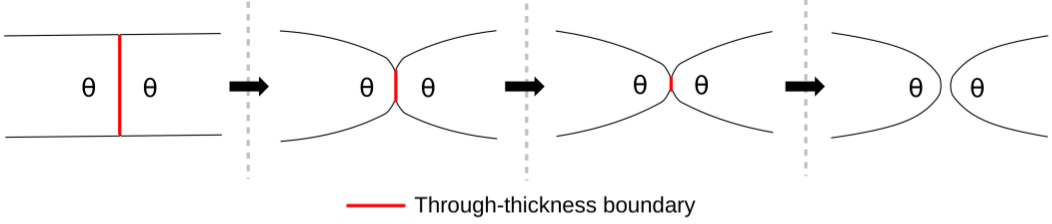}
    \end{tabular}
    \caption{ Thermal grooving in the through-thickness boundary leading to the fragmentation of the precipitate.
    \label{fig:boundary_split}}
\end{figure}

Another form of the instability which disrupts the morphological make-up of the microstructure is the boundary splitting.
The temporal change accompanying the boundary splitting is schematically shown in Fig.~\ref{fig:boundary_split}.
This form of instability is exclusively observed in the precipitate with through-thickness sub-boundary, a boundary that penetrates the entire thickness of the structure~\cite{poths2004effect,cabibbo2013loss,lin2013instability}.
Boundary splitting is initiated by another equilibration phenomenon known as thermal grooving~\cite{mullins1957theory}.
In a triple junction wherein three boundaries meet, a groove is introduced at high temperature which establishes an equilibrium between the interfaces based on their corresponding energy density ($\gamma$), as shown in Fig.~\ref{fig:boundary_split}.
The ridges which accompany the groove introduce a curvature difference and eventually vanish, owing to the mass transfer to the adjacent flat surfaces.
This disappearance of the ridge, disturbs the local equilibrium which is subsequently restored by the deepening of the groove.
The combination of these events in a plate-like structures results in boundary splitting as shown in Fig.~\ref{fig:boundary_split}.
The fragmentation of the seemingly infinitely long structures in metallic system is predominantly achieved by the boundary splitting~\cite{rastegari2015warm,wang2010prediction}.
In the process chain, the materials are deformed to increase the density of the through-thickness sub-boundary and thereby, accelerate the morphological change~\cite{lupton1972influence,kaspar1988hot,wu2011dynamic,ma2012kinetics}.

\begin{figure}
    \centering
      \begin{tabular}{@{}c@{}}
      \includegraphics[width=0.75\textwidth]{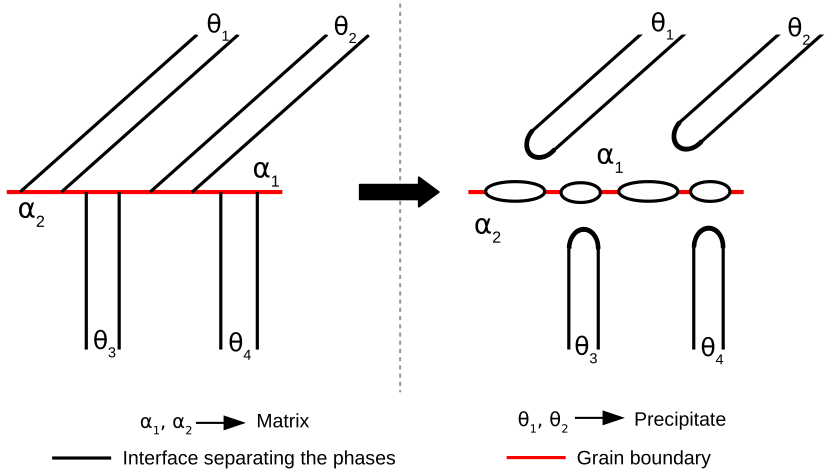}
    \end{tabular}
    \caption{ The shape-change in the lamellar arrangement of the phases initiated by its fragmentation from the grain boundary. 
    \label{fig:GB_ovulation}}
\end{figure}

As opposed to conventional theoretical consideration, a precipitate in a microstructure frequently interacts with the grain boundary.
For instance, the orientation and continuity of the lamellar arrangement is confined to a grain in a polycrystalline structure~\cite{yamaguchi2000high,parthasarathy1997observations}.
Therefore, the role of the grain boundary in the curvature-driven transformation cannot be overlooked.
Recently, experimental investigations have unraveled that a seemingly continuous structure fragments near the grain boundary~\cite{arruabarrena2014influence}, as shown in Fig.~\ref{fig:GB_ovulation}.
Although the driving force for this shape-change is not the inherent curvature-difference of the shape, the structure breaks-up in adherence to the Young's law~\cite{young1805iii}, which dictates the equilibrium profile of the interfaces at a triple junction.

Apart from the different forms, the shape instabilities are also distinguished by the dominant mode of mass transfer which establishes the morphological change.
Correspondingly, the shape transformation can either be governed by the surface or volume diffusion.
Several critical aspects of the evolution including rate and mechanism are significantly influenced by the mass-transfer mode which dominates~\cite{nichols1965surface}.
Broadly, it is assumed that the instabilities in the fluids are dictated by the surface diffusion and the evolutions are numerically analysed accordingly~\cite{cussler2009diffusion}.
However, in solid-state system, particularly in metallic systems, this governing factor vary with the chemical make-up.
The difference introduced by the principal mode of the mass transfer can be elegantly described by considering the respective analytical framework.

Irrespective of the dominant mode of the mass transfer, or even the form, the shape instabilities are governed by the Gibbs-Thomson relation~\cite{gibbs1928collected,gibbs1957collected}.
The influence of the curvature on the chemical potential, based on this relation, is expressed as
\begin{align}\label{eq:gbthom_rel}
\mu(k)=\gamma V_{\text{m}} k,
\end{align}
where $V_{\text{m}}$is the molar volume and $\gamma$ is the interfacial energy density.
\nomenclature{$V_{\text{m}}$}{Molar volume}%
\nomenclature{$\gamma$}{Interfacial energy density}%
The curvature $k$ is the ascertained from the principal radii of the shape $R_1$ and $R_2$ by $k\approx(1/R_{1}+1/R_{2})$.
\nomenclature{$k$}{Curvature}%
\nomenclature{$R_1$ and $R_2$}{Principal radii}%
From Eqn~\ref{eq:gbthom_rel}, it is evident that any difference in the curvature introduces a gradient in the chemical potential.
The chemical-potential gradient consequently induces a mass transfer. 
The velocity of the atoms ($v_{\text{a}}$) migrating in response to the potential gradient is written as
\nomenclature{$v_{\text{a}}$}{Velocity of diffusing atoms}%
\begin{align}\label{eq:gbthom_vel}
v_{\text{a}}=-\frac{D}{\kappa T}\nabla\mu=-\frac{D\gamma V_{\text{m}}}{\kappa T}\nabla k,
\end{align}
where temperature and Boltzmann's constant are respectively represented by $T$ and $\kappa$.
\nomenclature{$T$}{Temperature}%
\nomenclature{$\kappa$}{Boltzmann's constant}%
The initial distinction, based on the mode of mass transfer, is introduced in the analytical formulation through the diffusion co-efficient or the diffusivity $D$.
\nomenclature{$D$}{Diffusion co-efficient}%
In volume-diffusion governed transformation, volume diffusivity $D_\text{v}$ is involved in Eqn.~\ref{eq:gbthom_vel} while the surface diffusivity $D_\text{s}$ is adopted wherein surface diffusion is dominant.
\nomenclature{$D_\text{v}$}{Volume diffusivity}%
\nomenclature{$D_\text{s}$}{Surface diffusivity}%

Consider a binary two-phase system of $\alpha$ and $\theta$ wherein the precipitate-$\theta$, which undergoes the morphological evolution, is enveloped by the matrix-$\alpha$.
Owing to the concentration, $c_{\text{eq}}\p{\alpha}$ and $c_{\text{eq}}\p{\theta}$, which represent the equilibrium composition, the phases are in chemical equilibrium.
\nomenclature{$c_{\text{eq}}\p{\alpha}$}{Equilibrium concentration of phase-$\alpha$}%
\nomenclature{$c_{\text{eq}}\p{\theta}$}{Equilibrium concentration of precipitate-$\theta$}%
Under this condition, if $c_{\text{eq}}\p{\theta}\gg c_{\text{eq}}\p{\alpha}$ as in cementite and ferrite, the atomic flux ($J$) which establishes the shape change can be written as
\nomenclature{$J$}{Atomic flux}%
\begin{align}\label{eq:gbthom_flux}
 J=v_{\text{a}}c_{\text{eq}}\p{\theta}=-\frac{D\gamma V_{\text{m}}c_{\text{eq}}\p{\theta}}{\kappa T}\nabla k.
\end{align}
In the surface-diffusion governed evolution, the curvature difference and mass transfer are considered exclusively in the surface.
Therefore, based on Nernst-Einstein relation~\cite{nernst1888kinetik,einstein1905z,einstein1905einstein}, Eqn.~\ref{eq:gbthom_flux} for the shape instabilities dictated by the surface diffusion, the surface flux is expressed as
\begin{align}\label{eq:gbthom_flux_surface}
 J_s=-\frac{D_s\gamma V_{\text{m}}c_{\text{eq}}\p{\theta}}{\kappa T}\nabla_s k,
\end{align}
where $\nabla_s k$ is the surface gradient of the curvature.
In contrast, for the volume-diffusion dominant transformation, the curvature gradient in the entire system is considered through $\nabla k$, with  volume diffusivity replacing $D$ in Eqn.~\ref{eq:gbthom_flux}.
\nomenclature{$\nabla_s k$}{Surface gradient of the curvature}%
\nomenclature{$\nabla k$}{Curvature gradient}%

Owing to the confined consideration of the setup in the surface-diffusion governed transformation, the amount of mass transferred per unit area and time is determined from the surface gradient of the flux, $-J$~\cite{mullins1957theory}.
Correspondingly, the migration rate of an surface element in the normal direction reads
\begin{align}\label{eq:gbthom_rate_surface}
\frac{\partial \eta}{\partial t}= -V_{\text{m}}\nabla_s J_s=\frac{D_s\gamma V_{\text{m}}\p2c_{\text{eq}}\p{\theta}}{\kappa T}\Delta_s k,
\end{align}
where $\nabla_s\p2k$ is expressed as the laplacian of the curvature, $\Delta_s k$.
\nomenclature{$\Delta_s k$}{Surface laplacian of the curvature}%
However, since the analytical formulation of the mass transfer dictated by volume-diffusion considers the entire system, the normal migration is expressed as
\begin{align}\label{eq:gbthom_rate_volume}
 \frac{\partial \eta}{\partial t}=\frac{D_v\gamma V_{\text{m}}\p2c_{\text{eq}}\p{\theta}}{\kappa T}\nabla k.
\end{align}

From the above elucidation, it is conceivable that the assumption of the surface-diffusion governed shape-change, wherein the morphological evolution is solely governed by the atomic migration along the surface, simplifies the theoretical approach.
In other words, while numerically analysing the microstructural transformation wherein the evolution is exclusively determined by the surface diffusion, the focus shifts from the entire domain to the interface.
Correspondingly, the theoretical investigation on the stability of an isolated three-dimensional structure, under the consideration of the surface-diffusion dominance,  is confined to the interface, which is a two-dimensional plane.
Although these considerations are appropriate for the systems wherein the surface diffusion is physically and extensively dominant, like fluid flow, it cannot be directly adopted for the metallic system for two fundamental reason.
One, in solid-state systems, the governing mode of mass transfer various with the chemical composition of the system~\cite{mehrer1990diffusion}.
Moreover, even if one mode is dominant, the contribution of the other mode cannot be overlooked, in its entirety.
Two, a critical assumption in the theoretical treatment of the surface-diffusion governed evolution is that the mass enclosed by the interface (or surface) offers absolutely no resistance to the morphological change in the structure.
Though such conditions are prevalent in fluids, in metallic systems it is hardly observed.
Therefore, a more definitive approach which encompasses the entire system should be adopted to examine the shape instabilities in solid-state systems.



\chapter{Phase-field modelling}\label{chap:pfi}

In understanding any natural phenomenon, theoretical studies have always complemented the experimental investigations. 
With the exponential increase in the computational resources, numerous techniques with unique theoretical framework have been developed to untangle the intricacies associated with any given physical process. 
The resulting enhanced understanding has enabled engineers and scientists to optimally employ the energy resources and render an efficient product.

The phase-field modelling is one such computational technique which is notably gaining ground for the past couple of decades. 
The adoption of this approach in diverse avenues of research including material sciences~\cite{provatas2011phase}, fluid dynamics~\cite{kim2012phase}, magnetism~\cite{koyama2008phase},  classical and quantum mechanics~\cite{ammar2009combining,steinbach2017quantum} vindicates its versatility. 
The characteristic features of the phase-field modelling which distinguishes it from its counterparts is of two fold.
One, in addition to the thermodynamic variables, a scalar state-variable referred to as \lq phase field\rq \thinspace is introduced. 
The phase field assumes a constant value in the bulk phases while smoothly varying across the region separating the phases, called the interface.
This unique spatial dependence of the phase field leads to the second characteristic feature of the phase-field modelling, which is the introduction of the diffuse interface replacing the sharp one.  

In this chapter, as an introduction, the evolution of the theoretical formulations leading to the phase-field model is presented.
The phase-field technique can broadly be introduced from two different standpoints. 
In the \lq top-down\rq \thinspace scheme, the phase-field technique is setup as an aspect of a larger numerical problem, namely \textit{free-boundary problem}.
Whereas, the \lq bottom-up\rq \thinspace framework introduces this simulation technique based on its motivation from the statistical treatment of physical transformations.
In the current work, a meaningful preface to the phase-field modelling is rendered through the bottom-up scheme.

\section{Order parameter and Landau free energy}

Electrical conductivities of most metals are inversely proportional to the temperature.
Therefore, below a critical temperature, a metal exhibits superconductivity by the absolute suspension of its electrical resistivity.
Overlooking the microscopic aspect, Landau proposed a phenomenological approach to analyse the transition from normal to superconductivity by considering it to be thermodynamically analogous to the second-order or continuous phase transformation~\cite{cyrot1973ginzburg}.
Although later is was shown that this consideration poses a serious limitation to the model, for the present discussion such constraints are ineffective.

Consider a system exhibiting a combination of two states of conductivity. 
In a simple lattice model, the total energy (Hamiltonian) $E$ of such a system with both normal and superconducting fractions is expressed as
\nomenclature{$E$}{Total energy of the system (Hamiltonian)}%
\begin{align}\label{eq:hamiltonian}
E\{\sigma_{i}\} = -\tilde{J}\sum_{i=1}^{N}\sum_{j\neq i}^{v} \sigma_{i}\sigma_{j}-B\sum_{i=1}^{N}\sigma_{i},
\end{align}
where $\sigma_{i}$ and $v$ correspondingly represent the conducting state of the lattice point (normal or superconducting) and co-ordination number, which accounts for the number of neighbouring lattice points.
\nomenclature{$v$}{Co-ordination number}%
The interaction energy between the lattices $i$ and $j$ is quantified by the constant $\tilde{J}$, and $B$ is the external magnetic field.
\nomenclature{$\tilde{J}$}{Interaction energy}%
\nomenclature{$B$}{External magnetic field}%
The first terms in Eqn.~\ref{eq:hamiltonian} encompasses all possible interaction between the lattices while the second term represents the contribution of the individual lattice.
The partition function of the system consisting of normal and superconducting fractions is expressed as
\begin{align}\label{eq:partition}
 \mathcal{Z} = \prod_{i}^{N}\sum_{\sigma_{i}}\exp(-{\beta}_{T} E\{\sigma_{i}\}),
\end{align}
where $N$ is the total lattice points in the system and $\beta_{T}$ is the thermodynamic beta, which is the reciprocal of the $\kappa T$ with the $\kappa$ and $T$ denoting the Boltzmann constant and temperature.
\nomenclature{$\beta_{T}$}{Thermodynamic beta (Reciprocal of $\kappa T$)}%

Landau postulated that the transition to the superconductivity can be \textit{phenomenologically} viewed as the increase in the fraction of the superconducting material or \textit{growth} of the superconducting \textit{phase}.
Therefore, to theoretically investigate the transition he distinguished the superconducting phase from the rest of the system, by introducing a scalar state-variable.

The scalar state-variable, referred to as \lq \textit{order parameter} ($\phi$)\rq \thinspace, numerically distinguishes the normal and superconducting fractions of the material by assuming a definite value within a given phase.
\nomenclature{$\phi$}{Phase field (or Order Parameter)}%
In the framework of the lattice model, the order parameter can be considered as the average of a local set of lattice points $\tilde{N}$, expressed as
\begin{align}\label{eq:order}
 \phi = \frac{1}{\tilde{N}}\Big<\sum_{1}^{\tilde{N}}\sigma_{i}\Big> \equiv \Big< \sigma_{i} \Big>_{\tilde{N}}.
\end{align}
In this \textit{coarse-graining} approach, the local set is assumed to be large enough to encompass sufficient lattice points, while significantly small when compared to the geometric scale of the domain.

By treating the order parameter as the fundamental variable, the partition function in Eqn.\ref{eq:partition} can be written as
\begin{align}\label{eq:partition_order}
 \mathcal{Z} = \int_{-\infty}^{\infty} \diff \phi \exp(-\beta_{T} Vf(\phi)),
\end{align}
where $V$ is the volume of the system and $f(\phi)$ is the order-parameter based free-energy density.
\nomenclature{$f(\phi)$}{Order-parameter based free-energy density}%
By adopting a thermodynamical description, the free energy density reads
\begin{align}\label{eq:free_energy}
 f(\phi) = \frac{F(\phi)}{N} = \frac{U(\phi)}{N} - \frac{TS(\phi)}{N}
\end{align}
where $U(\phi)$ and $S(\phi)$ are internal energy and entropy, respectively.
\nomenclature{$U(\phi)$}{Internal energy}%
\nomenclature{$S(\phi))$}{Entropy}%
To ascertain these thermodynamic entities ($U(\phi)$ and $S(\phi)$) which dictate the free energy density $f(\phi)$, the \textit{Mean-field approximations}, as in Braggs-Williams approach~\cite{hill1985bragg}, is employed.

Assuming the interaction energy $\tilde{J}$ to be constant, the internal energy can be written analogous to Eqn.~\ref{eq:hamiltonian}.
Correspondingly, the order-parameter dependent internal-energy $U(\phi)$ reads
\begin{align}\label{eq:internal_energy}
 \frac{U}{N}=-\tilde{J}\sum_{<i,j>}\phi_{i}\phi_{j}-B\sum_{i}\phi_{i}=-\frac{v}{2}\tilde{J}\phi^{2}-B\phi,
\end{align}
where $v$ is the co-ordination number.
The entropy $S(\phi)$, in Eqn.~\ref{eq:free_energy}, which is predominantly dictated by the configuration, is determined from the density of states, $\bar{\Omega}$.
For $N_s$ superconducting lattices, the total number of configurations reads
\begin{align}\label{eq:configuration}
\bar{\Omega} = \frac{N!}{N_s!(N-N_s)!}. 
\end{align}
Therefore, the contribution of the entropy, given by $S = \log\bar{\Omega}$, using Stirling's approximation can be written as
\begin{align}\label{eq:entropy}
 \frac{S(\phi)}{N}\approx \log2 - \frac{1}{2}(1+\phi)\log(1+\phi) - \frac{1}{2}(1-\phi)\log(1-\phi).
\end{align}

By substituting Eqns.~\ref{eq:internal_energy} and ~\ref{eq:entropy} in Eqn.~\ref{eq:free_energy}, the order-parameter based free-energy density is written as
\begin{align}\label{eq:free_energy2}
 f(\phi) = -\frac{v}{2}\tilde{J}\phi^{2}-B\phi-T\Big[\log2 - \frac{1}{2}(1+\phi)\log(1+\phi) - \frac{1}{2}(1-\phi)\log(1-\phi)\Big].
\end{align}
It is evident that the above formulation of the free energy density is devoid of the characteristic critical temperature which governs the transition, under Landau's consideration.
Therefore, Eqn.~\ref{eq:free_energy2} is solved for specific condition to incorporate the critical temperature.
Under the extremum condition, $\frac{\partial f}{\partial \phi}=0$, the aforementioned formulation of the free energy density yields 
\begin{align}\label{eq:extremum}
 -\tilde{J}\phi v-B-\frac{T}{2}\log\Big(\frac{1+\phi}{1-\phi}\Big) = 0.
\end{align}
Correspondingly, from Eqn.~\ref{eq:extremum}, the order parameter is expressed as
\begin{align}\label{eq:order}
 \phi = \text{tanh}\Big(\frac{\tilde{J}\phi v+B}{T}\Big).
\end{align}
In the absence of the external magnetic field ($B=0$), and marginally above the critical temperature, wherein the system assumes a homogeneous state with the order parameter, $\phi=0$,  the critical temperature can be written as $T_{c}=\tilde{J}v$ from Eqn.~\ref{eq:order}.
\nomenclature{$T_{c}$}{Critical Temperature}%
Expanding the free energy density in Eqn.~\ref{eq:free_energy2} upto fourth order, and incorporating the critical temperature yields
\begin{align}\label{eq:free_energy_functional}
 f(\phi)=-A-B\phi+\frac{\phi^2}{2}(T-\tilde{J}v)+\frac{1}{12}T\phi^4,
\end{align}
where $A$ is a constant.
This order parameter based polynomial expansion of the free energy density is referred to as Landau free energy.

The prominent feature of the Landau free energy is the influence of the critical temperature on the nature of the $f(\phi)$.
When temperature is greater than the critical temperature ($T>\tilde{J}v$), both the second order (quadratic) and the fourth order (quartic) terms in Eqn.~\ref{eq:free_energy_functional} are positive. 
Correspondingly, only one minimum exists for the free energy density.
However, below the critical temperature, since only the quartic term is positive, the free energy density exhibits two mimimas, which is conventionally referred to as \textit{double well}.

The well-known Landau theory is characteristically formulated for the transition to the superconductivity.
Therefore, in order render a coherent description of the theoretical approach which led to the phase-field formulation, a deviation that conforms to \textit{the Ising model}~\cite{ising1925beitrag} is adopted in the above elucidation.

\section{Ginzburg-Landau free energy functional}

Although the description in the previous section begins with the consideration of the entire system, the approach is simplified by involving constant interaction energy $\tilde{J}$ and expressing the internal energy for a fraction in Eqn.~\ref{eq:internal_energy}.
Therefore, the resulting free energy function in Eqn.~\ref{eq:free_energy_functional} is confined to the bulk volume-fraction of the system which can either be normal or superconducting.
However, in addition to the bulk phases, the system includes the regions separating the phases which are referred to as interfaces.
To formulate a free energy which encompasses the interface, the internal energy must be extensively defined.
Correspondingly, the internal energy now reads
\begin{align}\label{eq:hamiltonian_interface}
U\equiv E\{\sigma_{i}\} = -\sum_{i=1}^{N}\sum_{j\neq i}^{v}[\tilde{J}_{ss}\sigma_{i}\sigma_{j} + \tilde{J}_{s\tilde{s}}(1-\sigma_{i})\sigma_{j} & + \tilde{J}_{\tilde{s}s}(1-\sigma_{j})\sigma_{i} \\ \nonumber
&+ \tilde{J}_{\tilde{s}\tilde{s}}(1-\sigma_{i})(1-\sigma_{j})],
\end{align}
where the interaction between similarly conducting lattice is quantified by $\tilde{J}_{ss}$ and $\tilde{J}_{\tilde{s}\tilde{s}}$, with the former considering the superconducting lattice while the latter represents the normal fractions of the system. 
The interaction between lattices with different conductivities is represented by $\tilde{J}_{s\tilde{s}}$.
Furthermore, it is important to note that the formulation in Eqn.~\ref{eq:hamiltonian_interface} entails the consideration that a lattice can either be normal or superconducting.

The internal energy expression in Eqn.~\ref{eq:hamiltonian_interface} can be simplified, without compromising on the intent to include the interface contribution, by assuming $\tilde{J}_{ss} = \tilde{J}_{\tilde{s}\tilde{s}}$ and temporarily suspending the role of any constants.
Under these assumptions and introducing the order parameters, Eqn.~\ref{eq:hamiltonian_interface} is written as
\begin{align}\label{eq:internal_ex}
 E(\phi)=\frac{1}{2}\sum_{i=1}^{N}\sum_{j\neq i}\tilde{J}_{ij}\phi_{i}(\vec{x})(1-\phi_{j}(\vec{x})),
\end{align}
where the spatial dependency of the order parameters are included.
Although theoretically, by considering the interaction between the phases, Eqn.~\ref{eq:internal_ex} encompasses both the bulk phases and interface, the contribution of the interface is not separate and explicit in the expression. 
Therefore, the order parameter term in Eqn.~\ref{eq:internal_ex} is algebraically split as
\begin{align}\label{eq:ab_ex}
 \phi_{i}(\vec{x})(1-\phi_{j}(\vec{x}))=\frac{1}{2}\Big\{[\phi_{i}(\vec{x})-\phi_{j}(\vec{x})]^2-[(\phi_{i}(\vec{x})^2+\phi_{j}(\vec{x})^2]+2\phi_{i}(\vec{x})\Big\},
\end{align}
such that first term on the right hand side represents the contribution across the phases while the other terms correspond to the individual phases.
In a two-dimensional setup wherein the co-ordination number, $v=4$, the interaction of $\phi_{i}$ is restricted to the phases at its right ($\phi_{R}$), left ($\phi_{L}$),  top ($\phi_{T}$) and bottom ($\phi_{B}$).
Therefore, the first term in Eqn.~\ref{eq:internal_ex}, can be expanded as
\begin{align}\label{eq:neighbour}
 \sum_{j\neq i}(\phi_{i}-\phi_{j})^2 = a^{2}\Big[\Big(\frac{(\phi_{i}-\phi_{R})^2}{a^{2}}+\frac{(\phi_{i}-\phi_{T})^2}{a^{2}}\Big)+\Big(\frac{(\phi_{i}-\phi_{L})^2}{a^{2}}+\frac{(\phi_{i}-\phi_{B})^2}{a^{2}}\Big)\Big],
\end{align}
where $a$ is the infinitesimal distance separating the phases.
The order-parameter terms on the right hand side of Eqn.~\ref{eq:neighbour} can be viewed as the gradients of the $\phi_{i}$.
Correspondingly, the interaction between the phases which represents the contribution form the interface can be expressed as
\begin{align}\label{eq:gradient}
 \sum_{j\neq i}(\phi_{i}-\phi_{j})^2 \approx a^{2}|\nabla \phi_{i}(\vec{x})|^{2}.
\end{align}
Substituting Eqn.~\ref{eq:gradient} in Eqn.~\ref{eq:ab_ex}, the total internal energy of the system which encompasses the interface is written as
\begin{align}\label{eq:int_grad}
  E(\phi)= \frac{1}{2} a^{2}|\nabla \phi_{i}(\vec{x})|^{2}+\frac{1}{2}\tilde{J}[\phi(\vec{x})(1-\phi(\vec{x}))].
\end{align}
It is important to note that the second term on the right hand side of the above expression is the simplified form its corresponding term from Eqn.~\ref{eq:ab_ex}, which accounts for the contribution of the bulk phase.
In addition to this contribution of the bulk phase, the role of the entropy ($S(\phi(\vec{x}),T(\vec{x}))$) should be included to translate the internal energy into free energy.  
Conventionally, mean field approximation is employed to formulate the entropy term.
After the inclusion of the entropy, if the free energy contribution of the bulk phase is represented by $f(\phi(\vec{x}),T(\vec{x}))$, the resulting free energy functional is expressed as
\begin{align}\label{eq:free_functional}
 F(T, \phi,\nabla \phi) = \int_{V}\Big\{ \bar{K}|\nabla \phi(\vec{x})|^{2}+f(\phi(\vec{x}),T(\vec{x}))\Big\}\diff V,
\end{align}
where $V$ is the volume of the system and $\bar{K}$ is the \textit{gradient energy} co-efficient which includes the interaction energy parameter. 
\nomenclature{$\bar{K}$}{Gradient energy co-efficient}%
This form of the functional is referred to as Ginzburg-Landau functional~\cite{ginsburg1950k}, wherein the order parameter spatially vary continuously.
Chronologically, the Ginzburg-Landau functional was adopted to Ising's model around 1970's~\cite{ferrell1969field,thouless1969critical}, implying that Landau's theory precedes Ising' model in their contribution to continuum-field (or phase-field) approach.

\section{Cahn-Hilliard equation}

The order parameters, both in Landau theory and Ising model, possess a characterising property which aides in distinguishing this approach from other similar numerical treatments.
For instance, in Landau theory, the order parameter which represents the fraction of the system exhibiting a particular conductivity (normal or superconducting), changes with its conductivity.
Therefore, considering a transformation wherein the normal conducting system changes entirely to superconducting, the order parameter pertaining to the original system completely vanishes. 
In other words, the order parameters elucidated in the previous sections are \textit{non-conserved}. 
Consequently, it becomes unphysical to directly relate this order parameter with the concentration.
However, in their seminal work, Cahn and Hilliard derived an expression similar to the Landau free energy with concentration replacing the order parameter~\cite{cahn1958free}.  

It is important to note that the starting points of these two approaches, Landau and Cahn-Hilliard, are significantly different.
While framework of the Landau's theory is aimed at treating the transition to the superconductivity as the second-order phase transformation, Cahn-Hilliard formulated a free energy for the physically realized diffuse interface~\cite{woodruff1973solid,hoyt2003atomistic}.
It is interesting to note that such treatment of the interface, as a diffuse region instead conventionally perceived sharp division between the phases, dates back to Van der Waal~\cite{vand1894,van1979thermodynamic}.
Therefore, it is reasonable to state that, despite the prevalent understanding, the sharp interface model make comparable assumptions like the phase-field approach to treat the interface.
In other words, assumption pertaining to the width of the interface is not exclusively confined to the phase-field models.

As described in the previous section, the interface contribution in the Landau approach was introduced by considering the interaction of the order parameters.
In this section, region separating chemically-distinct phases is treated as a diffuse interface wherein the concentration exhibits a spatial variation.
Consider a closed system of a binary alloy with components $A$ and $B$, let $c$ be the independent concentration variable representing the mole fraction of $B$.
\nomenclature{$c$}{Independent concentration in mole fraction}%
If the system comprises of $A$- and $B$-rich phases in equilibrium, the overall free-energy of the system can be expressed as the summation of the free energies wherein the concentration is homogeneous, representing the bulk phases, and inhomogeneous, characterising the interface. 
This consideration entails the assumption that the free energy is continuous across the interface.
Furthermore, the concentration in the inhomogeneous region is determined by adopting the coarse grain approximation.
Correspondingly, the overall free-energy of the system can be written as
\begin{align}\label{eq:free_CH}
 f(c, \nabla c, \nabla^{2}c,\dots)=f_{\text{bulk}}(c)+f_{\text{intf}}(\nabla c, \nabla^{2}c,\dots),
\end{align}
where $f_\text{bulk}(c)$, expressed as $f_\text{bulk}(c)=f_{\alpha}(c)+f_{\beta}(c)$, is the free energy contribution from the bulk phases with homogeneous concentration $c$ with $\alpha$ and $\beta$ representing $A$- and $B$-rich phases, respectively. 
\nomenclature{$f_\text{bulk}(c)$}{Concentration based free-energy (Cahn-Hilliard)}%
The free energy contribution from the interface is represented by $f_{\text{intf}}(\nabla c, \nabla^{2}c,\dots)$.
\nomenclature{$f_{\text{intf}}(\nabla c, \dots)$}{Interface contribution (Cahn-Hilliard)}%
Using Taylor's expansion for the overall free-energy $f(c, \nabla c, \nabla^{2}c,\dots)$, Eqn.~\ref{eq:free_CH} transforms to 
\begin{align}\label{eq:CH_taylor}
 f(c, \nabla c, \nabla^{2}c,\dots)= f_{o}(c)+\Bigg\{\frac{\partial f}{\partial \nabla c}\nabla c & +\frac{\partial f}{\partial \nabla^{2} c}\nabla^{2} c+\frac{\partial f}{\partial \nabla_{x}\nabla_{y} c}\nabla_{x}\nabla_{y} c+\dots \\ \nonumber
 & +\frac{1}{2}\Big[\frac{\partial^{2} f}{\partial (\nabla c)^{2}}(\nabla c)^{2}+\frac{\partial^{2} f}{\partial (\nabla^{2} c)^{2}}(\nabla^{2} c)^{2}\\ \nonumber
 &+\frac{\partial^{2} f}{\partial (\nabla_{x}\nabla_{y} c)^{2}}(\nabla_{x}\nabla_{y} c)^{2}+\dots\Big]+\dots\Bigg\}
\end{align}
with $f_{o}(c)\equiv f_\text{bulk}(c)$, the terms enclosed by $\{\dots\}$ in the above Eqn.~\ref{eq:CH_taylor} relate to the interface contribution $f_{\text{intf}}(\nabla c, \nabla^{2}c,\dots)$ in Eqn.~\ref{eq:free_CH}.
Assuming the system to be configurationally isotropic, Eqn.~\ref{eq:CH_taylor} can be simplified by eliminating the terms that are equivalent and mutually-reversed owing to the symmetry.
By confining to the fourth order, the simplified form of Eqn.~\ref{eq:CH_taylor} reads
\begin{align}\label{eq:CH_simplified}
 f(c, \nabla c, \nabla^{2}c,\dots)= f_{o}(c)+\frac{\partial f}{\partial \nabla^{2} c}\nabla^{2} c+\frac{1}{2}\frac{\partial^{2} f}{\partial (\nabla c)^{2}}(\nabla c)^{2}.
\end{align}
Invoking Stokes theorem, the overall free-energy of the system can be expressed as functional of the form
\begin{align}\label{eq:CH_functional}
 F(c,\nabla c)=\int_{V}\Big\{ f_{o}(c)+ \tilde{K}|\nabla c(\vec{x})|^{2}\Big\}\diff V,
\end{align}
with $V$ representing the volume and constant $\tilde{K}$, the gradient energy co-efficient.
\nomenclature{$\tilde{K}$}{Gradient energy co-efficient (Cahn-Hilliard)}%
If the order parameters are defined based on the concentration, particularly for the second-order phase transformations like spinodal decomposition, the Cahn-Hilliard functional in Eqn.~\ref{eq:CH_functional} would resemble Eqn.~\ref{eq:free_functional}, the Ginzburg-Landau functional.

\section{Phase field}

From the above elucidation it is evident that the order parameter is versatile and can be conveniently defined based on the thermodynamical setup of the system.
For instance, in Ginzburg-Landau approach the order parameter represents the conductivity while in Ising's model, it is defined based on the magnetic spin.
Moreover, the spatially-dependent concentration variable in the Cahn-Hilliard formulation can be replaced by the order parameter by defining it appropriately.
Despite the difference in the nature of the thermodynamic variable which is employed to define the order parameter, certain features characterise these order parameters.
The order parameter assume a definite value in the bulk phases while spatially vary in the region separating the phases.
Furthermore, it is evident from Eqns.~\ref{eq:free_functional} and~\ref{eq:CH_functional} that the order parameter represent a continuum field in the system.
However, the nature of the order parameter, \textit{conserved or non-conserved}, plays a significant role when the system is in non-equilibrium condition.

Under a non-equilibrium condition, in addition to spatial-dependency, the order parameters exhibits temporal-dependency.
In accordance with the \textit{linear response theory}, the temporal evolution of the order parameter is related to the variation derivative of the functional expressed in Eqn.~\ref{eq:CH_functional} (or Eqn.~\ref{eq:free_functional}).
However, the form adopted by this relation depends explicitly on the nature of the order parameter.
For the non-conserved order parameter, defined by the variables like conductivity, magnetic spin or density, the evolution equation is of the form
\begin{align}\label{eq:GL_evolution}
 \frac{\partial \phi(\vec{x},t)}{\partial t}=-\bar{M}\frac{\delta F}{\delta \phi(\vec{x},t)},
\end{align}
where $\bar{M}$ is a constant which governs the \textit{mobility} of the interface.
Eqn.~\ref{eq:GL_evolution} is referred to as time-dependent Ginzburg-Landau equation~\cite{schmid1966time}.
Later, Allen and Cahn treated derived similar expression from the Cahn-Hilliard functional, which is now known as Allen-Cahn equation~\cite{allen1979microscopic}.
Apart from its applicability, the Allen-Cahn and time-dependent Ginzburg-Landau equation are numerically analogous.
The order parameter when defined based on conserved variables like concentration, the form of the evolution equation changes to 
\begin{align}\label{eq:CH_evolution}
 \frac{\partial \phi(\vec{x},t)}{\partial t}=\tilde{M}\nabla^{2}\frac{\delta F}{\delta \phi(\vec{x},t)}.
\end{align}
The aforementioned form is the simplified representation evolution equation adopted for numerically treating spinodal decomposition.
Although it has been shown that both these evolution equations can be derived from the same probabilistic master equation using path-integral approach~\cite{metiu1976derivation}, the disparity in Eqns.~\ref{eq:GL_evolution} and~\ref{eq:CH_evolution} is significant.

Physically, a phase transformation encompasses both conserved and non-conserved components.
For instance, during solidification, the state of the system is non-conserved, as the liquid state completely vanishes, whereas the concentration remains conserved.
Advancements in the \textit{continuum} approach, elucidated in previous sections, enabled numerical treatment of both conserved and non-conserved parameters in a single framework~\cite{halperin1974renormalization,langer1986directions}.
In this framework, the thermodynamical description of the non-conserved order parameter is subtle and conventionally, it is understood to represent the nature of the phase-fraction.
Such minimization of the thermodynamic attributions to the order parameter,  relaxes the constraint on the values assumed by the order parameters in the bulk phases.
Therefore, by appropriate formulation, simpler (handle-able) values are assigned to the order parameters for efficient numerical treatment.
This state variable, with or without any thermodynamic significance, but restricted to the condition $\sum_{i}^{N}\phi_{i}=1$ is referred as \textit{phase field}.
In other words, phase fields are order parameters with relaxed thermodynamical description and confined to the criterion that $\phi_{\alpha}=1$ within phase-$\alpha$ and $\phi_{\alpha}=0$ in the remnant of the system.

\section{Interface properties}\label{sec:inter_prop}

Based on the above elucidation of the phase field, under non-equilibrium condition, the functional which governs its evolution can be expressed as
\begin{align}\label{eq:Pf_functional}
 F(\cdot,\phi,\nabla \phi)=\int_{V}\{f_{\text{drv}}(\cdot)+\frac{K}{2}|\nabla \phi(\vec{x})|^{2}+Hf_{\text{dw}}(\phi)\}\diff V,
\end{align}
where $K$ and $H$ are constants, and $f_{\text{drv}}(\cdot)$ is the \lq driving force\rq \thinspace which governs the temporal change in the phase field.
Since the primary focus of this section is to delineate the interface, an detailed explication of the term $f_{\text{drv}}(\cdot)$ will be resumed in the next section.
However, its vital to note that the driving force influences the nature of the double-well function $f_{\text{dw}}(\phi)$, which is motivated by the Landau free-energy formulation in Eqn.~\ref{eq:free_energy_functional}. 
\nomenclature{$f_{\text{dw}}(\phi)$}{Double-well function}%
\nomenclature{$K$}{Gradient energy co-efficient (Allen-Cahn)}%
Owing to the phase-field constraint $\sum_{i}^{N}\phi_{i}=1$, the function $f_{\text{dw}}(\phi)$ consists of two minimas at $0$ and $1$.
Depending on the $f_{\text{drv}}(\cdot)$, one of these minimas turns global.

At equilibrium, when $f_{\text{drv}}(\cdot)=0$, the free energy functional in Eqn.~\ref{eq:Pf_functional} gets simplified with the contributions restricted to gradient energy term ($|\nabla \phi(\vec{x})|^{2}$) and double-well function $f_{\text{dw}}(\phi)$.
Since the energy of the system is solely dictated by the free energy of the interface under equilibrium, a corollary of this condition is that the interface contribution based on the function in Eqn.~\ref{eq:Pf_functional} is expressed as
\begin{align}\label{eq:Pf_interface}
 f_{\text{intf}}(\phi,\nabla \phi(\vec{x}))=Hf_{\text{dw}}(\phi)+\frac{K}{2}|\nabla \phi(\vec{x})|^{2}.
\end{align}
Owing to the lack of phase transformation under equilibrium, irrespective of the nature of the order parameter, $\frac{\partial \phi(\vec{x},t)}{\partial t}=0$.
Consequently, at equilibrium, the variational derivative of the functional $F(\phi)$ is written as
\begin{align}\label{eq:functional_derivative}
 \frac{\delta F}{\delta \phi}=\frac{\partial f_{\text{intf}}(\phi,\nabla \phi(\vec{x}))}{\partial \phi}-\Big[\nabla\frac{\partial f_{\text{intf}}(\phi,\nabla \phi(\vec{x}))}{\partial \nabla\phi(\vec{x})}\Big]=0.
\end{align}
Assuming that the spatial dependency of phase field $\phi$ is restricted to X-direction, and considering a simplified form of the double-well function $f_{\text{dw}}(\phi)=\frac{1}{4}(1-\phi^2)^2$, Eqn.~\ref{eq:Pf_interface} yields
\begin{align}\label{eq:functional_simple}
 K\frac{\partial^2 \phi(x)}{\partial x^2}-H\frac{\partial f_{\text{dw}}(\phi)}{\partial \phi}=K\frac{\partial^2 \phi(x)}{\partial x^2}-H(\phi-\phi^3)=0.
\end{align}
From Eqn.~\ref{eq:functional_simple}, the profile of the interface can be written as
\begin{align}\label{eq:phi_profile}
 \phi(x)=\text{tanh}\Big(\frac{x}{\epsilon\sqrt{2}}\Big),
\end{align}
where the constant $\epsilon=\sqrt{{K}/{H}}$.
Here, it is important to note that the relation expressed in Eqn.~\ref{eq:phi_profile} is significantly influenced by the choice of $f_{\text{dw}}(\phi)$.
Therefore, with the change in the double-well function $f_{\text{dw}}(\phi)$, the profile of the interface varies.

The free energy density $f_{\text{intf}}(\phi(\vec{x}))$ in Eqn.~\ref{eq:Pf_interface} is of the dimension of energy per unit volume.
Since $f_{\text{dw}}(\phi)$ is dimensionless, the dimension of the constant $H$ should be identical to $f_{\text{intf}}(\phi(\vec{x}))$.
For the similar reason, $K$ is of the dimension, energy per unit length.
Now, the relation $\epsilon=\sqrt{{K}/{H}}$ yields that dimension of $\epsilon$ is length and thus, relates the width of the diffuse interface.

In principle, having decoupled the interface contribution from the rest of the system in Eqn.~\ref{eq:Pf_interface}, the interfacial energy $\gamma$ can be related to $f_{\text{intf}}(\phi(\vec{x}))$ as
\begin{align}\label{eq:gamma}
 \gamma=\int_{-\infty}^{\infty}f_{\text{intf}}(\phi,\nabla \phi(\vec{x}))\diff x =H\int_{-\infty}^{\infty}\Big[\frac{1}{2}\epsilon^2|\nabla \phi(\vec{x})|^2+f_{\text{dw}}(\phi)\Big] \diff x.
\end{align}
Non-dimensionalising the variable $x$ using the length parameter $\epsilon$ by $\tilde{x}=\frac{x}{\epsilon}$ and expressing Eqn.~\ref{eq:gamma} yields
\begin{align}\label{eq:gamma_nonD}
 \gamma=\sqrt{KH}\int_{-\infty}^{\infty}|\nabla_{\tilde{x}} \phi(\vec{x})|^2+f_{\text{dw}}(\phi) \diff x.
\end{align}
To solve the terms enclosed by the integrals, Eqn.~\ref{eq:functional_simple} is non-dimensionalised which results in the condition
\begin{align}\label{eq:solve_condition}
 \frac{\partial^2 \phi(\tilde{x})}{\partial \tilde{x}^2}=\frac{\partial f_{\text{dw}}(\phi)}{\partial \phi}.
\end{align}
Using the above condition, the integral term in Eqn.~\ref{eq:gamma_nonD} is evaluated, which for the convenience is referred to as $I$.
Therefore, from Eqn.~\ref{eq:gamma_nonD}, the interfacial energy can be written as
\begin{align}\label{eq:gamma_cons}
 \gamma=I\sqrt{KH}=\epsilon I H.
\end{align}
Based on the above relation, for a given interface width $W$, the interfacial energy $\gamma$ can be recovered in a phase-field model by manipulating the constants $H$ and $K$.
\nomenclature{$W$}{Interface width}%
It is evident from Eqn.~\ref{eq:Pf_interface} that the constant $H$ governs the shape, or particularly the amplitude, of the double-well function $f_{\text{dw}}(\phi)$.
Therefore, as represented in Eqn.~\ref{eq:gamma_cons}, the width of the diffuse interface ($\epsilon$) and the product of the energy barrier ($H$) phenomenologically relates to interfacial energy ($\gamma$).

\section{Thermodynamic formulation of the functional}\label{sec:thermodynamic_function}

The relaxation of the thermodynamic attributions to the phase field by the inclusion of a more coherent variable, widens the applicability of the approach.
Therefore, since the preliminary works of Langer~\cite{langer1986directions} and Hohenberg et al~\cite{halperin1974renormalization}, numerous phase-field models have been increasingly reported.
Despite being substantially different in its thermodynamical consideration, the similarity in the numerical framework of the different phase-field models is one of the main reason for the growing recognition of this approach.

The framework of the phase-field modelling begins with the formulation of the functional.
The thermodynamical makeup of the system is defined in the model by the choice of an appropriate functional.
Moreover, the variation of this functional dictates the resulting phase-field simulation.
In this section, the progressive changes in the formulation of the functional, which led to the model adopted in the present work, is concisely presented.

The driving force which govern the phase transformation is encompassed in the functional.
Therefore, in Eqn.~\ref{eq:Pf_functional}, in addition to the double-well and the gradient energy term a function, $f_{\text{drv}}(\cdot)$, is included.
To simulate the solidification of a pure material, the driving force is governed by the temperature.
Correspondingly, the functional can be expressed as
\begin{align}\label{eq:solid_functional1}
 F(T,\phi,\nabla \phi)=\int_{V}\left[ f_{\text{dw}}(\phi)+\frac{\epsilon^{2}}{2}|\nabla \phi|^2+\lambda U(T)\phi\right] \diff V,
\end{align}
where $\lambda$ is a dimensionless constant.
The term $U(T)$ governs the driving force through the relation $U(T)=C_{p}\left(\frac{T-T_{m}}{L}\right)$, where $T_{m}$, $C_{p}$ and $L$ are the melting temperature, the specific heat and the latent heat, respectively. 
\nomenclature{$T_{m}$}{Interface width}%
\nomenclature{$C_{p}$}{Interface width}%
\nomenclature{$L$}{Interface width}%
Based on the sharp interface description of the enthalpy, a function $H(U(T))$ is defined which reads
\nomenclature{$H(T)$}{Enthalpy}%
\begin{align}\label{eq:drive_phi}
 H(U(T))=U(T)+\frac{l}{2}\phi 
\end{align}
with $l$ being a constant.
This numerical formulation was adopted in the early works of Caginalp~\cite{caginalp1986analysis} and Collins et al~\cite{collins1985diffuse}.

For the functional in Eqn.~\ref{eq:solid_functional1}, the temporal evolution which ensures the minimization of the overall free energy is written as
\begin{align}\label{eq:intro_phi_ev1}
 -\tau\frac{\partial \phi}{\partial t}=\frac{\delta F(T,\phi,\nabla \phi)}{\delta \phi}=f_{\text{dw}}'(\phi)+\epsilon^{2}\nabla^2\phi+\lambda U(T),
\end{align}
where $\tau$ is a relaxation constant and the phase-field derivative of the double-well function is represented by $f_{\text{dw}}'(\phi)$.
\nomenclature{$\tau$}{Relaxation constant}%
Furthermore, owing to the conserved nature of the term $U(T)$, its temporal evolution is expressed as
\begin{align}\label{eq:intro_u_ev1}
 \frac{\partial U}{\partial t} = {K_{o}}\nabla^2 U+\frac{l}{2}\frac{\partial \phi}{\phi t},
\end{align}
where the constant $K_{o}$ accounts for the conductivity.
\nomenclature{$K_{o}$}{Conductivity}%

The monotonic decrease in the free energy is evident in the time-dependent variation of the functional expressed in Eqn.~\ref{eq:GL_evolution} and~\ref{eq:CH_evolution}.
However, for the functional in Eqn.~\ref{eq:solid_functional1}, the inclusion of the term which dictates the driving force does not explicitly affirm this temporal behaviour.
Recognizing this influence of the additional term, Penrose and Fife formulated a different functional which ensures the monotonic change~\cite{penrose1990thermodynamically}. 
The formulation of the functional begins with the thermodynamical description of the free energy as a \textit{Legendre transform} expressed as 
\begin{align}\label{eq:entropy_leg}
 f(T)=e(T)-Ts(e(T)),
\end{align}
where $e$ and $s(e)$ are the energy and entropy density of the system.
\nomenclature{$e$}{Overall energy density of the system}%
\nomenclature{$s(e)$}{Entropy density of the system}%
Therefore, by augmenting the dependency of the free energy $f(T)$ on the phase field through the conventional double-well function, the Eqn.~\ref{eq:entropy_leg} yields
\begin{align}\label{eq:entropy_phase}
 s(e,\phi)=-\frac{f(T,\phi)}{T}+\frac{e}{T}
\end{align}
where $s(e(T),\phi)$, and $f(T,\phi)$ are phase field dependent entropy and free-energy density of the system.
From the relation in Eqn.~\ref{eq:entropy_leg}, the energy density of the system can be expressed as
\begin{align}\label{eq:energy_en}
 e=\frac{\partial(f(T(e),\phi)/T)}{\partial(1/T)}.
\end{align}
Based on this thermodynamical consideration, a functional based on the entropy density can be expressed as
\begin{align}\label{eq:ent_functional_intro}
 S(e,\phi)=\int_{V}\left[s(e,\phi)-\frac{\epsilon^2}{2}|\nabla \phi|^2\right] \diff V.
\end{align}
In this formulation, it is important to note that the dependence of the energy density on the phase field is through its relation to the free energy in Eqn.~\ref{eq:energy_en}. 
Therefore, subsequently, an attempt was made to improve the thermodynamical consistency of this approach~\cite{wang1993thermodynamically}.
Correspondingly, the energy density of the system is written, by distinguishing it between the phases, as
\begin{align}\label{eq:energy_en_phi}
 e(T,\phi)=e_{S}(T)h(\phi)+e_{L}(T)(1-h(\phi)),
\end{align}
where $e_{S}(T)$ and $e_{L}(T)$ are the energy densities of solid and liquid phases of the system, respectively.
The function $h(\phi)$, referred to as \textit{interpolation function}, changes monotonically across the interface and satisfies the condition
\nomenclature{$h(\phi)$}{Interpolation function}%
\begin{align}\label{eq:inter_condition}
 h(\phi)=
 \begin{cases}
  1,&  \phi=1 ~(\text{solid}),\\
  0,&  \phi=0 ~(\text{liquid}),
 \end{cases} 
\end{align}
Furthermore, in both the bulk phases ($\phi=0,1$), the phase-field derivative of the interpolation function must be $h'(\phi)=0$. 
This interpolation function is determined from the double-well function $f_{\text{dw}}(\phi)$  through
\begin{align}\label{eq:inter_double}
 h(\phi)=\frac{\int_{0}^{\phi}f_{\text{dw}}(\phi)\diff\phi}{\int_{0}^{1}f_{\text{dw}}(\phi)\diff\phi}.
\end{align}
Since the phase field assume $\phi=1$ in solid and $\phi=0$ in liquid, the double-well function $f_{\text{dw}}(\phi)=\phi^{2}(1-\phi)^{2}$ is considered and the interpolation is ascertained accordingly.
The phase-field evolution equation for this entropy based formulation can be expressed as variation of the functional defined in Eqn.~\ref{eq:ent_functional_intro}  as 
\begin{align}\label{eq:entropy_evolution}
 \tau\frac{\partial \phi}{\partial t}=\frac{\delta S(e,\phi) }{\delta \phi}.
\end{align}
Additionally, by appropriately defining the mobility $\tilde{M}$ which encompasses the relation $\frac{\partial s(e,\phi)}{\partial e}=\frac{1}{T(e,\phi)}$, the evolution of the energy density is written as
\begin{align}\label{eq:enr_evolution}
 \frac{\partial e}{\partial t}=\tilde{M}\nabla^{2}T.
\end{align}
The thermodynamic consistency of this formulation for the solidification of the pure material is ensured by the monotonic increase in the overall entropy density.

The numerical descriptions of the functional so far have been confined to the phase transformation of a pure material.
Wheeler, Beottinger and Mcfadden pioneered the attempt of simulating the transformation of the binary alloy in an isothermal condition, which is commonly referred to as WBM model~\cite{wheeler1992phase}. 
This initial approach is an extension of the numerical treatment adopted for the unary system~\cite{kobayashi1993modeling}.
The free energy of the components, $A$ and $B$, constituting the solid and liquid phase in the binary system is defined analogous to the unary system as
\begin{align}\label{eq:wbm_pureA}
 f_{A}(T,\phi)=W_{A}\int_{0}^{\phi}p(p-1)[p-\frac{1}{2}-\beta_{A}(T)]\diff p,
\end{align}
and 
\begin{align}\label{eq:wbm_pureB}
 f_{B}(T,\phi)=W_{B}\int_{0}^{\phi}p(p-1)[p-\frac{1}{2}-\beta_{B}(T)]\diff p,
\end{align}
respectively.
The function $\beta_{\{A,B\}}(T)$ manipulates the double-well function depending on the temperature and its relation to the melting temperature of the corresponding material ($A$ or $B$) and $W_{\{A,B\}}$ is a constant.
Owing to the binary nature of the system, the chemical setup of the phases can be expressed by one independent concentration variable $c$.
Therefore, invoking the thermodynamical relation, the free energy density of the entire system, based on Eqns.~\ref{eq:wbm_pureA} and~\ref{eq:wbm_pureB}, is be expressed as
\begin{align}\label{eq:wbm_fre}
 f(c,T,\phi)=cf_{B}(T,\phi)+(1-c)f_{A}(T,\phi)+\frac{RT}{V_m}[c\ln c+(1-c)\ln(1-c)],
\end{align}
where $c$ is the mole fraction of component $B$ with $R$ and $V_m$ representing universal gas constant and molar volume, respectively.
\nomenclature{$R$}{Universal gas constant}%
The formulation in Eqn.~\ref{eq:wbm_fre} assumes the binary system to be ideal solution.
However,  from Eqns.~\ref{eq:wbm_pureA} and~\ref{eq:wbm_pureB}, free energy description can be extended for regular solutions as well.
Based on the above thermodynamical considerations, the free energy functional of the system is written as
\begin{align}\label{eq:wbm_func}
 F(c,T,\phi,\nabla\phi)=\int_{V}\left[f(c,T,\phi)+\frac{\epsilon^2}{2}|\nabla \phi|^2\right] \diff V.
\end{align}
The temporal evolution of the phase field is determined from the variation of the functional $F(c,T,\phi,\nabla\phi)$.
\nomenclature{$F$}{Free energy functional}%
Furthermore, assuming isothermal solidification, the evolution of the independent concentration variable is 
\begin{align}\label{eq:wbm_concV}
 \frac{\partial c}{\partial t} = \nabla \Big\{[c(1-c)]\tilde{M}\nabla\frac{\delta F(c,\phi)}{\delta c}\Big\},
\end{align}
where $\tilde{M}$ is the mobility which encompasses the diffusivity $D$ of the component.
The coefficient $c(1-c)$ is included in the above Eqn.~\ref{eq:wbm_concV} to ensure that the mobility is constant with in a phase and independent of the concentration.

Several advancements have been made to this initial WBM model to relax certain restrictions and improve the applicability of the approach.
One major limitation of this model, which was identified almost immediately, pertains to the very high driving force.
With increase in the driving force governing the phase transformation the interface velocity correspondingly increases.
However, as indicated in Eqn.~\ref{eq:wbm_concV}, the concentration evolution is predominantly governed by the diffusivity which is a constant at a given temperature.
Therefore, at a very high interface velocity, the diffusing component gets segregated in the interface, which is referred to \textit{solute trapping}.
To address this limitation, Karma~\cite{karma1994phase} as well as the authors of the WBM model~\cite{wheeler1993phase} independently extended the free functional to 
\begin{align}\label{eq:wbm_func2}
 F(c,\nabla c,\phi,\nabla\phi)=\int_{V}\left[f(c,T,\phi)+\frac{\epsilon^2}{2}|\nabla \phi|^2+\frac{\tilde{\epsilon}^2}{2}|\nabla c|^2\right] \diff V,
\end{align}
by including the gradient concentration term, $\frac{\tilde{\epsilon}^2}{2}|\nabla c|^2$.
Another advancements involved the redefining the entire framework to investigate the non-isothermal solidification of the binary alloys.
In this model, while the concentration based description of the free energy remains unchanged, the approach of Penrose and Fife~\cite{penrose1990thermodynamically}, as well Wang et al.~\cite{wang1993thermodynamically}, is adopted to formulate entropy density based functional~\cite{conti1997solidification}.

In the WBM model, it is evident from the numerical description that a single concentration field is considered for both the solid and liquid phase.
Such consideration entails that the concentration field is continuous across the interface and exhibit a spatial dependency.
At each point in the interface, the concentration can be assumed to be the average of the both the phases.
Therefore, this formalism is, at times, referred to coarse grained approach.
Although the framework of the model  appears to be thermodynamically consistent, Tiaden et al. have shown that, in equilibrium, an excess energy is attributed to the interface, particularly due to the nature concentration field~\cite{tiaden1998multiphase}.
This introduces a non-physical interaction between the free energy bulk phases with the interface and imposes constraint on the interface parameter.
Furthermore, it has been identified that, since in interface the concentration adheres to the condition $c^{S}=c^{L}=c$, where $c^{S}$ and $c^{L}$ concentration in respective liquid and solid phases, a chemical potential gradient is introduced within the diffuse region which should be equal under equilibrium~\cite{kim1998interfacial}.
To circumvent this limitation and to render an efficient decoupling of the interface and bulk free energy contribution, a different approach was rendered by Kim, Kim and  Suzuki, which is referred to as KKS model~\cite{kim1999phase}.

In their work, Tiaden et al.~\cite{tiaden1998multiphase}, while theoretically unravelling the limitation of the WBM model, presented a groundwork for the KKS model.
This initial attempt involves replacing the continuous concentration variable $c$ with a phase-dependent entity $c^{\alpha}$, where $\alpha$ can be solid, liquid or any phase,  and relating the concentration of the individual phases through a correlation factor.
\nomenclature{$c^{\alpha}$}{Concentration in phase-$\alpha$ (mole fraction)}%
Since this approach was confined to the dilute solutions and thermodynamical consistency was not completely explicit, it was subsequently superseded by the KKS model.

The theoretical framework of the KKS model is motivated by the sharp interface description of the similar problem.
Correspondingly, the free energy density comprising of the contributions from the bulk phases is expressed as
\begin{align}\label{eq:kks_bulk_fre}
 f(c^{\alpha},c^{\beta},\phi)=h(\phi)f^{\alpha}(c^{\alpha})+(1-h(\phi))f^{\beta}(c^{\beta}),
\end{align}
where $h(\phi)$ is the interpolation function that satisfies the condition as in previous models, Eqn.~\ref{eq:inter_condition}.
The free energy density pertaining to the phases $\alpha$ and $\beta$ is represented by $f^{\alpha}$ and $f^{\beta}$, respectively.
\nomenclature{$f^{\alpha}$}{Free-energy density of phase-$\alpha$}%
Here, it is important note this formulation of the bulk free-energy density in Eqn.~\ref{eq:kks_bulk_fre} enables the introduction of  physical free energy from the database like CALPHAD.
Furthermore, the continuous concentration $c$ in the WBM model is replaced by the phase-dependent concentrations $c^{\alpha}$ and $c^{\beta}$.
However, instead of the correlation factor, these concentrations are related through the interpolation function as
\begin{align}\label{eq:conc_inter}
 c(\phi)=h(\phi)c^{\alpha}+(1-h(\phi))c^{\beta}.
\end{align}
At the interface, the phases are assumed be in equilibrium characterised by the condition
\begin{align}\label{eq:chempot_inter}
 \frac{\partial f^{\alpha}(c^{\alpha}(\vec{x},t))}{\partial c^{\alpha}}=\frac{\partial f^{\beta}(c^{\beta}(\vec{x},t))}{\partial c^{\beta}}=\mu_{eq}(\vec{x},t),
\end{align}
where the equilibrium chemical potential is represented by $\mu_{eq}$.
\nomenclature{$\mu$}{Chemical potential}%

Considering the aforementioned numerical and thermodynamical description, the functional under this formalism is expressed as
\begin{align}\label{eq:kks_functional}
 F(c^{\alpha},c^{\beta},\phi,\nabla\phi)=\int_{V}\left[f(c^{\alpha},c^{\beta},\phi)+f_{\text{dw}}(\phi)+\frac{\epsilon^2}{2}|\nabla \phi|^2\right ]\diff V.
\end{align}
The phase-field evolution, governed by variation of functional in Eqn.~\ref{eq:kks_functional}, can be written as
\begin{align}\label{eq:kks_pf_ev}
 -\tau\frac{\partial \phi}{\partial t}=h'(\phi)[f^{\alpha}(c^{\alpha})-f^{\beta}(c^{\beta})-\mu_{eq}(c^{\alpha}-c^{\beta})]-f_{\text{dw}}'(\phi)-\epsilon^2\nabla^2\phi.
\end{align}
The temporal evolution of the concentration is similarly expressed as
\begin{align}\label{eq:kks_con_ev}
 \frac{\partial c}{\partial t}=\nabla\Big[\frac{\tilde{M}(\phi)}{f_{cc}}\nabla\frac{\partial f^{\alpha}(c^{\alpha})}{\partial c^{\alpha}}\Big]
\end{align}
where the constant $f_{cc}=\frac{\partial^{2}f^{\alpha}(c^{\alpha})}{\partial c^{\alpha^2}}$ plays the role of the coefficient $c(1-c)$ in the WBM model and ensures that the mobility $\tilde{M}(\phi)$, which is dictated by the diffusivity, is constant within a phase and is independent of the concentration.
\nomenclature{$f_{cc}$}{Second derivative of free energy}%

\section{Grand potential}\label{sec:ther_gran}

The thermodynamic variable which dictates the phase transformation in the solidification of the pure material is temperature.
This fundamental field becomes constant throughout the system, irrespective of the phases, once the equilibrium is attained.
In the solidification of the binary alloy, under the isothermal condition, the governing role of the temperature is assumed by the independent concentration variable.
Consequently, the equilibrium condition is ascertained by the phase-dependent concentration reaching a characteristic value, $c^{\alpha}_{eq}$ and $c^{\beta}_{eq}$ , defined by the free energy plot.
It is worth noting that, unlike the temperature which is an intensive variable, these concentrations are densities of the extensive variable and are not equal at equilibrium, $c^{\alpha}_{eq}\neq c^{\beta}_{eq}$.
Although there is absolutely no thermodynamical inconsistencies in the phase-dependency of the concentration variable, this nature of the fundamental-field compromises the computational efficiency.
This compromise is evident from the condition, Eqn.~\ref{eq:chempot_inter}, adopted in the KKS model.
Since the KKS model, assumes that the co-existing phases are in equilibrium at the interface, the phase-dependent concentrations are determined by solving the non-linear relation in Eqn.~\ref{eq:chempot_inter} for each phase and component.
In a polycrystalline setup, the amplitude of the calculation increases proportionately with increase in the number of components and phases.
The computational load introduced by the phase-dependent field can be averted by considering an intensive variable which is independent of the phases.
Therefore, a phase-field model wherein the chemical potential $\mu$ acts as the dynamic fundamental-variable is more efficient than the KKS model, particularly for a multicomponent consideration.

In order to employ the chemical potential as the fundamental field which replaces the concentration, the thermodynamic function defining the functional needs to be re-formulated.
Conventionally, in both WBM and KKS model, the functional is defined based on the free energy density $f^{\alpha}(c^{\alpha})$.
The phase-dependent concentration variable in the function $f^{\alpha}(c^{\alpha})$ can be replaced by the intensive variable $\mu$ by considering the \textit{Legendre transform} of the free energy which yields
\begin{align}\label{eq:leg_f}
 \Psi^{\alpha}(\mu)=f^{\alpha}(c^{\alpha})-\mu c^{\alpha}.
\end{align}
The function $\Psi^{\alpha}(\mu)$ in the above Eqn.~\ref{eq:leg_f} is referred to as \textit{grand potential density} in the statistical thermodynamic.
\nomenclature{$\Psi^{\alpha}(\mu)$}{Grand potential density of phase-$\alpha$}%
Furthermore, from Eqn.~\ref{eq:leg_f} the concentration $c^{\alpha}$ can be expressed
\begin{align}\label{eq:leg_c}
 c^{\alpha}=-\frac{\partial \Psi^{\alpha}(\mu)}{\partial \mu}
\end{align}
Here it is important to note that this description assumes phase transformation in isothermal condition and hence, the temperature dependency is not indicated.

The nature of the grand potential density $\Psi^{\alpha}(\mu)$ at equilibrium can be elucidated in a thermodynamical framework.
Consider a extensive variant of the grand potential which depends on the volume of the phase, $\tilde{\Psi}^{\alpha}(\mu,V^{\alpha})$.
Eqn.~\ref{eq:leg_f} can now be written as
\begin{align}\label{eq:leg_fv}
 \tilde{\Psi}^{\alpha}(\mu,V^{\alpha})=\tilde{f}^{\alpha}(c^{\alpha},V^{\alpha})-\mu c^{\alpha},
\end{align}
Under this condition, since the free energy $\tilde{f}^{\alpha}(c^{\alpha},V^{\alpha})$ can be expressed as the 
\begin{align}\label{eq:leg_fg}
 \tilde{f}^{\alpha}(c^{\alpha},V^{\alpha})=G^{\alpha}-PV^{\alpha},
\end{align}
Eqn.~\ref{eq:leg_fv} reads
\begin{align}\label{eq:leg_fvV}
 \tilde{\Psi}^{\alpha}(\mu,V^{\alpha})=G^{\alpha}-PV^{\alpha}-\mu c^{\alpha},
\end{align}
where $P$ is the pressure.
\nomenclature{$P$}{Pressure}%
Invoking the Gibbs-Duhem relation, $G^{\alpha}=\mu c^{\alpha}$, the relation in Eqn.~\ref{eq:leg_fvV} yields,
\begin{align}\label{eq:leg_fvP}
 P=-\frac{\tilde{\Psi}^{\alpha}(\mu,V_{\alpha})}{V_{\alpha}}=-\Psi^{\alpha}(\mu).
\end{align}
Since the pressure $P$ is equal at equilibrium, correspondingly, the nature of the grand potential density at equilibrium is written as
\begin{align}\label{eq:grand_eq}
 \Psi^{\alpha}(\mu)=\Psi^{\beta}(\mu)
\end{align}
Considering the above description of the grand potential and the its fundamental variable, chemical potential, a relation between the model formulated based on this thermodynamical function and solidification of the pure material can be deduced.
Free energy of the solid and liquid phase during the solidification are equal at equilibrium.
This equality is not sustained in the two component system.
However, as elucidated through Eqn.~\ref{eq:grand_eq}, similar to free energy in pure system, the grand potential density loses the phase-dependency at equilibrium and reaches an equal value.
Furthermore, akin to the temperature in the solidification of unary melt, the chemical potential in the binary system assumes an identical value across the interface in the equilibrium.
Despite these similarity with the single component solidification, a phase-field model can be formulated based on the grand potential density which does not compromise in the specific quantitative aspects like the WBM model.



\newpage\null\thispagestyle{empty}\newpage

\newpage
\thispagestyle{empty}
\vspace*{8cm}
\begin{center}
 \Huge \textbf{Part II} \\
 \Huge \textbf{Grand-potential formalism}
\end{center}

\newpage\null\thispagestyle{empty}\newpage

\chapter{Grand potential based multiphase-field model}~\label{sec:grand_potentialM}

A phase-field model formulated around the grand potential functional was initially proposed for binary two-phase system~\cite{plapp2011unified} and subsequently extended to the multicomponent multiphase system~\cite{choudhury2012grand}.
Since the simulations analysed in the present work emerge from the grand potential density based phase-field model, a concise description of the model is rendered in this section.
Considering the nature of the theoretical investigations pursued in the following sections, the model is derived for binary multiphase systems.
Furthermore, since the morphological evolutions reported in this work pertains to the isothermal treatment, the dependence of temperature on the thermodynamic parameters is overlooked. 

As elucidated in the previous sections, the formulation of the phase-field model stems from the description of the functional which encompasses the contribution from bulk phases and the corresponding interface.
The double-well function, or its variant, which penalises the phase field digressing from the defined value, along with the gradient energy term can be categorized as the energetic contribution of the interface.
While, the bulk-phase contribution is formulated based on a thermodynamic function which is associated with a fundamental dynamic variable.
The approach adopted to numerically treat the function and its corresponding variable bring about a significant differences in the model.
In the present model, the functional $\Omega(\mu,\vphi,\n \vphi)$ is formulated as
\nomenclature{$\Omega$}{Grand potential density based functional}%
\begin{align}\label{eq:GP_functional}
\Omega(\mu,\vphi,\n \vphi)=\int_{V}\Big[\underbrace{\Psi(\mu,\vphi)}_{\mathclap{\text{bulk}}}+\underbrace{\epsilon a(\vphi,\n \vphi) +\dfrac{1}{\epsilon}w( \vphi)}_{\mathclap{\text{interface}}}\Big]dV,
\end{align}
where the contribution from the bulk phases is dictated by the corresponding grand potential density $\Psi(\mu,\vphi)$.
\nomenclature{$\Psi(\mu,\vphi)$}{Homogenised grand potential density}%
Owing to the multiphase formulation of the model, in Eqn.~\ref{eq:GP_functional}, the phase field is vector-valued continuous state variable of $N$ components, $\vphi=\{\phi_{\alpha}, \phi_{\beta},\dots,\phi_{N}\}$, with $N$ representing total number of phases.
\nomenclature{$\vphi$}{Continuous vector of $N$ components}%
\nomenclature{$N$}{Total number of phases}%
Whereas, the chemical potential is expressed as $\mu$, since the system comprises of a single independent component.

In a multiphase, the bulk-phase contribution $\Psi(\mu,\vphi)$ is expressed as the interpolation of individual contributions,
\begin{align}\label{eq:interpolation_grandchem}
\Psi_{\text{bulk}}\left(\mu,\vphi\right) =
\sum_{\alpha=1}^N\Psi^{\alpha}\left(\mu\right)h_{\alpha}\left(\vphi\right),
\end{align}
where $h_{\alpha}\left(\vphi\right)$ is the interpolation function which satisfies the condition discussed in Sec.~\ref{sec:thermodynamic_function} and Eqn.~\ref{eq:inter_condition}.
The grand potential density $\Psi^{\alpha}\left(\mu\right)$ of the phase-$\alpha$ adheres to the thermodynamic description render in the previous Sec.~\ref{sec:ther_gran}.

In the current multiphase description, the gradient energy density, $\epsilon {a}(\bm{\phi},\n \bm{\phi})$,  is defined as the summation of the pairwise interaction between the adjoining phases~\cite{nestler2005multicomponent}.
\nomenclature{$\epsilon {a}(\bm{\phi},\n \bm{\phi})$}{Gradient energy term}%
\nomenclature{$\epsilon$}{Length scale parameter which dictates interface width}%
The corresponding formulation of the gradient energy density reads
\begin{align}\label{eq:gradient_energy}
\epsilon {a}(\bm{\phi},\n \bm{\phi})  =  \epsilon\displaystyle\sum_{\alpha<\beta} \gamma_{\alpha\beta}{[a_{\alpha\beta}(\textbf{q}_{\alpha\beta})]}^{2} {|\textbf{q}_{\alpha\beta}|}^{2},
\end{align}
where $\gamma_{\alpha\beta}$ is the interface energy density between phase-$\alpha$ and-$\beta$.
The length-scale parameter $\epsilon$ dictates the interface width.
The fundamental variable in Eqn.~\ref{eq:gradient_energy}, which is replaced by phase-field gradient in a two-phase system,  is the gradient vector, $\textbf{q}_{\alpha\beta}$.
\nomenclature{$\textbf{q}_{\alpha\beta}$}{Gradient vector}%
This gradient vector is expressed as
\begin{align}\label{eq:qab}
\textbf{q}_{\alpha\beta} & = {\phi_{\alpha} {\n}{\phi}_{\beta}} - {\phi_{\beta} {\n}{\phi}_{\alpha}}.
\end{align}
The adjunct parameter $a_{\alpha\beta}$, which is written as the function of the gradient vector in Eqn.~\ref{eq:gradient_energy}, definitively influences the interface energy density and enables the introduction of anisotropy.

Conventionally, the double-well function along with the gradient energy density renders the interface contribution.
However, it has been identified that replacing the well-function with the \textit{obstacle-type potential} enhances the stability and numerical efficiency of the simulation, particularly in a multiphase setup~\cite{oono1988study,garcke1999multiphase}.
Therefore, the obstacle-type potential $\omega(\vphi)$ replaces the well-function in Eqn.~\ref{eq:GP_functional}.
This potential density $\omega(\vphi)$ is expressed as
\nomenclature{$\frac{1}{\epsilon}\omega(\vphi)$}{Obstacle type potential}%
\begin{align}\label{eq:omega}
  \frac{1}{\epsilon}\omega(\vphi) &= \frac{16}{\epsilon\pi^2} \sum_{\alpha, \beta > \alpha} \gab \phia \phib + \frac{1}{\epsilon}\sum_{\alpha, \beta > \alpha, \delta > \beta} \gamma_{\alpha \beta \delta} \phia \phib \phi_{\delta}.
\end{align}
While the first term in Eqn.~\ref{eq:omega} accommodates all possible interfaces, the second terms ensures the stability of the interface between the two phases by preventing the formation of the third-spurious phases.
The obstacle potential penalises the phase field in combination with the Gibbs simplex $\mathcal{G}$, which reads
\begin{align}\label{eq:simplex}
\mathcal{G} = \left\{ \bm{\phi} \in \mathbb{R}^N: \sum_\alpha \phi_\alpha = 1, \phi_\alpha \geq 0 \right\},
\end{align}
by imposing the criterion that $\omega(\vphi)$ becomes infinity ($\omega(\vphi)\to\infty$), if $\vphi$ is not confined to the $\mathcal{G}$.
The thermodynamical consistency of adopting this combination of $\omega(\vphi)$ and $\mathcal{G}$ has already been extensively reported~\cite{hotzer2016calibration}.

The temporal evolution of the phase field, as indicated in Eqn.~\ref{eq:GL_evolution}, is dictated by the variation of the functional.
Accordingly, the phase-field evolution equation for the present model is expressed as
\begin{align}\label{eq:equation_evolution_phi}
\tau\epsilon\dfrac{\partial \phi_{\alpha}}{\partial t}=\underbrace{ \epsilon \left(\n
 \cdot \dfrac{\partial {a}\left(\vphi,\n \vphi\right)}{\partial \n
 \phi_{\alpha}}- \dfrac{\partial{a}\left(\vphi,\n
 \vphi\right)}{\partial \phi_\alpha}\right) -\dfrac{1}{\epsilon}\dfrac{\partial
 {w}\left(\vphi\right)}{\partial \phi_\alpha}-\dfrac{\partial
 \Psi\left(\mu, \vphi\right)}{\partial \phi_\alpha}}_{\mathclap{=:\text{rhs}_{\alpha}}} - \Lambda,
\end{align}
where $\tau$ is the relaxation constant.
For a binary system, in order to ensure the zero (vanishing) interface kinetics, this constant is defined as
\begin{align}\label{eq:relaxation}
  \tau=\frac{(c^{\alpha}_{eq}-c^{\beta}_{eq})^2(M_c+F_c)}{D\dfrac{\partial c^{\alpha}}{\partial \mu}},
\end{align}
where $c^{\alpha}_{eq}$ and $c^{\beta}_{eq}$ are the respective equilibrium composition of phase-$\alpha$ and-$\beta$, and $D$ is the diffusivity.
The concentration are expressed as the mole fraction of the independent component.
In this work the diffusivities of the both phases are assumed to be equal ($D^{\alpha}=D^{\beta}=D$).
Furthermore,  the constants, $M_c$ and $F_c$, are influence by the interpolation function.
For interpolation function $h_{\alpha}\left(\vphi\right)=\phi_{\alpha}^2(3-2\phi_{\alpha})$, the summation of the constants is $M_c+F_c\approx0.223$.

In the two-phase system, the sum of the phase field is retained at $1$, apparently, through the double-well function or its variant.
However, in a multiphase setup, the constraint $\sum_{\alpha=1}^N \phi_\alpha =1$ which characterises the multiphase-field approach is ensured by introducing $\Lambda$, the \textit{Lagrange parameter} in Eqn.~\ref{eq:equation_evolution_phi}.
\nomenclature{$\Lambda$}{Lagrange multiplier to ensure $\sum_{\alpha=1}^N \phi_\alpha =1$}%
This parameter is written as $\frac{1}{\tilde{N}}\sum_{\alpha=1}^{\tilde{N}} \text{rhs}_{\alpha}$, where $\tilde{N}$ is the number of active phases.
A phase is considered as active only when a gradient exists in the phase-field. 

A relation between the independent concentration and the grand potential can be expressed by considering Eqn.~\ref{eq:leg_fv}.
Correspondingly, the independent concentration of the system can be expressed as
\begin{align}\label{eq:conc_as_diff}
c(\vphi) = -\dfrac{\partial \Psi\left(\mu,\vphi\right)}{\partial \mu}.
\end{align}
Substituting Eqn.~\ref{eq:interpolation_grandchem} in the above Eqn.~\ref{eq:conc_as_diff} yields
\begin{align}\label{eq:conc_interpolated}
c(\vphi) = -\Big( \sum_{\alpha=1}^N \dfrac{\partial \Psi^{\alpha}\left(\mu,\vphi\right)}{\partial \mu}h_{\alpha}\left(\vphi\right)\Big),
\end{align}
which, based on the Eqn.~\ref{eq:leg_c}, can be simplified as
\begin{align}\label{eq:conc_interpolated1}
 c(\vphi) = \sum_{\alpha=1}^N c^{\alpha}h_{\alpha}\left(\vphi\right).
\end{align}
The invertibility between the concentration and chemical potential, as represented by the Eqns.~\ref{eq:conc_as_diff} and~\ref{eq:conc_interpolated1}, forms the basis of the model.

The evolution of the fundamental variable, the chemical potential $\mu$, can be derived by considering the temporal evolution of the concentration described in Eqn.~\ref{eq:conc_interpolated1}.
Owing to the invertibility, the concentration in Eqn.~\ref{eq:conc_interpolated1} can be represented by $c(\mu,\vphi)$.
Therefore, the time derivative of the independent concentration variable is written as
\begin{align}\label{eq:conc_time}
\dfrac{\partial c}{\partial t}= \dfrac{\partial c(\mu,\vphi)}{\partial \mu} \dfrac{\partial \mu}{\partial t}+ \dfrac{\partial c(\mu,\vphi)}{\partial \vphi} \dfrac{\partial \vphi}{\partial t}.
\end{align}
However, the temporal evolution of the concentration, when solved separately as the conserved order parameter (Eqn.~\ref{eq:CH_evolution}), yields
\begin{align}\label{eq:conc_evolution}
 \dfrac{\partial c}{\partial t}= 
\n \cdot \Big( \vv{M}(\vphi) \n \mu\Big),
\end{align}
where $\vv{M}(\vphi)$ is the mobility matrix for the multiphase system.
\nomenclature{$\vv{M}(\vphi)$}{Mobility matrix}%
The mobility matrix $\vv{M}(\vphi)$ is written as interpolation of the individual phase mobilities,
\begin{align}\label{eq:M_mobility}
 \vv{M}(\vphi) = \sum_{\alpha=1}^N D\vv{\mathcal{X}}^{\alpha} g_{\alpha}(\vphi),
\end{align} 
with $g_{\alpha}(\vphi)$ representing the interpolation function.
\nomenclature{$g_{\alpha}(\vphi)$}{Mobility interpolation function}%
In Eqn.~\ref{eq:M_mobility}, $\vv{\mathcal{X}}^{\alpha}$ is a constant which ensures that the diffusivity $D$ does not vary within a given phase.
This constant is defined as $\frac{\partial c^{\alpha}(\mu)}{\partial \mu}$.
Generally, the interpolation of the diffusivities are different from that of the concentration or grand potential density.
In other words, often $h(\vphi)\neq g(\vphi)$.
However, in view of the thermodynamical description of the present system and microstructural evolution to be investigated, in this work, no distinction is made between the interpolation functions ($h(\vphi)= g(\vphi)$).

Having considered the evolution of the concentration, in Eqn.~\ref{eq:conc_time}, and phase field in Eqn.~\ref{eq:equation_evolution_phi}, it is evident that the temporal dependency of these  variable is influenced by the chemical potential.
To ascertain the evolution of the chemical potential, based on Eqn.~\ref{eq:conc_interpolated1}, the derivative of the concentration with respect to $\mu$, which is written as 
\begin{align}\label{eq:chem_mu}
\dfrac{\partial c(\mu,\vphi)}{\partial \mu} = \sum_{\alpha=1}^N \dfrac{\partial c^{\alpha}(\mu)}{\partial \mu} h_{\alpha}\left(\vphi\right),
\end{align}
is considered.
Furthermore, from the same Eqn.~\ref{eq:conc_interpolated1}, the second term in the right hand of the Eqn.~\ref{eq:conc_time} can be expressed
\begin{align}\label{eq:chem_time}
\dfrac{\partial c(\mu,\vphi)}{\partial t} = \sum_{\alpha=1}^N  c^{\alpha}(\mu) \dfrac{\partial h_{\alpha}\left(\vphi\right)}{\partial t}.
\end{align}
Substituting Eqns~\ref{eq:conc_evolution},~\ref{eq:chem_mu} and ~\ref{eq:chem_time} in Eqn.~\ref{eq:conc_time} and re-arranging the terms, the temporal evolution in the chemical potential reads
\begin{align}\label{eq:ChemPot_eqn}
\dfrac{\partial \mu}{\partial t} = \Big[\n \cdot \Big( \vv{M}(\vphi) \n \mu\Big) - \sum_{\alpha=1}^N  c^{\alpha}(\mu) \dfrac{\partial h_{\alpha}\left(\vphi\right)}{\partial t} \Big]  \Big[ \sum_{\alpha=1}^N \vv{\mathcal{X}}^{\alpha} h_{\alpha}\left(\vphi\right) \Big]^{-1}.
\end{align}
The term $\sum_{\alpha=1}^N \vv{\mathcal{X}}^{\alpha} h_{\alpha}\left(\vphi\right)=\vv{\mathcal{X}}(\vphi)$, in the above Eqn.~\ref{eq:ChemPot_eqn}, is referred to as \textit{susceptibility matrix}~\cite{plapp2011unified}.
\nomenclature{$\vv{\mathcal{X}}$}{Susceptibility matrix}%

\chapter{Quantitative treatment of the grand potential model}\label{chap:quant}

\section{Recovering Gibbs-Thomson relation through the asymptotic analysis}\label{sec:asym}

Experimental observations unravel that the interface, particularly during a phase transformation like solidification, is diffuse and of the order of few atomic distances~\cite{woodruff1973solid,hoyt2003atomistic}. 
This nature of the interface seemingly vindicates the diffuse interface model over the sharp interface approach, wherein the region separating the bulk phases is considered to be of zero thickness~\cite{stefan1889diffusion,stefan1889uber,stefan1891theorie,stefan1890verdampfung,gupta2017classical}.
However, in view of the dimensions of the emerging microstructure, the sharp interface consideration is exceedingly complaint with the physical conditions than models adopting finitely diffuse interface.
Consequently, the solutions of the sharp interface models are generally acclaimed to be quantitative.
Furthermore, as elucidated in the early sections, the phase-field approach renders only a phenomenological view with the state variable phase field holding no (real) thermodynamical significance.
Therefore, in the phase-field modelling, the quantitative aspect is ensured when the sharp interface solutions are recovered, thereby indicating that the diffuse interface does not influence the overall physics of the problem. 
To that end, a variant of the boundary-layer method, referred to as \textit{matched asymptotic analysis}~\cite{nayfeh2011introduction}, is employed to demonstrate that the model adheres to the sharp interface solutions and the physical laws.

In principle, the asymptotic analysis investigates the mathematical problem on two different length scales through the matching conditions~\cite{caginalp1986analysis}.
Corresponding, in the phase-field modelling, this analysis examines the evolution of the variables, which emerges from the model formulation, at the inner and outer scale.
While inner scale pertains to the interface, the dimensions of the microstructure (or bulk) is encompassed by the outer scale.

The asymptotic analysis accompanying the early phase-field models were confined to the \textit{sharp interface limit} wherein
the model was analysed for the condition that the interface width tends to zero ($\epsilon\to0$)~\cite{wheeler1992phase}.
This analysis unravels the compliance of the phase-field approach to the physical laws, particularly the Gibbs-Thomson relation~\cite{gibbs1928collected,gibbs1957collected}, in addition to other sharp interface solutions.
Although the asymptotic analysis was subsequently extended to included \textit{finite interface limit}~\cite{karma1996phase,karma1998quantitative}, since this section primarily aims to prove the adherence of the present model to the Gibbs-Thomson relation, the following analysis is restricted to the sharp interface limit.

For the grand-potential phase-field model, an extensive asymptotic analysis accompanying both sharp and finite interface limit has already been reported~\cite{choudhury2012grand}.
However, fundamentally oriented towards phase transformation, specifically solidification, this asymptotic analysis was reasonably confined to the one-dimensional setup.
Since the influence of the curvature is absolutely trivial in one-dimension, in this section, the asymptotic analysis is extended to attest the adherence to the Gibbs-Thomson relation.

For the asymptotic analysis of the present model, a simplified functional with the double-well function $f_\text{dw}(\phi)$ and the interface width governing parameter $\text{W}_{\phi}$ is considered.
The corresponding phase-field evolution equation is written as
\begin{align}\label{eq:evophi_simp}
 \tau \frac{\partial \phi}{\partial t}=-\frac{\delta \Omega(\mu,\phi,\nabla\phi)}{\delta \phi}=\text{W}_{\phi}\p{2}\nabla \p{2}\phi-\frac{df_\text{dw}(\phi)}{d\phi}-\frac{\partial \Psi(\mu,\phi)}{\partial \phi}.
\end{align}
The notations like $\text{W}_{\phi}$ are adopted to facilitate compliance with the conventional expressions associated with the asymptotic analysis.

The asymptotic analysis begins by distinguishing the system as \textit{inner} and \textit{outer} regions.
In principle, the inner region corresponds to the interface while the outer to the bulk phases.
Since the length scales associated with interface width and microstructure are significantly differently, the scale pertaining to the inner and outer regions are appropriately defined. 
While in the inner region, the length scale is characterised by $x^{\text{in}}\ll\text{W}_{\phi}$, the length scale in the outer region considered to be $x\p{\text{o}}\gg d_{o}$.
\nomenclature{$x^{\text{in}}$}{Inner region of interface length scale}%
\nomenclature{$x\p{\text{o}}$}{Outer region of interface length scale}%
The parameter $d_{o}$, referred to as capillary length, is expressed as $d_{o}=\frac{D}{v_n}$, where $D$ and $v_n$ are diffusivity and normal interface velocity, respectively.
\nomenclature{$d_{o}$}{Capillarity length}%
\nomenclature{$v_n$}{Normal velocity}%

The parameters involved in the evolution equation, Eqn.~\ref{eq:evophi_simp}, are non-dimensionalised appropriately.
The non-dimensional terms replacing the parameter in Eqn.~\ref{eq:evophi_simp} can be summarized as
\begin{align}\label{eq:nonDimen1}
 &\text{W}_{\phi}  \to  \epsilon=\frac{\text{W}_{\phi}}{d_{o}}\\ \nonumber
 &t  \to  \bar{t}=t\Big(\frac{d_{o}\p{2}}{D}\Big)\p{-1}\\ \nonumber
 &\tau  \to  \bar{D}=\tau \Big(\frac{\text{W}_{\phi}\p{2}}{D}\Big)\p{-1}\\ \nonumber
\end{align}
Adopting the non-dimensional parameters the evolution equation is written as
\begin{align}\label{eq:nonD_ev}
 \bar{D}\epsilon\p{2}\frac{\partial \phi}{\partial \bar{t}}=\epsilon\p{2}\bar{\nabla}\p{2}\phi-\frac{df_\text{dw}(\phi)}{d\phi}-\epsilon\frac{\partial \Psi(\mu,\phi)}{\partial \phi},
\end{align}
where $\bar{\nabla}$ is the gradient in the dimensionless scale.
\nomenclature{$\bar{\nabla}$}{Gradient in the dimensionless scale}%

The asymptotic expansion of the fundamental variables $\phi$ and $\mu$ in the outer and inner region of the system is expressed as
\begin{align}\label{eq:outer_sol}
 \phi\p\text{o}=\phi\p\text{o}_0+\epsilon \phi\p\text{o}_1+\epsilon\p{2} \phi\p\text{o}_2+\dots \\ \nonumber
 \mu\p\text{o}=\mu\p\text{o}_0+\epsilon \mu\p\text{o}_1+\epsilon\p{2} \mu\p\text{o}_2+\dots
\end{align}
and 
\begin{align}\label{eq:inner_sol}
 \phi\p\text{in}=\phi\p\text{in}_0+\epsilon\phi\p\text{in}_1+\epsilon\p{2}\phi\p\text{in}_2+\dots \\ \nonumber
 \mu\p\text{in}=\mu\p\text{in}_0+\epsilon\mu\p\text{in}_1+\epsilon\p{2}\mu\p\text{in}_2+\dots,
\end{align}
respectively.
Although inner and outer solutions are written in the scaling powers of $\epsilon$, these solutions are distinguished by the length scales associated with the respective regions.
Therefore, before establishing the matching conditions, the length scales are correspondingly non-dimensionalised.
Since the inner region is related to the interface while the outer region is identified with the microstructure, in the curvilinear co-ordinate system (refer Appendix.~\ref{sec:app1}), the normal $u$ is made dimensionless through $\xi=u/\text{W}_{\phi}$ and $\eta=u/d_{o}$ in the inner and outer region, respectively.
The matching conditions are ascertained by extending the interface to the bulk and retracting the outer to the inner region.
Considering a system with phases $\alpha$ and $\beta$ which occupy the left and right side of the domain, the matching conditions for the phase field is written as
\begin{align}\label{eq:matching}
 \lim_{\eta \to 0\p{+}}\phi\p\text{o}_0(\eta)=\lim_{\xi \to -\infty}\phi\p\text{in}_0(\xi)=\phi_{\alpha}\\
 \lim_{\eta \to 0\p{-}}\phi\p\text{o}_0(\eta)=\lim_{\xi \to +\infty}\phi\p\text{in}_0(\xi)=\phi_{\beta}
\end{align}

Having established the matching conditions, the phase-field evolution equation, in Eqn.~\ref{eq:evophi_simp}, in the outer region is asymptotically expanded as
\begin{align}\label{eq:outer_phiev}
 \bar{D}\epsilon\p{2}\frac{\partial \phi\p\text{o}_0}{\partial \bar{t}}= & \epsilon\p{2}\bar{\nabla}\p{2}\phi\p\text{o}_0- \frac{df_\text{dw}(\phi\p\text{o}_0)}{d \phi}-\epsilon\frac{\partial \Psi(\mu\p\text{o}_0,\phi\p\text{o}_0)}{\partial \phi}
  -\epsilon \frac{d\p{2}f_\text{dw}(\phi\p\text{o}_0)}{d \phi\p{2}}\phi\p\text{o}_1\\ \nonumber
& -\epsilon\p{2}\frac{\partial \p{2} \Psi(\mu\p\text{o}_0,\phi\p\text{o}_0)}{\partial \phi\p{2}}\phi\p\text{o}_1
-\epsilon\p{2}\frac{\partial \p{2} \Psi(\mu\p\text{o}_0,\phi\p\text{o}_0)}{\partial \mu \partial \phi}\mu\p\text{o}_1
-\epsilon\p{2}\frac{d\p{2}f_\text{dw}(\phi\p\text{o}_0)}{d \phi\p{2}}\phi\p\text{o}_2-\dots
\end{align}
To resemble the expression in Eqn.~\ref{eq:outer_sol}, the above Eqn.~\ref{eq:outer_phiev} can be re-arranged to
\begin{align}\label{eq:outer_phiev_re}
\epsilon\p{2}\Bigg[\bar{D}\frac{\partial \phi\p\text{o}_0}{\partial \bar{t}}-\bar{\nabla}\p{2}\phi\p\text{o}_0&+\frac{\partial \p{2} \Psi(\mu\p\text{o}_0,\phi\p\text{o}_0)}{\partial \phi\p{2}}\phi\p\text{o}_1+\frac{\partial \p{2} \Psi(\mu\p\text{o}_0,\phi\p\text{o}_0)}{\partial \mu \partial \phi}\mu\p\text{o}_1+\frac{d\p{2}f_\text{dw}(\phi\p\text{o}_0)}{d \phi\p{2}}\phi\p\text{o}_2+\dots\Bigg] \\ \nonumber
&=\epsilon\p{0}\Bigg[-\frac{df_\text{dw}(\phi\p\text{o}_0)}{d \phi}\Bigg]
-\epsilon\Bigg[\frac{\partial \Psi(\mu\p\text{o}_0,\phi\p\text{o}_0)}{\partial \phi}+\frac{d\p{2}f_\text{dw}(\phi\p\text{o}_0)}{d \phi\p{2}}\phi\p\text{o}_1\Bigg].
\end{align}
In Eqn.~\ref{eq:outer_phiev_re}, it evident that the asymptotic expansion is confined to the second order. 
The equations pertaining to each order of the asymptotic expansion in Eqn.~\ref{eq:outer_phiev_re} separately read
\begin{align}\label{eq:outer_order_sep}
 &\mathcal{O}(\epsilon\p{0}):\frac{df_\text{dw}(\phi\p\text{o}_0)}{d \phi}=0 \\ \nonumber
 &\mathcal{O}(\epsilon):\frac{\partial \Psi(\mu\p\text{o}_0,\phi\p\text{o}_0)}{\partial \phi}+\frac{d\p{2}f_\text{dw}(\phi\p\text{o}_0)}{d \phi\p{2}}\phi\p\text{o}_1=0 \\ \nonumber
 &\mathcal{O}(\epsilon\p{2}):\bar{D}\frac{\partial \phi\p\text{o}_0}{\partial \bar{t}}-\bar{\nabla}\p{2}\phi\p\text{o}_0+\frac{\partial \p{2} \Psi(\mu\p\text{o}_0,\phi\p\text{o}_0)}{\partial \phi\p{2}}\phi\p\text{o}_1+\frac{\partial \p{2} \Psi(\mu\p\text{o}_0,\phi\p\text{o}_0)}{\partial \mu \partial \phi}\mu\p\text{o}_1+\frac{d\p{2}f_\text{dw}(\phi\p\text{o}_0)}{d \phi\p{2}}\phi\p\text{o}_2=0.
\end{align}
The equation of the zeroth order can be interpreted as the corollary of the criterion that the phase field acquires a constant value depending on the phase.
Accordingly, this lowest order solution yields 
\begin{align}\label{eq:outer_order_first}
 \phi\p\text{o}_0=
 \begin{cases}
  \phi_{\alpha}&  \eta \to -\infty\\
  \phi_{\beta}&  \eta \to +\infty.
 \end{cases} 
\end{align}
Furthermore, in the outer region, far from the interface, the phase field remain homogeneous in a given a phase.
Therefore, the driving force in the respective phase, away from the interface, exhibit no significant change.
Consequently, $\frac{\partial \Psi(\mu\p\text{o}_0,\phi\p\text{o}_0)}{\partial \phi}$ becomes $0$ in the first order term in Eqn.~\ref{eq:outer_order_sep}.
Extending similar considerations to the second order equation, it can be deduced that $\phi\p\text{o}_1=0$ and $\phi\p\text{o}_2=0$.

The role of the co-ordinate system in the outer region is marginal, owing to the condition that the phase field assumes homogeneous constant value in the bulk phase.
However, in the inner region, which is associated with the interface, the evolution equation should be appropriately expanded to encompass the curvilinear co-ordinate system.
To that end, the second order asymptotic expansion of the Eqn.~\ref{eq:evophi_simp} is written as
\begin{align}\label{eq:inner_withD}
 \tau\Bigg(\frac{\partial \phi}{\partial t} - v_{n}\frac{\partial \phi}{\partial u} + \frac{\partial s}{\partial t} \frac{\partial \phi}{\partial s} \Bigg) =& \text{W}_{\phi}\Bigg[ \frac{\partial\p{2}\phi}{\partial u\p{2}} + \Bigg(\frac{1}{1+uk}\Bigg) k \frac{\partial \phi}{\partial u}
 + \Bigg(\frac{1}{1+uk}\Bigg)\p{2}\frac{\partial\p{2} \phi}{\partial s\p{2}} \\ \nonumber
& -  \Bigg(\frac{1}{1+uk}\Bigg)\p{3}u \frac{\partial k}{\partial s} \frac{\partial \phi}{\partial s} \Bigg] 
  - \frac{df_\text{dw}(\phi)}{d \phi} - \frac{\partial \Psi(\mu,\phi)}{\partial \phi}.
\end{align}
Although the above Eqn.~\ref{eq:inner_withD} encompasses the curvilinear system of $u$ and $s$, it is vital to recognize that the parameters retain their corresponding dimensions.
Therefore, the parameters are appropriately non-dimensionalised.
The normal velocity, $v_{n}$, is replaced by dimensionless $\bar{v}_{n}$ which is expressed as $\bar{v}_{n}=v_{n}\Big(\frac{D}{d_{o}}\Big)\p{-1}$.
\nomenclature{$\bar{v}_{n}$}{Dimensionless normal velocity}%
The length scales associated with Eqn.~\ref{eq:inner_withD} are separately non-dimensionalised considering their description in the system.
The length along normal direction of the interface, $u$, in the inner region, should be of the order of the interface width. 
And accordingly, it is non-dimensionalised through $\xi=\frac{u}{\text{W}_{\phi}}$.
In contrast, the scale of the curvature and entire interface (interface length as opposed to its width) should be  greater than $\frac{\text{W}_{\phi}}{d_{o}}$ and generally, in the order of capillary length.
Consequently, the curvature and the tangential scale is made dimensionless through $\bar{k}=\frac{k}{d_{o}}$ and $\Xi=\frac{s}{d_{o}}$, respectively.

The non-dimensionalised evolution equation for the inner region correspondingly reads
\begin{align}\label{eq:inner_withoutD}
 \bar{D}\epsilon\p{2}\frac{\partial \phi}{\partial \bar{t}}-\bar{D}\epsilon v_{n}\frac{\partial \phi}{\partial \xi}+\bar{D}\epsilon\p{2}\frac{\partial \Xi}{\partial \bar{t}}\frac{\partial \phi}{\partial \Xi}&=\Bigg[ \frac{\partial\p{2}\phi}{\partial \xi\p{2}}+\Bigg( \frac{\epsilon\bar{k}}{1+\epsilon\xi\bar{k}} \Bigg) \frac{\partial \phi}{\partial \xi} + \Bigg( \frac{\epsilon}{1+\epsilon\xi\bar{k}} \Bigg)\p{2}\frac{\partial\p{2} \phi}{\partial \Xi \p{2}} \Bigg]\\ \nonumber
 &- \frac{df_\text{dw}(\phi)}{d \phi} -\epsilon \frac{\partial \Psi(\mu,\phi)}{\partial \phi}.
\end{align}
The above Eqn.~\ref{eq:inner_withoutD} can further be simplified by considering that 
\begin{align}\label{eq:inner_simply}
  \frac{\epsilon\bar{k}}{1+\epsilon\xi\bar{k}} \approx \epsilon\bar{k}-\xi(\epsilon\bar{k})\p{2} &&\text{and}&&
  \Bigg( \frac{\epsilon}{1+\epsilon\xi\bar{k}} \Bigg)\p{2}\approx \epsilon\p{2}.
\end{align}
Substituting Eqn.~\ref{eq:inner_simply} in Eqn.~\ref{eq:inner_withoutD} yields
\begin{align}\label{eq:inner_simWitoutD}
 \frac{df_\text{dw}(\phi)}{d \phi}+\epsilon \frac{\partial \Psi(\mu,\phi)}{\partial \mu}&=\frac{\partial\p{2}\phi}{\partial \xi\p{2}}+ \epsilon\Bigg(\bar{k} \frac{\partial \phi}{\partial \xi} + \bar{D}v_{n}\frac{\partial \phi}{\partial \xi}\Bigg)\\ \nonumber
 &+\epsilon\p{2}\Bigg( \frac{\partial\p{2} \phi}{\partial \Xi \p{2}}-\xi\bar{k}\p{2}\frac{\partial \phi}{\partial \xi}-\bar{D}\frac{\partial \phi}{\partial \bar{t}}- \bar{D}\frac{\partial \Xi}{\partial \bar{t}}\frac{\partial \phi}{\partial \Xi}\Bigg).
\end{align}
Above Eqn.~\ref{eq:inner_simWitoutD} is expressed in a such a way that the asymptotic expansion of the terms on the right hand side are straightforward.
Therefore, the remnant terms on the left hand side of the Eqn.~\ref{eq:inner_simWitoutD} are separately expanded.

For the asymptotic expansion, the double well function and grand potential density can be expressed as
\begin{align}\label{eq:inner_leftexpan}
 &f_\text{dw}(\phi)=f_\text{dw}(\phi\p\text{in}_0+\delta\phi\p\text{in})\\
 &\Psi(\mu,\phi)=\Psi(\mu\p\text{in}_0+\delta\mu\p\text{in},\phi\p\text{in}_0+\delta\phi\p\text{in})
\end{align}
where $\delta\mu\p\text{in}=\epsilon\mu\p\text{in}_{1}+\epsilon\p{2}\mu\p\text{in}_{2}+\cdots$ and $\delta\phi\p\text{in}=\epsilon\phi\p\text{in}_{1}+\epsilon\p{2}\phi\p\text{in}_{2}+\cdots$. 
The terms on the left hand side of the Eqn.~\ref{eq:inner_simWitoutD} can, therefore, be asymptotically expanded as
\begin{align}\label{eq:inner_sepExpan}
 &\frac{\partial f_\text{dw}(\phi\p\text{in}_0+\delta\phi\p\text{in})}{\partial \phi}+ \epsilon \frac{\Psi(\mu\p\text{in}_0+\delta\mu\p\text{in},\phi\p\text{in}_0+\delta\phi\p\text{in})}{\partial \phi} =\frac{df_\text{dw}(\phi\p\text{in}_0)}{\partial \phi} \\ \nonumber
 &+ \epsilon \Bigg( \frac{\partial \Psi(\mu\p\text{in}_0,\phi\p\text{in}_0)}{\partial \phi} + \frac{d\p{2}f_\text{dw}(\phi\p\text{in}_0)}{\partial \phi\p{2}}\phi\p\text{in}_1\Bigg)
 +\epsilon\p{2}\Bigg( \frac{\partial\p{2} \Psi(\mu\p\text{in}_0,\phi\p\text{in}_0)}{\partial \phi\p{2}} \phi\p\text{in}_1
 + \frac{\partial\p{2} \Psi(\mu\p\text{in}_0,\phi\p\text{in}_0)}{\partial \phi \partial \mu} \mu\p\text{in}_1 \Bigg).
\end{align}
From Eqn.~\ref{eq:inner_sepExpan}, the terms pertaining to the different orders are separately written as
\begin{align}\label{eq:inner_sepExpan1}
\mathcal{I}(\epsilon\p{0}):\frac{\partial\p2 \phi\p\text{in}_0}{\partial\xi\p2}-\frac{df_\text{dw}(\phi\p\text{in}_0)}{\partial \phi}=0
\end{align}
\begin{align}\label{eq:inner_sepExpan2}
\mathcal{I}(\epsilon):-\bar{D}\bar{v}_{n_0}\frac{\partial \phi\p\text{in}_0}{\partial \xi}=\bar{k}\frac{\partial \phi\p\text{in}_0}{\partial \xi}+\frac{\partial\p2 \phi\p\text{in}_1}{\partial \xi\p2}-\frac{d\p2 f_\text{dw}(\phi\p\text{in}_0)}{\partial \phi\p2}\phi\p\text{in}_1- \frac{\partial \Psi(\mu\p\text{in}_0,\phi\p\text{in}_0)}{\partial \phi}
\end{align}
\begin{align}\label{eq:inner_sepExpan3}
\mathcal{I}(\epsilon\p{2}):\frac{\partial\p2 \phi\p\text{in}_2}{\partial \xi\p2} & -\frac{d\p2 f_\text{dw}(\phi\p\text{in}_0)}{\partial \phi\p2}\phi\p\text{in}_2 =
  \bar{D}\frac{\partial \phi\p\text{in}_0}{\partial \bar{t}}
 - \frac{\partial\p2 \phi\p\text{in}_0}{\partial \Xi\p2}
 -(\bar{D}\bar{v}_{n_0}+\bar{k})\frac{\partial \phi\p\text{in}_1}{\partial \xi}\\ \nonumber
 & -(\bar{D}\bar{v}_{n_1}-\xi\bar{k}\p2)\frac{\partial \phi\p\text{in}_0}{\partial \xi}
   +\frac{\partial\p2 \Psi(\mu\p\text{in}_0,\phi\p\text{in}_0)}{\partial \phi\p2}\phi\p\text{in}_1
   +\frac{\partial\p2 \Psi(\mu\p\text{in}_0,\phi\p\text{in}_0)}{\partial \phi \partial \mu}\mu\p\text{in}_1.
\end{align}

From the order based expansion of the inner solutions, it is evident that the lowest order equation handling curvature $\bar{k}$ is $\mathcal{I}(\epsilon)$ in Eqn.~\ref{eq:inner_sepExpan2}.
Therefore, this inner solution is further analysed to understand the role of curvature $\bar{k}$ in influencing the fundamental variable.
\nomenclature{$\bar{k}$}{Dimensionless curvature}%
Eqn.~\ref{eq:inner_sepExpan2} can be re-arranged and expressed as
\begin{align}\label{eq:inner_order1}
 \frac{\partial\p2 \phi\p\text{in}_1}{\partial \xi\p2}-\frac{d\p2 f_\text{dw}(\phi\p\text{in}_0)}{\partial \phi\p2}\phi\p\text{in}_1
 =-(\bar{D}\bar{v}_{n_0}+\bar{k})\frac{\partial \phi\p\text{in}_0}{\partial \xi}+\frac{\partial \Psi(\mu\p\text{in}_0,\phi\p\text{in}_0)}{\partial \phi}
\end{align}
Above Eqn.~\ref{eq:inner_order1} is multiplied with $\frac{\partial \phi\p\text{in}_0}{\partial \xi}$ and integrated as
\begin{align}\label{eq:inner_order1_re}
 \int_{-\infty}\p{+\infty}\frac{\partial \phi\p\text{in}_0}{\partial \xi}\Bigg[ \frac{\partial\p2 \phi\p\text{in}_1}{\partial \xi\p2}- \frac{d\p2 f_\text{dw}(\phi\p\text{in}_0)}{\partial \phi\p2}\phi\p\text{in}_1\Bigg]\diff\xi  = &-(\bar{D}\bar{v}_{n_0}+\bar{k}) \int_{-\infty}\p{+\infty}\Bigg( \frac{\partial \phi\p\text{in}_0}{\partial \xi}\Bigg)\p2 \diff\xi \\ \nonumber
 & + \int_{-\infty}\p{+\infty}\frac{\partial \Psi(\mu\p\text{in}_0,\phi\p\text{in}_0)}{\partial \phi}\frac{\partial \phi\p\text{in}_0}{\partial \xi} \diff\xi.
\end{align}
Integrating the left hand side of the Eqn.~\ref{eq:inner_order1_re} by parts, and considering $\frac{\partial\p2 \phi\p\text{in}_0}{\partial\xi\p2}=\frac{df_\text{dw}(\phi\p\text{in}_0)}{\partial \phi}$ from Eqn.~\ref{eq:inner_sepExpan1}, yields
\begin{align}\label{eq:inner_order1_lhs}
 \int_{-\infty}\p{+\infty}\frac{\partial \phi\p\text{in}_0}{\partial \xi}\mathcal{L}(\phi\p\text{in}_1)\diff\xi=0,
\end{align}
where $\mathcal{L}=\frac{\partial\p2 }{\partial \xi\p2}-\frac{d\p2 f_\text{dw}(\phi\p\text{in}_0)}{\partial \phi\p2}$.
The term governing the thermodynamic driving force, which is second term in the right hand side of Eqn.~\ref{eq:inner_order1_re}, can be written as
\begin{align}\label{eq:inner_order1_rhs}
 \int_{-\infty}\p{+\infty}\frac{\partial \Psi(\mu\p\text{in}_0,\phi\p\text{in}_0)}{\partial \phi}\frac{\partial \phi\p\text{in}_0}{\partial \xi} \diff\xi = \int_{-\infty}\p{+\infty}\frac{\partial \Psi(\mu\p\text{in}_0,\phi\p\text{in}_0)}{\partial \xi}\diff\xi - \int_{-\infty}\p{+\infty}
 \frac{\partial \Psi(\mu\p\text{in}_0,\phi\p\text{in}_0)}{\partial \mu}\frac{\partial \mu\p\text{in}_0}{\partial \xi}\diff\xi.
\end{align}
In order to solve the term,
$\int_{-\infty}\p{+\infty} \frac{\partial \Psi(\mu\p\text{in}_0,\phi\p\text{in}_0)}{\partial \mu}\frac{\partial \mu\p\text{in}_0}{\partial \xi}\diff\xi$,
in the above Eqn.\ref{eq:inner_order1_rhs}, it is vital to asymptotically expand the other fundamental variable, chemical potential ($\mu$).
However, since the present model treats chemical potential as the dynamic variable instead of the concentration, which is conventionally adopted, a workaround is permitted.
 
Gibbs-Thomson relation which is predominantly governed by the curvature operates both in the presence and in the absence of the chemical driving force that dictates the phase transformation.
If a thermodynamical condition is assumed wherein the chemical driving force is absent, in other words the phases are equilibrium across the \textit{flat} interface, the chemical potentials in the bulk much farther from the interface is equal.
This condition can be expressed as $\mu\p{\alpha}=\mu\p{\beta}=\mu_{\text{eq}}$, where $\mu_{\text{eq}}$ is the equilibrium chemical potential across the flat interface separating phases $\alpha$ and $\beta$.

The second term in the right hand side of Eqn.~\ref{eq:inner_order1_rhs} ultimately yields the difference in the chemical potential assumed by the bulk phases far from the interface, $\Delta\mu=\mu|_{\infty}-\mu|_{-\infty}$.
Assuming the phases to be chemical equilibrium, the term becomes 
\begin{align}\label{eq:inner_order1_lhs}
\int_{-\infty}\p{+\infty}
\frac{\partial \Psi(\mu\p\text{in}_0,\phi\p\text{in}_0)}{\partial \mu}\frac{\partial \mu\p\text{in}_0}{\partial \xi}\diff\xi=0.
\end{align}
In order to solve the remnant term in Eqn.~\ref{eq:inner_order1_rhs}, which relates to the difference $\Delta\Psi = \Psi_{\alpha}(\mu,\phi)-\Psi_{\beta}(\mu,\phi)$, the fundamental description of the grand chemical potential function is revisited.
The grand chemical potential of a given phase $\Theta$ can be linearly expanded as
\begin{align}\label{eq:inner_order1_lhs1}
 \Psi^{\Theta}(\mu,\phi)=\Psi^{\Theta}(\mu_\text{eq},\phi)+\frac{\partial \Psi^{\Theta}(\mu,\phi)}{\partial \mu}\Bigg|_{\mu_{\text{eq}}}(\mu-\mu_{\text{eq}}),
\end{align}
where $\Theta$ could be $\alpha$ or $\beta$.
Furthermore, $\mu$ is the chemical potential that is influenced by the curvature of the interface.
In the absence of the curvature, at the chemical equilibrium, since $\Psi^{\alpha}(\mu_\text{eq})=\Psi^{\beta}(\mu_\text{eq})$, the difference in the grand chemical potential is expressed as
\begin{align}\label{eq:dr_mu_psi}
 \Delta\Psi=\Bigg[\frac{\partial \Psi^{\alpha}(\mu,\phi)}{\partial \mu}\Bigg|_{\mu_{\text{eq}}}- \frac{\partial \Psi^{\beta}(\mu,\phi)}{\partial \mu}\Bigg|_{\mu_{\text{eq}}}\Bigg](\mu-\mu_{\text{eq}}).
 \end{align}
 Since $\frac{\partial \Psi^{\Theta}(\mu,\phi)}{\partial \mu}=-c^{\Theta}$, the remnant term in Eqn.~\ref{eq:inner_order1_rhs} becomes
  \begin{align}\label{eq:dr_mu_overal}
 \int_{-\infty}\p{+\infty}\frac{\partial \Psi(\mu\p\text{in}_0,\phi\p\text{in}_0)}{\partial \xi}\diff\xi=[c\p{\beta}(\mu_{\text{eq}},\phi)-c\p{\alpha}(\mu_{\text{eq}},\phi)](\mu-\mu_{\text{eq}}).
 \end{align}
Substituting Eqns.~\ref{eq:inner_order1_lhs} and ~\ref{eq:dr_mu_overal} in Eqn.~\ref{eq:inner_order1_re} yields the relation
\begin{align}\label{eq:Gib_pen}
 [c\p{\alpha}(\mu_{\text{eq}},\phi)-c\p{\beta}(\mu_{\text{eq}},\phi)](\mu-\mu_{\text{eq}})=(\bar{D}\bar{v}_{n_0}+\bar{k}) \int_{-\infty}\p{+\infty}\Bigg( \frac{\partial \phi\p\text{in}_0}{\partial \xi}\Bigg)\p2 \diff\xi.
\end{align}
Now, considering $\displaystyle\int_{-\infty}\p{+\infty}\left( \frac{\partial \phi\p\text{in}_0}{\partial \xi}\right)\p2 \diff \xi=\gamma_{\phi}$, the change in the chemical potential introduced by the curvature through Eqn.~\ref{eq:Gib_pen} can be written as
\begin{align}\label{eq:Gib_tho}
 (\mu-\mu_{\text{eq}})=\frac{(\bar{D}\bar{v}_{n_0}+\bar{k})\gamma_{\phi}}{c\p{\alpha}(\mu_{\text{eq}},\phi)-c\p{\beta}(\mu_{\text{eq}},\phi)}.
\end{align}
Above Eqn.~\ref{eq:Gib_tho} indicates that the Gibbs-Thomson relation is recovered in the grand potential phase-field model.

\section{CALPHAD based free-energy description}

Depending on the system and its corresponding evolution, the thermodynamic function which dictates the driving force of the resulting microstructural changes is appropriately chosen to formulate the phase-field model. 
In addition to the adoption of the suitable function, with the advancements in the modelling, the description of these functions have been progressively enhanced to render a quantitative simulation.
Considering phase transformations in binary alloys, the evolution is dictated by the chemical composition of the system and particularly, the nature of the components involved.
In other words, the driving force governing the kinetics and mechanism of the microstructural transformation depend significantly on the alloy system considered.
Therefore, attempts are made to thermodynamically describe the system based on the corresponding phase diagrams of the components involved.
Incorporating CALPHAD, CALculation of PHAse Diagrams~\cite{saunders1998calphad,spencer2008brief}, data in the simulation has proven to be an effective tool in quantitatively defining the system. 
Furthermore, several approaches have been employed to couple the quantitative data with the simulation setup.
Since the quantitative data essentially comprises of Gibbs energy, chemical potential and equilibrium conditions, one approach involves devising a file through an external CALPHAD software, which consists of these thermodynamic parameters, and augmenting it to the modelling scheme~\cite{ode2007prediction,wang2008coarsening}.
Moreover, while some modelling techniques adopt locally linearised phase diagrams by parallely running the CALPHAD software~\cite{grafe2000coupling,eiken2006multiphase}, others \lq directly\rq\thinspace couple the quantitative data by formulating the phase-field model to accommodate the entire CALPHAD database~\cite{chen2004quantitative,zhang2015incorporating}.

\begin{figure}
    \centering
      \begin{tabular}{@{}c@{}}
      \includegraphics[width=0.5\textwidth]{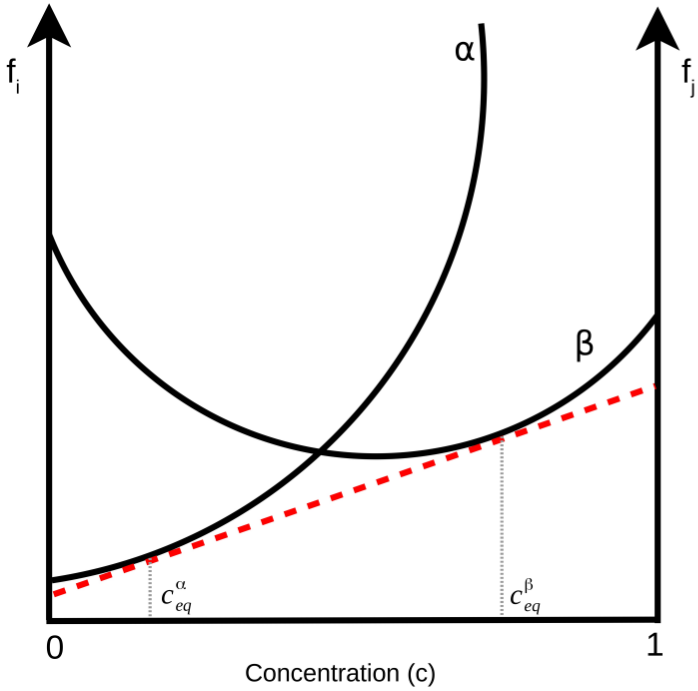}
    \end{tabular}
    \caption{ Schematic illustration of the free energy density vs concentration plot pertaining to a binary system of components $i$ and $j$.
    \label{fig:free_energy_plot}}
\end{figure}

Much different from the aforementioned methodologies, in the current study a simpler approach is extended to incorporate the CALPHAD data.
In this approach, the geometric scheme in the relation between the Gibbs free-energy density and the concentration is exploited to represent the thermodynamic data in the form of a parabolic equation~\cite{redlich1948algebraic,wu2004simulating}. 
The choice of this simpler, yet effective, technique can be justified by realizing that the microstructural evolution simulated in the present work occur in a chemical equilibrium.

Free energy density of a phase-$\alpha$, $f^{\alpha}(c_i,c_j)$, in a binary system with components $i$ and $j$ can be expressed as
\begin{align}\label{eq:f_energy}
 f^{\alpha}(c_i,c_j)=c_if_i+c_jf_j+RT[c_i\ln c_i+c_j\ln c_j],
\end{align}
where $c_i$ and $c_j$ are the mole fractions of components $i$ and $j$, respectively.
Temperature and universal gas constant are correspondingly represented  by $T$ and $R$ in Eqn.~\ref{eq:f_energy}.
The respective free energies of the pure $i$ and $j$ are $f_i$ and $f_j$.
Owing to the constraint $c_i+c_j=1$, the number of independent concentration becomes one and if $c_j$ is represented by $c$, Eqn.~\ref{eq:f_energy} becomes
\begin{align}\label{eq:f_energy2}
 f^{\alpha}(c)=(1-c)f_i+cf_j+RT[(1-c)\ln (1-c)+c\ln c].
\end{align}
Upon considering the influence of concentration $c$ on the free energy density, particularly the geometric trend in the dependence of the $f_i$ and $f_j$ on $c$, as schematically shown in Fig.~\ref{fig:free_energy_plot}, Eqn.~\ref{eq:f_energy2} can be approximated as
\begin{align}\label{eq:free_energy_app}
 f^{\alpha}(c)=A^{\alpha}c^{2}+B^{\alpha}c+D^{\alpha},
\end{align}
where $A^{\alpha}$, $B^{\alpha}$ and $D^{\alpha}$ are the phase-dependent coefficients.
Since the free energy density $f_{\alpha}(c)$ is significantly influenced the temperature, the phase-dependent coefficients vary with the temperature.
However, owing to the isothermal consideration of the present study, the dependency of the coefficients on the temperature is not explicitly mentioned.

\begin{figure}
    \centering
      \begin{tabular}{@{}c@{}}
      \includegraphics[width=0.7\textwidth]{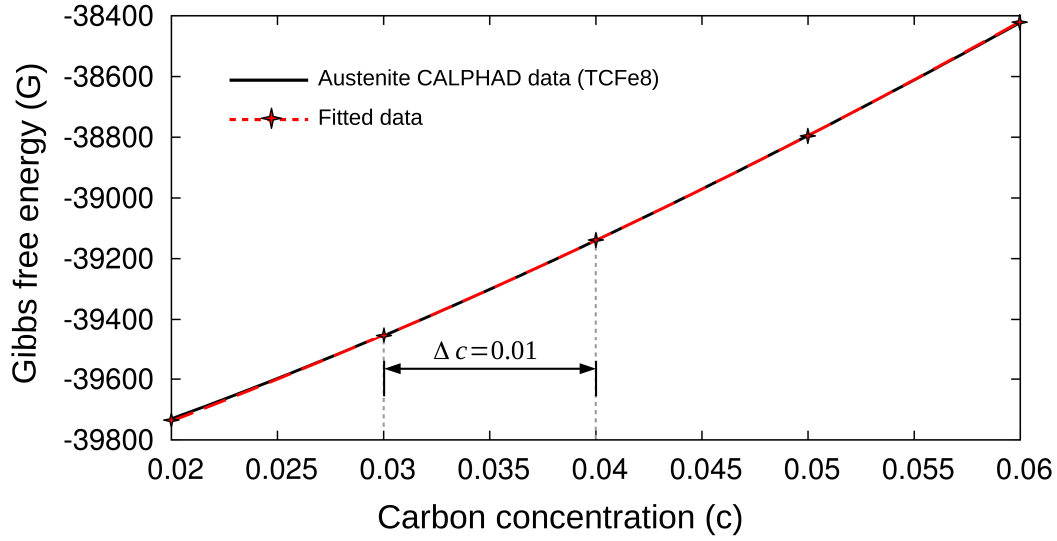}
    \end{tabular}
    \caption{ Gibbs free-energy density dependence of austenite on carbon concentration as rendered by the CALPHAD database (TCFe8) is compared to the fitted curve.
    \label{fig:CALPHAD}}
\end{figure}

CALPHAD data provides the Gibbs free-energy density of the phase for a given concentration and temperature.
However, the model is conventionally formulated based on the Helmholtz free-energy density.
While the Gibbs free-energy density is expressed in term of per mole, the Helmholtz free-energy density is ascertained for unit volume.
Therefore, assuming the molar volume of the phases to be equal ($V_m^{\alpha}=V_m^{\beta}=V_m$), the Gibbs and Helmholtz free-energy density can be related through $f^{\alpha}(c)=\frac{1}{V_m}G^{\alpha}(c)$.
\nomenclature{$G^{\alpha}(c)$}{Gibbs free energy of phase-$\alpha$}%
Correspondingly, the differentiation of the approximated free energy in Eqn.~\ref{eq:free_energy_app} yields 
\begin{align}\label{eq:diff_fre1}
 \frac{\partial f^{\alpha}}{\partial c}=2A^{\alpha}c+B^{\alpha}=\frac{1}{V_m}\frac{\partial G^{\alpha}}{\partial c}=\mu^{\alpha}
\end{align}
\begin{align}\label{eq:diff_fre2}
 \frac{\partial^2 f^{\alpha}}{\partial c^2}=2A^{\alpha}=\frac{1}{V_m}\frac{\partial^2 G^{\alpha}}{\partial c^2}.
\end{align}
In order to couple the CALPHAD data with the parabolic approximation in Eqn.~\ref{eq:free_energy_app}, a small concentration step, $\Delta c$, is defined.
The Gibbs free-energy density at a base concentration $c$ is extracted from the CALPHAD database through an external software.
If the Gibbs energy at $c$ is designated as $G^{\alpha}_0$ and for the neighbouring compositions as
\begin{align}\label{eq:designate_g}
 c+\Delta c \to G^{\alpha}_1 && c+2\Delta c \to G^{\alpha}_2 \\ \nonumber
 c-\Delta c \to G^{\alpha}_{\bar{1}} && c-2\Delta c \to G^{\alpha}_{\bar{2}} \nonumber
\end{align}
by employing finite-difference approximation, using five-point stencil scheme, the derivatives of the Gibbs free-energy density can be calculated through
 \begin{align}\label{eq:five_point1}
 \frac{\partial^2G^\alpha}{\partial c_i^2}&=\frac{-G^{\alpha}_2+16G^{\alpha}_1-30G_{\alpha:0}+16G^{\alpha}_{\bar{1}}-G^{\alpha}_{\bar{2}}}{12(\Delta c_i)^2}
 \end{align}
 and
 \begin{align}\label{eq:five_point2}
 \frac{\partial G^\alpha}{\partial c_i}&=\frac{-G^{\alpha}_2+8G^{\alpha}_1-8G^{\alpha}_{\bar{1}}+G^{\alpha}_{\bar{2}}}{12\Delta c_i}.
\end{align}
The coefficients $A^{\alpha}$, $B^{\alpha}$ and $D^{\alpha}$ can be ascertained by combining the Eqns.~\ref{eq:five_point1} and ~\ref{eq:five_point2} with Eqns.~\ref{eq:diff_fre1} and ~\ref{eq:diff_fre2}.
Upon determining the coefficients which incorporates the quantitative data to the model, its accuracy is affirmed by ensuring that the equilibrium conditions are replicated.
By imposing the constrain that chemical potential between the phases is equal at equilibrium, $\frac{\partial f_{\text{eq}}^{\alpha}}{\partial c}=\frac{\partial f_{\text{eq}}^{\beta}}{\partial c}$, using \textit{Newton-Raphson iteration} technique, it is ensured that the coefficients yield equilibrium compositions as rendered by the CALPHAD data.
Furthermore, from Eqn.~\ref{eq:diff_fre2}, it is evident that the susceptibility matrix $\mathcal{X^{\alpha}} h_{\alpha}(\vphi)$ is the interpolation of the reciprocal of $2A^{\alpha}$, in a multiphase system.

For all the simulations in the present work, the thermodynamic parameters are extracted from the CALPHAD-database \textit{TCFe8} pertaining to the binary Fe-C system.
The data is accessed through external commercial software, \textit{Pandat} (v8.1).
The Gibbs energy of the austenite phase at 973K as described by the CALPHAD database is plotted in Fig.~\ref{fig:CALPHAD}.
Additionally, the curve resulting the aforementioned parabolic approximation technique is included. 
It is evident from Fig.~\ref{fig:CALPHAD}, the \textit{fitted} data exhibits a noticeable compliance with the quantitative data.

Although this parabolic approximation approach can be extended to all phases irrespective of the alloy system, additional care must be taken for two definitive cases.
One, when equilibrium composition of the phase is very low and close to zero.
Ferrite is a prime example for this case in Fe-C system.
Under this condition, a casual approximation would yield unphysical negative equilibrium concentration, thus impairing the entire thermodynamical consideration.
This numerical inaccuracy can be averted by choosing the equilibrium concentration as the base concentration ($c=c_{\text{eq}}^{\alpha}$) and imposing the condition, $c_{\text{eq}}^{\alpha}-2\Delta c > 0$.
The other case which demands attention while adopting the fitting approach is the introduction of the stoichiometric phases.
Since the stoichiometric phases defy the geometric trend in Fig.~\ref{fig:free_energy_plot}, which forms the basis of the parabolic approach, an polynomial expression cannot be attained directly.
However, based on the information from the CALPHAD database which pertains to a point (or a small line), a sharp curve is numerically constructed which satisfies the equilibrium condition.
This numerically built sharp curve ensures that the $G^{\theta} \to \infty$ when the concentration is not the $c_{\text{eq}}^{\theta}$, where $\theta$ is the stoichiometric phase.
In Fe-C system, this approach is adopted for the cementite phase.

\chapter{Chemo-elastic phase-field model}\label{chap:chm_elast}

Phase transformations along with the morphological evolution ensuing the chemical equilibrium, particularly in metallic systems, are significantly influenced by the crystallographic features of the phases involved~\cite{watson1973crystallography,tian1987mechanisms}.
Generally, these crystallographic features include crystal structures and the lattice parameters.
The crystal structure, particularly, is one of the primary factors which distinguishes the phases in a solid-state system.
Therefore, the spatial arrangement of atoms, besides the chemical composition, independently introduces an additional component to the energy densities governing the phase evolution.
The energy expended to establish the change in the crystallographic arrangement of the atoms in the growing phase and the resulting misfits in the lattice parameters can be categorized as the elastic free-energy density.
This description of the elastic free-energy density entails the assumption that no irreversible change is introduced in the crystallographic features due to the additional component.
In other words, it is considered that the difference in the crystal structure between the parent and child phase does not induce any plastic deformation during the evolution.
Furthermore, since the crystallographic arrangement of atoms inherently exhibits anisotropy, an appropriate description of the elastic free-energy density would automatically influence the morphology of the growing phase.

\section{A multiphase-field model}

In this section, a multiphase-field model which encompasses chemo-elastic contribution is presented for the binary system.
The free energy functional which includes the elastic contribution is expressed as
\begin{align}\label{eq:functional}
  \mathcal{F}(\vphi, \n \vphi, c, \veps) = \int_V f_\text{intf}(\vphi, \n \vphi) + f_\text{el}(\vphi, \veps) + f_\text{ch}(\vphi, c) \diff{V},
\end{align}
where $f_\text{el}(\vphi, \veps)$ is the elastic free-energy density.
\nomenclature{$f_\text{el}(\vphi, \veps)$}{Elastic free-energy density}%
\nomenclature{$\veps$}{Total elastic strain}%
In Eqn.~\ref{eq:functional}, it is important to note that the chemical contribution is coupled as the free energy density $f_\text{ch}(\vphi, c)$, while the grand potential density was directly introduced in the functional $\Omega(\mu,\vphi,\n \vphi)$ in Eqn.~\ref{eq:GP_functional}. 
Therefore, it is agreeable that this formalism is not explicitly built on grand potential but adopts the undergirding framework in handling the chemical variables, concentration and chemical potential which would be evident in the following derivation.
The contribution of the interface, $f_\text{intf}$, conforms to the formulation elucidated in Chapter~\ref{sec:grand_potentialM}.

The elastic and chemical free-energy density governing the phase transformation can be expressed as the summation of the pairwise interaction between the phases.
As described during the asymptotic analysis in Sec.~\ref{sec:asym}, the pairwise interaction is considered as the driving force for the evolution.
Correspondingly, the phase-field evolution based on the driving force can be expressed as
\begin{align}\label{eq:phase_field_evolution}
  \frac{\partial \phia}{\partial t} = - \frac{1}{\varepsilon} \frac{1}{\tilde{N}}\sum^{\tilde{N}}_{\beta \neq \alpha} M_{\alpha \beta} \l[\frac{\delta f_\text{intf}}{\delta \phia} - \frac{\delta f_\text{intf}}{\delta \phi_{\beta}} + \frac{8 \sqrt{\phia \phi_{\beta}}}{\pi} \l( \Delta^{\alpha \beta}_\text{ch} + \Delta^{\alpha \beta}_\text{el} \r) \r],
\end{align}
where $M_{\alpha \beta}$ is the mobility of the interface pertaining to the region separating the phase-$\alpha$ and -$\beta$.
\nomenclature{$M_{\alpha \beta}$}{Mobility of the interface separating the phase-$\alpha$ and -$\beta$}%
Furthermore, in the above expression $\Delta^{\alpha \beta}_\text{ch}$ and $\Delta^{\alpha \beta}_\text{el}$ correspond to the chemical and elastic driving force, where $\Delta^{\alpha \beta}_{i}=\left(\frac{\delta}{\delta \phia}+\frac{\delta}{\delta \phi_{\beta}} \right)f_{i}$ with  $i\in\{\text{ch},\text{el}\}$.
\nomenclature{$\Delta^{\alpha \beta}_\text{ch}$}{Chemical driving-force between phase-$\alpha$ and -$\beta$}%
\nomenclature{$\Delta^{\alpha \beta}_\text{el}$}{Elastic driving-force between phase-$\alpha$ and -$\beta$}%
In Eqn.~\ref{eq:phase_field_evolution}, $\tilde{N}$ is the active number of phases, wherein $\tilde{N}\leq N$.

The overall chemical free-energy density is expressed as
\begin{align}\label{eq:bar_W_chem}
 f_\text{ch}(\vphi, c)=\sum_\alpha f^{\alpha}_{\text{ch}} \phi_{\alpha},
\end{align}
where the free energies of the individual phases, $f_{\text{ch}}^\alpha$, are interpolated through the corresponding phase field instead of a separate interpolation function, $h_{\alpha}(\vphi)$.
From Eqn.~\ref{eq:bar_W_chem}, the phase-field derivative of the chemical free-energy density is written as
\begin{align}\label{eq:d_bar_W_chem}
  \frac{\partial f_\text{ch}}{\partial \phia} = f^{\alpha}_{\text{ch}} + \sum_\beta \frac{\partial f^{\beta}_{\text{ch}}}{\partial c^\beta}\frac{\partial c^\beta}{\partial \phia}\phi_{\beta}.
\end{align}
Considering that the phases are equilibrated across the interface, the chemical potentials become phase independent and equal, 
\begin{align}\label{eq:d_chem}
 \frac{\partial f^{\alpha}_{\text{ch}}}{\partial c\p \alpha}=\frac{\partial f^{\beta}_{\text{ch}}}{\partial c\p \beta}=\cdots=\mu.
\end{align}
Furthermore, the concentration $c$ can expressed as $c=\sum_\beta c^\beta \phi_{\beta}$, analogous to Eqn.~\ref{eq:conc_interpolated1}.
Consequently, owing to the mass constraint, the following relation can be derived
\begin{align}\label{eq:conc_der}
 \frac{\partial c^\beta}{\partial \phia}\phi_{\beta}=-c\p\alpha.
\end{align}
Substituting Eqns.~\ref{eq:d_chem} and ~\ref{eq:conc_der}, the chemical factor governing the evolution of the phase field, which is expressed in Eqn.~\ref{eq:d_bar_W_chem} transforms to
\begin{align}\label{eq:ind_gran}
 \frac{\partial f_\text{ch}}{\partial \phia} = f^{\alpha}_{\text{ch}}-\mu c\p\alpha.
\end{align}
From Sec.~\ref{sec:ther_gran}, the left hand side of Eqn.~\ref{eq:ind_gran} is the \textit{Legendre} transform of the chemical free-energy density $f^\alpha_\text{ch}(c^\alpha)$ and is the grand potential density $\Psi(\mu)$.
Therefore, having established that the chemical driving force of the chemo-elastic model is dictated the grand potential density, the derivation of the model elucidated in Sec.~\ref{sec:grand_potentialM} can be adopted and the evolution of the dynamic variable $\mu$ is written as
\begin{align}\label{eq:ChemPot_eqn}
\dfrac{\partial \mu}{\partial t} = \Big[\n \cdot \Big( M(c,\vphi) \n \mu\Big) - \sum_{\alpha=1}^N  c^{\alpha}(\mu) \dfrac{\partial h_{\alpha}\left(\vphi\right)}{\partial t} \Big]  \Big[ \sum_{\alpha=1}^N \vv{\mathcal{X}}^{\alpha} h_{\alpha}\left(\vphi\right) \Big]^{-1}.
\end{align}

The elastic driving force for the present model is formulated based on the mechanical jump condition~\cite{schneider2015phase,schneider2018small}.
This introduces a dichotomy wherein the phases are assumed to be in chemical equilibrium at the interface to establish a continuity in the chemical potential, while the jump condition is involved in the formulation of the elastic free-energy density.
Particularly, the mechanical jump condition is adopted in the calculation of the stresses and the corresponding driving force.

Before explicitly formulating the elastic driving force, a homogenized normal vector is defined based on the scalar field as
 \begin{align}\label{eq:n_aus_M}
M(\vv\phi)=\sum_{\alpha<\beta}\phia\phi_{\beta} \quad \Rightarrow \quad \vv n(M(\vv\phi)) = \frac{\nabla M(\vv\phi)}{|\nabla M(\vv\phi)|}.
\end{align}
The stresses and strains are transformed to the orthonormal basis $\vv{B}=\{\vv{n},\vv{t},\vv{s}\}$, which comprises of the homogenized normal vector formulated in Eqn.~\ref{eq:n_aus_M}.
Subsequently, the fundamental elastic variables are expressed in the Voigt notation as 
\begin{align}\label{eq:define_sigma_epsilon_n_t}
  \begin{split}
  \vv\sigma_B^\alpha(\vv{n}) :=&{\Big(
  \sigma_{nn},\sigma_{nt},\sigma_{ns},
  \sigma_{tt}^\alpha,\sigma_{ss}^\alpha,\sigma_{ts}^\alpha \Big)}^\T={\l(\vv\sigma_n, \vv\sigma_t^\alpha \r)}^\T, \\
   \vv\epsilon_B^\alpha(\vv{n}) :=&{\Big(
  \epsilon_{nn}^\alpha,2 \epsilon_{nt}^\alpha,2 \epsilon_{ns}^\alpha,
  \epsilon_{tt},\epsilon_{ss},2 \epsilon_{ts}\Big)}^\T=
  {\left(\vv\epsilon_n^\alpha,\vv\epsilon_t\right)}^\T.
  \end{split}
\end{align}

With $\ljump \cdot \rjump$ representing the jump of the variables, according to the force balance $\ljump \vv\sigma_n \rjump = \vv 0$ and the Hadamard kinematic compatibility condition $\ljump \vv\epsilon_t \rjump = \vv 0$, the jump of the stresses $\vv{\sigma}_n$ and strains $\vv{\epsilon}_t$  vanishes across the interface.
\nomenclature{$\ljump \cdot \rjump$}{Jump in a variable}%
\nomenclature{$\vv{\sigma}_n$}{Normal component of stress}%
\nomenclature{$\vv{\epsilon}_t$}{Tangential component of strain}%
Therefore, the continuous contributions of these fundamental variables, for an infinitesimal deformation imposed on a single plane, can be written as 
\begin{align}\label{eq:stress_strain_cont}
 \vv{\sigma}_n := {(\sigma_{nn},\sigma_{nt},\sigma_{ns})}, \\ 
 \vv{\epsilon}_t := {(\epsilon_{tt},\epsilon_{ss},2 \epsilon_{ts})}.
\end{align}
Correspondingly, the discontinuous components of the stresses and strains are expressed as 
\begin{align}\label{eq:stress_strain_discont}
 \vv{\sigma}^\alpha_t &:= {(\sigma_{tt}^\alpha,\sigma_{ss}^\alpha,\sigma_{ts}^\alpha)}, \\
 \vv\epsilon_n^\alpha &:= {(\epsilon_{nn}^\alpha,2 \epsilon_{nt}^\alpha,2 \epsilon_{ns}^\alpha)},
\end{align}
respectively.
The notation $\alpha$ in the formulation of the discontinuous variables indicates that the particular stresses and strains are phase-dependent.

The local strain is ascertained from the gradient of the displacement field as
\begin{align}\label{eq:local_strain}
 \vv\epsilon = (\vv\nabla\vv u + (\vv\nabla\vv u)^{T})/2,
\end{align}
wherein the displacement-field gradient, using the Einstein summation convention, is expressed as $(\nabla \vv u )_{ij} = \partial u_i/\partial x_j$.
\nomenclature{$\vv u$}{Displacement}%

The stiffness tensor is defined based on the orthonormal basis $\vv{B}$.
For the ease of handling in the subsequent treatment, the matrix is fragmented into smaller matrices (blocks) as
\begin{align}\label{eq:C_nt}
\C^\text{v}_B &=
\l(\begin{array}{ccc|ccc}
\mathcal{C}_{nnnn} & \mathcal{C}_{nnnt} & \mathcal{C}_{nnns} & \mathcal{C}_{nntt} & \mathcal{C}_{nnss} & \mathcal{C}_{nnts}\\
\mathcal{C}_{ntnn} & \mathcal{C}_{ntnt} & \mathcal{C}_{ntns} & \mathcal{C}_{nttt} & \mathcal{C}_{ntss} & \mathcal{C}_{ntts}\\
\mathcal{C}_{nsnn} & \mathcal{C}_{nsnt} & \mathcal{C}_{nsns} & \mathcal{C}_{nstt} & \mathcal{C}_{nsss} & \mathcal{C}_{nsts}\\ \hline
\mathcal{C}_{ttnn} & \mathcal{C}_{ttnt} & \mathcal{C}_{ttns} & \mathcal{C}_{tttt} & \mathcal{C}_{ttss} & \mathcal{C}_{ttts}\\
\mathcal{C}_{ssnn} & \mathcal{C}_{ssnt} & \mathcal{C}_{ssns} & \mathcal{C}_{sstt} & \mathcal{C}_{ssss} & \mathcal{C}_{ssts}\\
\mathcal{C}_{tsnn} & \mathcal{C}_{tsnt} & \mathcal{C}_{tsns} & \mathcal{C}_{tstt} & \mathcal{C}_{tsss} & \mathcal{C}_{tsts}
\end{array} \r)
=:
\begin{pmatrix}
\C_{nn} & \C_{nt}\\
\C_{tn} & \C_{tt}
\end{pmatrix},
\end{align}
where $\C_{nn}$ and $\C_{tt}$ are $3 \times 3$ symmetrical matrices, while $\C_{nt}$ and $\C_{tn}$ are similar matrice variables which satisfy the condition $\C_{tn} = \C^\T_{nt}$.
\nomenclature{$\C$}{Stiffness tensor}%
This scheme of formulating a tensor is extended to construct the compliance tensor $\vv{\mathcal{S}}^\alpha$.
\nomenclature{$\vv{\mathcal{S}}$}{Compliance tensor}%

By exclusively considering the continuous variables, the stresses are calculated as
\begin{align}\label{eq:sigma_B_multi}
\vv{\bar{\sigma}}_{B} = 
 \underbrace{
 \begin{pmatrix}
  - \bar{\TT}_{nn}^{-1}                & -\bar{\TT}_{nn}^{-1} \bar{\TT}_{nt} \\
  - \bar{\TT}_{tn} \bar{\TT}_{nn}^{-1} & \bar{\TT}_{tt} - \bar{\TT}_{tn} \bar{\TT}_{nn}^{-1} \bar{\TT}_{nt}
  \end{pmatrix} 
}_{\bar{\C}^\text{v}_B (\vv{\phi})}
 \begin{pmatrix} 
 \vv{\epsilon}_n \\ \vv{\epsilon}_t 
 \end{pmatrix}
  + \underbrace{
  \begin{pmatrix}
  \bar{\TT}_{nn}^{-1}                & \vv{O} \\
  \bar{\TT}_{tn} \bar{\TT}_{nn}^{-1} & - \vv{I}
  \end{pmatrix} 
 \begin{pmatrix} 
 \tilde{\vv{\chi}}_n \\ \tilde{\vv{\chi}}_t 
 \end{pmatrix}
 }_{\tilde{\vv{\sigma}}_{B}},
\end{align}
where the normal and the tangential components of the inelastic strains $\tilde{\vv\epsilon}^\alpha$  are correspondingly denoted by 
$\tilde{\vv{\chi}}_n$ and $\tilde{\vv{\chi}}_t$.
\nomenclature{$\tilde{\vv\epsilon}$}{Inelastic strain}%
These normal and tangential parts of the $\tilde{\vv\epsilon}^\alpha$ are written as
\begin{align}\label{eq:chi_n_t_poly}
  \tilde{\vv\chi}_n = \sum_\alpha \l(\tilde{\vv{\epsilon}}^\alpha_n + \TT^\alpha_{nt}\tilde{\vv{\epsilon}}^\alpha_t \r) \phia,  \quad   \tilde{\vv\chi}_t = \sum_\alpha \TT^\alpha_{tt} \tilde{\vv{\epsilon}}^\alpha_t \phia,
\end{align}
respectively. 
Furthermore, in Eqn.~\ref{eq:sigma_B_multi}, $\bar{\TT}$ represents the locally averaged contribution of the proportionality matrix.
\nomenclature{$\bar{\TT}$}{Averaged proportionality matrix}%
This averaged contribution between the continuous and discontinuous variables is expressed as
\begin{align}
\bar{\TT}_{nn} &:= \sum_\alpha \TT^\alpha_{nn} \phia := - \sum_\alpha {(\C^\alpha_{nn})}^{-1} \phia\label{eq:T_nn}, \\ 
\bar{\TT}_{nt} &:= \sum_\alpha \TT^\alpha_{nt} \phia :=  \sum_\alpha {(\C^\alpha_{nn})}^{-1} \C^\alpha_{nt} \phia \label{eq:T_nt}, \\
\bar{\TT}_{tt} &:= \sum_\alpha \TT^\alpha_{tt} \phia := 
\sum_\alpha \l(\C^\alpha_{tt} - \C^\alpha_{tn} {(\C^\alpha_{nn})}^{-1} \C^\alpha_{nt} \r) \phia.
\end{align}

By formulating the transformation matrices $\vv{M}_\epsilon$ and $\vv{M}_\sigma$ as in Ref.~\cite{schneider2015phase,schneider2018small}, in the Cartesian coordinate system, the resulting stresses in the Voigt notation can be determined by 
\begin{align}
 \vv{\bar{\sigma}}^\text{v}(\vv{\phi})  = \bar{\C}^\text{v}(\vv{\phi}) \bar{\vv{\epsilon}}^\text{v} + \tilde{\vv\sigma}^\text{v}(\vv{\phi}),
\end{align}
wherein the transformations $\bar{\C}^\text{v}(\vv{\phi}) = \vv{M}_\epsilon^\T \C^\text{v}_B (\vv{\phi}) \vv{M}_\epsilon$ and $\tilde{\vv{\sigma}}^\text{v}(\vv{\phi}) = \vv{M}_\sigma^\T \tilde{\vv\sigma}_{B}(\vv{\phi})$ is adopted.
For the displacement field $\vv{u}$, the momentum balance $\n \cdot \bar{\vv{\sigma}}(\vv{\phi}) = \vv{0}$ is solved by transforming the resulting stress $\vv{\bar{\sigma}}^\text{v}(\vv{\phi})$ into matrix notation.
Accordingly, the stresses in the bulk region of phase-$\alpha$ are calculated by 
\begin{align}
  \sigma_{ij}^{\alpha} = (\C^\alpha[\vv\epsilon -\tilde{\vv\epsilon}^\alpha])_{ij} =\mathcal{C}_{ijkl}^\alpha(\epsilon_{kl} -\tilde{\epsilon}^\alpha_{kl}).
\end{align}

Ultimately, the mechanical driving-force dictated by the derivative of the elastic free-energy density is written as
\begin{align}
\frac{\partial f_\text{el}(\vv \phi, \vv{\epsilon}_{\vv B})}{\partial\phia} = \sum_\alpha\frac{\partial p^\alpha(\vv{\sigma}_n, \vv{\epsilon}_t) \phia}{\partial\phia}, 
\end{align}
wherein, 
\begin{align}\label{eq:p_alpha}
p^\alpha(\vv{\sigma}_n, \vv{\epsilon}_t) 
&=\frac{1}{2} \left( 
\begin{pmatrix} 
\vv{\sigma}_n \\ \vv{\epsilon}_t 
\end{pmatrix} 
\cdot 
\begin{pmatrix} 
  \TT^\alpha_{nn} & \TT^\alpha_{nt} \\
  \TT^\alpha_{tn} & \TT^\alpha_{tt}   
\end{pmatrix}
\begin{pmatrix}
  \vv{\sigma}_n \\ \vv{\epsilon}_t
\end{pmatrix} 
  \right) -\left( 
\begin{pmatrix} 
  \vv{\sigma}_n \\ \vv{\epsilon}_t 
\end{pmatrix} 
  \cdot 
\begin{pmatrix} 
  \vv{I} & \TT^\alpha_{nt}        \\
  \vv{O} & \TT^\alpha_{tt}   
\end{pmatrix}
\begin{pmatrix} 
  \tilde{\vv\epsilon}^\alpha_n \\ \tilde{\vv\epsilon}^\alpha_t 
\end{pmatrix} 
  \right) \\ 
  &+\frac{1}{2}\left( \tilde{\vv\epsilon}^\alpha_t \cdot \TT^\alpha_{tt} \tilde{\vv\epsilon}^\alpha_t \right).\nonumber
\end{align}
The components of proportionality matrix $\TT^\alpha$, which contributes to the mechanical driving-force through $p^\alpha(\vv{\sigma}_n, \vv{\epsilon}_t)$ in Eqn.~\ref{eq:p_alpha}, are expressed as
\begin{align}\label{eq:T_parts_multi}
\TT^\alpha_{nn}  &:= -  \S^\alpha_{nn}, \\ 
\TT^\alpha_{nt}  &:=   \S^\alpha_{nn} \C^\alpha_{nt}, \\
\TT^\alpha_{tt}  &:=  \C^\alpha_{tt}- \C^\alpha_{tn} \S^\alpha_{nn} \C^\alpha_{nt}.
\end{align}

\section{Phase-field model distinguishing interstitial and substitutional diffusion}

In phase-field modelling, the interface separating the two phases ($\alpha$ and $\beta$) are replaced by definite diffuse-region wherein the scalar variable, phase field ($\phi$), varies smoothly.
Therefore, the entire system is treated as the combination of the interface and the remnant bulk regions.
Thermodynamically, such a system is described through a functional ($\mathcal{F}$) as
\begin{align}\label{eq:functional1}
  \mathcal{F}(\phi, \n \phi,\vv{S}) = \mathcal{F}_\text{intf}(\phi, \n \phi) + \mathcal{F}_\text{bulk}(\phi, \vv{S}),
\end{align}
where $\mathcal{F}_\text{intf}$ and $\mathcal{F}_\text{bulk}(\phi, \vv{S})$ are the respective contributions from the diffuse interface and the bulk region, wherein the phase field ($\phi$) is constant.
Furthermore, in Eqn.~\ref{eq:functional1}, $\vv{S}$ is an independent thermodynamic variable(s) which dictates the contribution of the bulk phases.
Depending on the energetics considered, the nature of the variable $\vv{S}$ changes.
The framework of the present model follows Ref.~\cite{nestler2005multicomponent}.
However, since the model is derived for a two-phase system ($\alpha$ and $\beta$), the phase field is written as a scalar, $\phi$.
Furthermore, $\phi$ between the phase-$\alpha$ and -$\beta$ are related $\phi_{\alpha}+\phi_{\beta}=1$.

The Ginzburg-Landau type functional is often adopted to describe a system with diffuse interface.
Correspondingly, the contribution of the interface reads
\begin{align}\label{eq:interface}
  \mathcal{F}_\text{intf}(\phi, \n \phi) =\int_{V} \left[\epsilon a(\n \phi) + \frac{1}{\epsilon}\omega(\phi)\right]\diff V,
\end{align}
where $\epsilon a(\n \phi)$ is the gradient-energy term while $\omega(\phi)/\epsilon$ is the penalising potential.
The length parameter $\epsilon$ dictates the width of the diffuse interface.
With $\gamma$ representing the energy density of the interface separating two phases, $\alpha$ and $\beta$, the gradient-energy term is written as
\begin{align}\label{eq:Aterm}
  \epsilon a(\n \phi) = \epsilon \gamma|\n \phi|^2.
\end{align}
As opposed to the double-well potential, which is conventionally used for penalising the state variable, the double-obstacle potential is employed in this formulation.
The double-obstacle potential, $\omega(\phi)$, is expressed as 
\begin{align}\label{eq:Wterm}
 \omega(\phi)=  \gamma\frac{16}{\pi^2}\phi(1-\phi),  \quad  \phi\in[0,1].
\end{align}
The role of the obstacle-type potential in enhancing the numerical and computational efficiency has already been discussed in Refs.~\cite{oono1988study,garcke1999multiphase}.

For the present analysis, the elastic and the chemical contributions of the bulk phases are considered.
Therefore, the functional which represents the bulk-phase contribution extends to
\begin{align}\label{eq:bulk_free}
  \mathcal{F}_\text{bulk}(\phi, \vv{S}) = \mathcal{F}_\text{el}(\phi, \vv{S}_\text{el})+\mathcal{F}_\text{ch}(\phi, \vv{S}_\text{ch}),
\end{align}
where $\mathcal{F}_\text{el}(\phi, \vv{S}_\text{el})$ and $\mathcal{F}_\text{ch}(\phi, \vv{S}_\text{ch})$ represent the elastic and the chemical contributions of the phases, respectively.

\subsection{Elastic model for displacive transformation}

The elastic contribution from the bulk phases in formulated as the Helmholtz free energy.
For a consistent derivation, wherein the bulk phase and the interface are energetically decoupled, the approach in Ref.~\cite{schneider2015phase} is employed.

In a solid-state system with singular surface, the traction vectors are equal across the interface.
This force balance between the two phases, $\alpha$ and $\beta$, reads
\begin{align}\label{eq:force_balance}
  \vv{\sigma}^{\alpha}\vv{n}=\vv{\sigma}^{\beta}\vv{n},  
\end{align}
where $\vv{n}$ is the normal vector to the interface separating $\alpha$ and $\beta$.
The normal vector in the diffuse interface is expressed as
\begin{align}\label{eq:normal}
 \vv{n}=\frac{\n \phi}{|\n \phi|},
\end{align}
where $|\n \phi|=\sqrt{\n \phi\cdot\n \phi}$.
The jump condition which results from the force balance, in Eqn.~\ref{eq:force_balance}, is written as
\begin{align}\label{eq:force_balance1}
 \llbracket \vv{\sigma}\rrbracket{\vv{n}} = \vv{0},
\end{align}
where $\llbracket \cdot \rrbracket$ represents the jump in the variable across the interface.
Furthermore, according to Hadamard compatibility condition~\cite{silhavy2013mechanics}, the continuity of the displacement vector ($\vv{u}$) is expressed as
\begin{align}\label{eq:hadamard_jump}
  \llbracket \n \vv{u} \rrbracket = \vv{a} \vv{n}^{T},
\end{align}
where $\vv{a}$ is an arbitrary vector and $(\vv{a} \vv{n}^{T})_{ij}=a_{i}n_{j}$ is the corresponding dyadic product.
The corollary of Eqn.~\ref{eq:hadamard_jump}, which describes the jump of the displacement vector in the normal direction, is
\begin{align}\label{eq:hadamard_jump2}
 \llbracket \n \vv{u} \rrbracket \vv{t}= \vv{0},
\end{align}
where $\vv{t}$ is a tangential vector.

For the present derivation, the total strain, $\vv{\epsilon}$, is treated as the summation of elastic strain, $\vv{\epsilon}_{\text{el}}$, and inelastic eigenstrain, $\vv{\tilde{\epsilon}}$.
This total strain is related to the displacement vector as
\begin{align}\label{eq:strain_displacement}
 \vv{\epsilon} = \vv{\epsilon}_{\text{el}} +  \vv{\tilde{\epsilon}} = \frac{1}{2}\left[ \n \vv{u} + (\n \vv{u})^{T} \right].
\end{align}
From Eqns.~\ref{eq:hadamard_jump2} and ~\ref{eq:strain_displacement}, the total strain along the tangential direction of the $\alpha\beta$-interface can be expressed as
\begin{align}\label{eq:tangential_strain}
(\vv{\epsilon}_{\text{el}}^{\alpha} +  \vv{\tilde{\epsilon}}^{\alpha}) \vv{t} &= (\vv{\epsilon}_{\text{el}}^{\beta} +  \vv{\tilde{\epsilon}}^{\beta}) \vv{t},
\end{align}
where $\vv{\epsilon}_{\text{el}}^{\alpha}$ and $\vv{\tilde{\epsilon}}^{\alpha}$ are phase-dependent elastic and inelastic strain.
Based on Eqn.~\ref{eq:tangential_strain}, a jump condition for the tangential strain across the interface can be defined as
\begin{align}\label{eq:baseB}
\llbracket \vv{\epsilon}_{\text{el}:{\vv{t}}}+  \vv{\tilde{\epsilon}}_{\vv{t}} \rrbracket &= \vv{0}.
\end{align}

In base $\vv{B}=\{ \vv{n}, \vv{s}, \vv{t} \}$, with fixed normal vector $\vv{n}$ and two tangential vectors, $\vv{s}$ and $\vv{t}$, the jump in the total strain across the interface is expressed as
\begin{align}\label{eq:strainB}
\llbracket \vv{\epsilon}_{\text{el}:{\vv{B}}}+  \vv{\tilde{\epsilon}}_{\vv{B}} \rrbracket=
 \begin{pmatrix} 
  \llbracket \epsilon_{\text{el}:{nn}}+  \tilde{\epsilon}_{nn} \rrbracket & \llbracket \epsilon_{\text{el}:{ns}}+  \tilde{\epsilon}_{ns} \rrbracket & \llbracket \epsilon_{\text{el}:{nt}}+  \tilde{\epsilon}_{nt} \rrbracket \\
  \llbracket \epsilon_{\text{el}:{ns}}+  \tilde{\epsilon}_{ns} \rrbracket & 0 & 0 \\
  \llbracket \epsilon_{\text{el}:{nt}}+  \tilde{\epsilon}_{nt} \rrbracket & 0 & 0
\end{pmatrix}.
\end{align}
The phase-dependent normal components of the strain are distinguished from the phase-independent tangential components, using Voigt notation as
\begin{align}\label{eq:strainB_voigt}
 \vv{\epsilon}_{\text{el}:{\vv{B}}}^{\alpha}+  \vv{\tilde{\epsilon}}_{\vv{B}}^{\alpha} =[ \underbrace{(\epsilon_{\text{el}:{nn}}^{\alpha}+  \tilde{\epsilon}_{nn}^{\alpha}), 2(\epsilon_{\text{el}:{ns}}^{\alpha}+  \tilde{\epsilon}_{ns}^{\alpha}), 2(\epsilon_{\text{el}:{nt}}^{\alpha} +  \tilde{\epsilon}_{nt}^{\alpha})}_{:= \vv{\epsilon}_{\text{el}:\vv{n}}^{\alpha}+  \vv{\tilde{\epsilon}}_{\vv{n}}^{\alpha} =\vv{\epsilon}_{\vv{n}}^{\alpha}},\\ \nonumber \underbrace{(\epsilon_{\text{el}:{ss}}+  \tilde{\epsilon}_{ss}), (\epsilon_{\text{el}:{tt}}+  \tilde{\epsilon}_{tt}), 2(\epsilon_{\text{el}:{st}}+  \tilde{\epsilon}_{st})}_{:= \vv{\epsilon}_{\text{el}:\vv{t}}+  \vv{\tilde{\epsilon}}_{\vv{t}} =\vv{\epsilon}_{\vv{t}}}  ]^{T}
\end{align}
Similarly, based on jump condition in Eqn.~\ref{eq:force_balance1}, the phase-dependent and -independent components of the stresses are realised as
\begin{align}\label{eq:stress_nt} 
 \vv{\sigma}_{\vv{B}}^{\alpha}=(\underbrace{\sigma_{nn}, \sigma_{ns}, \sigma_{nt}}_{\vv{\sigma}_{\vv{n}}},\underbrace{ \sigma_{ss}^{\alpha}, \sigma_{tt}^{\alpha}, \sigma_{st}^{\alpha}}_{\vv{\sigma}_{\vv{t}}^{\alpha}} )^{T}=(\vv{\sigma}_{\vv{n}},\vv{\sigma}_{\vv{t}}^{\alpha})^{T}.
\end{align}

Despite the elegant formulation, in-consistencies are introduced in the phase-field model when the bulk and interface are not efficiently decoupled. 
It has been shown that by the appropriate choice of the continuous variables, that smoothly vary across the interface, the unphysical contribution of the bulk phases to the interface can be averted~\cite{kim1999phase,plapp2011unified}.
Based on the jump conditions, which facilitate the distinction of the phase-dependent and -independent components in Eqns.~\ref{eq:strainB_voigt} and ~\ref{eq:stress_nt}, the tangential strain ($\vv{\epsilon}_{\vv{t}}$) and the normal stress ($\vv{\sigma}_{\vv{n}}$) are considered as the continuous variables in the present model.

The elastic free-energy in phase-$\alpha$, using elastic strain ($\vv{\epsilon}_{\text{el}:{\vv{B}}}^{\alpha}$), is expressed as
\begin{align}\label{eq:el_free} 
 f_{\text{el}}^{\alpha}(\vv{\epsilon}_{\text{el}:{\vv{B}}}^{\alpha})=\frac{1}{2}\left( \vv{\epsilon}_{\text{el}:{\vv{B}}}^{\alpha} \cdot \mathcal{C}_{\vv{B}}^{\alpha}\vv{\epsilon}_{\text{el}:{\vv{B}}}^{\alpha} \right),
\end{align}
where $\mathcal{C}_{\vv{B}}^{\alpha}$ is the stiffness matrix of the phase-$\alpha$.
By encompassing the inelastic eigenstrain, $\vv{\tilde{\epsilon}}_{\vv{B}}^{\alpha}$, the free energy formulation transforms to
\begin{align}\label{eq:totalEL_free} 
 f_{\text{el}}^{\alpha}(\vv{\epsilon}_{\vv{B}}^{\alpha})=\frac{1}{2}\left[ (\vv{\epsilon}_{\vv{B}}^{\alpha}-\vv{\tilde{\epsilon}}_{\vv{B}}^{\alpha})\cdot \mathcal{C}_{\vv{B}}^{\alpha}(\vv{\epsilon}_{\vv{B}}^{\alpha}-\vv{\tilde{\epsilon}}_{\vv{B}}^{\alpha}) \right].
\end{align}
For the above formulation, it has been shown that the phase-field evolution is governed by the difference in the \textit{Legendre transform} of the free energy~\cite{wheeler1992phase}.
Therefore, the elastic driving force, based on Eqn.~\ref{eq:totalEL_free}, is expressed as
\begin{align}\label{eq:elastic_potential}
 p^{\alpha}(\vv{\epsilon}_{\vv{t}},\sigma_{\vv{n}})=\frac{1}{2}\left\{ \left[(\sigma_{\vv{n}} \cdot \vv{\epsilon}_{\vv{t}}-\vv{\tilde{\epsilon}}^{\alpha}_{\vv{t}}) \cdot \vv{\mathcal{T}}^{\alpha} 
 \begin{pmatrix}
  \sigma_{\vv{n}}\\
  \vv{\epsilon}_{\vv{t}}-\vv{\tilde{\epsilon}}^{\alpha}_{\vv{t}}
 \end{pmatrix}
  \right] \right\}-(\sigma_{\vv{n}}\cdot\vv{\tilde{\epsilon}}^{\alpha}_{\vv{n}}),
\end{align}
where tensor $\vv{\mathcal{T}}^{\alpha}$ encompasses all material parameter.
Above Eqn.~\ref{eq:elastic_potential}, which is the \textit{Legendre transform} of the free energy, while describing the driving force, renders its formulation based on the continuous variables, $\vv{\epsilon}_{\vv{t}}$ and  $\vv{\sigma}_{\vv{n}}$.
In this section, aspects of the model which is pertinent to the present analysis is elucidated, for a comprehensive understanding the readers are directed to Refs.~\cite{schneider2015phase}.

\subsection{Model for diffusion}

For simulating diffusion-governed (reconstructive) phase transformation, the quantitative information including the equilibrium concentration are recovered from the CALPHAD database.
Often, these data are not inherently compatible for the formulation of a multi-physics model.
For instance, the CALPHAD database provides the concentration based (chemical) free-energy is the form of the Gibbs free-energy, while, even in the present model, the elastic contribution is formulated based on Helmholtz free-energy.
Therefore, by assuming that the molar volume of the phases are equal ($v_m$), the Gibbs free-energy is converted to Helmholtz free-energy by
\begin{align}\label{eq:chfree_energy}
 f_{\text{ch}}^{\alpha}(\vv{c}^{\alpha})=\frac{1}{v_m}G^{\alpha}(\vv{c}^{\alpha}),
\end{align}
where $\vv{c}^{\alpha}$ is a continuous vector which represents the concentration of each component in mole fraction, $\vv{c}^{\alpha}=\{c^{\alpha}_{0}, c^{\alpha}_{1}, c^{\alpha}_{2}, \dots\}$.
\nomenclature{$\vv{c}^{\alpha}$}{Continuous concentration vector representing the composition of multicomponent phase-$\alpha$}%
Although the free energy formulated in Eqn.~\ref{eq:chfree_energy} encompasses multicomponent systems, the mole-fraction based concentration expression restricts its applicability~\cite{cha2001phase}.
Particularly, the substitutional and interstitial diffusion, which are prevalent in a multicomponent system, cannot be distinguished in a free energy formulation based on mole fraction.
Furthermore, mole-fraction based formulation cannot be employed to model transformation under para-equilibrium conditions, wherein diffusion occurs selectively in certain components.
By adopting number density, number of moles per volume, to represent concentration, these limitations have been considerably addressed~\cite{yeon2001phase,cha2005phase}.
Moreover, recently, site fraction has been employed to analyse the phase transformations in system with several sub-lattices~\cite{zhang2015incorporating}.
In the present model, site fraction is adopted in the framework of the grand-potential formulation to distinguish the interstitial and substitutional diffusion~\cite{plapp2011unified}.

Consider a system of two sub-lattices.
One corresponding to the regular lattices, which is henceforth referred to substitutional, while the other is associated with the inherent voids formed by the crystallographic arrangement of atoms, interstitial.
Since the sizes of the substitutional lattice is significantly different from the interstitial site, a component is often limited to one of the sub-lattices~\cite{wang2015atomic}.
Therefore, the concentration is expressed through two continuous vectors, $\vv{y}^{\alpha}_{\text{int}}=\{y^{\alpha}_{\text{int}:0}, y^{\alpha}_{\text{int}:1}, \dots, y^{\alpha}_{\text{int}:j}\}$ and $\vv{y}^{\alpha}_{\text{sub}}=\{y^{\alpha}_{\text{sub}:0}, y^{\alpha}_{\text{sub}:1}, \dots, y^{\alpha}_{\text{sub}:l}\}$, with each representing a sub-lattice.
\nomenclature{$\vv{y}^{\alpha}_{\text{int}}$}{Continuous vector representing the site fractions of interstitial solutes in phase-$\alpha$}%
\nomenclature{$\vv{y}^{\alpha}_{\text{sub}}$}{Continuous vector representing the site fractions of solvent and substitutional solutes in phase-$\alpha$}%
\nomenclature{$y^{\alpha}_{\text{int}:j}$}{Site fraction of interstitial solute $j$ in phase-$\alpha$}%
\nomenclature{$y^{\alpha}_{\text{sub}:l}$}{Site fraction of substitutional solute $l$ in phase-$\alpha$}%
Site fraction of a component $i$, which occupies interstitial site, in phase-$\alpha$ is expressed as
\begin{align}\label{eq:site_fraction}
 y^{\alpha}_{\text{int}:i}=\frac{n^{\alpha}_{\text{int}:i}}{\sum_{i=0}^{j}n^{\alpha}_{\text{int}:i}}=\frac{N^{\alpha}_{\text{int}:i}}{\sum_{i=0}^{j}N^{\alpha}_{\text{int}:i}},
\end{align}
where $n^{\alpha}_{\text{int}:i}$ and $\sum_{i=0}^{j}n^{\alpha}_{\text{int}:i}$ are number of $i$ atoms in $\alpha$ and total number of interstitial sites, respectively.
By incorporating Avagadro's number, as in Eqn.~\ref{eq:site_fraction}, site fraction can be defined as the ratio of number of moles of $i$ ($N^{\alpha}_{\text{int}:i}$) and the total interstitial sites in moles ($\sum_{i=0}^{j}N^{\alpha}_{\text{int}:i}$).
From Eqn.~\ref{eq:site_fraction} it follows that $\sum_{i=0}^{j}y^{\alpha}_{\text{int}:i}(=\sum_{k=0}^{l}y^{\alpha}_{\text{sub}:k})=1$.
Furthermore, based on Eqn.~\ref{eq:site_fraction}, site fraction and mole fraction can be related as
\begin{align}\label{eq:site_mole}
 y^{\alpha}_{\text{int}:i}&=c_{i}^{\alpha}\left(\frac{N^{\alpha}}{\sum_{i=0}^{j}N^{\alpha}_{\text{int}:i}}\right),
\end{align}
where $N^{\alpha}$ summation of number of moles of individual components ($N^{\alpha}_i$) in $\alpha$.
\nomenclature{$N^{\alpha}$}{Number of moles of all components in phase-$\alpha$}%
Since $N^{\alpha}$ and $\sum_{i=0}^{j}N^{\alpha}_{\text{int}:i}$, which is total number of interstitial sites in $\alpha$, are constant, Eqn.~\ref{eq:site_mole} yields the relation $c_{i}^{\alpha} = a^{\alpha}_{\text{int}}y^{\alpha}_{\text{int}:i}$, where $a^{\alpha}_{\text{int}}$ is the number of interstitial sites per atom in $\alpha$.
Similar relation can be derived for the substitutional alloying elements.
However, it should be noted that the constants, $a^{\alpha}_{\text{int}}$ and $a^{\alpha}_{\text{sub}}$, depend extensively on the crystal structure of the phases.
\nomenclature{$a^{\alpha}_{\text{int}}$}{Number of interstitial sites per atom in phase-$\alpha$}%
\nomenclature{$a^{\alpha}_{\text{sub}}$}{Number of substitutional sites per atom in phase-$\alpha$}%

The free energy of the phase-$\alpha$, using the conjugate pairs, is expressed as
\begin{align}\label{eq:gibbs_chem}
 G^{\alpha}=\sum_{i=0}^{l+2}\mu_{i}^{\alpha}N_{i}^{\alpha},
\end{align}
where $\mu_{\text{int}:i}^{\alpha}$ is the chemical potential and $N_{i}^{\alpha}$ is the number of mole of $i$ in $\alpha$.
By associating the components to their respective sub-lattices, and correspondingly transforming the chemical potential, the free energy reads
\begin{align}\label{eq:gibbs_chem2}
 G^{\alpha} =\sum_{i=0}^{j}\mu_{\text{int}:i}^{\alpha}N_{\text{int}:i}^{\alpha}+\sum_{k=0}^{l}\mu_{\text{sub}:k}^{\alpha}N_{\text{sub}:k}^{\alpha}.
\end{align}
From above Eqn.~\ref{eq:gibbs_chem2}, the free energy based on the site fractions is written as
\begin{align}\label{eq:gibbs_site1}
 G^{\alpha}(\vv{y}^{\alpha}_{\text{int}},\vv{y}^{\alpha}_{\text{sub}})=\sum_{i=0}^{j}N^{\alpha}_{\text{int}:i}\left[ \sum_{i=0}^{j}\mu^{\alpha}_{\text{int}:i}y^{\alpha}_{\text{int}:i} \right] + \sum_{k=0}^{l}N^{\alpha}_{\text{sub}:k}\left[ \sum_{k=0}^{l}\mu^{\alpha}_{\text{sub}:k}y^{\alpha}_{\text{sub}:k} \right],
\end{align}
where $y^{\alpha}_{\text{int}:i}$ and $y^{\alpha}_{\text{sub}:k}$ are site fractions of $i$ and $k$ in $\alpha$, respectively.
By separating one component from each sub-lattices, the free energy can be written as
\begin{align}\label{eq:gibbs_site2}
 G^{\alpha}(\vv{y}^{\alpha}_{\text{int}},\vv{y}^{\alpha}_{\text{sub}})=\sum_{i=0}^{j}N^{\alpha}_{\text{int}:i}\left[ \mu^{\alpha}_{\text{int}:0}y^{\alpha}_{\text{int}:0} + \sum_{i=1}^{j}\mu^{\alpha}_{\text{int}:i}y^{\alpha}_{\text{int}:i} \right] + \sum_{k=0}^{l}N^{\alpha}_{\text{sub}:k}\left[ \mu^{\alpha}_{\text{sub}:0}y^{\alpha}_{\text{sub}:0} +  \sum_{k=0}^{l}\mu^{\alpha}_{\text{sub}:k}y^{\alpha}_{\text{sub}:k} \right].
\end{align}
Since $\sum_{i=0}^{j}y^{\alpha}_{\text{int}:i}(=\sum_{k=0}^{l}y^{\alpha}_{\text{sub}:k})=1$, Eqn~\ref{eq:gibbs_site2} transforms to
\begin{align}\label{eq:gibbs_site3}
 G^{\alpha}(\vv{y}^{\alpha}_{\text{int}},\vv{y}^{\alpha}_{\text{sub}})=\sum_{i=0}^{j}N^{\alpha}_{\text{int}:i}\left[ \mu^{\alpha}_{\text{int}:0} + \sum_{i=1}^{j}\left(\tilde{\mu}^{\alpha}_{\text{int}:i}\right)y^{\alpha}_{\text{int}:i} \right] + \sum_{k=0}^{l}N^{\alpha}_{\text{sub}:k}\left[ \mu^{\alpha}_{\text{sub}:0} +  \sum_{k=1}^{l}\left(\tilde{\mu}^{\alpha}_{\text{sub}:k}\right)y^{\alpha}_{\text{sub}:k} \right]
\end{align}
where $\tilde{\mu}^{\alpha}_{\text{int}:i}$ and $\tilde{\mu}^{\alpha}_{\text{sub}:k}$ are the diffusional potentials that read
\begin{align}\label{eq:diffusion_potential1}
 \tilde{\mu}^{\alpha}_{\text{int}:i} = \mu^{\alpha}_{\text{int}:i}-\mu^{\alpha}_{\text{int}:0} \\
 \tilde{\mu}^{\alpha}_{\text{sub}:k} = \mu^{\alpha}_{\text{sub}:k}-\mu^{\alpha}_{\text{sub}:0}.
\end{align}
\nomenclature{$\tilde{\mu}^{\alpha}_{\text{int}:i}$}{Diffusion potential of interstitial solute $i$ in phase-$\alpha$}%
\nomenclature{$\tilde{\mu}^{\alpha}_{\text{sub}:k}$}{Diffusion potential of component $k$ in phase-$\alpha$}%

In a multicomponent system with more than one sub-lattice, thermodynamically~\cite{hillert2008phase,hillert2001compound}, the chemical potential is estimated by
\begin{align}\label{eq:potential_site1}
 \mu^{\alpha}_{\text{int}:i}=\frac{\partial G^{\alpha}}{\partial N_{\text{int}:i}^{\alpha}} + \frac{\partial G^{\alpha}}{\partial y_{\text{int}:i}^{\alpha}}\frac{\partial y_{\text{int}:i}^{\alpha}}{\partial N_{\text{int}:i}^{\alpha}} + \displaystyle\sum_{j} \frac{\partial G^{\alpha}}{\partial y_{\text{int}:j}^{\alpha}}\frac{\partial y_{\text{int}:j}^{\alpha}}{\partial N_{\text{int}:i}^{\alpha}}.
\end{align}
Solving the partial derivatives in the second and third terms on the right-hand side of Eqn.~\ref{eq:potential_site1} yields the relation
\begin{align}\label{eq:moles_site}
 \frac{\partial y_{\text{int}:i}^{\alpha}}{\partial N_{\text{int}:i}^{\alpha}}=\frac{1-y^{\alpha}_{\text{int}:i}}{\sum_{i=0}^{j}N^{\alpha}_{\text{int}:i}} &&  \text{and} && \frac{\partial y_{\text{int}:j}^{\alpha}}{\partial N_{\text{int}:i}^{\alpha}}=\frac{-y^{\alpha}_{\text{int}:j}}{\sum_{i=0}^{j}N^{\alpha}_{\text{int}:i}}.
\end{align}
By substituting Eqn.~\ref{eq:moles_site} in Eqn.~\ref{eq:potential_site1}, the chemical potential of a component in the interstitial site is expressed as
\begin{align}\label{eq:potential_site2}
 \mu^{\alpha}_{\text{int}:i} = G^{\alpha}_m + \frac{1}{\sum_{i=0}^{j}N^{\alpha}_{\text{int}:i}}\left[ \frac{\partial G^{\alpha}}{\partial y_{\text{int}:i}^{\alpha}} - \sum_{j}y_{\text{int}:j} \frac{\partial G^{\alpha}}{\partial y_{\text{int}:j}^{\alpha}} \right],
\end{align}
where $G^{\alpha}_m$ is the molar free-energy of phase-$\alpha$.
Similarly, the chemical potential of a substitutional component can be written as
\begin{align}\label{eq:potential_site5}
 \mu^{\alpha}_{\text{sub}:k} = G^{\alpha}_m + \frac{1}{\sum_{i=0}^{j}N^{\alpha}_{\text{sub}:k}}\left[ \frac{\partial G^{\alpha}}{\partial y_{\text{sub}:k}^{\alpha}} - \sum_{l} y_{\text{sub}:l} \frac{\partial G^{\alpha}}{\partial y_{\text{sub}:l}^{\alpha}} \right].
\end{align}
From Eqns.~\ref{eq:potential_site2} and ~\ref{eq:potential_site5}, the diffusion potentials of the interstitial and substitutional components are expressed as
\begin{align}\label{eq:diffusion_potential2}
 \tilde{\mu}^{\alpha}_{\text{int}:i} &= \frac{1}{\sum_{i=0}^{j}N^{\alpha}_{\text{int}:i}}\left[ \frac{\partial G^{\alpha}}{\partial y_{\text{int}:i}^{\alpha}} - \frac{\partial G^{\alpha}}{\partial y_{\text{int}:0}^{\alpha}} \right]\\
 \tilde{\mu}^{\alpha}_{\text{sub}:k} &= \frac{1}{\sum_{k=0}^{l}N^{\alpha}_{\text{sub}:k}}\left[ \frac{\partial G^{\alpha}}{\partial y_{\text{sub}:k}^{\alpha}} - \frac{\partial G^{\alpha}}{\partial y_{\text{sub}:0}^{\alpha}} \right],
\end{align}
respectively.
Substituting the above relations in Eqn.~\ref{eq:gibbs_site3}, the free energy can be expressed in terms of site fractions and diffusion potentials.
Furthermore, based on Eqn.~\ref{eq:chfree_energy}, the Gibbs free-energy is transformed to Helmholtz free-energy.

When the phase-$\alpha$ and -$\beta$ are in chemical equilibrium, the chemical (diffusion) potential of a component across the interface are equal.
Therefore, at equilibrium, Eqn.~\ref{eq:gibbs_site3}, yields the relation  
\begin{align}\label{eq:equilibrium}
 \frac{\partial G^{\alpha}}{\partial y_{\text{int}:i}}=\frac{\partial G^{\beta}}{\partial y_{\text{int}:i}}& = \tilde{\mu}_{\text{int}:i} \left(= \mu_{\text{int}:i} \right) \\
 \frac{\partial G^{\alpha}}{\partial y_{\text{sub}:k}}=\frac{\partial G^{\beta}}{\partial y_{\text{sub}:k}}&=\tilde{\mu}_{\text{sub}:k}.
\end{align}
In Eqn.~\ref{eq:gibbs_site2}, which introduces the diffusion potential, the solvent is conventionally considered as the 0th component in the substitutional sub-lattice.
In the interstitial sub-lattice, since the entire sites are rarely filled in any alloy-system, the vacant-sites are treated as 0th component.
The interstitial vacancies render zero contribution of the thermodynamic nature of the system. 
Therefore, the diffusion potential of a component in the interstitial sub-lattice is considered equal to its respective chemical potential.
This equivalence in the chemical and diffusion potential of the interstitial components is stated in Eqn.~\ref{eq:equilibrium}.

Based on Eqn.~\ref{eq:equilibrium}, the chemical and the diffusion potential of the corresponding interstitial and substitutional component are treated as the continuous variables~\cite{kim1999phase,plapp2011unified}.
The continuous variables are adopted as the dynamic variables by considering the \textit{Legendre transform} of the chemical free-energy, which results in the grand chemical-potential density.
The grand chemical-potential for the phase-$\alpha$ with interstitial and substitutional lattice is written as
\begin{align}\label{eq:grandchem_site}
 \Psi^{\alpha}(\vv{\mu}_{\text{int}},\vv{\mu}_{\text{sub}}) = f_{\text{ch}}^{\alpha}(\vv{y}^{\alpha}_{\text{int}} (\vv{\mu}_{\text{int}}),\vv{y}^{\alpha}_{\text{sub}} (\vv{\mu}_{\text{sub}})) - \left[ \sum_{i=1}^{j} \mu_{\text{int}:i} y^{\alpha}_{\text{int}:i} (\vv{\mu}_{\text{int}}) + \sum_{k=1}^{l} \tilde{\mu}_{\text{sub}:k} y^{\alpha}_{\text{sub}:k} (\vv{\mu}_{\text{sub}}) \right],
\end{align}
where $\vv{\mu}_{\text{int}}$ is a continuous vector representing the chemical potential of $j$ interstitial components, while $\vv{\mu}_{\text{sub}}$ corresponds to the diffusional potential of $l$ substitutional components.
By using the interpolation function $h(\phi)$, the grand potential density of a two-phase system is expressed as
\begin{align}\label{eq:grand_inter}
 \Psi(\vv{\mu}_{\text{int}},\vv{\mu}_{\text{sub}})=\Psi^{\alpha}(\vv{\mu}_{\text{int}},\vv{\mu}_{\text{sub}})h(\phi)+ \Psi^{\beta}(\vv{\mu}_{\text{int}},\vv{\mu}_{\text{sub}})(1-h(\phi)).
\end{align}
The formulation of the grand chemical potential in Eqn.~\ref{eq:grandchem_site} yields the relation
\begin{align}\label{eq:grand_site}
 \frac{\partial \Psi^{\alpha}(\vv{\mu}_{\text{int}},\vv{\mu}_{\text{sub}})}{\partial \mu_{\text{int}:i}}=-y^{\alpha}_{\text{int}:i}
\end{align}
Therefore, based on Eqns.~\ref{eq:grand_inter} and ~\ref{eq:grand_site}, the continuity of the interstitial and substitutional components is expressed as
\begin{align}\label{eq:site_inter}
 y_{\text{int}:i}(\phi)&=y^{\alpha}_{\text{int}:i}h(\phi)+y^{\beta}_{\text{int}:i}(1-h(\phi))\\
 y_{\text{sub}:k}(\phi)&=y^{\alpha}_{\text{sub}:k}h(\phi)+y^{\beta}_{\text{sub}:k}(1-h(\phi)).
\end{align}
Since the aforementioned continuity is derived from the equilibrated chemical (diffusion) potential, in Eqn.~\ref{eq:equilibrium}, the bulk phases and interface are efficiently decoupled ~\cite{plapp2011unified}.

\subsection{Evolution equations}

The evolution of the phase field is phenomenologically governed by the minimization of the overall free energy, which includes chemical and elastic component.
Correspondingly, the phase-field evolution, ascertained from the variational derivative of the functional defined in Eqn.~\ref{eq:functional1}, reads
\begin{align}\label{eq:phase_field}
 \tau \epsilon \frac{\partial \phi}{\partial t}=-\frac{\delta F(\phi, \n \phi,\vv{S})}{\delta \phi}=-\left[\frac{\partial F}{\partial \phi}-\n\cdot\frac{\partial F}{\partial \n\phi} \right].
\end{align}
In the above Eqn.~\ref{eq:phase_field}, $\tau$ is the inverse of the interface mobility~\cite{nestler2005multicomponent}.
Consistency of the present formulation is recovering the capillarity has already been elucidated in Ref.~\cite{tschukin2019elasto}

For the evolution of the continuous variables associated with the elastic free-energy, the momentum balance equation is solved~\cite{schneider2015phase}.
However, during the phase transformation, which is predominantly governed by the concentration and inelastic eigenstrain, the system is in mechanical equilibrium.
Therefore, the evolution equation of the elastic variables is determined by solving
\begin{align}\label{eq:elastic}
\rho \frac{\partial^2 \vv{u}}{\partial t^2}=\n\cdot\vv{\sigma}=\vv{0},
\end{align}
where $\rho$ is the mass density and the divergence of stress is $(\n\cdot\vv{\sigma})_{i}=\partial \sigma_{ij}/\partial x_j$.

The evolution of the chemical (diffusional) potential is derived by considering the temporal evolution of the concentration.
The change in the concentration, in site fractions, of an interstitial component $i$ with time is written as
\begin{align}\label{eq:int_potential}
 \frac{\partial y_{\text{int}:i}}{\partial t} = \frac{\partial y_{\text{int}:i}(\mu_{\text{int}:i},\phi)}{\partial \mu_{\text{int}:i}}\frac{\partial \mu_{\text{int}:i}}{\partial t}+\frac{\partial y_{\text{int}:i}(\mu_{\text{int}:i},\phi)}{\partial \phi}\frac{\partial \phi}{\partial t}.
\end{align}
By incorporating the conventional formulation of the concentration evolution, the temporal change in the interstitial chemical potential is expressed as
\begin{align}\label{eq:ipotential_evolution}
 \frac{\partial \mu_{\text{int}:i}}{\partial t}&=\left\{ \underbrace{\n\cdot\left[\vv{M}_{\text{int}:ij}(\phi)\n \mu_{\text{int}:i}\right]}_{:=\frac{\partial y_{\text{int}:i}}{\partial t}}-\left[ y_{\text{int}:i}^{\alpha}\frac{\partial h(\phi)}{\partial t} + y_{\text{int}:i}^{\beta}\frac{\partial (1-h(\phi))}{\partial t} \right] \right\}\\ \nonumber
 &\left[ \vv{\mathcal{X}}^{\alpha}_{\text{int}:ij}h(\phi)+\vv{\mathcal{X}}^{\beta}_{\text{int}:ij}(1-h(\phi))\right]^{-1},
\end{align}
where $\vv{M}_{\text{int}:ij}(\phi)$ is the mobility which encapsulates the diffusivity of the component.
The mobility $\vv{M}_{\text{int}:ij}(\phi)$ is written as
\begin{align}\label{eq:int_mobility}
 \vv{M}_{\text{int}:ij}(\phi)= \vv{D}^{\alpha}_{\text{int}:ij}\vv{\mathcal{X}}^{\alpha}_{\text{int}:ij}h(\phi)+\vv{D}^{\beta}_{\text{int}:ij}\vv{\mathcal{X}}^{\beta}_{\text{int}:ij}(1-h(\phi)),
\end{align}
where $ \vv{D}^{\alpha}_{\text{int}:ij}$ and $\vv{D}^{\beta}_{\text{int}:ij}$ is the diffusivity matrix pertaining to phase-$\alpha$ and -$\beta$, respectively.
Furthermore, in Eqns.~\ref{eq:ipotential_evolution} and ~\ref{eq:int_mobility}, the susceptibility matrix which ensures constant diffusivity in a given phase is represented by $\vv{\mathcal{X}}^{\alpha}_{\text{int}:ij}$. 
The susceptibility can be ascertained from the well-known Darken factor~\cite{plapp2011unified,godreche1991solids,darken1948diffusion}.

Similar to the interstitial chemical-potential in Eqn.~\ref{eq:ipotential_evolution}, the evolution of the diffusion potential is derived, and expressed as
\begin{align}\label{eq:spotential_evolution}
 \frac{\partial \tilde{\mu}_{\text{sub}:k}}{\partial t}&=\left\{ \underbrace{\n\cdot\left[\vv{M}_{\text{sub}:kl}(\phi)\n \tilde{\mu}_{\text{sub}:k}\right]}_{:=\frac{\partial y_{\text{sub}:k}}{\partial t}}-\left[ y_{\text{sub}:k}^{\alpha}\frac{\partial h(\phi)}{\partial t} + y_{\text{sub}:k}^{\beta}\frac{\partial (1-h(\phi))}{\partial t} \right] \right\} \\ \nonumber
 &\left[ \vv{\mathcal{X}}^{\alpha}_{\text{sub}:kl}h(\phi)+\vv{\mathcal{X}}^{\beta}_{\text{sub}:kl}(1-h(\phi))\right]^{-1}.
\end{align}
The substitutional mobility ($\vv{M}_{\text{sub}:kl}(\phi)$) is of the form $\vv{M}_{\text{int}:ij}(\phi)$ in Eqn.~\ref{eq:int_mobility}.
However, the mobility $\vv{M}_{\text{sub}:kl}(\phi)$ involves diffusivity of substitutional components ($\vv{D}^{\alpha}_{\text{sub}:ij}$ and $\vv{D}^{\beta}_{\text{sub}:ij}$) and the corresponding susceptibility matrices, $\vv{\mathcal{X}}^{\alpha}_{\text{sub}:kl}$ and $\vv{\mathcal{X}}^{\beta}_{\text{sub}:kl}$.

Generally, a definite smooth-function(s) is adopted to interpolate the variables across the diffuse interface~\cite{boettinger2002phase,tschukin2019elasto}.
However, recently, it has been shown that employing phase field ($\phi$) as the interpolation function for the chemo-elastic model enhances the computational efficiency while being thermodynamically consistent~\cite{amos2018chemo}.
Therefore, for the present model, the phase field is considered as the interpolation function ($h(\phi)=\phi$).

\afterpage{\blankpage}

\newpage
\thispagestyle{empty}
\vspace*{8cm}
\begin{center}
 \Huge \textbf{Part III} \\
 \Huge \textbf{Volume-diffusion governed shape-instabilities}
\end{center}

\chapter{Fault migration in lamellar arrangement of phases}\label{chap:fault}

An \lq ideal\rq \thinspace lamellar microstructure, with the alternating arrangement of the constituent phases, is characterised by the morphology of its precipitate.
It is assumed that the distribution of the precipitate is homogeneous and extends uniformly across the matrix.
Correspondingly, such ideal configuration of the phases is considered to be stable~\cite{nakagawa1972stability,mazurski1995effect}.
However, as mentioned in introduction, curvature difference in these seemingly ideal microstructure is introduced in two ways.
One, due to non-uniformity in the cross-section of the precipitate and the other, owing to the presence of the termination.
Since the non-uniformity in the structure is relatively less prevalent in the solid-state metallic systems, the threat to the stability imposed by the termination.
Particularly, in the form of the discontinuous and finite \lq faulty\rq \thinspace structure.
Therefore, in this chapter the morphological changes introduced by the lamella-fault is extensively investigated.

\section{Analytical framework}\label{sec:ana}

\begin{figure}
    \centering
      \begin{tabular}{@{}c@{}}
      \includegraphics[width=0.25\textwidth]{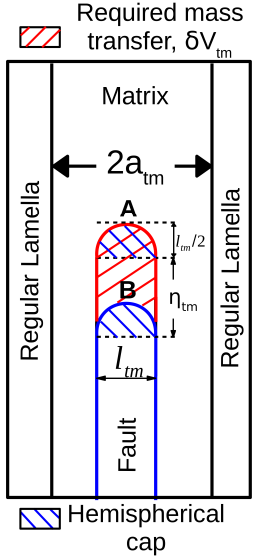}
    \end{tabular}
    \caption{ A representative depiction of the fault migration for a finite duration of $t_{\text{tm}}$ in discontinuous lamellar arrangement of phases.
    \label{fig:analytical}}
\end{figure}

The movement of the discontinuous structure accompanying the termination migration has already been schematically introduced in previous chapter.
An analytical relation for the recession of the faulty lamella can be derived by considering an elementary configuration of the phases as shown in Fig.~\ref{fig:analytical}.
Similar setup has been adopted in the previous works, wherein the evolution is investigated by involving \textit{ln cosh-cylinder} co-ordinate system~\cite{ho1974coarsening,tian1987kinetics}.

As illustrated in Fig.~\ref{fig:analytical}, the two-dimensional setup comprises of a discontinuous finite structure positioned between infinite regular-structures.
The interlamellar spacing, which is the distance between the precipitates, is represented by $a_{\text{tm}}$.
\nomenclature{$a_{\text{tm}}$}{Interlamellar spacing}%
The morphological change during the fault migration is characterised by the recession of the termination.
Therefore, for finite duration of $t_{\text{tm}}$, the fault migrates over a definite distance, $\eta_{\text{tm}}$.
For the present analytical treatment, magnitude of $t_{\text{tm}}$ is considered to be sufficiently small so that the regular structures remain visibly flat despite the mass transfer from the fault.
Furthermore, in order to quantify the curvature difference, the tip of the finite lamella is encapsulated with a hemisphere of radius $l_{\text{tm}}/2$, where $l_{\text{tm}}$ is the width of the precipitate.
\nomenclature{$l_{\text{tm}}$}{Width of the discontinuous precipitate}%

The presence of the termination introduces the curvature difference in the lamellar setup.
Consequently, a gradient in the chemical potential, proportional to the curvature difference, is induced which facilitates mass transfer from the tip to the surrounding flat surfaces.
If the area available for mass transfer from the tip is represented by $A_{\text{tm}}$, then the volume transferred during the finite interval $t_{\text{tm}}$ can be expressed as
\nomenclature{$A_{\text{tm}}$}{Area available for mass transfer at termination}%
\begin{align}\label{eq:time}
   \delta V_{\text{tm}} = |J A t|_{\text{tm}},
\end{align}
where $J_{\text{tm}}$ is the flux of the atoms per unit area.

The atomic flux $J_{\text{tm}}$ is ascertained from the velocity of the migrating species, $v_{\text{tm}}$.
\nomenclature{$J_{\text{tm}}$}{Atomic flux accompanying termination migration}%
As introduced in the early chapter, for a volume-diffusion governed transformation, the velocity of the atoms migrating in response to the potential gradient ($\nabla \mu$) induced by the curvature difference reads
\begin{align}\label{eq:V_d}
  v_{\text{tm}} = \frac{D V_m}{\kappa T} \nabla \mu,
\end{align}
where $D$ and $V_m$ correspond to the volume diffusivity and the molar volume, while $\kappa$ and $T$ are Boltzmann's constant and temperature, respectively.
Therefore, from Eqn.~\ref{eq:V_d}, the atomic flux is expressed as
\begin{align}\label{eq:flux_v}
  J_{\text{tm}} = v_{\text{tm}} c_{eq}^{\theta} =\frac{D V_m c_{eq}^{\theta}}{\kappa T} \nabla \mu,
\end{align}
where $c_{eq}^{\theta}$ is the equilibrium composition of the precipitate-$\theta$.
Eqn.~\ref{eq:flux_v} entails the assumption that $c_{eq}^{\theta}>>c_{eq}^{\alpha}$, where equilibrium concentration of matrix-$\alpha$ is represented by $c_{eq}^{\alpha}$.

\begin{figure}
    \centering
      \begin{tabular}{@{}c@{}}
      \includegraphics[width=0.25\textwidth]{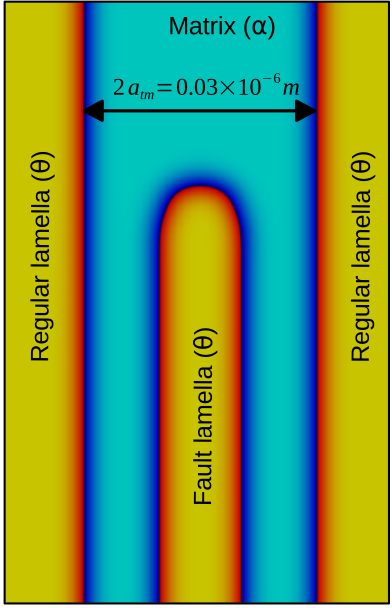}
    \end{tabular}
    \caption{ Two-dimensional domain setup considered for the simulation of the termination migration.
    \label{fig:pf_domain}}
\end{figure}

Owing to the elementary nature of the setup, the chemical potential gradient can be written as
\begin{align}\label{eq:delta_mu}
  \nabla \mu = \frac{\gamma V_m\left[\left(\frac{1}{R_1}+\frac{1}{R_2}\right)_{s}-\left(\frac{1}{R_1}+\frac{1}{R_2}\right)_{\tilde{s}}\right]}{a}=\frac{2\gamma V_m}{a_{\text{tm}} l_{\text{tm}}},
\end{align}
where, if the hemispherical tip and the flat surfaces in Fig.~\ref{fig:analytical} are respectively distinguished as source ($s$) and sinks ($\tilde{s}$), $R_1$  and $R_2$ are their corresponding principal radii of curvature~\cite{courtney1989shape,semiatin2005prediction,park2012mechanisms}. 
Based on Eqn.~\ref{eq:delta_mu}, the atomic flux is expressed as
\begin{align}\label{eq:flux_v}
  J_{\text{tm}} = \frac{2D V_m^2 \gamma c_{eq}^{\theta}}{\kappa T} \left ( \frac{1}{a_{\text{tm}} l_{\text{tm}}} \right).
\end{align}

Considering the morphology of the faulty structure, the tip area which exclusively acts as the source for the mass transfer is written as $A_{\text{tm}} = \frac{\pi l_{\text{tm}}}{2}$.
Furthermore, from the depiction in Fig.~\ref{fig:analytical}, the amount of the mass transferred during the interval $t_{\text{tm}}$ reads
\begin{align}\label{eq:volume}
  \delta V_{\text{tm}}= \eta_{\text{tm}} l_{\text{tm}}.
\end{align}
Therefore, by incorporating the expressions for the volume of the mass transferred $\delta V_{\text{tm}}$, the available area $A_{\text{tm}}$ and the flux $J_{\text{tm}}$, the finite duration can be written as
\begin{align}\label{eq:tau}
  \tau_t = \frac{\eta_{\text{tm}} l_{\text{tm}} \kappa T}{D  V_m^2\gamma c_{eq}^{\alpha}}a_{\text{tm}}.
\end{align}
The corollary of the above Eqn.~\ref{eq:tau} is that the time taken for the transfer of a definite volume $\delta V_{\text{tm}}$ remains unchanged with time.
Therefore, the volume fraction of the faulty structure should decrease linearly exhibiting no change in slope with time.

\begin{table*}
  \caption{Parameters used in the present work.} \label{tab:table_1}
 \centering
 \begin{tabular}{c c c c}
 
  Parameter & Value & Units \\ [0.5ex]
  \hline
  Temperature ($T$) & 973 & $\text{K}$ \\
  Interfacial Energy $(\gamma)$ & 0.49 & $\text{Jm}^{-2}$\\
  Diffusivity in ferrite($D_{\alpha}$) & 2 x $10^{-9}$ & $\text{m}^{2}\text{s}^{-1}$\\
  Diffusivity in cementite($D_{\theta}$) & 2 x $10^{-9}$   & $\text{m}^{2}\text{s}^{-1}$\\	
  Molar volume ($V_m$) & 7 x$10^{-6}$  & $\text{m}^{3}$/mole \\
  Equilibrium concentration($c^{\theta}_{eq}$) & 0.25 &  mole fraction\\
  Equilibrium concentration($c^{\alpha}_{eq}$) & 0.00067 &  mole fraction\\
  \end{tabular} 
  \end{table*} 

Existing studies~\cite{ho1974coarsening,tian1987kinetics} relate the interlamellar spacing $a_{\text{tm}}$ with the rate of migration through
\begin{align}\label{eq:termination_migration}
  \frac{\partial \eta_{\text{tm}}}{\partial t} \propto a_{\text{tm}}^{-2}.
\end{align}
The above expression indicates that for a given interlamellar spacing, the migration velocity of the lamella-fault is constant.
Correspondingly, similar to the volume ($\delta V_{\text{tm}}$), the tip of the faulty lamella recedes at a constant rate, exhibiting no dependence on time.
In order to verify these multifaceted claims of the present and previous theoretical studies, the termination migration is simulated by employing the phase-field approach.

\section{Two-dimensional simulation of the fault migration}

\subsection{Domain configuration}

A simulation domain as shown in Fig.~\ref{fig:pf_domain}, is set up to investigate the morphological changes accompanying the termination migration.
The evolution equations characterising the present phase-field model is discretised using finite difference approach on uniform numerical grid. 
These equations are subsequently solved by Euler's forward marching scheme.
An equidistant grid of dimension $\Delta \text{x}=\Delta \text{y}(=\Delta \text{z})=2 \times 10\p{-9} $ is considered for the entirety of the work.
The interface width is defined by assigning a  definite value to the length parameter $\epsilon$.
A width of approximately six grid points is established by affixing the length parameter at $\epsilon = 2.5 \times \Delta \text{x}$.
The simulations are efficiently performed by decomposing the larger domains into smaller fragments through Message Passing Interface (MPI) standard.
Additionally, the numerically accuracy of the numerical treatment is enhanced by non-dimensionalising the parameters involved.
The chemical equilibrium between the matrix-$\alpha$ and precipitate-$\theta$ is established by incorporating CALPHAD-based parameters.

In addition to the present analysis, the aforementioned numerical scheme is adopted for all the theoretical investigations elucidated in this work.
Moreover, the thermodynamical parameters adopted in all the numerical and analytical treatments are tabulated in Table.~\ref{tab:table_1}.
Since the present analyses exclusively examine the volume-diffusion governed morphological changes, the equal diffusivities are assigned to both the phases, matrix  and precipitate, to ensure the dominance of volume diffusion.

\subsection{Mechanism and kinetics of the evolution}

\begin{figure}
    \centering
      \begin{tabular}{@{}c@{}}
      \includegraphics[width=0.80\textwidth]{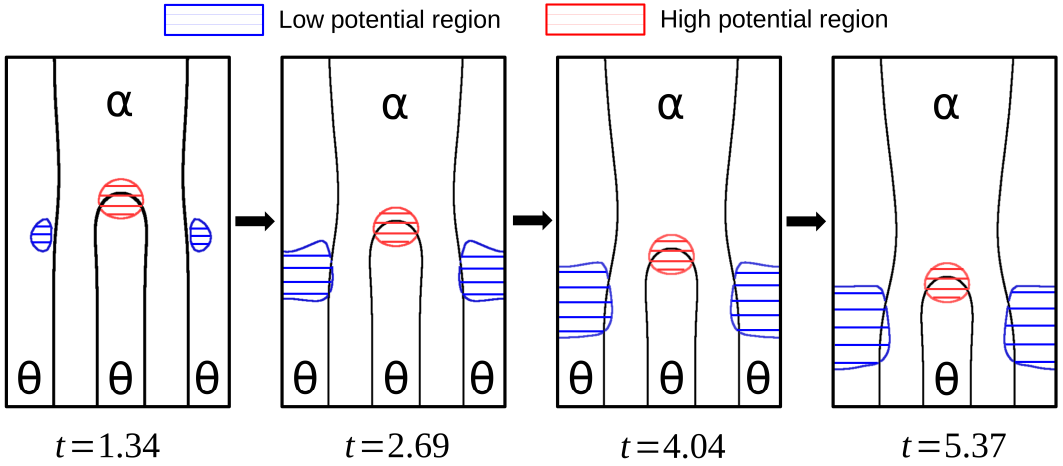}
    \end{tabular}
    \caption{ The morphological evolution in the lamellar precipitate accompanying the recession of the faulty structure. The distribution of the high and low  potential which govern the mass transfer is overlayed.
    \label{fig:term_mig1}}
\end{figure}

The recession exhibited by the discontinuous \lq faulty\rq \thinspace lamella sandwiched between the regular structures, which are separated by a distance of $2a_{\text{tm}}=0.003\times 10\p{-6}$m, is illustrated in Fig.~\ref{fig:term_mig1}.
This graphical depiction is an isoline representation wherein phase field of value $\phi_{\theta}=0.5$ is considered to schematically distinguish the phases by acting as the interface.
Depending on the distribution of the chemical potential, certain regions of the microstructural setup (Fig.~\ref{fig:pf_domain}) act as \lq source\rq \thinspace or \lq sink\rq \thinspace.
Regions wherein the potential is relatively higher become source, since they lose mass to the low potential regions, which are sinks.
Employing chemical potential as the dynamic variable enables the visualization of these sources and sinks.
In Fig.~\ref{fig:term_mig1}, these active regions are highlighted in red and blue.

The time $t$ involved in all forthcoming discussion is a dimensionless quantity which is normalised by the parameter $\tau'=\frac{a_{\text{tm}}\p3 \kappa T}{D \gamma V_m\p2 c\p{\theta}_{eq}}$.
\nomenclature{$\tau'$}{Time non-dimensionalising constant ($s^{-1}$)}%
This non-dimensionalising parameter is formulated based on the analytical framework elucidated in the early chapter and is coherent with existing studies~\cite{courtney1989shape,park2012mechanisms,semiatin2005prediction}.

Consistent with the existing consideration, owing the presence of the termination, the tip of the fault becomes the source of the evolution, as illustrated in Fig.~\ref{fig:term_mig1} at $t=1.34$.
Moreover, the flat nature of the neighboring structures introduces the curvature difference which consequently establishes a potential gradient.
Governed by the disparity in the chemical-potential distribution, the mass transfers from the high-potential source to the low-potential sinks.
While the mass transferred to the adjacent sinks disturbs the homogeneity of the regular lamellae, the tip position of the discontinuous precipitate progressively varies.
This recession of the faulty structure is referred to as the fault migration.

\begin{figure}
    \centering
      \begin{tabular}{@{}c@{}}
      \includegraphics[width=0.8\textwidth]{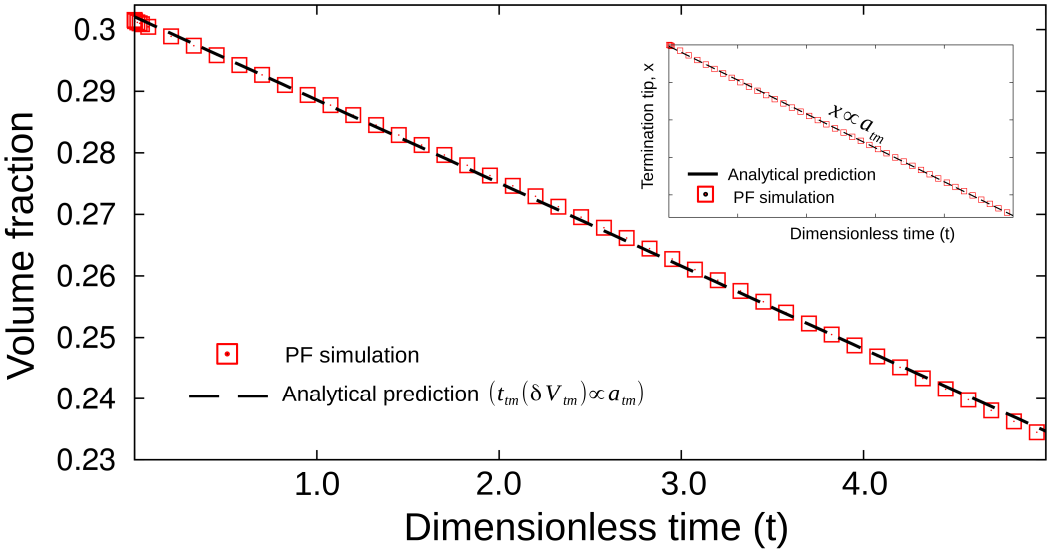}
    \end{tabular}
    \caption{ The temporal change in the volume of the discontinuous lamellar-fault. The position of the tip during the fault migration is monitored and included as subset.
    \label{fig:velocity_volume}}
\end{figure}

Pioneering numerical treatments of the volume-diffusion governed termination-migration, involving \textit{ln cosh-cylinder} co-ordinate system~\cite{ho1974coarsening,tian1987kinetics}, indicate that the profile of the receding fault tip is preserved during the transformation.
The morphological evolution rendered by the phase-field simulation, as shown in Fig.~\ref{fig:term_mig1}, adheres to this theoretical claim for the given interlamellar spacing, $2a_{\text{tm}}=0.003\times 10\p{-6}$m.

In order to verify the analytical relation in Eqn.~\ref{eq:tau}, the volume of the discontinuous precipitate is measured during the evolution.
The temporal change in the volume of the faulty structure is illustrated in Fig.~\ref{fig:velocity_volume}.
A linear monotonic decrease in the volume, as observed in Fig.~\ref{fig:velocity_volume}, asserts that the time taken to transfer a definite volume from the faulty to the regular precipitate is constant throughout the transformation.
This trend in the temporal change of the volume expounds the consistency of the evolution with Eqn.~\ref{eq:tau}.
Furthermore, to examine Eqn.~\ref{eq:termination_migration}, the position of the tip during the transformation is monitored and plotted as a subset of Fig.~\ref{fig:velocity_volume}.
Similar linear monotonic change in the tip position during the termination migration indicates that the recession rate, as expressed in Eqn.~\ref{eq:termination_migration}, remains constant for a definite interlamellar spacing.

\subsection{Influence of the interlamellar spacing}

\begin{figure}
    \centering
      \begin{tabular}{@{}c@{}}
      \includegraphics[width=0.8\textwidth]{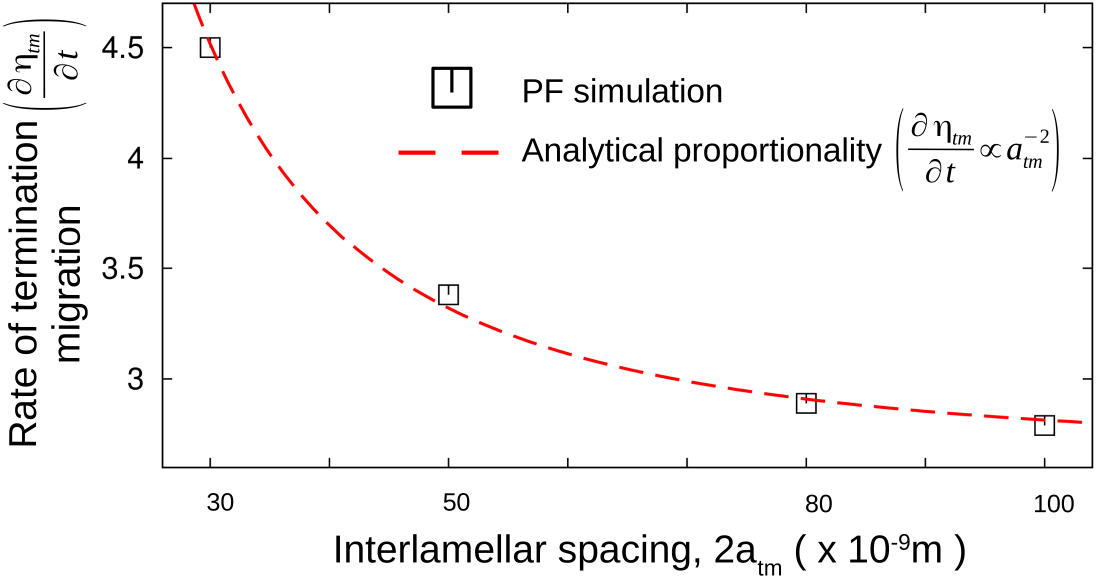}
    \end{tabular}
    \caption{ Change in the kinetics of the termination migration under the influence of the interlamellar spacing.
    \label{fig:a_velocity}}
\end{figure}

The spacing between the precipitates in a lamellar microstructure which result from the co-operative growth of the phases is influence by the thermodynamic conditions.
To understand the role of the interlamellar spacing on the evolution of the discontinuous precipitate, termination migration in different setups, akin to Fig.~\ref{fig:pf_domain}, with various spacing ($2a_{\text{tm}}$) is analysed.
The recession rate under different interlamellar spacing is ascertained and the graphically depicted in Fig.~\ref{fig:a_velocity}.
A convincing agreement with the analytical proportionality in Eqn.~\ref{eq:termination_migration} is evident in Fig.~\ref{fig:a_velocity}, which emphasises the consistency of the approach in simulating the curvature-governed evolution in a multiphase system.
The observed decrease in the migration rate with the increase in the interlamellar spacing is due to the proportional increase in the time taken for the mass transfer from the source to the neighbouring sinks.

\begin{figure}
    \centering
      \begin{tabular}{@{}c@{}}
      \includegraphics[width=0.72\textwidth]{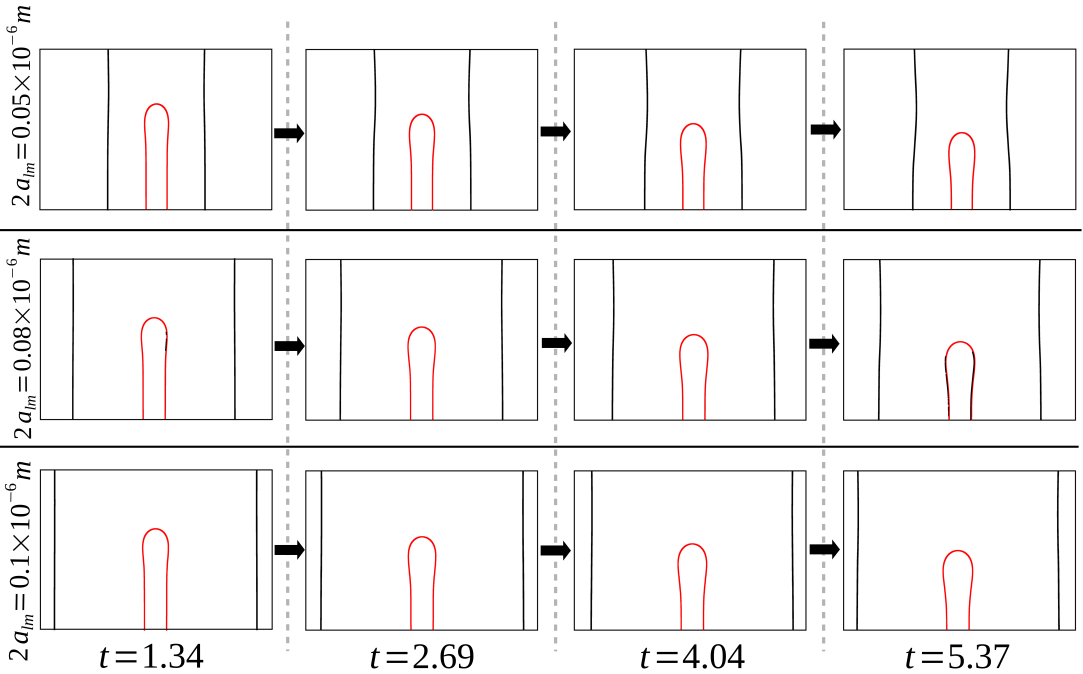}
    \end{tabular}
    \caption{ The influence of the spacing between the regular structures on the morphological evolution of the finite precipitate. Three different domain setups with varying spacing, $2a_{\text{tm}}=0.05\times 10\p{-6}$, ~$0.08\times 10\p{-6}$ and ~$0.1\times 10\p{-6}$m, are considered.
    \label{fig:term_diff}}
\end{figure}

\begin{figure}
    \centering
      \begin{tabular}{@{}c@{}}
      \includegraphics[width=0.8\textwidth]{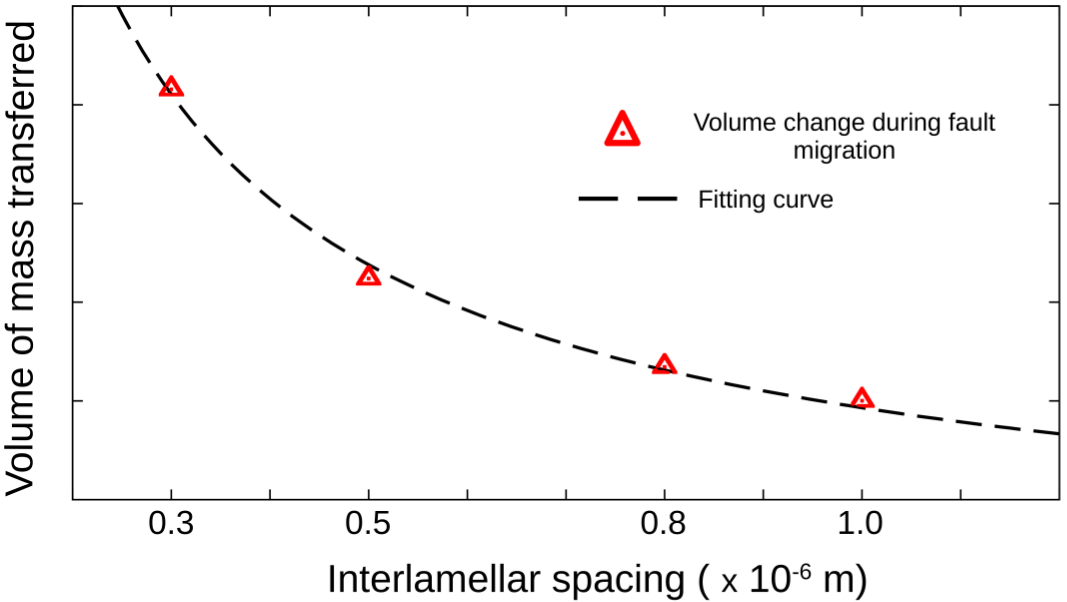}
    \end{tabular}
    \caption{ The influence of the interlamellar distances, $2a_{\text{tm}}=0.05\times 10\p{-6}$, $0.08\times 10\p{-6}$ and $0.1\times 10\p{-6}$m, on the amount of mass transferred to the surrounding lamellar structures.
    \label{fig:volume_spacing}}
\end{figure}

Isoline representation of the morphological evolution accompanying the termination migration, under different interlamellar spacing, is shown in Fig.~\ref{fig:term_diff}.
Independent of the spacing between the regular structures, the faulty lamella, which is distinguished in red, recedes.
However, it is interesting to note that, in contrast to Fig.~\ref{fig:term_mig1} where $2a_{\text{tm}}=0.003\times 10\p{-6}$m, the profile of the tip changes during the evolution in Fig.~\ref{fig:term_diff}.
In all the interlamellar conditions, $2a_{\text{tm}}=0.05\times 10\p{-6}$, $0.08\times 10\p{-6}$ and $0.1\times 10\p{-6}$m, the well-defined tip transforms to a ridge (or perturbation) during the migration.
Furthermore, it is noticeable in Fig.~\ref{fig:term_diff} that with the increase in the interlamellar spacing the size of the perturbation progressively intensifies.

The upsetting of the shape of the fault tip can be elucidated by considering the morphology of the discontinuous precipitate.
The shape of the lamella-fault comprises of the inherent flat surfaces leading up to the tip. 
Therefore, with increase in the interlamellar spacing, sinks begin to develop within the flat surfaces of the evolving precipitate, in addition to the surrounding regular structures.
In other words, as the neighbouring structures become more distant, the mass from the tip gets transferred to its own flat surface resulting in the distortion of its profile.
With increase in the interlamellar spacing the size of the ridge formed at the tip correspondingly increase.
Although this behaviour is apparently similar to the evolution of the finite structure~\cite{nichols1976spheroidization}, the mechanism of mass transfer from the fault to the regular structure introduces a significant difference.

The increase in the magnitude of the tip perturbation with the interlamellar spacing indicates that the amount of mass transferred progressively decreases with $2a$.
In order to investigate the influence of the interlamellar distance on the atomic flux to the adjacent structures, the volume change in the fault precipitate, for a definite duration,  is ascertained.
As shown in Fig.~\ref{fig:volume_spacing}, the amount of mass transferred to the infinite lamellae, during the termination migration, noticeably reduces with the spacing.
In other words, the characteristic mass transfer which differentiates the fault migration from the evolution of the isolate structure decreases with the distance $2a$.
Therefore, in a coarse microstructure, wherein the lamellar structures are considerably distant, the role of the neighbors in the evolution of the precipitates is minimal.
Correspondingly, the precipitate evolves like an isolated structure, despite the complexity of the microstructure.

\section{Three-dimensional simulation of the fault migration}

\begin{figure}
    \centering
      \begin{tabular}{@{}c@{}}
      \includegraphics[width=0.8\textwidth]{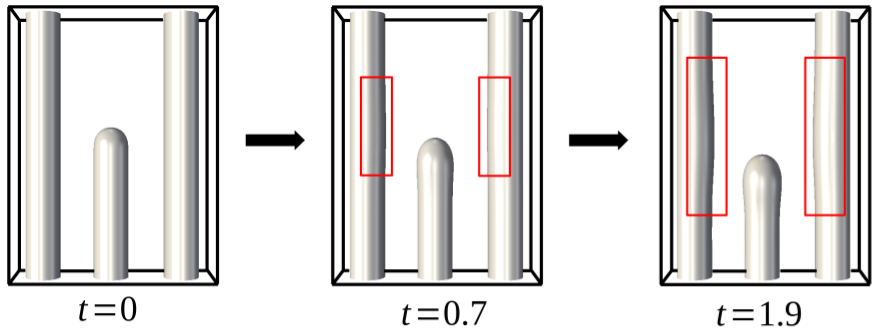}
    \end{tabular}
    \caption{ The migration of the three-dimensional faulty precipitate and the consequent change in the morphology of the regular lamellar structures.
    \label{fig:3d_temination}}
\end{figure}

The theoretical treatment of the fault migration, particularly governed by the volume diffusion, are often restricted to the two-dimension.
As discussed earlier, unlike the surface-diffusion governed evolution, since the analysis of the volume-diffusion dictated phenomenon demands the consideration of the entire domain, the 2-dimensional setup is computationally favoured.
However, owing to the efficiency of the present phase-field approach, the investigation of the fault migration is extended to the three-dimension.

Akin to the previous studies, a simple three-dimension system is configured to analyse the evolution of the discontinuous precipitate. 
The elementary setup with the interlamellar spacing of $2a_{\text{tm}}=0.04\times 10^{-6}$m, and the respective evolution of the lamella-fault is shown in Fig.~\ref{fig:3d_temination}.
Although the dimension of the system restricts the choice the minimum distance between the structures ($2a_{\text{tm}}$), the morphological changes broadly remain unaltered.
As illustrated in Fig.~\ref{fig:3d_temination}, the three-dimensional faulty precipitate recedes by losing it mass to the neighbouring regular structures, which in-turn disturbs the uniformity of its cross-section (highlighted in red).

\begin{figure}
    \centering
      \begin{tabular}{@{}c@{}}
      \includegraphics[width=0.8\textwidth]{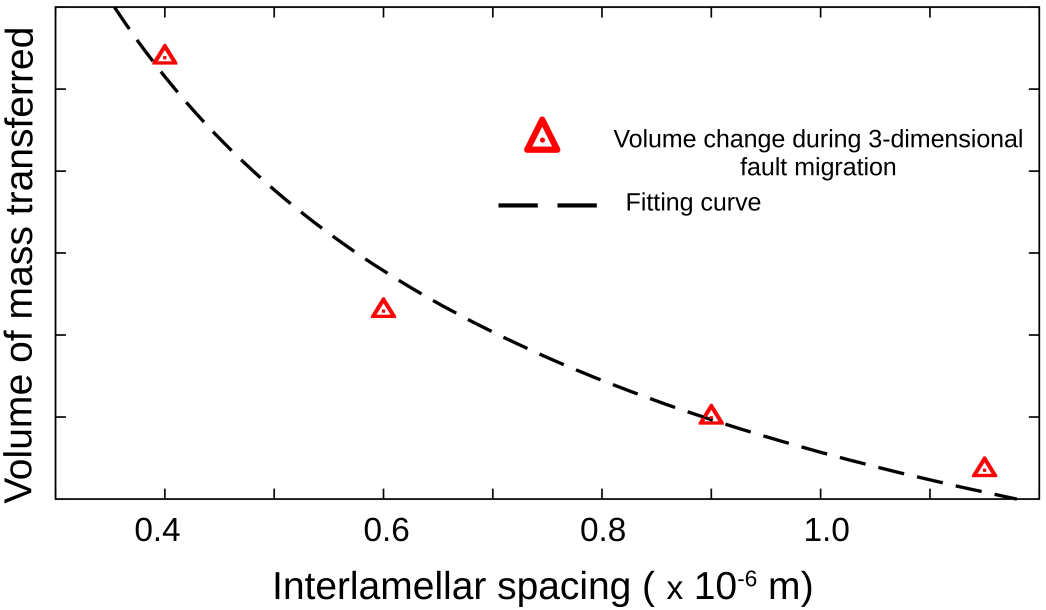}
    \end{tabular}
    \caption{The change in the volume of the three-dimensional discontinuous structure, for a given period of time, in different lamellar arrangements with varying interlamellar spacing.
    \label{fig:3dvolume_spacing}}
\end{figure}

Despite the proximity of the neighbouring precipitates, a slight deviation is introduced in the tip profile of the evolving precipitate in Fig.~\ref{fig:3d_temination}.
The change in the morphology of the tip during the recession suggests that two factors contribute to the evolution of the fault.
These factors, as observed in the coarse two-dimensional setup, include the volume lost by the fault to the regular lamellar and the mass transferred to its own low potential region (flat surface).
In order to quantify the mass transferred to the regular structures in relation to the interlamellar spacing, the volume change accompanying the evolution under different setups are presented in Fig.~\ref{fig:3dvolume_spacing}.
Similar to the two-dimensional migration, as shown in Fig.~\ref{fig:volume_spacing}, the amount of the mass lost to the regular precipitate progressively decreases with increase in the distance.
Moreover, Fig.~\ref{fig:3dvolume_spacing} indicates that, in the three-dimensional systems, the mass gained by the lamellar precipitates drastically reduces with increase in spacing.
Therefore, it is conceivable that beyond a critical distance the influence of the neighbours are negligible and the precipitate evolves like as isolated structure governed by the inherent curvature-difference in its shape.

\begin{figure}
    \centering
      \begin{tabular}{@{}c@{}}
      \includegraphics[width=0.8\textwidth]{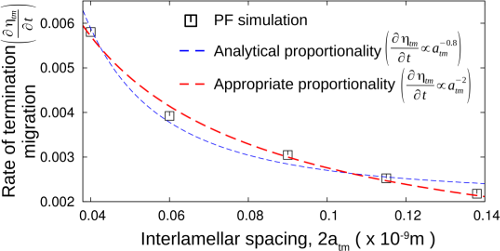}
    \end{tabular}
    \caption{ The role of the distance separating the regular structures ($2a_{\text{tm}}$) in governing the kinetics of the fault migration in three-dimensional setup.
    \label{fig:3velocity_vs_spacing}}
\end{figure}

Of the two factors contributing to the recession of the faulty structure, the prominent role is assumed by the mass transferred from the discontinuous to the regular structures.
Since, as shown in the Fig.~\ref{fig:3dvolume_spacing}, the mass lost by the fault precipitate to the surrounding lamellae decreases significantly with the increase in spacing, the migration rate noticeably reduces with interlamellar distance.
In Fig.~\ref{fig:3velocity_vs_spacing}, the influence of the spacing on the kinetics of the evolution is illustrated.
Consistent with the volume-change in the discontinuous precipitate, the recession kinetics reduces substantially with increase in the interlamellar spacing ($2a_{\text{tm}}$).
The relation in Eqn.~\ref{eq:termination_migration} is included in Fig.~\ref{fig:3velocity_vs_spacing}.
However, it is evident that a noticeable deviation from this relation is observed in Fig.~\ref{fig:3velocity_vs_spacing}.
The lack of complete agreement with the predicted relation can be attributed to two critical components of the present investigation.
One, unlike the analytical treatment, the present work considers the fault migration in a three-dimensional setup.
Two, the mass transfer from the fault tip to its own flat surfaces is not explicitly considered in the existing studies.
These two factor introduce the deviation from the analytical relation.
By fitting the recession rates resulting from the simulation, a different relation is derived which can be expressed as
\begin{align}\label{eq:termination_migration1}
  \frac{\partial \eta_{\text{tm}}}{\partial t} \propto a_{\text{tm}}^{-0.8}.
\end{align}
The above expression includes mass transfer to the neighbouring precipitates and the inherent flat surfaces during the fault migration in the three-dimensional system.

\section{Conclusion}

The lamellar microstructure is plainly described as a alternating arrangement of the seemingly continuous phases, precipitate and matrix.
During the phase transformation, which yields the lamellar microstructure, the continuity of the precipitate is often disturbed due to factors like dislocation density~\cite{bramfitt1973transmission}.
The discontinuity, in an otherwise ideal arrangement of the phases, introduces a curvature difference owing to the presence of the termination.
Consequently, a difference in the chemical potential distribution is introduced which actuates a mass transfer from the region of high potential to the low potential.
Since, owing to the curvature, the region of low potential is assumed by the flat surface of the surrounding continuous structures, the evolution of the faulty discontinuous precipitate is analysed in relation to its neighbours.
In the present analysis, the morphological evolution of the faulty precipitate, referred to as fault migration, is numerically analysed using phase-field approach in both two- and three-dimension.
This investigation affirms that the recession of the discontinuous structure is predominantly governed by the mass transferred to the adjacent regular precipitates.
Furthermore, it is shown the volume lost by the evolving precipitate, for a definite duration, decreases with the interlamellar spacing.
Consequently, a proportionate decrease in the kinetics of the migration is observed with increase in the distance between the precipitates.

Present analysis unravel that the profile of the fault tip is preserved during the migration exclusively in the finer lamellar microstructure, wherein the interlamllar spacing is small.
With increase in the spacing between the regular structures, an additional mass transfer path is introduced.
This additional mass transfer from the fault tip to its own flat surface occurs when the lamellar structures are separated by larger distance.
The inherent accumulation of the mass in the flat surface of the discontinuous structure disturbs the profile of the tip and introduces a ridge (or perturbation).
The size of this ridge increases with the interlamellar spacing indicating that the evolution is predominantly governed by the inherent mass diffusion in widely spaced system.
In other words, owing to the increased distance between the regular structures, the discontinuous precipitate transforms like an isolates structure in a coarse microstructure.
Ultimately, the vital corollary of this investigation is that, even though the microstructure typically comprises of numerous precipitate distributed in the matrix, the understanding of the evolution of an isolated structure is pivotal, since the role of the neighbours is significantly influenced by its distance.
Correspondingly, the analysis of the shape instabilities in the upcoming chapters are largely confined to the isolate structure.

\chapter{Cylinderization of the ribbon-like structures}\label{chap:ribbon}

Conventional micrographs depict the lamellar microstructure as the alternating arrangement of the constituent phases~\cite{zhang2006microstructural,zherebtsov2011spheroidization}.
These two-dimensional images, however, do not render a comprehensive description of the precipitate morphology.
During the phase transformation, the shape adopted by the precipitate in a given lamellar microstructure is governed by several factors including the chemical makeup of the material~\cite{loretto1998influence}, transformation temperature~\cite{ridley1984review} and microscopic homogeneity~\cite{zelin2002microstructure}.
While the eutectic transformation inherently yields rod-like precipitates~\cite{mesa2011microstructure}, the cementite in the pearlite exhibit plate morphology~\cite{kral2000three}.
In some cases, it has been identified that the precipitate assumes a ribbon-like morphology wherein the length seemingly extends infinitely, while the cross-section is definite~\cite{go1991strengthening}.
Moreover, the ribbon-like shapes (or semi-infinite plates) form an integral segment of the morphological evolution of complex structures~\cite{sandim2004annealing,hardwick1993effect}.
A schematic representation of the shape change observed during the transformation of the semi-infinite plates has already been introduced in Fig.~\ref{fig:cylinderization}.
This morphological evolution, referred to as the cylinderization~\cite{kampe1989shape}, is extensively analysed through the phase-field simulations in the present chapter.

\section{Domain setup}

\begin{figure}
    \centering
      \begin{tabular}{@{}c@{}}
      \includegraphics[width=0.70\textwidth]{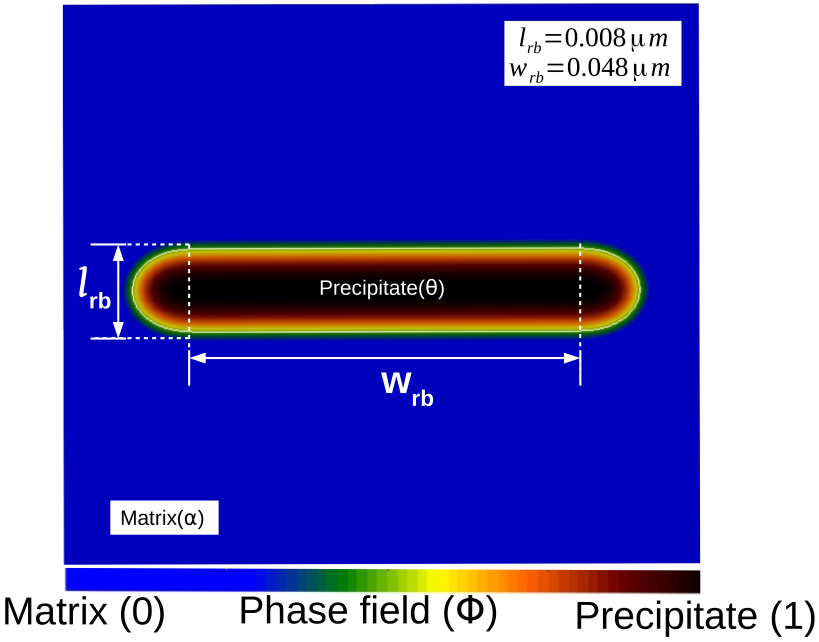}
    \end{tabular}
    \caption{The domain setup considered for the investigation of the cylinderization. The precipitate-$\theta$ of aspect ratio ($w_\text{rb}/l_\text{rb}=$) 6 is embedded onto the matrix $\alpha$.
    \label{fig:1a}}
\end{figure}

Numerical investigations of the fault migration, in the previous chapter, indicate that the influence of the neighbouring structures decreases significantly with increase in the distance separating the precipitates.
Moreover, since an inherent curvature-difference is often introduced in an isolate structure owing to its shape, an understanding on the stability of the individual shapes is critical for describing the stability of the microstructure.
Therefore, for the present analysis, a two-dimensional setup as shown in Fig.~\ref{fig:1a} is considered, wherein precipitate ($\theta$) is in chemical equilibrium with the matrix.
The observed morphology of the precipitate is the cross-section of the infinite ribbon-like structure.

A pivotal difference in the analytical treatment of the stability of the infinite and finite structure lies in the parameters pertinent to the morphology, which govern the transformation.
In the investigation of the infinite structure, the geometrical nature of the introduced perturbation influences the evolution of the precipitate, whereas the ratio of the width to thickness, called the aspect ratio, is the governing shape-factor for finite structures.
In addition to the numerical studies, this shape factor has been realised in the experimental investigations as well~\cite{gupta1978instability,santala2008rayleigh}.
Correspondingly, the thickness of the precipitate, shown in Fig.~\ref{fig:1a}, is set at $l_\text{rb} = 0.008 \times \mu$m~\cite{arruabarrena2014influence} and the width $w_\text{rb}$ is varied to accommodate the desired aspect ratio.
\nomenclature{$l_\text{rb}$}{Thickness of the ribbon-like structure}%
\nomenclature{$w_\text{rb}$}{Width of the ribbon-like structure}%
Furthermore, in the spirit of the existing studies~\cite{courtney1989shape}, hemispherical caps are augmented on both the ends of the plate for the elegant description of the curvature difference.
The radius of the hemispherical ends is affixed at $l_\text{rb}/2$.
The thermodynamical parameters involved in the simulations are tabulated in Table~\ref{tab:table_1}.
No explicit distinction is made between the different modes of mass transfer, surface and volume diffusion.
Additionally, the equal diffusivities in both the matrix and precipitate ensures that the volume diffusion is the operating mode of diffusion.

\section{Cylinderization of capped ribbon-like structure}

\subsection{Morphological evolution}

\begin{figure}
    \centering
      \begin{tabular}{@{}c@{}}
      \includegraphics[width=0.4\textwidth]{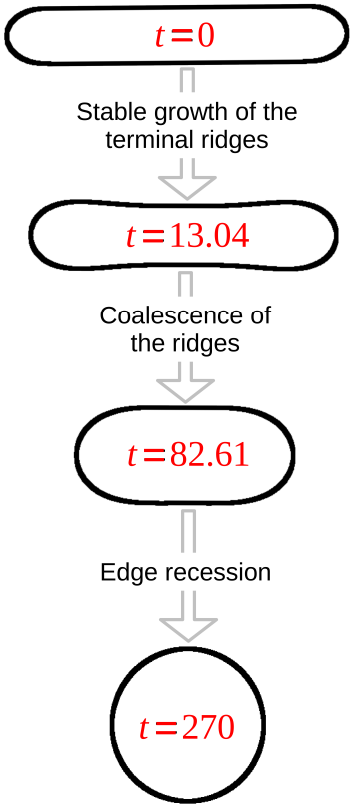}
    \end{tabular}
    \caption{An isoline representation ($\phi = 0.5$) of the temporal change in the morphology of the precipitate of aspect ratio 6 during curvature-driven transformation.
    \label{fig:2}}
\end{figure}

Fig.~\ref{fig:2} shows the morphological changes accompanying the cylinderization of the precipitate-$\theta$ of aspect ratio ($w_\text{rb}/l_\text{rb}=$) 6.
Akin to the fault migration, the presence of termination(s) introduces a curvature difference in the system.
However, unlike the multiphase setup wherein the curvature difference established between the faulty and surrounding regular structures predominantly dictates the fault migration, in cylinderization, the driving force established within the individual precipitate exclusively governs the evolution.

As shown in Fig.~\ref{fig:2} at $t=0$, where $t$ is the non-dimensionalised time, the morphology of the precipitate encompasses two terminations owing to the finitude of its cross-section.
The presence of these terminations introduces a difference in curvature between the capped ends and the flat surface of the precipitate.
Owing to this curvature difference, a gradient in the chemical potential is induced, in accordance with the Gibbs-Thomson relation.
Governed by the potential gradient, the mass begins to transfer from the region of the high potential to the low potential.
In other words, the curved ends of the plates act as the source for the mass transfer.
Furthermore, since the  atomic flux prefers shortest diffusion path, the flat surfaces adjacent to the capped ends become the sink.
Consequently, the mass from the termination gets deposited on the immediate flat regions which leads to the formation of the terminal perturbations or ridges.
The formation of the ridges, as shown in  Fig.~\ref{fig:2} at $t=13.04$, disturbs the morphology of the terminations.
Despite the change in the termination profile, the ends continue to lose mass to the adjacent low potential region.
The progressive mass transfer, governed by the potential gradient, facilitates the stable growth of the ridges.

The stable growth of the ridges, as illustrated in Fig.~\ref{fig:2}, occurs at the expense of the inherent flat surfaces.
Therefore, the corresponding morphological evolution appears as the shrinking of the precipitate.
With time, the continually growing terminal perturbations coalesce and transform the cross-section of the precipitate into capsule-like structure, as shown in Fig.~\ref{fig:2} at $t=82.61$.
The capsule morphology resembles the initial configuration of the precipitate with significantly low aspect-ratio.
Therefore, the subsequent temporal change follows the mechanism identical to evolution of the precipitate in the initial stages of the cylinderization.
However, since the amount of the flat surface resulting from the coalescence of the ridges are visibly much lower, the mass gets deposited in the central region of the capsule-like precipitate instead of forming the secondary longitudinal perturbations.
Owing to the mass transfer to the central region, the precipitate assumes a elliptical structure.
The elliptical precipitate, governed by the potential gradient induced by the inherent curvature-difference, ultimately transforms into a spherical structure, as shown in Fig.~\ref{fig:2} at $t=270$.
The entire scheme of the morphological changes leading to the cylinderization can be summarized by considering the incremental shift in the low potential region.
Particularly, from its terminal proximity in the initial stages to the central position following the coalescence of the ridges.

\subsection{Analytical investigation of the cylinderization kinetics}

\subsubsection{Analytical framework}

The pioneering attempt to analytically investigate the kinetics of the cylinderization is attributed to the Courtney and Kampe~\cite{courtney1989shape}.
The approach begins by identifying three specific stages in the evolution, namely initial, midpoint and final. 
The driving force at each of these stages of evolution is ascertained, in a simplified form, from the corresponding geometry of the precipitate.
In the initial stages, since the geometry of the precipitate, particularly the curvature difference, is numerically well-defined, the driving force is consistently expressed.
Moreover, the evolution ends with the complete elimination of the curvature difference, and correspondingly, the driving force at the final stage is fittingly considered to be zero.
Despite these appropriate treatment of the driving force, the curvature difference at the midpoint of the cylinderization is assumed in its entirety, owing to the lack of a complete description of the shape-change.
The geometry of the precipitate at the midpoint is assumed to be the average of the initial and final state, and the concentration (potential) gradient is defined accordingly.

The phase-field model adopted to numerically examine the cylinderization considers chemical potential as the dynamic variable.
Since the gradient of the chemical potential which, in accordance with the curvature difference, enables the mass transfer that ultimately results in the cylinderization, the temporal evolution of the driving force can be monitored directly.
Accordingly, the difference in the highest and lowest potential at a given time of the transformation, $\Delta \mu(x,t) = \mu^+(x)|_t - \mu^-(x)|_t$, is calculated from the simulation.
The temporal change in the chemical-potential difference ($\Delta \mu(x,t)$) is subsequently ascertained.
\nomenclature{$\Delta \mu(x,t)$}{Difference in the chemical potential}%

\begin{figure}
    \centering
      \begin{tabular}{@{}c@{}}
      \includegraphics[width=0.7\textwidth]{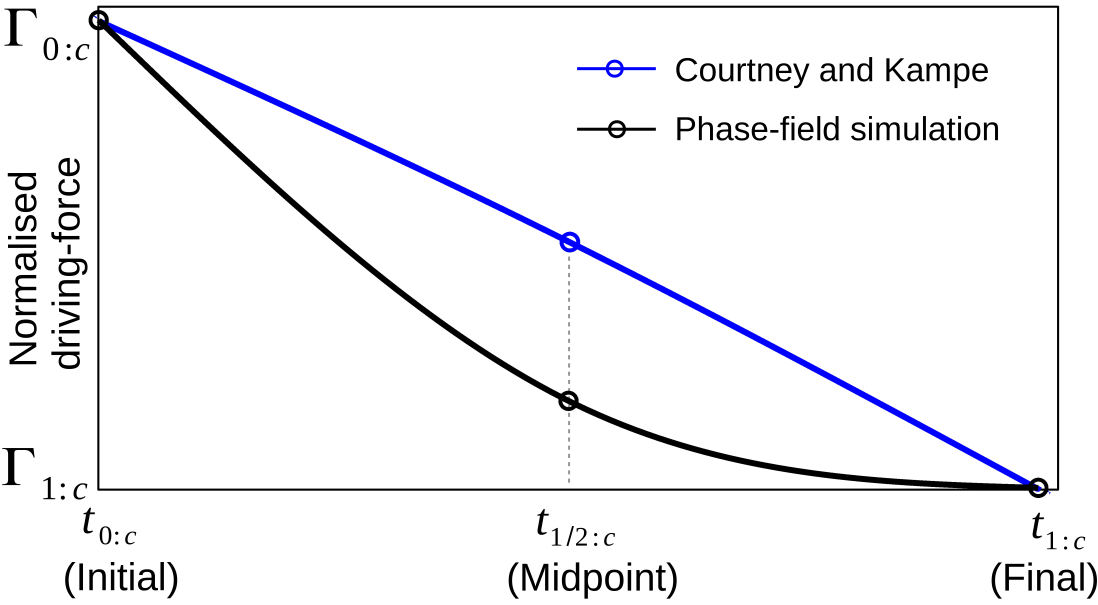}
    \end{tabular}
    \caption{The temporal change in the driving force which governs the simulated transformation is compared with the analytically calculated driving force.
    \label{fig:3}}
\end{figure}

The evolution of the potential difference accompanying cylinderization is plotted along with the analytically formulated driving force in Fig.~\ref{fig:3}.
Since the analytical calculations are confined to the three specific points, initial, midpoint and final, the curve is numerically fitted for this representation.
Furthermore, both the driving force and time are normalised for the comparative depiction in Fig.~\ref{fig:3}.

The graphical representation of the temporal change in the driving force, illustrated in Fig.~\ref{fig:3}, unravels that the existing theoretical treatment predicts a nearly linear decrease in the driving force with time.
This un-physical evolution of the curvature difference (driving force) can be attributed to the geometrical assumptions  made concerning the morphology of the precipitate at the midpoint of the cylinderization.
On the other hand, it is evident from Fig.~\ref{fig:3} that the phase-field simulation yields relatively more coherent evolution of the potential difference.
Therefore, the analytical approach is revisited with the aide of the present simulations.

\subsubsection{Revisiting the theoretical treatment}

The morphology of the precipitate, of aspect ratio ($w_\text{rb}/l_\text{rb}=$) 6, at the initial ($t_{0:\text{c}}$), midpoint ($t_{\frac{1}{2}:\text{c}}$) and final ($t_{1:\text{c}}$) stage of the cylinderization is illustrated in Fig.~\ref{fig:fig_4}.
In the present analysis, the final stage ($t_{1:\text{c}}$) is characterised by the aspect ratio of the precipitate, such that $t_{1:\text{c}}$ is time taken to complete the cylinderization ($w_\text{rb}/l_\text{rb}=1$).
As predicted by the existing studies, the precipitate assumes a elliptical shape at the midpoint of the transformation.

Assuming that the driving force encompasses the instantaneous morphology of the precipitate, the overall driving force of the cylinderization, $\bar{\Gamma}_{\text{c}}$, can be expressed as 
\nomenclature{$\bar{\Gamma}_{\text{c}}$}{Overall driving-force determining the cylinderization rate}%
\begin{align}\label{eq:driving_force1}
 \bar{\Gamma}_{\text{c}}= \frac{1}{3}\left[\Gamma_{0:\text{c}}\left(A_{0:\text{c}},\left(\frac{\delta c}{\delta x}\right)_{0:\text{c}}\right)
+\Gamma_{\frac{1}{2}:\text{c}}\left(A_{\frac{1}{2}:\text{c}},\left(\frac{\delta c}{\delta x}\right)_{\frac{1}{2}:\text{c}}\right)+\Gamma_{1:\text{c}}\left(A_{1:\text{c}},\left(\frac{\delta c}{\delta x}\right)_{1:\text{c}}\right)\right],
\end{align}
where $\Gamma_{0:\text{c}}$, $\Gamma_{\frac{1}{2}:\text{c}}$ and $\Gamma_{1:\text{c}}$ are the respective driving forces are initial, midpoint and final stage of the transformation.
\nomenclature{$\Gamma_{0:\text{c}}$}{Driving force at initial stages of cylinderization}%
\nomenclature{$\Gamma_{\frac{1}{2}:\text{c}}$}{Driving force at midpoint of cylinderization}%
\nomenclature{$\Gamma_{1:\text{c}}$}{Driving force at final stages of cylinderization}%
The driving force which is expressed as the function of $A_{i:\text{c}}$, area available for the mass transfer,  difference in the equilibrium concentration induced by the curvature ($\delta c_{i:\text{c}}$) and the diffusion distance ($\delta x_{i:\text{c}}$) indicates the transitory nature of $\Gamma_{i}$, with $i \in \{ 0, \frac{1}{2}, 1\}$.
\nomenclature{$A_{i:\text{c}}$}{Area available for diffusion enabling cylinderization}%
\nomenclature{$\delta c_{i:\text{c}}$}{Difference introduced by ribbon shape on equilibrium concentration}%
\nomenclature{$\delta x_{i:\text{c}}$}{Diffusion distance for cylinderization}%
Furthermore, separating the concentration gradient as $\delta c_{i:\text{c}}$ and $\delta x_{i:\text{c}}$ renders a straightforward solution for the driving force. 
Since the cylinderization halts with the driving force becoming zero, $\Gamma_{1:\text{c}} \to 0$, Eqn.~\ref{eq:driving_force1} becomes
\begin{align}\label{eq:driving_force1_2}
 \bar{\Gamma}_{\text{c}}={\frac{1}{3}\left[\Gamma_{0:\text{c}}\left(A_{0:\text{c}},\left(\frac{\delta c}{\delta x}\right)_{0:\text{c}}\right) +\Gamma_{\frac{1}{2}}\left(A_{\frac{1}{2}:\text{c}},\left(\frac{\delta c}{\delta x}\right)_{\frac{1}{2}:\text{c}}\right)\right]}.
\end{align}

\begin{figure}
    \centering
      \begin{tabular}{@{}c@{}}
      \includegraphics[width=0.9\textwidth]{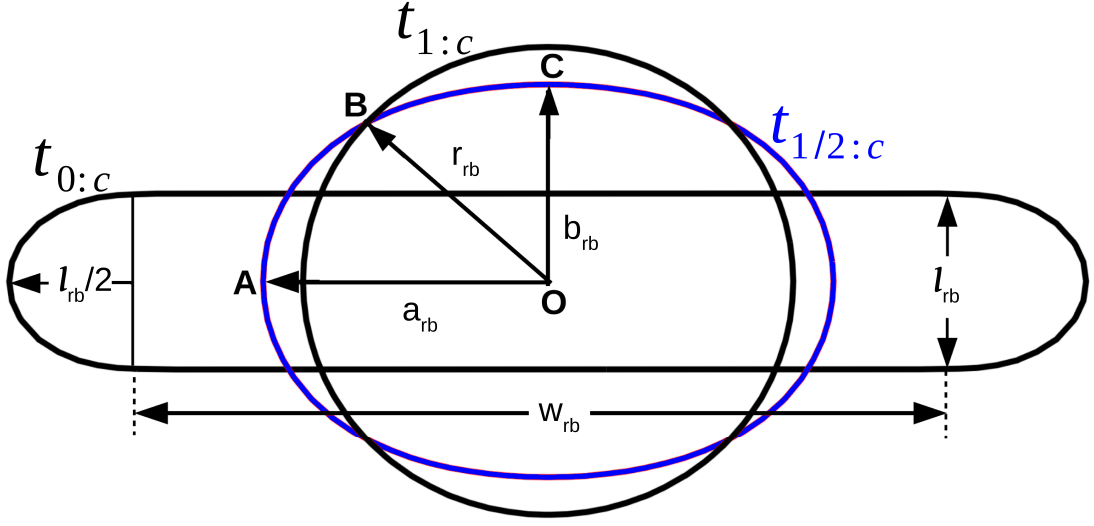}
    \end{tabular}
    \caption{ An isoline representation of the precipitate morphology at the initial ($t_{0:\text{c}}$), midpoint ($t_{\frac{1}{2}:\text{c}}$) and final ($t_{1:\text{c}}$) stage of the cylinderization simulated using the phase-field model.
    \label{fig:fig_4}}
\end{figure}

Based on the termination profile, the area available for the diffusion at the initial stages of the evolution can be considered as $A_{0:\text{c}} \approx n_d(\lambda \frac{l_\text{rb}}{2})$.
The pre-factor $n_d$ in the area $A_{0:\text{c}}$ refers to the number of diffusion paths, which is defined as $n_d=4$, owing to the symmetry of the present configuration.
Furthermore, the seemingly infinite length of the plate, which is along viewing angle of Fig.~\ref{fig:1a} is assumed to be $\lambda$.
In accordance with the Gibbs-Thomson relation, the change in the equilibrium concentration introduced by the curvature is written as
\begin{align}\label{eq:c_0}
 \delta c_{i:\text{c}}={\frac{c^{\theta}_{eq}\gamma V_\text{m}}{\kappa T}\Big[\Big(\frac{1}{R_1}+\frac{1}{R_2}\Big)_{\text{sources}}-\Big(\frac{1}{R_3}+\frac{1}{R_4}\Big)_{\text{sinks}}\Big]},
\end{align}
where $R_1$, $R_2$, $R_3$ and $R_4$ are the principal radii of high-potential sources and low-potential sinks, respectively.
The thermodynamical constants involved in Eqn.~\ref{eq:c_0} are interfacial energy density ($\gamma$), molar volume ($V_\text{m}$), Boltzmann's constant ($\kappa$) and absolute temperature ($T$).
At the initial stages the principal radius of the source $R_1$ is well-defined owing to the hemispherical cap.
Furthermore, since the sinks are flat surfaces, $\frac{1}{R_3}+\frac{1}{R_4} = 0$.
Therefore, the concentration difference at the initial stage of the cylinderization reads
\begin{align}\label{eq:c0_0}
 \delta c_{0:\text{c}}=\frac{2c^{\theta}_{eq}\gamma V_\text{m}}{\kappa Tl_\text{rb}}.
\end{align}
From the morphology of the precipitate at $t_0$, the diffusion length can be written as
\begin{align}\label{eq:x0_0}
 \delta x_{0:\text{c}}=\frac{1}{4}\left(2w+\pi l_\text{rb}-4r_{\text{rb}} \right),
\end{align}
where $r_{\text{rb}}$ is the radius of the rod emerging from the cylinderization process.
\nomenclature{$r_{\text{rb}}$}{Radius of cylinder}%
The driving force based on the diffusion area and the concentration gradient is expressed as
\begin{align}\label{eq:x0_1}
 \Gamma_{i:\text{c}}\left(A_{i:\text{c}},\left(\frac{\delta c}{\delta x}\right)_{i:\text{c}}\right)=D V_\text{m} A_{i:\text{c}} \left(\frac{\delta c_{i:\text{c}}}{\delta x_{i:\text{c}}}\right).
\end{align}
Therefore, by substituting Eqns.~\ref{eq:c0_0} and ~\ref{eq:x0_0} and incorporating the area available for volume diffusion in Eqn.~\ref{eq:x0_1}, the driving force at the initial stage of the cylinderization is determined by
 \begin{align}\label{eq:dr_0}
 \Gamma_{0:\text{c}} =\frac{DV_\text{m}^{2}c^{\theta}_{eq}\gamma}{\kappa T}\frac{16\lambda }{l_\text{rb}(2w_\text{rb}+\pi l_\text{rb}-4r_{\text{rb}})}.
\end{align}

Existing analysis assumes that the geometry of the structure at the midpoint of the cylinderization is the average of the dimension at the initial and final stage~\cite{courtney1989shape}.
Under such consideration, the distance relating the concentration gradient is expressed as
\begin{align}\label{eq:x_half_old}
 \delta x_{\frac{1}{2}:\text{c}}=\frac{l_\text{rb}}{4}\left[ \frac{w_\text{rb}}{l_\text{rb}} + (\pi - r_{\text{rb}})\frac{r_{\text{rb}}}{l_\text{rb}} \right],
\end{align}
and correspondingly, the concentration difference reads
\begin{align}\label{eq:c_half_old}
 \delta c_{\frac{1}{2}:\text{c}}=\frac{c^{\theta}_{eq} \gamma V_\text{m}}{\kappa T r_{\text{rb}}}.
\end{align}
However, Fig.~\ref{fig:3} indicates that the aforementioned assumption, and the resulting expressions in Eqns.~\ref{eq:x_half_old} and ~\ref{eq:c_half_old}, render a thermodynamically inaccurate description of the temporal evolution of the driving force.
Particularly, owing to the over-estimation of the driving force at the midpoint, the progressive decrease in the curvature difference appears to be linear.
In order to rectify this misappropriation, the geometrical description of the precipitate at the midpoint is recovered from the phase-field simulations.

\begin{figure}
\centering
   \begin{subfigure}[b]{1.0\textwidth}
   \includegraphics[width=1.0\linewidth]{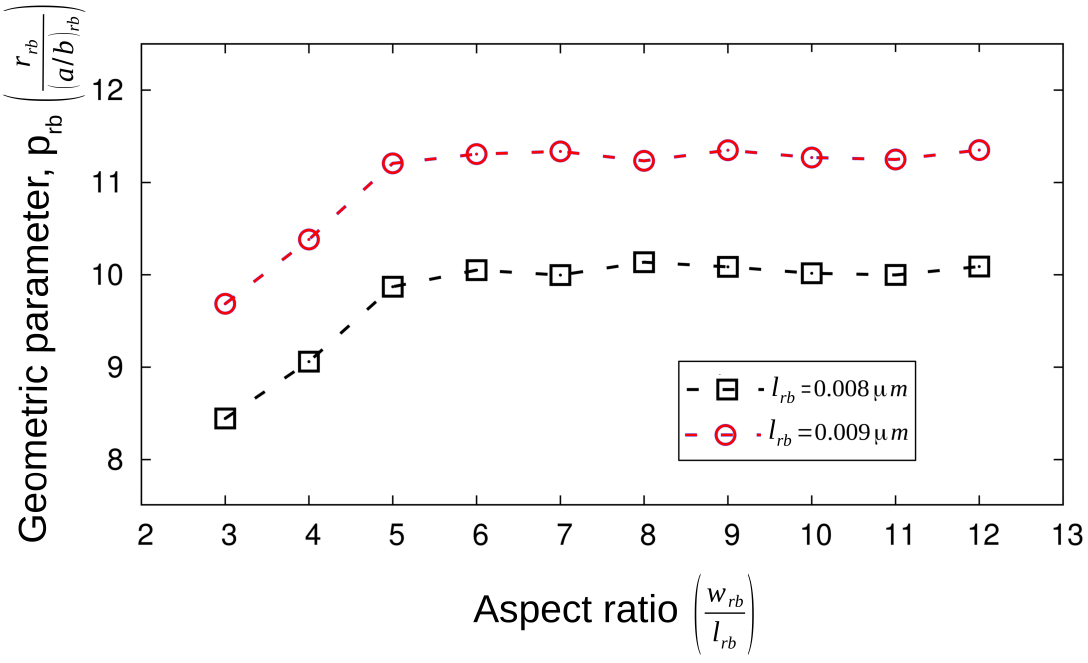}   
   \caption{Parameter $p_\text{rb}$ at the midpoint of cylinderization is ascertained from the simulation of plates with different aspect ratios. The influence of the thickness $l_\text{rb}$ on the $p_\text{rb}$ is illustrated by considering initial structures of two different thickness ($l_\text{rb} = 0.008, 0.009\mu$m).}
   \label{fig:fig_4a}
\end{subfigure}

\begin{subfigure}[b]{1.0\textwidth}
   \includegraphics[width=1.0\linewidth]{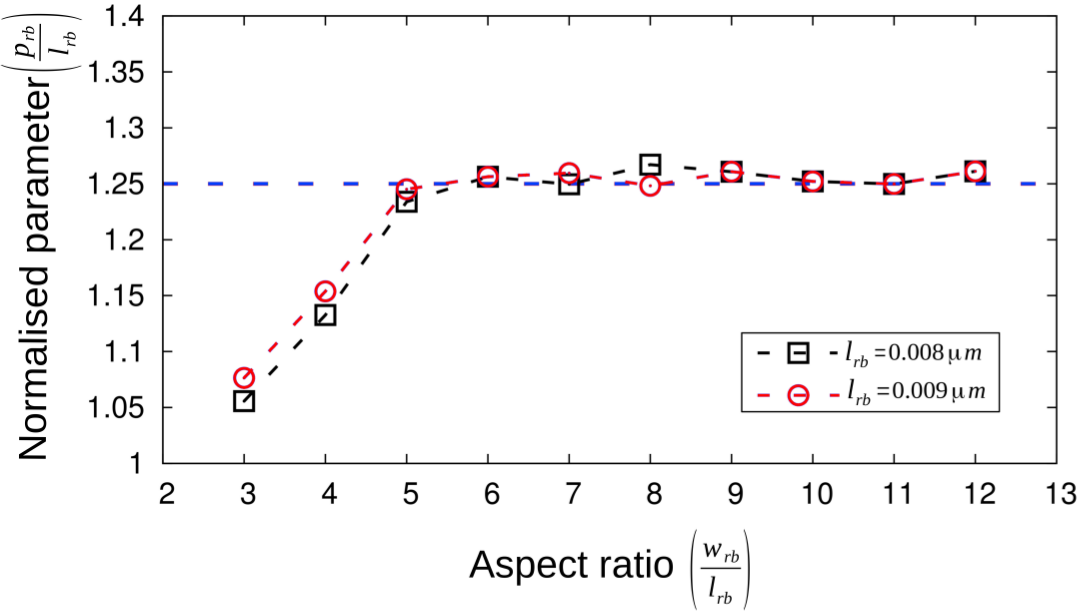}  
   \caption{The dimensionless geometric variable ($\frac{p_\text{rb}}{l_\text{rb}}$) is determined and presented for various aspect ratios.}
   \label{fig:fig_4b}
\end{subfigure}

\end{figure}

The morphology of the precipitate at the midpoint of the evolution is presented in Fig.~\ref{fig:fig_4}, distinguished in blue.
Owing to the elliptical shape of the precipitate two geometrical parameters, semi-major ($a_{\text{rb}}$) and -minor ($b_{\text{rb}}$) length, are introduced.
The area available for the volume diffusion at this stage of the cylinderization can be considered as $A_{\frac{1}{2}:\text{c}}\approx 4(\lambda b_{\text{rb}})$.
Geometrical treatments, Ref.~\cite{timoshenko1959theory}, postulate that the principal radii of the elliptical structures can be calculated as
\begin{align}\label{eq:curvature_ellipse}
 & R_1= \frac{a_{\text{rb}}^2b_{\text{rb}}^2}{(a_{\text{rb}}^2\text{sin}^2\theta+b_{\text{rb}}^2\text{cos}^2\theta)^{\frac{3}{2}}} \\ \nonumber
&R_2= \frac{a_{\text{rb}}^2}{(a_{\text{rb}}^2\text{sin}^2\theta+b_{\text{rb}}^2\text{cos}^2\theta)^{\frac{1}{2}}}.
\end{align}
By adopting Eqn.~\ref{eq:curvature_ellipse}, the curvature at the source and sink of the elliptical precipitate is ascertained.

The concentration difference introduced by the curvature, based on the geometrical parameters $a_{\text{rb}}$ and $b_{\text{rb}}$,  is written as
\begin{align}\label{eq:c_1/2}
 \delta c_{\frac{1}{2}:\text{c}}=\frac{c^{\theta}_{eq}\gamma V_\text{m}}{\kappa T}\Big(\frac{a_{\text{rb}}^3+a_{\text{rb}}b_{\text{rb}}^2-2b_{\text{rb}}^3}{a_{\text{rb}}^2b_{\text{rb}}^2}\Big).
\end{align}
Furthermore, since the perimeter of the ellipse is calculated as $2\pi\Big(\frac{a_{\text{rb}}^2+b_{\text{rb}}^2}{2}\Big)^{\frac{1}{2}}$, the distance which the diffusing mass needs to transverse to accomplish the cylinderization is expressed as
\begin{align}\label{eq:x_1/2}
 \delta x_{\frac{1}{2}:\text{c}}\approx \frac{1}{2\sqrt{2}}\big(\pi(a_{\text{rb}}^2+b_{\text{rb}}^2)^{\frac{1}{2}}-2\sqrt{2}(r_{\text{rb}}^2-b_{\text{rb}}^2)^{\frac{1}{2}}\big).
\end{align}

Unlike the conventional phase transformations, since the constituent phases are in chemical equilibrium, the corresponding volume fractions remain unchanged during cylinderization.
Therefore, the radius of the cylinder, which results from the morphological transformation, can be directly estimated from the initial dimensions of the plate by
\begin{align}\label{eq:x_1/2i}
 r_{\text{rb}}=\left(\frac{4w_{\text{rb}}l_\text{rb}+\pi l_\text{rb}^2}{4\pi}\right)^{\frac{1}{2}}.
\end{align}
Intuitively, the geometrical parameters of the elliptical structure, $a_{\text{rb}}$ and $b_{\text{rb}}$, formed at the midpoint of the cylinderization can be expressed in terms of the radius $r_{\text{rb}}$.
Exploiting the inter-dependencies of the geometrical parameters, the analytical treatment is simplified by introducing a variable $p_{\text{rb}}=\frac{r_{\text{rb}}}{a_{\text{rb}}/b_{\text{rb}}}$.
\nomenclature{$p_{\text{rb}}$}{Aspect ratio of cylinderization-midpoint ellipse}%
The parameter $p_{\text{rb}}$ while encompassing the geometry of the midpoint precipitate, reduces the number of variables involved in this analytical treatment.
In order to the numerically quantify  $p_{\text{rb}}$, the dimensions of the elliptical structure at midpoint of the evolution is ascertained from the simulation, for different plates, and plotted in Fig.~\ref{fig:fig_4a}.

Interestingly, beyond the aspect ratio $4$, the geometric parameters appears to be independent of the initial aspect-ratio of the precipitate.
However, in order encompass the influence of the thickness ($l_\text{rb}$), plates with increased thickness ($l_\text{rb}=0.009\mu$m) is numerically investigated and included in Fig.~\ref{fig:fig_4a}.
Since the volume of the precipitate increase with the increase in the thickness, the midpoint variable $p_{\text{rb}}$ correspondingly increases.
Furthermore, it is vital to note that, since $p_{\text{rb}}$ is expressed as the ratio $r_{\text{rb}}$ and $a_{\text{rb}}/b_{\text{rb}}$, the parameter assumes length unit.
Therefore, to normalise the variable and to eliminate its dependencies on the initial configuration of the precipitate, the ratio of $p_{\text{rb}}$ and thickness ($l_\text{rb}$) is subsequently considered.

Fig.~\ref{fig:fig_4b} shows the numerical nature of the normalised parameter $p_{\text{rb}}/l_\text{rb}$ (dimensionless) for plates of different aspect ratios.
Evidently, in the plates with aspect ratio $5$ and above, the geometric term $p_{\text{rb}}/l_\text{rb}$ is independent of the initial size and thickness of the precipitate.
Furthermore, based on Fig.~\ref{fig:fig_4b}, dimensions of the precipitate at the midpoint of the cylinderization can be ascertained through
\begin{align}\label{eq:degree}
\frac{a_{\text{rb}}}{b_{\text{rb}}}\approx\frac{r_{\text{rb}}}{1.25l_\text{rb}}. 
\end{align}
In addition to enabling the calculation of the curvature difference at the midpoint, the above relation can be employed to estimate the degree of cylinderization.
For instance, the ratio of the major- and minor-axis length will be less than the definite value $\frac{r_{\text{rb}}}{1.25l_\text{rb}}$, if the cylinderization is beyond its midpoint.

By incorporating the parameter $p_{\text{rb}}$, the concentration difference introduced at the midpoint of the cylinderization is written as
\begin{align}\label{eq:ce_1/2}
 \delta c_{\frac{1}{2}:\text{c}} = & \frac{c^{\theta}_{eq}\gamma V_\text{m}}{\kappa T}\left(\frac{r_{\text{rb}}}{p_{\text{rb}}}\right)^{\frac{3}{2}}\left(\frac{r_{\text{rb}}^3+p_{\text{rb}}^2-2p_{\text{rb}}^3}{r_{\text{rb}}^4}\right),
 \end{align}
while the corresponding diffusion-length is expressed as
 \begin{align} \label{eq:xe_1/2}
 \delta x_{\frac{1}{2}:\text{c}} = \left(\frac{r_{\text{rb}}}{8p_{\text{rb}}}\right)^{\frac{1}{2}}\left\{\pi(r_{\text{rb}}^2+p_{\text{rb}}^2)^{\frac{1}{2}}-2\left[2p_{\text{rb}}(r_{\text{rb}}-p_{\text{rb}})\right]^{\frac{1}{2}}\right\}.
\end{align}
Substituting the Eqns.~\ref{eq:ce_1/2} and ~\ref{eq:xe_1/2} in Eqn.~\ref{eq:x0_1}, the driving force at the midpoint of the cylinderization reads
\begin{align}\label{eq:dr_1/2}
 \Gamma_{\frac{1}{2}:\text{c}} =\frac{DV_\text{m}^{2}c^{\theta}_{eq}\gamma}{\kappa T}\frac{\left(8\sqrt{2}\lambda b_{\text{rb}}\right) \left(r_{\text{rb}}^3+p_{\text{rb}}^2-2p_{\text{rb}}^3\right)}{r_{\text{rb}}^2\left\{\pi(r_{\text{rb}}^2+p_{\text{rb}}^2)^{\frac{1}{2}}-2\left[2p(r_{\text{rb}}-p_{\text{rb}})\right]^{\frac{1}{2}}\right\}}
\end{align}
By combining the driving forces at the initial and midpoint of the cylinderization, as in Eqn.~\ref{eq:driving_force1_2}, the overall driving-force is expressed as
\begin{align}\label{eq:overall}
 \bar{\Gamma}_{\text{c}}=\frac{DV_\text{m}^{2}c^{\theta}_{eq}\gamma}{\kappa T}\Big(\frac{8\sqrt{2}\lambda}{3}\Big)\left\{\frac{\sqrt{2} }{l_\text{rb}(2w_{\text{rb}}+\pi l_\text{rb}-4r_{\text{rb}})}
  +\frac{b_{\text{rb}}\left(r_{\text{rb}}^3+p_{\text{rb}}^2-2p_{\text{rb}}^3\right)}{r_{\text{rb}}^2\left[\pi(r_{\text{rb}}^2+p_{\text{rb}}^2)^{\frac{1}{2}}-2\left(2p_{\text{rb}}(r_{\text{rb}}-p_{\text{rb}})\right)^{\frac{1}{2}}\right]}\right\}.
\end{align}
From the overall driving-force the time taken for the cylinderization is estimated by the relation
\begin{align}\label{eq:cy_time}
 t_{1:\text{c}}=\frac{\delta V_{\text{c}}}{\bar{\Gamma}_{\text{c}}},
\end{align}
where $\delta V_{\text{c}}$ is the amount of volume transfer required for the cylinderization of the plate.
\nomenclature{$V_{\text{c}}$}{Amount of volume transfer required for the cylinderization}%
\nomenclature{$t_{1:\text{c}}$}{Time taken for cylinderization}%
Since the volume of the precipitate is preserved during the morphological evolution, the required mass transfer is expressed as
\begin{align}\label{eq:req_mass}
 \delta V_{\text{c}}=r_{\text{rb}}^2\left\{\pi-\frac{l_\text{rb}}{r_{\text{rb}}}\left[1-\left(\frac{l_\text{rb}}{2r_{\text{rb}}}\right)^2\right]-2\text{sin}^{-1}\left(\frac{l_\text{rb}}{2r_{\text{rb}}}\right)\right\}.
\end{align}
Combining Eqns.~\ref{eq:overall} and ~\ref{eq:req_mass} in Eqn.~\ref{eq:cy_time}, the dimensionless time taken for the cylinderization is written as
\begin{align}\label{eq:tau}
 \frac{t_{1:\text{c}}}{\tau'}=\frac{3}{8\sqrt{2}} \left(\frac {r_{\text{rb}}}{l_\text{rb}}\right)^2 \left\{\frac{\pi-\frac{l_\text{rb}}{r_{\text{rb}}}\left[1-\left(\frac{l_\text{rb}}{2r_{\text{rb}}}\right)^2\right]-\left[2\text{sin}^{-1}\left(\frac{l_\text{rb}}{2r_{\text{rb}}}\right)\right]}
 {\frac{l_\text{rb}\sqrt{2} }{l_\text{rb}(2w+\pi l_\text{rb}-4r_{\text{rb}})}+\frac{p_{\text{rb}}l_\text{rb}\left(r_{\text{rb}}^3+p_{\text{rb}}^2-2p_{\text{rb}}^3\right)}{r_{\text{rb}}^2\left[\pi \left(r_{\text{rb}}^2+p_{\text{rb}}^2\right)^{\frac{1}{2}}-2\left(2r_{\text{rb}}p_{\text{rb}}-2p_{\text{rb}}^2)\right)^{\frac{1}{2}}\right]}}\right\},
\end{align}
where $\tau'=\frac{l_\text{rb}^3 \kappa T}{DV_\text{m}^{2}c^{\theta}_{eq}\gamma}$ is the non-dimensionalising factor.

\begin{figure}
    \centering
      \begin{tabular}{@{}c@{}}
      \includegraphics[width=0.9\textwidth]{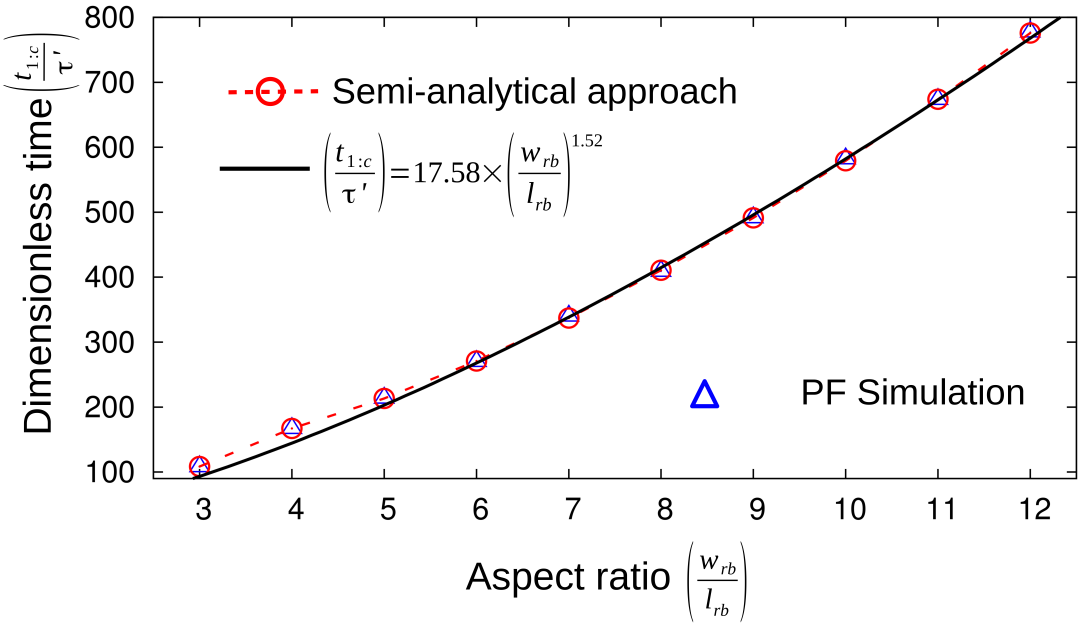}
    \end{tabular}
    \caption{ The cylinderization kinetics predicted by the  simulation-aided semi-analytical treatment (Eqn.~\ref{eq:tau}) is compared with the outcomes of the phase-field simulations.
    \label{fig:fig_5}}
\end{figure}

The time taken for the cylinderization of plates with different aspect ratios is calculated based on the semi-analytical treatment elucidated above.
The contribution of the simulation to the present theoretical formulation, though pivotal, is confined to the precipitate geometry at the midpoint ($t_{\frac{1}{2}}$).
Therefore, the simulation results are compared to the predictions of the analytical treatment in Fig.~\ref{fig:fig_5}.
Evidently, the results of the phase-field simulations noticeably agree with the predictions of the present semi-analytical approach.
Here it is vital to note that, as shown in Figs.~\ref{fig:fig_4a} and ~\ref{fig:fig_4b}, since the smaller plates of aspect ratios $3$ and $4$ do not exhibit the geometrical consistency at the midpoint, for the comparison in Fig.~\ref{fig:fig_5}, appropriate relation which is similar to Eqn.~\ref{eq:degree} is adopted.
By fitting the data points, an expression for the influence of the aspect ratio on the kinetics of the cylinderization is obtained which reads
\begin{align}\label{eq:predict}
 \frac{t_{1:\text{c}}}{\tau'}=17.58\left(\frac{w_\text{rb}}{l_\text{rb}}\right)^{1.52}.
\end{align}
The above relation indicates that with increase in the aspect ratio, the time taken for cylinderization increases smoothly and monotonically.
The predominant factor contributing to this monotonic increase in the cylinderization time is the increase in the volume required to be transferred ($\delta V_{\text{c}}$) with the aspect ratio of the plate.

\section{Cylinderization of faceted ribbon}\label{sec:facet}

During the phase transformation, several factors govern the morphology of the precipitate in the lamellar arrangement.
Amongst these factors, the crystallographic relation between the phases renders a pivotal influence on the shape of the precipitate.
Correspondingly, in specific alloy systems, precipitates assume a faceted morphology which are characterised by sharp edges~\cite{cline1970effect}.
Since, the principal curvatures along the sharp regions of the faceted precipitates are analytically ill-defined, hemispherical caps are augmented to facilitate the theoretical treatment.
Irrespective geometry of the shape, it has been shown that the phase-field approach elegantly handles the curvature while recovering the sharp interface solutions~\cite{karma1996phase,karma1998quantitative}.
Therefore, the kinetics and mechanism of the cylinderization exhibited by the faceted ribbons are analysed in this section.

\begin{figure}
    \centering
      \begin{tabular}{@{}c@{}}
      \includegraphics[width=0.9\textwidth]{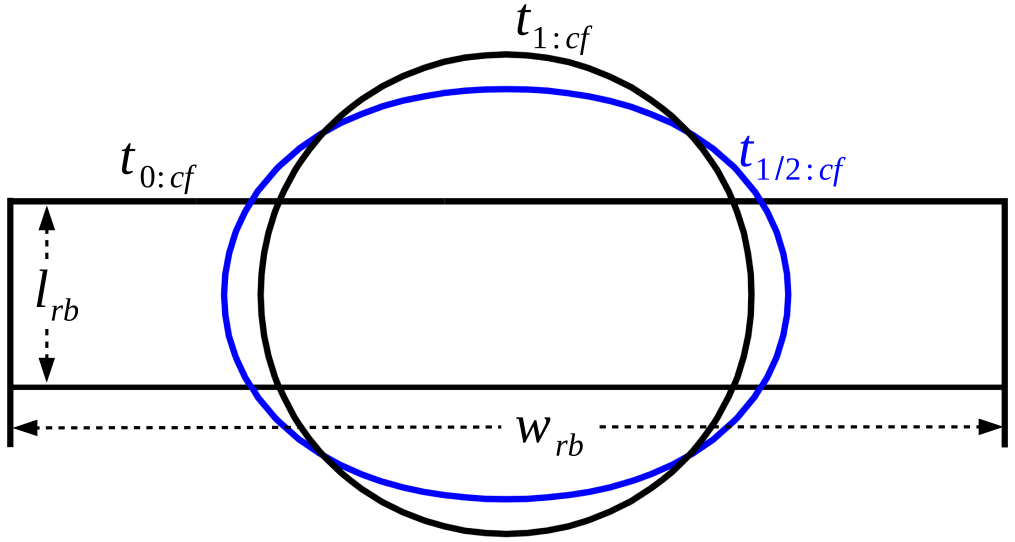}
    \end{tabular}
    \caption{ Morphology of the faceted precipitate of aspect ratio $6$ at the initial, midpoint and end of the cylinderization.
    \label{fig:fig_6}}
\end{figure}

In Fig.~\ref{fig:fig_7}, the shape of the faceted precipitate at the initial, midpoint and final stages of the cylinderization is shown.
Similar to the capped ribbons, the faceted structures assume elliptical shape at the midpoint of the evolution.
Analytical treatment of the cylinderization, considered in the previous section, cannot be extended to the faceted shape, since the curvature difference which govern the driving force is analytical ill-defined.
Therefore, the evolution rates are entirely determined from the phase-field simulations.

Change in the cylinderization time with increase in the aspect ratio of the plates is graphically presented in Fig.~\ref{fig:fig_6}.
Irrespective of the morphology of the plates, since the transformation mechanism is unaltered, the time taken for the cylinderization monotonically increases with time.
With the increase the size (aspect ratio) of the faceted ribbon, the mass transferred to cylinderise the precipitate proportionately increases.
Correspondingly, as shown in Fig.~\ref{fig:fig_7}, the transformation time increases.

\section{Comparing the cylinderization of the capped and faceted ribbons}

\subsection{The rate of transformation}

The numerical investigations unravel that the influence of the aspect ratio on the cylinderization time of the faceted rods ($t_{1:\text{cf}}$) can be expressed as
\begin{align}\label{eq:predict1}
 \frac{t_{1:\text{cf}}}{\tau'}=11.22\left(\frac{w_\text{rb}}{l_\text{rb}}\right)^{1.57}.
\end{align}
The above relation along with the outcomes of the simulations are presented in Fig.~\ref{fig:fig_7}.
In addition to the cylinderization kinetics of the faceted structures, the time taken  for the transformation of the capped precipitates are included in Fig.~\ref{fig:fig_7}.
Two factors contribute to the noticeable disparity in the cylinderization kinetics of the capped and faceted plates.
One, for a given aspect ratio, owing to the capped ends, the amounts of mass transfer ($\delta V_{\text{c}}$) required to transform the capped structure is greater than the faceted precipitate.
Two, since the termination profile of the capped and faceted precipitates are substantially different, the initial driving force correspondingly vary.
This disparity in the initial curvature-difference additionally contributes to the difference in the evolution rate shown in Fig.~\ref{fig:fig_7}.

\begin{figure}
    \centering
      \begin{tabular}{@{}c@{}}
      \includegraphics[width=0.9\textwidth]{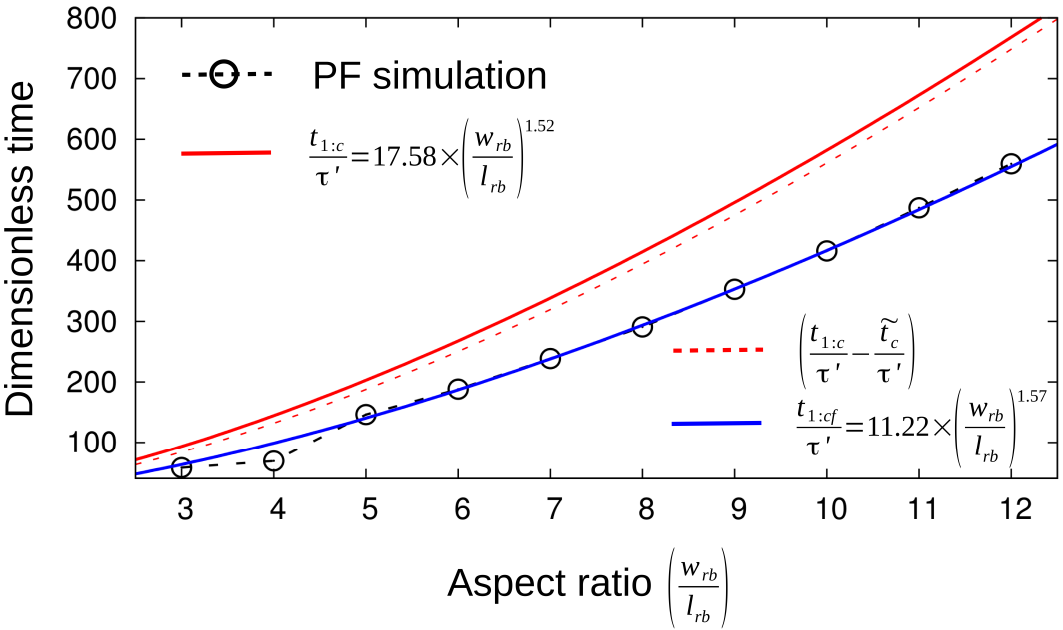}
    \end{tabular}
    \caption{ Increase in the time taken for the cylinderization of the faceted structure with increase in the aspect ratio. The influence of size on the cylinderization kinetics of the capped ribbons are included to render a comparative investigation.
    \label{fig:fig_7}}
\end{figure}

The curvature at the ends of the plate characterises the morphology of the precipitate.
While, in the capped precipitate, the curvature is well-defined, the principal curvature is zero in the faceted structure.
In order to distinguish the role of the termination geometry on the kinetics of the cylinderization, the influence of the additional volume introduced by the hemispherical caps is delineated.

As elucidated above, the amount of mass transfer which is required to cylinderise the plate is not the same for the capped and faceted structure, despite the equal aspect ratio.
In the capped ribbons, the hemispherical caps add to $\delta V_{\text{c}}$ , and consequently, prolong the cylinderization time.
Therefore, the time taken for the migration of the additional volume is estimated from the  required mass-transfer pertaining to the capped ($\delta V_{\text{c}}$) and faceted precipitate ($\delta V_{\text{cf}}$).
Adopting the overall driving force formulation in Eqn.~\ref{eq:overall}, the time taken for the transfer of the additional volume, which is introduced by the hemispherical ends in the capped structures, is estimated by
\begin{align}
\frac{\tilde{t}_{\text{c}}}{\tau'}=& \frac{\delta V_{\text{c}}-\delta V_{\text{cf}}}{\bar{\Gamma}_{\text{c}}} \nonumber
\end{align}
\begin{align}\label{eq:terminal}
& = \frac{3}{32\sqrt{2}}\frac{\pi l_\text{rb}^2}{\left\{\frac{\sqrt{2} }{l_\text{rb}(2w_{\text{rb}}+\pi l_\text{rb}-4r_{\text{rb}})}+\frac{b_{\text{rb}}\left(r_{\text{rb}}^3+p_{\text{rb}}^2-2p_{\text{rb}}^3\right)}{r_{\text{rb}}^2\left[\pi(r_{\text{rb}}^2+p_{\text{rb}}^2)^{\frac{1}{2}}-2\left(2p_{\text{rb}}(r_{\text{rb}}-p_{\text{rb}})\right)^{\frac{1}{2}}\right]}\right\}}. 
\end{align}

The time taken for the migration of the \lq cap-volume\rq \thinspace can be eliminated from the cylinderization time of the capped structures, thereby ensuring that, for a given aspect ratio, the required mass-transfer ($\delta V_{\text{c}}$) are equal in capped and faceted structures.
The transformation kinetics after discarding the role of the additional cap-volume is included in Fig.~\ref{fig:fig_7}.
Conceivably, the cylinderization rate of the capped structures is apparently enhanced by the elimination of the time required for the transfer of the cap-volume.
Furthermore, despite the increase in the aspect ratio of the plate, since the cap-volume remains unchanged, the decrease in the cylinderization time is seemingly independent of the precipitate size.
In Fig.~\ref{fig:fig_7}, the observed difference between the transformation kinetics of  the capped (discontinuous red lines) and faceted (solid blue lines) is exclusively governed by the characteristic morphology of the termination.
Moreover, it is evident from Fig.~\ref{fig:fig_7} that the faceted termination accelerates the morphological evolution, when compared to the smooth ends of the capped precipitates.

\subsection{Curvature-enhanced kinetics}

\begin{figure}
    \centering
      \begin{tabular}{@{}c@{}}
      \includegraphics[width=0.9\textwidth]{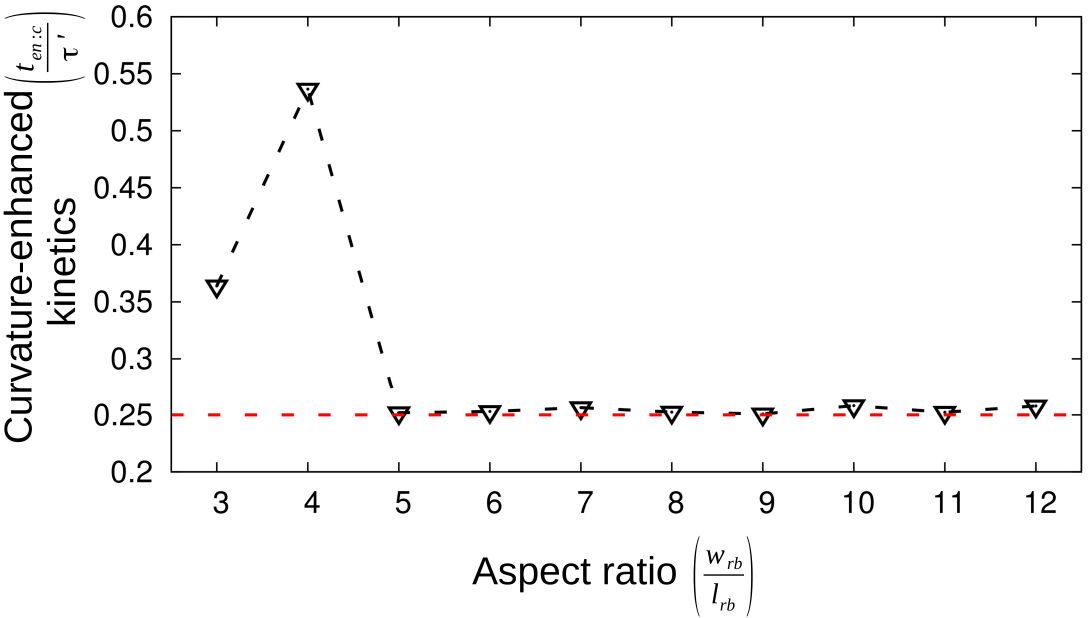}
    \end{tabular}
    \caption{ The change in the cylinderization rate induced solely by the morphology of the terminations which is ascertained through the Eqn.~\ref{eq:time-difference}.
    \label{fig:fig_9}}
\end{figure}

Despite the diminution of the cylinderization time pertaining to the capped precipitate, through the Eqn.~\ref{eq:terminal}, Fig.~\ref{fig:fig_7} indicates that the transformation rate is considerably higher in the faceted precipitate.
Since the required volume-transfer is made identical for faceted and capped precipitates by eliminating the contribution of the cap-volume, the sole factor contributing to the disparity in the kinetics is the curvature at the ends of the precipitate.
Particularly, it  is evident that the sharp edges of the faceted ribbon enhance the cylinderization kinetics when compared to the curved terminations of the capped structures.
The increased rate of transformation, which is governed by the initial difference in the curvature, can be quantified as
\begin{align}\label{eq:time-difference}
\frac{t_{\text{en:c}}}{\tau'}=\frac{1}{\tau'}\left[ \left(t_{1:\text{c}}-\tilde{t}_{\text{c}}\right)-t_{1:\text{cf}} \right].
\end{align}
In the above Eqn.~\ref{eq:time-difference}, $t_{1:\text{c}}$ and $\tilde{t}_{\text{c}}$ correspond to the time taken for the cylinderization of the capped precipitate and migration of the cap-volume.
Therefore, the term $t_{1:\text{c}}-\tilde{t}_{\text{c}}$~ on the right hand side of Eqn.~\ref{eq:time-difference} represents the time taken for the mass transfer in the capped precipitate, which is equalised in relation to its corresponding faceted ribbon.
Furthermore, the cylinderization time of the faceted plate is represented by $t_{1:\text{cf}}$.
Irrespective of the aspect ratio of the plates, since the curvature difference between the capped and faceted structures in the initial stages of the cylinderization is identical, a similarity in the enhanced kinetics is intuitively expected.
To examine the curvature-driven acceleration of the transformation rate, the enhanced kinetics ($t_{\text{en:c}}$) is numerically estimated for the precipitates of different aspect ratios.
Fig.~\ref{fig:fig_9} shows the influence on the ribbon size (aspect ratio) on the acceleration of the morphological transformation.
Consistent with the analytical claim, expect for the smaller ribbons of aspect ratio $3$ and $4$, the acceleration in the evolution rate ($t_{\text{en:c}}$) is independent of the initial aspect-ratios of the precipitate.
Furthermore, Fig.~\ref{fig:fig_9} indicates that the sharp edges reduces the time taken for the transfer of the required mass by $25\%$ when compared to the curved terminations of the capped ribbons.
The non-compliance exhibited by the smaller structures of aspect ratio $3$ and $4$ can be attributed to the comparable magnitude of the cap-volume and required mass transfer for transformation ($\delta V_{\text{c}}$).

\section{Absence of \lq Contra-diffusion \rq \thinspace}

\begin{figure}
    \centering
      \begin{tabular}{@{}c@{}}
      \includegraphics[width=1\textwidth]{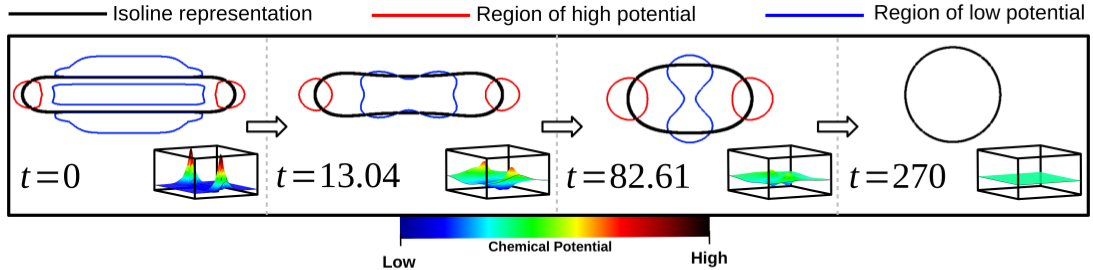}
    \end{tabular}
    \caption{ Distribution of the chemical potential and the resulting morphological changes exhibited by the capped ribbons during cylinderization.
    \label{fig:fig_8}}
\end{figure}

Extending the present numerical analysis to the structures of larger aspect ratio indicates that the transformation mechanism remains unchanged and the kinetics adheres to the relations expressed in Eqns.~\ref{eq:predict} or ~\ref{eq:predict1}, depending on the termination morphology of the precipitate.
However, theoretical treatment of the surface-diffusion governed morphological evolution of the capped finite-structure predict that, beyond a critical aspect-ratio, a substantial change is observed in the transformation mechanism~\cite{nichols1976spheroidization}.
Particularly, the larger ribbons are expected to disintegrate, through the process of \lq ovulation\rq \thinspace, to smaller ribbons before cylinderising.
In order to investigate the invariance of the transformation mechanism, in the volume-diffusion governed transformation, on the aspect ratio of the plates, the morphological transformation of the capped precipitate accompanying the cylinderization is re-examined.
 
The formation of neck in the central region of the precipitate is identified as the precursor for the shift in the transformation mechanism~\cite{nichols1976spheroidization}.
The necking process, which involves decrease in the thickness of the precipitate, demands mass transfer from the central flat-region to the receding ends of the precipitate.
In other words, as opposed to the conventional mass transfer from the terminations to the adjacent region (Fig.~\ref{fig:2}), the \lq contra-diffusion \rq \thinspace, which results in necking, is established by the central region becoming the source of the mass transfer.
Correspondingly, the contra-diffusion is actuated only when the potential in the central region of the precipitate is higher than its surrounding.
Fig.~\ref{fig:fig_8} illustrates the distribution of the chemical potential, especially the region of the high and low potential, induced during the volume-diffusion governed cylinderization of the capped plates.
Evidently, all through the morphological evolution, the central region of the finite structure remains the region of low potential, thus preventing the introduction of contra-diffusion and consequently, averting the shift in the transformation mechanism.

It is conceivable that during cylinderization, wherein the termination ridges stably grow in size with time, an appropriate curvature-difference can be established between the central region and perturbations which turns the flat region into the source of mass transfer.
However, in volume-diffusion governed evolution, introduction of such curvature-difference is precluded by the change in the distribution of the low potential region.
As shown in Fig.~\ref{fig:fig_8}, the profile of the low potential region progressively changes with the increase in the size of the termination ridges, while remaining confined to the central region.
This change in distribution of the low potential region directs the mass transfer accordingly and prevents the central region from turning into a source for mass transfer.
In other words, unlike the surface-diffusion governed evolution, the contra-diffusion, which induces a change in the transformation mechanism, is averted in the volume-diffusion governed cylinderization by the progressive change in the profile of the low potential region and restricting it to the central region (Fig.~\ref{fig:fig_8}).

\section{Conclusion}

The dimensions of the precipitate emerging from the co-operative growth of phases, a prominent phase transformation which yields lamellar microstructure, often appear relatively infinite in the growth direction while remaining finite in the other orthogonal directions~\cite{cline1970effect}.
The curvature-induced transformation of such precipitates involves progressive morphological change in the cross-section.
The shape-change induced in the cross-section of the precipitate, owing to the inherent difference in the curvature, is referred to as cylinderization.
In this chapter, the volume-diffusion governed cylinderization is numerically investigated through phase-field simulations.

Owing to the lack of in-situ information, the morphology of the precipitate at the midpoint of the cylinderization is assumed for the analytical treatment~\cite{courtney1989shape}.
This assumption is relaxed by tracking the morphological changes and the existing theoretical approach is revisited.
Considerable agreement is noticed between the revisited semi-analytical treatment and the outcomes of the phase-field simulations.
Furthermore, the mechanism of the cylinderization has been shown to be consistent with the existing studies~\cite{courtney1989shape}.

For the elegant treatment of the driving force (curvature difference), a hemispherical cap is augmented on the either ends (terminations) of the cross-section.
Since such smooth caps are rarely observed in a microstructure, the numerical investigation is extended to faceted precipitates with sharp edges.
Analytical treatment of the sharp edges are cumbersome as the principal curvature is zero at the termination.
However, as elucidated in Sec.~\ref{sec:facet}, the present phase-field approach convincingly handles the curvature difference associated with the sharp terminations.

The role of the termination morphology in governing the kinetics of the cylinderization is distinguished.
It is realised that, when compared to the capped ribbons, the shape edges of the faceted precipitate reduces the cylinderization time by $25\%$, approximately.
Moreover, it is identified that, irrespective of the size of the precipitate, the transformation mechanism remains unchanged for the volume-diffusion governed cylinderization.

\chapter{Spheroidization of finite three-dimensional rods}\label{chap:rods}

The onset of the shape-instability in a semi-infinite structure like ribbons is predominantly confined to the cross-section.
Therefore, the numerical treatments which analyse the morphological evolution of the ribbon-like precipitate, which is referred to cylinderization, is confined to two-dimension, as elucidated in the previous chapter.
Despite the transformation of the semi-infinite ribbon to cylindrical structure, the precipitate is hardly stable towards the curvature-induced driving forces~\cite{surek1976theory}.

In addition to the cylindrical structures that are formed during the morphological evolution of other structures, in certain alloy system phase transformation yields rod-like precipitates~\cite{ryum1975precipitation}.
Moreover, in composite materials, relatively high-strength precipitate rods are embedded onto the ductile matrix to the enhance the mechanical properties~\cite{crossman1971unidirectionally,williams1975microstructural}.
Owing to the wide applicability of these materials, and since the mechanical properties are significantly influenced by the shape adopted by the phases, comprehensive investigations are made to understand the morphological stability of the precipitates~\cite{ringer1994precipitate,miyazaki1982formation}.

Considering the significance of the curvature-driven transformation on the applicability of the huge spectrum of materials, in this chapter the stability of the rod-like precipitates are extensively analysed.
Unlike the ribbon-like precipitates, the entire morphology of the rods changes in response to the shape-instability. 
Therefore, the present phase-field study is extended to three-dimension.
Often, the curvature-driven evolution of the rods are analysed by considering infinitely-long cylindrical structures~\cite{sharp1983overview,abarzhi2010review}.
However, it has been experimentally identified that, owing to the intricacies of the microstructure, the seemingly infinite cylindrical precipitate are predominantly disintegrated into finite rods, through different processes like boundary splitting, at the very outset of the morphological evolution~\cite{allen2004microstructural,fan2011deformation,xu2003using}.
Therefore, in present chapter, the temporal change in the morphology of the finite three-dimensional rods are exclusively investigated.
In ribbon-like precipitate, the cylinderization is governed by the curvature difference established across the cross-section.
In contrast, the morphological transformation of the finite rods are influenced by the curvature difference established across the entire precipitate.
Since the resulting shape-change transforms the precipitate in spheroidal structure, this form of shape-instability is referred to as spheroidization~\cite{marich1970structural,li2011effects}.

\section{Domain setup}

Fig.~\ref{fig:fig_2} ~shows the simulation domain considered for the analysis of spheroidization of finite three-dimensional rods.
By incorporating CALPHAD data, the chemical equilibrium between the precipitate-$\theta$ and the matrix-$\alpha$ is established by assigning appropriate composition to the phases.
Moreover, for the elegant delineation of the curvature difference in the initial stages of the transformation, hemispherical caps are augmented in the either longitudinal ends of the rod.
The diameter of the \lq capped\rq \thinspace rod is fixed at $l_{\text{r}} = 0.012\mu$m, while the length $w_{\text{r}}$ is varied to accommodate the aspect ratio, $\frac{w_{\text{r}}}{l_{\text{r}}}$.
\nomenclature{$l_{\text{r}}$}{Diameter of three-dimensional capped rod}%
\nomenclature{$w_{\text{r}}$}{Length of three-dimensional capped rod}%
Owing to the definitive nature of the cross-section, the radius of the termination caps are affixed at $l_{\text{r}}/2$.
The simulation domain is chosen to avoid any influence of the boundary condition on the morphological evolution, while being computationally efficient.
Furthermore, the domain setup is consistently varied to encompass the rods of different aspect ratio.

\section{Edge recession-assisted spheroidization}

\subsection{Spheroidization of the capped rods}

\subsubsection{Mechanism}\label{lab:mech}

\begin{figure}
\centering
   \begin{subfigure}[b]{0.6\textwidth}
   \includegraphics[width=1\linewidth]{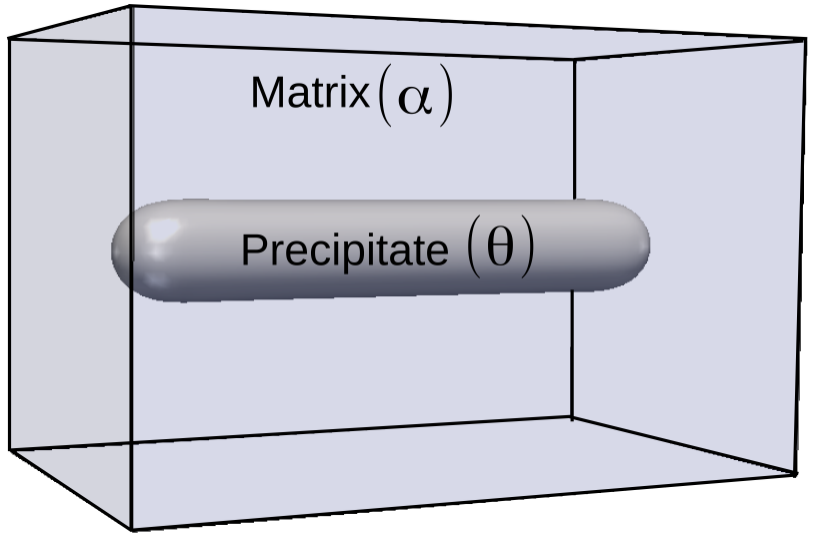}
   \caption{Three-dimensional domain setup adopted for analysing of spheroidization of rods.}
   \label{fig:fig_2} 
\end{subfigure}

\begin{subfigure}[b]{0.7\textwidth}
   \includegraphics[width=1\linewidth]{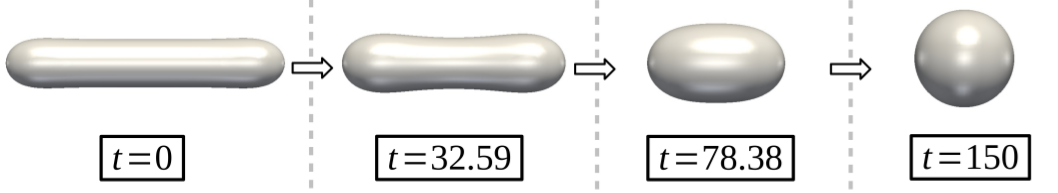}
   \caption{Temporal evolution of the capped rod of aspect ratio $5$ leading to the spheroidization of the precipitate, where $t$ is dimensionless time.}
   \label{fig:fig_3}
\end{subfigure}

\end{figure}

The morphological transformation of the capped rod of aspect ratio 5, governed by the inherent curvature-difference of its shape, is illustrated in Fig.~\ref{fig:fig_3}.
When compared to the cylinderization, significant resemblance in the mechanism of the evolution is evident.
The geometrically well-defined terminations of the capped rod introduce a difference in the curvature, particularly in relation to the abutting cylindrical structures.
The curvature difference induces a gradient in the chemical potential, which actuates the mass flow, ultimately, resulting the morphological transformation of the precipitate.

In the initial stage of the spheroidization the capped terminations act as the source of the mass transfer while the adjacent cylindrical body of the rod becomes sink.
The deposition of the mass in the adjoining regions of the termination caps results in the formation of the longitudinal ridges, as shown in Fig.~\ref{fig:fig_3} at $t=32.59$.
Since the source of the mass transfer, which facilitates the formation of the ridges, is the longitudinal terminations, the growth of the ridges is accompanied by the recession of the edges.
With the progressive migration of the mass from the terminations, the longitudinal perturbations continue to grow.
The recession of the edges and the corresponding growth of the perturbations disturbs the initial morphology of the rod.
While edge-recession appears to shrink the precipitate, the perturbations disturbs the cylindrical nature of the rod.
This scheme of evolution, particularly in the initial stages, indicates that the spheroidization mechanism of the rods are identical irrespective of the dominant mode of mass transfer, surface or volume diffusion~\cite{nichols1976spheroidization}.
Furthermore, the theoretical treatment which assumes that the rods retains morphological configuration all through the transformation is shown to be inaccurate~\cite{mclean1973kinetics}.

Akin to cylinderization, the longitudinal ridges continue to grow by consuming the adjacent cylindrical body and ultimately, coalesce.
The coalescence of the receding perturbation transforms the precipitate into capsule-like structure, Fig.~\ref{fig:fig_3} at $t=78.38$.
Owing to the shape of the precipitate following the coalescence of the ridges, the mass transferred from the edges gets deposited in the central region of the capsule.
Through the accumulation of mass in the central region, the precipitate assumes a three-dimensional ellipsoidal shape.
The subsequent evolution, which spheroidises the precipitate, is driven by the inherent curvature-difference induced by the ellipsoidal shape.
Since the entire spheroidization is predominantly governed by the recession of the terminations, which is caused by the mass transfer from the longitudinal edges, the transformation mechanism is referred to as termination migration- or edge recession-assisted spheroidization. 

\subsubsection{Kinetics}

Recently, Park et al extended the analytical treatment of cylinderization to investigate the spheroidization kinetics of the capped rods~\cite{park2012prediction}.
However, adopting the derivation to calculate the driving force at the initial and midpoint of the transformation, Eqns.9a and 13 in Ref.~\cite{park2012prediction},  it is realised that the driving force at the midpoint of the evolution is slightly higher than the initial driving-force.
Therefore, the theoretical treatment of the cylinderization is initially extended to three-dimensional finite rods and subsequently revisited with the aid of the phase-field simulation.

\paragraph{Adopting the cylinderization approach:}\label{sec:cylin}

As introduced in the previous chapter, the analytical approach to delineate the kinetics of the evolution begins with the identification of the three distinct stages of the transformation, which are referred to as initial ($t_{0:\text{sc}}$), midpoint ($t_{\frac{1}{2}:\text{sc}}$) and final ($t_{1:\text{sc}}$).
The transformation rate is subsequently ascertained by 
\begin{align}\label{eq:rate}
 t_{1:\text{sc}}=\frac{\delta V_{\text{sc}}}{\bar{\Gamma}_{\text{sc}}},
\end{align}
where $\delta V_{\text{sc}}$ is the amount of mass transferred to transform the precipitate completely.
\nomenclature{$t_{1:\text{sc}}$}{Time taken for the spheroidization of capped rod}%
\nomenclature{$\bar{\Gamma}_{\text{sc}}$}{Overall driving-force directing the spheroidization of capped rod}%
This required mass-transfer is dictated by the initial morphology of the precipitate~\cite{courtney1989shape,semiatin2005prediction}.
Furthermore, in Eqn.~\ref{eq:rate}, $\bar{\Gamma}_{\text{sc}}$ is the overall driving-force which is estimated as
\begin{align}\label{eq:rate}
 \bar{\Gamma}_{\text{sc}}=\frac{1}{3}\sum_{i\in\{0,\frac{1}{2},1\}}\Gamma_{i:\text{sc}},
\end{align}
where $\Gamma_{i:\text{sc}}$ is the instantaneous driving-force considered at initial, midpoint and end of the evolution.
However, since the transform halts with the exhaustion of the driving force, $\Gamma_{1:\text{sc}} = 0$.

\begin{figure}
    \centering
      \begin{tabular}{@{}c@{}}
      \includegraphics[width=0.9\textwidth]{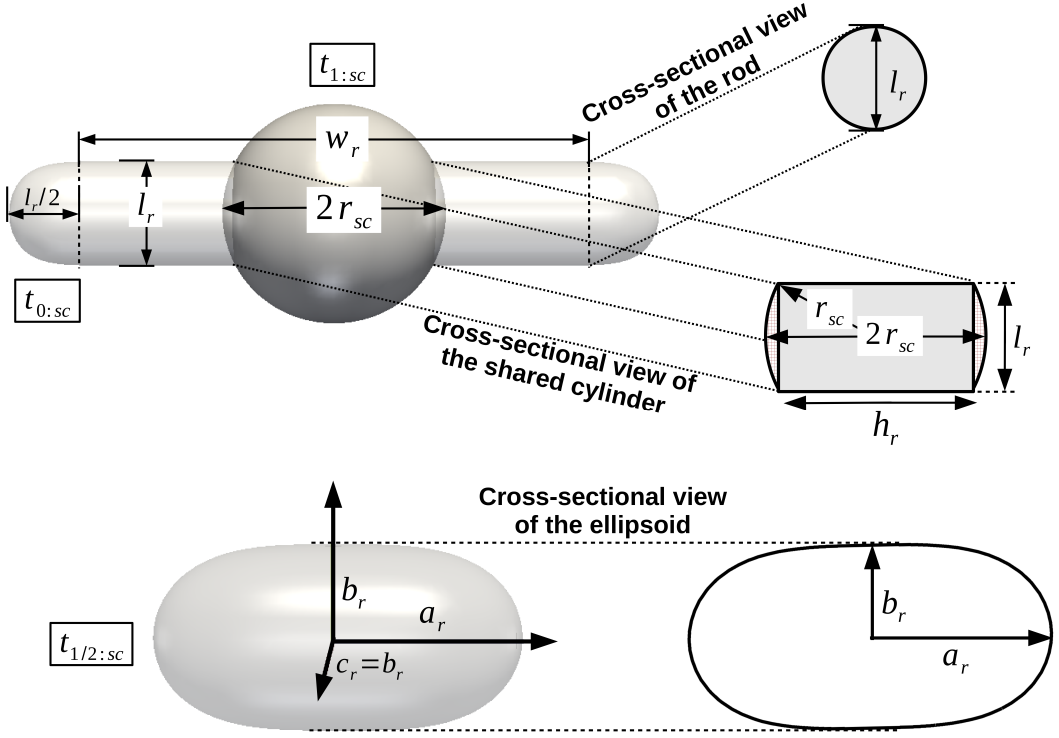}
    \end{tabular}
    \caption{ The morphology of the precipitate of aspect ratio $5$ at the initial, midpoint and final state of the spheroidization.
    \label{fig:fig_1}}
\end{figure}

The shape of the precipitate at the three distinct stages which are considered for the analytical treatment is shown in Fig.~\ref{fig:fig_1}.
The required mass-transfer is calculated by eliminating the volume shared by the rod and resulting spheroid from the overall volume of the precipitate.
The cross-section of the shared region is included in Fig.~\ref{fig:fig_1}.
To determine the volume of the shared precipitate, a parameter $h_{\text{r}}$ is introduced which holds the relation $h_{\text{r}}=\sqrt{\left(r_{\text{sc}}^2-\left(l_{\text{r}}/2\right)^2\right)}$.
Fig.~\ref{fig:fig_1} indicates that the volume of the overlapping region, is the sum of the cylinder with height $h_{\text{r}}$ and two spherical caps.
The volume of the spherical cap in the shared region can be expressed
\begin{align}\label{eq:cap_shared}
 V_{\text{cap}}=\frac{1}{3}\pi \left( r_{\text{sc}}-\frac{h_{\text{r}}}{2} \right)^2\left(2r_{\text{sc}}+\frac{h_{\text{r}}}{2}\right).
\end{align}
Correspondingly, the required mass-transfer for the spheroidization of the capped rod is determined by
\begin{align}\label{eq:req_mass}
 \delta V_{\text{sc}}= \frac{4}{3}\pi r_{\text{sc}}^3-\left\{\pi \left(\frac{l_{\text{r}}}{2}\right)^2h_{\text{r}}+2\left[ \frac{1}{3}\pi \left( r_{\text{sc}}-\frac{h_{\text{r}}}{2} \right)^2\left(2r_{\text{sc}}+\frac{h_{\text{r}}}{2}\right) \right] \right\}.
\end{align}
\nomenclature{$\delta V_{\text{sc}}$}{Required mass-transfer for the spheroidization of capped rod}%
The above expression can be simplified and the mass transfer required for the spheroidization can be written as
\begin{align}\label{eq:req_mass_sim}
 \delta V_{\text{sc}}= \pi r_{\text{sc}}^2h_{\text{r}}-\frac{\pi}{4}h_{\text{r}}\left(l_{\text{r}}^2+\frac{h_{\text{r}}^2}{3}\right).
\end{align}
Since the volume of the evolving phase-$\theta$ is preserved during the spheroidization, owing to the chemical equilibrium, a relation between the radius of the spheroid and initial dimension of the rod is derived, which is expressed as
\begin{align}\label{eq:radius_capped}
 r_{\text{sc}}=\left[\left(\frac{l_{\text{r}}}{4}\right)^2(3w_{\text{r}}+2l_{\text{r}}) \right]^{\frac{1}{3}}.
\end{align}

The two transitory parameters governing the driving force are the area available for the diffusion and the concentration (potential) gradient induced by the curvature.
As elucidated in Sec.~\ref{lab:mech}, during the termination migration-assisted spheroidization of the rods, the mass is transferred predominantly from the termination.
Therefore, since at the initial stage of the spheroidization, hemispherical caps are the exclusive source for mass transfer, the area available for diffusion is $A_{0:\text{sc}}=\pi l_{\text{r}}^2$.
\nomenclature{$A_{i:\text{sc}}$}{Diffusion-area available for the spheroidization in capped rod}%
The difference in the equilibrium concentration, which is introduced by the shape, is proportional to the curvature difference and is written as
\begin{align}\label{eq:c_o_capped}
 \delta c_{0:\text{sc}} \propto \Bigg[ \underbrace{\left(\frac{2}{l_{\text{r}}}+\frac{2}{l_{\text{r}}} \right)}_\mathcal{\text{source}}-\underbrace{\left( \frac{1}{\infty}+\frac{2}{l_{\text{r}}}\right)}_\mathcal{\text{sink}} \Bigg].
\end{align}
\nomenclature{$\delta c_{i:\text{sc}}$}{Difference in equilibrium concentration around capped rod}%
Furthermore, the diffusion distance associated with the concentration gradient at the outset of the spheroidization is analytically estimated as
\begin{align}\label{eq:x_o_capped}
 \delta x_{0:\text{sc}}=\frac{\pi}{4}l_{\text{r}}+\frac{w_{\text{r}}}{2}-r_{\text{sc}}.
\end{align}
\nomenclature{$\delta x_{i:\text{sc}}$}{Diffusion length in spheroidization of capped rod}%
By considering the aforementioned factors, the driving force at the beginning of the spheroidization can be expressed as
\begin{align}\label{eq:dr_0}
 \Gamma_{0:\text{sc}} \left( A_{0:\text{sc}},\left(\frac{\delta c}{\delta x}\right)_{0:\text{sc}}\right) \propto 2\pi l_{\text{r}}\left(\pi \frac{l_{\text{r}}}{4}+\frac{w_{\text{r}}}{2}-r_{\text{sc}}\right)^{-1}. 
\end{align}
Although the above Eqn.~\ref{eq:dr_0} adopts a simplified form of the concentration gradient, since no nonphysical approximation is made, the same relation is involved while revisiting the approach.

The diffusion area at the midpoint is approximated to be the average of the diffusion area at the initial and final state of the transformation.
In the existing theoretical study~\cite{park2012prediction}, the diffusion area at the final stage is assumed to the entire surface area of the precipitate.
However, it vital to note that the diffusion area is the region actively involved in the mass transfer, and predominantly associated with the source.
Therefore, the consideration that the surface area of the precipitate is the diffusion area at the final state, entails an implausible configuration with a lack of sink.
Moreover, such assumption can also be viewed as potential reason for the misappropriation of the midpoint driving-force, particularly, since the diffusion area significantly influences to the driving force.
Thus, in the present derivation, the diffusion area at the end of the transformation is considered to be a quadrant of the surface, which avoids a repetitive inclusion of the diffusion area. 
Correspondingly, the diffusion area at the midpoint of spheroidization is expressed as
\begin{align}\label{eq:a_o5_capped}
 A_{\frac{1}{2}:\text{sc}}=\frac{1}{2}\pi \left(l_{\text{r}}^2+r_{\text{sc}}^2 \right).
\end{align}

The principal radii of curvature at the midpoint is similarly ascertained as the average of the initial and final state.
Accordingly, the principal radii at the source and sink are approximated 
\begin{align}\label{eq:radii_mid}
 R_1=\frac{1}{2}\left(\frac{l_{\text{r}}}{2}+r_{\text{sc}}\right) &&& R_2=\frac{1}{2}\left(\frac{l_{\text{r}}}{2}+r_{\text{sc}}\right)
\end{align}
and
\begin{align}\label{eq:radii_mid}
 R_3=\frac{1}{2}\left(\frac{l_{\text{r}}}{2}+r_{\text{sc}}\right) &&& R_4=\frac{1}{2}\left(\infty+r_{\text{sc}}\right),
\end{align}
respectively.
By substituting these principal radii of curvature, the influence of curvature difference on the equilibrium concentration can be related as 
\begin{align}\label{eq:conc_mid}
 \delta c_{\frac{1}{2}:\text{sc}} \propto 2\left({\frac{l_{\text{r}}}{2}+r_{\text{sc}}}\right)^{-1}.
\end{align}
Including distance which quantifies the migration of diffusing atoms at midpoint, based on initial and final diffusion-distance, the midpoint driving force is expressed as
\begin{align}\label{eq:dis_mid}
  \Gamma_{\frac{1}{2}:\text{sc}} \left( A_{\frac{1}{2}:\text{sc}},\left(\frac{\delta c}{\delta x}\right)_{\frac{1}{2}:\text{sc}}\right) \propto \left[\frac{2\pi(l_{\text{r}}^2+r_{\text{sc}}^2)}{l_{\text{r}}+2r_{\text{sc}}}\right]\underbrace{\left\{ \frac{1}{2}\left[ \pi\frac{l_{\text{r}}}{4} + \frac{w_{\text{r}}}{2}+r_{\text{sc}}\left( \frac{\pi}{2}-1 \right)\right] \right\}^{-1}}_{:=\delta x_{\frac{1}{2}:\text{sc}}}
\end{align}
The driving force ascertained from Eqns.~\ref{eq:dr_0} and ~\ref{eq:dis_mid} along with Eqn.~\ref{eq:req_mass_sim} can be employed to determine the rate of spheroidization.
However, before presenting the outcomes of the aforementioned derivation, the approximation involving the average of the initial and final condition is relaxed and the treatment is revisited.

\paragraph{Revisiting the cylinderization approach:}\label{sec:re_cylin}

\begin{figure}
    \centering
      \begin{tabular}{@{}c@{}}
      \includegraphics[width=0.9\textwidth]{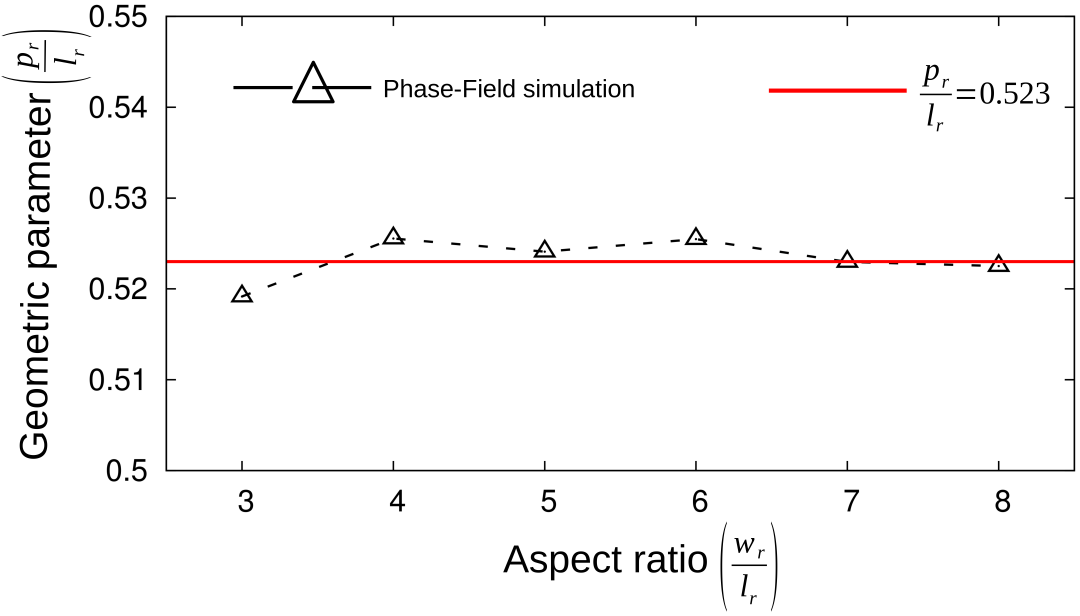}
    \end{tabular}
    \caption{ The parameter $p=\frac{r_{\text{sc}}}{\left(a/b\right)_{\text{r}}}$ is seemingly independent of the initial size of the three-dimensional rod.
    \label{fig:fig_4}}
\end{figure}

As mentioned earlier, the major aspect of the present treatment is to reformulate the driving force of the spheroidization at the midpoint.
As shown in Fig.~\ref{fig:fig_1}, the precipitate assumes a three-dimensional ellipsoidal shape at the midpoint.
Although an attempt was made to incorporate the curvature difference pertaining to the characteristic shape, Eqn.12a of Ref.~\cite{park2012prediction}, the description was confined to two-dimension.
Therefore, in this section three-dimensional extension is rendered while relaxing the geometric approximations.

Formulating the surface area of the three-dimensional structures is intrinsically not straightforward.
Therefore, a reasonable consideration employed in the existing analyses is adopted to describe the diffusion area at the midpoint.
Correspondingly, it is assumed that the precipitate along the major-axis of the ellipsoid exclusive acts as the source.
Although a rigorous treatment is rendered in Refs.~\cite{tee2005surface} and~\cite{rivin2007surface}, the diffusion area of the ellipsoidal precipitate is described in a simplistic way.

As shown in Fig.~\ref{fig:fig_1}, the ellipsoid formed at the midpoint of the evolution is characterised $a_{\text{r}}>b_{\text{r}}=c_{\text{r}}$, where $a_{\text{r}}$, $b_{\text{r}}$ and $c_{\text{r}}$ are semi-axes lengths along the orthogonal co-ordinates.
Such structures are referred to as \lq prolate spheroids\rq \thinspace.
In the present analysis, the overall surface-area of the midpoint precipitate is formulated through \textit{Knud Thomsen} approximation~\cite{klamkin1971elementary,klamkin1976corrections}.
Correspondingly, the surface area of the prolate spheroid is expressed as
\begin{align}\label{eq:dis_mid}
 S_{\text{pr}}=4\pi\left[\frac{2(a_{\text{r}}b_{\text{r}})^{\mathcal{P}}+b_{\text{r}}^{2\mathcal{P}}}{3}\right]^{\frac{1}{\mathcal{P}}},
\end{align}
where the constant $\mathcal{P}\approx 1.6$.
To estimate the diffusion area at the midpoint which is confined to the major axis, the area of the segment pertaining to the minor axes is removed from the overall area of the prolate spheroid.
Correspondingly, the midpoint diffusion-area is written as
\begin{align}\label{eq:area_mid}
 A_{\frac{1}{2}:\text{sc}}=S_{\text{pr}}-4\pi (a_{\text{r}}b_{\text{r}}^5)^{\frac{1}{3}}.
\end{align}
Although the above formulation assumes the spherical segment along the minor axes, when compared to the Eqn.~\ref{eq:a_o5_capped}, a relatively accurate approximation is rendered by Eqn.~\ref{eq:area_mid} for midpoint diffusion area.
Moreover, owing to the volume preservation, the radius of the ultimate spheroidal structure is related to the geometrical parameters of the midpoint ellipsoid as $r_{\text{sc}}=(a_{\text{r}}b_{\text{r}}^2)^{\frac{1}{3}}$.

\begin{figure}
    \centering
      \begin{tabular}{@{}c@{}}
      \includegraphics[width=0.9\textwidth]{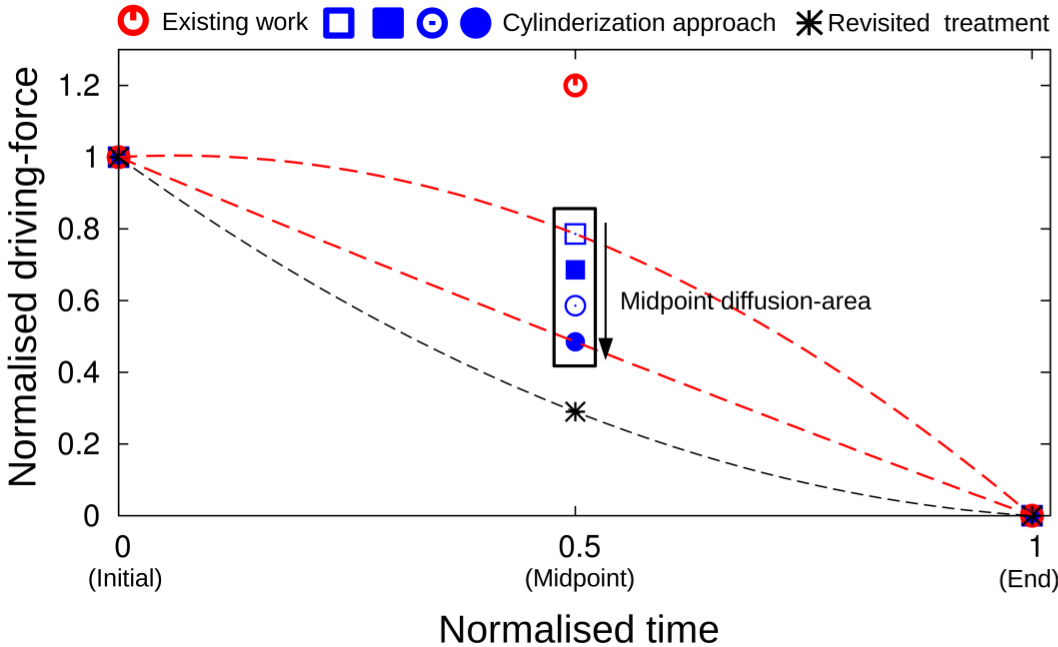}
    \end{tabular}
    \caption{ The transitory driving at the specific stages of the spheroidization which are calculated based on the existing work~\cite{park2012prediction}, present adoption and the simulation-aided treatment. The influence of the diffusion area on the kinetics for the cylinderization approach is included.
    \label{fig:fig_5}}
\end{figure}

The influence of the curvature difference, which is inherent to the prolate spheroid, on the equilibrium concentration can be expressed as
\begin{align}\label{eq:c_mid1}
\delta c_{\frac{1}{2}:\text{sc}}\propto\left[ H_{\text{sink}}-H_{\text{source}}\right],
\end{align}
where $H_{\text{sink}}$ and $H_{\text{source}}$ are the mean curvatures associated with the sink and source, respectively.
Reason for replacing the principal radii with the mean curvature is elucidated elsewhere (Appendix~\ref{sec:app2}).
The mean curvature for the ellipsoidal structure is defined based on the co-efficients of first and second fundamental forms as derived in the Appendix~\ref{sec:app3}.
Correspondingly, for the present prolate spheroid wherein $b=c$, the co-efficients of first fundamental form reads
\begin{align}\label{eq:c_mid2}
 E_{\text{r}}&=a_{\text{r}}^2\sin^2\theta+b_{\text{r}}^2\cos^2\theta \\ \nonumber
 F_{\text{r}}&=0\\ \nonumber
 G_{\text{r}}&=b_{\text{r}}^2\sin^2\theta.
\end{align}
\nomenclature{$E_{\text{i}}$, $F_{\text{i}}$, $G_{\text{i}}$}{Co-efficients of first fundamental form}%
Furthermore, for the prolate spheroid, the co-efficients characterising the second fundamental form is expressed as
\begin{align}\label{eq:c_mid3}
 \tilde{E}_{\text{r}}&=\frac{a_{\text{r}}b_{\text{r}}^2}{(b_{\text{r}}^4\cos^2\theta+a_{\text{r}}^2b_{\text{r}}^2\sin^2\theta)^\frac{1}{2}}\\ \nonumber
 \tilde{F}_{\text{r}}&=0\\ \nonumber
 \tilde{G}_{\text{r}}&=\frac{a_{\text{r}}b_{\text{r}}^2\sin^2\theta}{(b_{\text{r}}^4\cos^2\theta+a_{\text{r}}^2b_{\text{r}}^2\sin^2\theta)^\frac{1}{2}}.
\end{align}
\nomenclature{$\tilde{E}_{\text{i}}$, $\tilde{F}_{\text{i}}$, $\tilde{G}_{\text{i}}$}{Co-efficients of first fundamental form}%
Based on the above co-efficients, the mean curvature at a point on a prolate spheroid reads
\begin{align}\label{eq:mean_curv1}
 H=\frac{a_{\text{r}}(b_{\text{r}}^2+a_{\text{r}}^2\sin^2\theta+b_{\text{r}}^2\cos^2\theta)}{2(a_{\text{r}}^2\sin^2\theta+b_{\text{r}}^2\cos^2\theta)(b_{\text{r}}^4\cos^2\theta+a_{\text{r}}^2b^2\sin^2\theta)^\frac{1}{2}}.
\end{align}
From Eqns.~\ref{eq:c_mid2},~\ref{eq:c_mid3} and ~\ref{eq:mean_curv1}, the difference in the equilibrium concentration induced by the morphology of the precipitate at the midpoint can be written as
\begin{align}\label{eq:c_mid4}
\delta c_{\frac{1}{2}:\text{sc}}\propto\left( \frac{a_{\text{r}}}{b_{\text{r}}^2}-\frac{a_{\text{r}}^2+b_{\text{r}}^2}{2a_{\text{r}}^2b_{\text{r}}} \right).
\end{align}
The diffusion-length associated with curvature induced concentration-gradient, at the midpoint, is defined as 
\begin{align}\label{eq:x_mid1}
 \delta x_{\frac{1}{2}:\text{sc}}=\frac{\pi}{4}\left[ 4(2a_{\text{r}}^2+2b_{\text{r}}^2)^{\frac{1}{2}}-(a_{\text{r}}b_{\text{r}}^2)^{\frac{1}{3}} \right].
\end{align}

In order to quantify the fundamental geometric parameters $a_{\text{r}}$ and $b_{\text{r}}$, a variable $p_{\text{r}}$ which is defined as $p_{\text{r}}=\frac{r_{\text{sc}}}{a_{\text{r}}/b_{\text{r}}}$ is introduced.
The nature of this variable $p_{\text{r}}$ is determined by monitoring the morphological evolution of the precipitate.
Influence of the initial aspect-ratio on the parameter $p_{\text{r}}$,  which encompasses the dimensions of the midpoint ellipsoid, is plotted in Fig~\ref{fig:fig_4}.
This illustration unravels the relation that $\frac{p_{\text{r}}}{l_{\text{r}}}=0.523$, irrespective of the initial size of the finite rod.
Correspondingly, the fundamental parameters of the prolate spheroid can be related as
\begin{align}\label{eq:x_mid2}
a_{\text{r}}=1.95b_{\text{r}}.
\end{align}
Using the above relation, the driving force at the midpoint of the spheroidization can be calculated from Eqns.~\ref{eq:area_mid}, ~\ref{eq:c_mid4} and ~\ref{eq:x_mid1}.

\begin{figure}
    \centering
      \begin{tabular}{@{}c@{}}
      \includegraphics[width=0.9\textwidth]{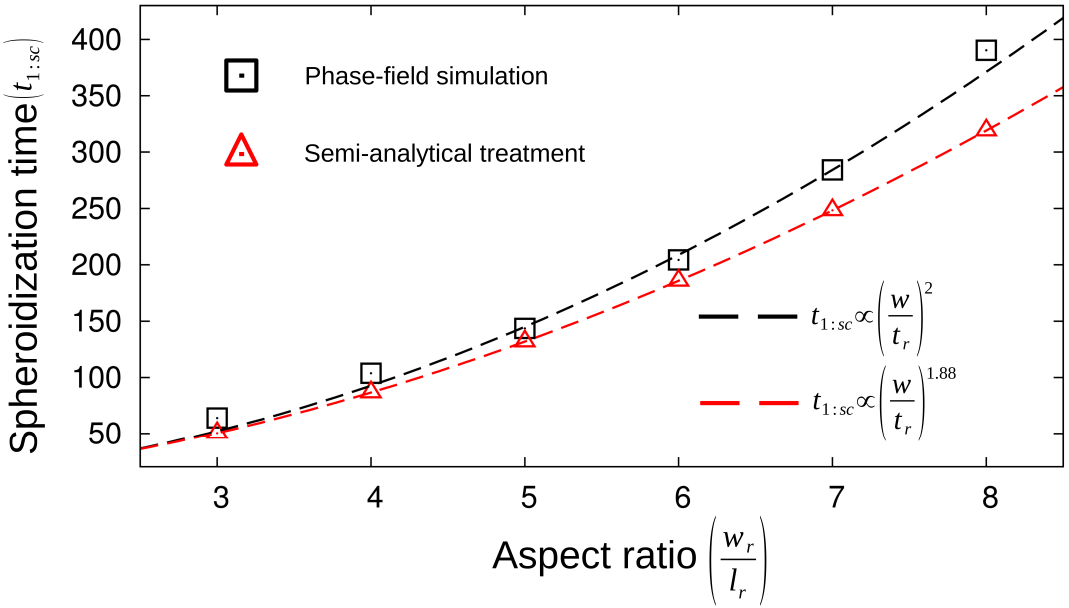}
    \end{tabular}
    \caption{ Time taken for the spheroidization of capped rods of different aspect ratio is compared with the predictions of the revisited analytical treatment.
    \label{fig:fig_6}}
\end{figure}

The driving forces governing the spheroidization of the capped rod of aspect ratio 5 is calculated at the three specific points of the evolution.
These driving forces estimated by the existing prediction~\cite{park2012prediction}, extension of the cylinderization treatment which is derived in Sec.~\ref{sec:cylin} and the semi-analytical treatment involving in-situ observation are compared in Fig.~\ref{fig:fig_5}.
Owing to some misappropriations, the existing studies render a driving force which is greater than the initial.
Earlier, it has been identified that the one factor contributing to this nonphysical representation of the driving force is the assumption that the diffusion area at the end of the transformation is the entire surface-area of the spheroid.
Therefore, the cylinderization approach is extended by relaxing this assumption.
As shown in Fig.~\ref{fig:fig_5}, the existing delineation with the appropriate formulation of the midpoint diffusion-area renders relatively consistent evolution of the driving force.
Furthermore, the diffusion area at the end of the evolution is progressively reduced, to unravel its influence on the driving force.
Evidently, with the decrease in the midpoint diffusion-area, the driving force correspondingly decreases and appears more thermodynamically consistent.
In Fig.~\ref{fig:fig_5}, the different points in the cylinderization approach respectively consider $S_\text{rq}$, $\frac{S_\text{rq}}{4}$, $\frac{S_\text{rq}}{2}$ and $\frac{3S_\text{rq}}{4}$ as the diffusion area at the end of the transformation, where $S_\text{rq}$ surface area of a quadrant of the resulting sphere.

The driving force ascertained through the revisited approach, which is enhanced by the simulation results, is included in Fig.~\ref{fig:fig_5}.
The revisited derivation, in Sec.~\ref{sec:re_cylin}, yields seemingly appropriate depiction of the temporal change in the driving force.
Based on the simulation-assisted formulation, the time taken for the transformation of the capped rods of different aspect-ratio is determined.
The kinetics predicted by this theoretical treatment is compared with the simulation result in Fig~\ref{fig:fig_6}.
The time taken for the spheroidization of the smaller capped rods, below aspect ratio 5, reasonably agrees with the postulated semi-analytical prediction.
However, despite the extensive geometrical treatment which augments in-situ information, noticeable disparity is introduced between the theoretical relation and simulation results.
It is interesting to note that the outcomes of the simulation progressively deviate from the analytical prediction with increase in aspect ratio.
Moreover, this significant difference between the analytical and simulation solutions contradicts the appreciable consistency observed in the cylinderization.

In order to recognise the factor(s) responsible for the observed difference between the analytical and numerical results, the temporal evolution of the driving force is examined.
Akin to the cylinderization, the driving force is estimated by considering the difference in the chemical potential, $\Delta \mu(x,t) = \mu(x)|^{+}_{t} - \mu(x)|^{-}_{t}$, where $\mu^+(x)_t$ and $\mu^-(x)_t$ are time-dependent maximum and minimum chemical potential in the domain, respectively.
The temporal change in the potential difference which governs the spheroidization of the capped rod of aspect ratio 5 is shown in Fig~\ref{fig:fig_7}.
For comparison, the evolution of the driving force accompanying the cylinderization of two-dimensional plate-like structures of the corresponding aspect ratio (5) is included. 

\begin{figure}
    \centering
      \begin{tabular}{@{}c@{}}
      \includegraphics[width=1.0\textwidth]{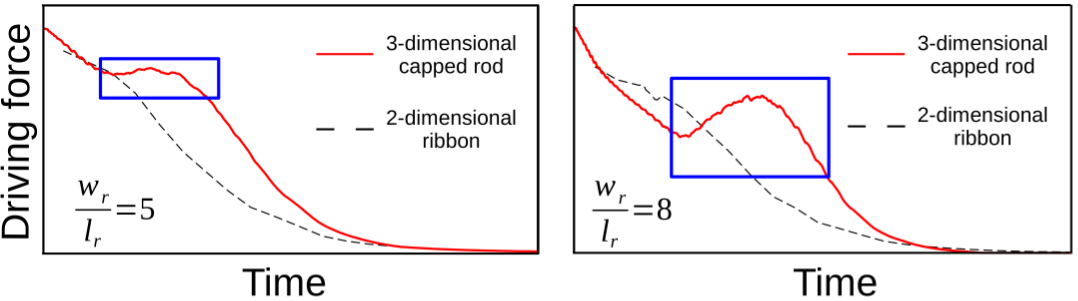}
    \end{tabular}
    \caption{ Temporal evolution of the curvature (potential) difference, $\Delta \mu(x,t) = \mu(x)|^{+}_{t} - \mu(x)|^{-}_{t}$, which dictate the morphological transformation of the capped rods of aspect ratio $5$ and $8$. The change in the driving force with time during the cylinderization of the two-dimensional ribbons of corresponding size is included for comparison.
    \label{fig:fig_7}}
\end{figure}

Analytical treatment of the curvature-driven transformation, inherently assumes that the potential difference decreases smoothly and monotonically with time.
As shown in Fig~\ref{fig:fig_7}, the driving force during the cylinderization of the two-dimensional plate of aspect ratio 5 decreases in the expected pattern.
Whereas, the potential difference accompanying the spheroidization of the three-dimensional capped rods,  experiences a period of stagnation, which disrupts the smooth-monotonic decrease in the driving force. 
The sluggish change in the curvature (potential) difference, which is highlighted in Fig~\ref{fig:fig_7}, prolongs the time taken for spheroidization.
Since the present theoretical formulation does not include the complete evolution the driving force, the characteristic temporal behaviour of the driving force predominantly contributes to the disparity observed in Fig~\ref{fig:fig_6}.

To explicate the progressive widening of the disparity between analytical and simulation results in Fig.~\ref{fig:fig_6}, the analysis of the potential difference is extended to larger rod of aspect ratio 8.
As shown in Fig~\ref{fig:fig_7}, the sluggish change in the curvature difference, observed in the rod of aspect ratio 5, is replaced by a definite period of increase and, subsequent, decrease in the driving force in larger structure.
This characteristic temporal evolution of the potential difference introduces a \lq hump\rq \thinspace, which contradicts the hitherto held assumption that the driving force decreases monotonically with time in shape-instabilities induced transformation.
The change in the transitional behaviour of the curvature difference, from a definite stagnation to a non-monotonic hump, with increase in aspect ratio prolongs the time taken for the spheroidization.  
Therefore, the influence of the size-dependent evolution of the driving force on the transformation rate is primarily responsible for the progressive disparity noticed in Fig~\ref{fig:fig_6}.
In other words, since the temporal change in the curvature difference during the spheroidization of the smaller rods, below aspect ratio 5, is smooth and monotonic, considerable agreement is seen between the analytical and numerical solutions.
However, with increase in the aspect ratio, the driving force characteristically evolve, in contradiction to the critical assumption of the analytical treatment, which introduces a significant difference in the outcomes of the simulation.
From Fig~\ref{fig:fig_7} it is evident that, although the smooth-monotonic decrease in curvature difference is a reasonable assumption for the evolution of two-dimensional structures, this consideration cannot be directly adopted for three-dimensional shapes.
Moreover, as expounded earlier, the temporal evolution of the driving force varies with the size, for a given morphology.

\subsubsection{Onset of \lq Contra-diffusion\rq \thinspace}\label{lab:contra}

\begin{figure}
    \centering
      \begin{tabular}{@{}c@{}}
      \includegraphics[width=0.6\textwidth]{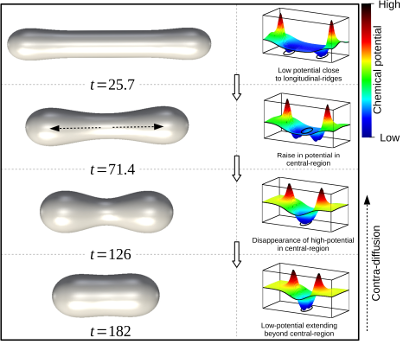}
    \end{tabular}
    \caption{ The transformation mechanism underpinning the spheroidization of the capped rod of  aspect ratio $8$. Three-dimensional depiction of the chemical-potential distribution which govern the shape change is included.
    \label{fig:fig_8}}
\end{figure}

Analysing the temporal change in the driving force associated with the cylinderization of the two-dimensional ribbons, in relation to the morphological evolution, indicates that the smooth-monotonic decrease in the curvature difference corresponds to the transformation mechanism wherein the longitudinal ridges grow stably before the coalescence.
However, the morphological behaviour pertaining to the stagnant or non-monotonic evolution of the potential difference is not entirely explicit. 
Therefore, in order to realise the shape-change accompanying the characteristic evolution of the driving force, the spheroidization mechanism of capped rod of aspect ratio $8$ is investigated.
The morphological changes leading to the spheroidization of the rod of aspect ratio $8$ is shown in Fig.~\ref{fig:fig_8}.
No significant difference is evident in the transformation mechanism of the rod in Fig.~\ref{fig:fig_8} when compared to the respective smaller structures of aspect ratio $5$.
Outwardly, the spheroidization is predominantly governed by the recession of the edges, through the stable growth of the longitudinal perturbations.
However, a closer examination of the change in the chemical-potential distribution unravels the morphological evolution governed by the characteristic evolution of the driving force.
A three-dimensional illustration of the chemical-potential distribution is included in Fig.~\ref{fig:fig_8}.

In the initial stages of the spheroidization, high potential is established around the terminations of the rod, while the regions adjacent to the edges assume low potential.
This potential distribution leads to mass transfer from the high-potential source to low-potential sink, which result in the formation of the longitudinal ridges.
Conventionally, as observed in the cylinderization, the chemical potential in the precipitate, but for the termination, continues to remain at low all through the morphological evolution.
However, during the spheroidization of the capped rod of aspect ratio $8$, the potential in the central region of the rod increases, as shown in Fig.~\ref{fig:fig_8} at $t=71.4$.
Accordingly, the potential at the central region is relatively higher than the potential at the foot of the longitudinal ridges (sinks).
The raise in the chemical potential at the midriff is attributed to the curvature difference established between the growing edge perturbations and the central region of the rod.
The relatively higher potential at the midriff of the precipitate transforms it into a source, which loses it mass to the surrounding sinks.
The unconventional transfer of mass from the central region to the receding ridges is referred to as Contra-diffusion.
The mass flow characterising the contra-diffusion is shown in Fig.~\ref{fig:fig_8}.
The contra-diffusion reduces the cross-sectional area of the rod in the central region, leading to the formation of necks.
However, since the evolving ridges progressively recede, the sinks get closer and the potential at the midriff continually decreases.
The continual decrease in the midriff-potential, which induces the the contra-diffusion, reduces the magnitude of mass transferred from the central region.
With time, as shown in Fig.~\ref{fig:fig_8} at $t=126$, the high potential in the central region completely disappears and transform it to sink.
Correspondingly, the resulting potential distribution, resumes the conventional mass flow from the ridges to the body of the precipitate.
By relating the spheroidization mechanism with Fig.~\ref{fig:fig_7}, it is evident that the non-monotonic evolution of the driving force pertains to the onset and subsequent disappearance of the contra-diffusion.

\begin{figure}
    \centering
      \begin{tabular}{@{}c@{}}
      \includegraphics[width=1.0\textwidth]{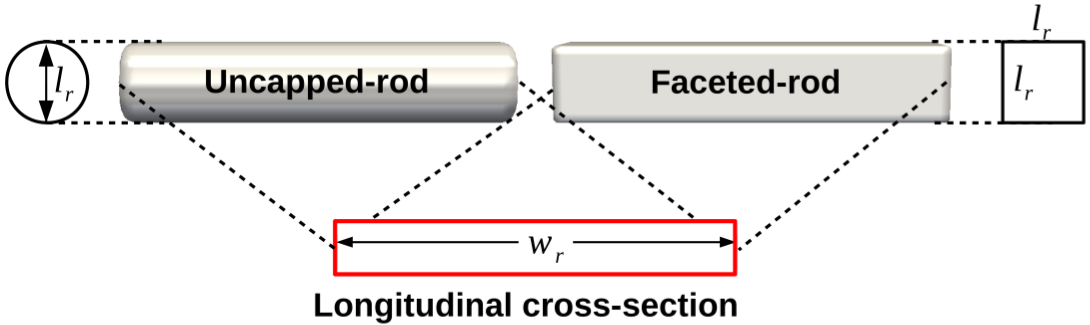}
    \end{tabular}
    \caption{ The geometrical description of the uncapped and faceted rods. The longitudinal and radial view of the cross-section is included.
    \label{fig:fig_9}}
\end{figure}

The present study unravels that the critical assumption in the existing theoretical treatment of the curvature-driven transformation, which is the smooth-monotonic decrease in the driving force, is due to the disjunction in the analysis of the kinetics and mechanism.
In other words, smooth and monotonic decrease in the curvature difference is assumed because the mechanism and the kinetics of the evolution are analysed separately.
Such separate consideration inherently overlooks the influence of mechanism on the kinetics.
However, the current investigation shows that the transformation mechanism, particularly in three-dimensional structures, significantly influences the kinetics of evolution.
By quantifying the amount of mass transferred due to the contra-diffusion, the analytical treatment can be extended.
Accordingly, as opposed to Eqn.~\ref{eq:rate}, the time taken for the spheroidization is expressed as
\begin{align}\label{eq:x_mid3}
  t_{1:\text{sc}}=\frac{\delta V_{\text{sc}}+2(\delta V_{\text{con}})}{\bar{\Gamma}},
\end{align}
\nomenclature{$t_{1:\text{sc}}$}{Time taken for the spheroidization of capped rod}%
where $V_{\text{con}}$ is the mass transferred through the contra-diffusion and pre-factor $2$ accounts for the characteristic non-monotonic hump.
Despite the theoretical framework, an extensive derivation is not pursued, since it is conceivable that the onset of contra-diffusion would introduce a change in the transformation mechanism in the larger rods.

\subsection{Spheroidization of the uncapped and faceted rods}

Hemispherical segments which characterises the capped rods are included to elegantly define the curvature difference at the initial stages of the spheroidization.
However, such morphologies are hardly observed in a microstructure.
Therefore, in this section, the spheroidization mechanism and the kinetics of the rods which are devoid of the hemispherical caps, henceforth referred to as uncapped rods, are analysed.
Lack of the termination caps, introduces a sharp edges on either longitudinal ends of the plate.
However, since it has already been shown in the previous chapter that the present model convincingly handles extreme curvature-difference, no geometrical manipulations are made to the circumvent the sharp edges.
Moreover, in certain material system, owing to the specific orientation relation between the phases, the precipitate assume a faceted morphology~\cite{sayir2000effect,dai1995synthesis}.
In an otherwise rod-like structure, the orientation relation introduces sharp edges and orthogonal corners which extend all through the length of the precipitate.
Similar to the uncapped rods, since the curvature of the faceted structures are analytical ill-defined, the theoretical study on the evolution of the faceted rods are limited.
Accordingly, the present numerical approach is adopted to understand the spheroidization of the faceted rods.

The morphological configuration of the uncapped and the faceted rods are shown in Fig.~\ref{fig:fig_9}.
The cross-section of the rods, both longitudinal and radial, are included in this illustration.
Similar to the capped rods, the aspect ratio of these rods is considered to be the ratio of the length ($w_{\text{r}}$) and diameter (or width, $l_{\text{r}}$).
For a given aspect ratio, although the longitudinal cross-sections of these rods are identical, the orthogonal corners that characterises the faceted rods establishes a significant change in the radial cross-section.

\subsubsection{Mechanism}

\begin{figure}
    \centering
      \begin{tabular}{@{}c@{}}
      \includegraphics[width=1.0\textwidth]{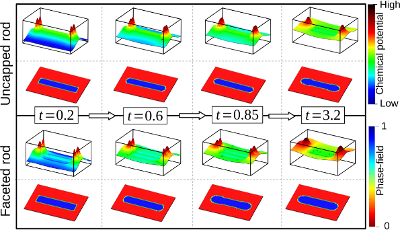}
    \end{tabular}
    \caption{ The continual change in the distribution of the chemical potential and respective change in the morphology of the capped and faceted rods of aspect ratio 8 in the initial stages of the spheroidization. The longitudinal cross-section of the rods are considered to illustrate the shape change. 
    \label{fig:fig_10}}
\end{figure}

The morphological difference in the uncapped and faceted rods, when compared to the capped rods, is likely to introduce a difference in the mechanism of the transformation.
From the outset of the spheroidization, these rods induce a unique driving force consistent with the characteristic morphology of the precipitate.
Therefore, to understand the influence of the shape, the initial stage of the spheroidization is extensively analysed.
Fig.~\ref{fig:fig_10} shows the morphological evolution of the uncapped and faceted rods of aspect ratio 8, along with the underpinning distribution of the chemical potential, in the initial stages of the transformation.
The initially identical longitudinal cross-section of the rods are considered to understand the morphological changes, comparatively.

Typifying the finitude of the precipitate, high potential is established around the terminations of the rods, irrespective of its geometrical configuration.
Although the high potential at the longitudinal ends of these rod are similar to the capped structures, the distribution of these high potential are visibly different.
In capped rods, as shown in Fig.~\ref{fig:fig_8}, the high potential at the edges are confined and are represented by a single peak.
However, in both uncapped and faceted rods, the distribution of the high potential comprises of two distinct peaks, as shown in Fig.~\ref{fig:fig_10} at $t=0.2$.
While these high-potential peaks are visible in the uncapped rods, they are definitively resolved in the faceted rods.
Furthermore, while the potential around the remnant body of the uncapped rod is relatively uniform, the orthogonal corners which extend across the length of the rod induces a significant difference in the potential distribution in the respective region of the faceted rod.
Therefore, prior to the mass transfer which is directed by the overall distribution of the chemical potential, the disparity within the high or low potential regions, seen as the peaks in Fig.~\ref{fig:fig_10}, actuates a responsive flux.

In the uncapped rod, the potential distribution is characterised by the peaks at the longitudinal ends and uniform low potential along the remnant precipitate. 
Consequently, the mass from the sharp termination-corners gets deposited in the region corresponding to the valley of the high-potential peaks.
The mass transferred from the sharp corners, as shown in Fig.~\ref{fig:fig_10} at $t=0.6$, transforms the longitudinal cross-section from a orthogonal shape to a smooth rectangle with curved ends.
The progressive migration of flux from the termination corners to the region between the high-potential peaks increasingly smoothens the edges of the rod.
The mass transfer within longitudinal termination, governed by the high-potential distribution, proceeds at the expense of the sharp corners.
Therefore, at $t=0.85$, when the disparity within the high potential vanishes, the rod assumes a shape of a capped rod with potential at the longitudinal ends represented by a single peak.
Owing to the resemblance in the distribution of the chemical potential, the subsequent morphological transformation of the rod follows the evolution of the capped structure, as shown in Fig.~\ref{fig:fig_10} at $t=3.2$.

Unlike the uncapped rods, in the faceted structures, the disparity within a specific potential (global maxima or minima) is not confined to terminations. 
The orthogonal corners, which is a distinctive feature of the faceted rod, induce difference in the potential distribution all through the length of the precipitate.
In other words, the potential distribution in the central region of the precipitate, which is generally smooth and low, are significantly inhomogeneous in the faceted rod, owing to the presence of the definite corners that characterises the radial cross-section. 
Therefore, in the initial stages of the spheroidization, the mass transfer is primarily governed by these internal disparities.
In the termination, the mass from the sharp corners migrate to the relatively low potential region within the termination.
This diffusing flux, similar to the uncapped rods, smoothens the longitudinal ends of the faceted precipitate by transforming it to a curved structure.
Similarly in the central region of the faceted rod, the corners act as the source for the mass transfer to the abutting flat surfaces.
The mass transfers in the early stages of the transformation, as shown in Fig.~\ref{fig:fig_10} at $t=0.6$ and $t=0.85$, collectively reduces the internal disparity in the potential distribution.
Consequently, the sharp termination and radial corners vanish with time, as the high-potential peaks coalesce.
At the end of this initial stage, as shown in Fig.~\ref{fig:fig_10} at $t=3.2$, the faceted rod loses it characteristic sharp corners, both in the termination and the remnant body, and transforms to a smooth structure, which resembles a capped rod.

The present study on the spheroidization mechanism of the three-dimensional structure accommodating sharp edges and corners indicate that, owing to the characteristic morphology of the precipitate, a inhomogeneity is introduced within the high- and low-potential distribution. 
The initial stage of the transformation is primarily governed by these internal disparity of the potential distribution.
The mass transfer induced by the internal disparity smoothens the sharp corners and edges by directing the flux to the adjacent low-potential region (relatively).
Therefore, at the end of the initial stage, the sharp terminations which characterise the morphology of the precipitate completely disappears.
Interestingly, in the spheroidization of the uncapped and faceted rods, the initial shape-change transform the rods to a familiar capped structure, which is expected to evolve in a manner discussed in the previous sections.

\subsubsection{Kinetics}

\begin{figure}
    \centering
      \begin{tabular}{@{}c@{}}
      \includegraphics[width=0.8\textwidth]{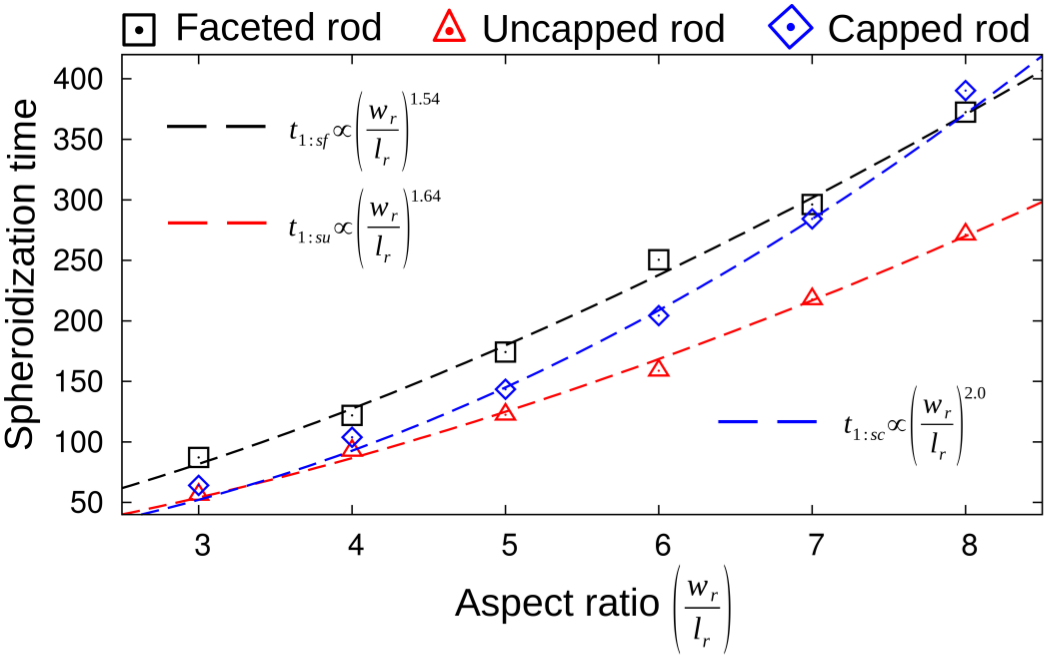}
    \end{tabular}
    \caption{ The influence of the aspect ratio on time taken for the spheroidization of the faceted and uncapped rods. The transformation kinetics of the capped rods are included for comparison. The spheroidization time of capped, uncapped and faceted rods are represented by $t_{1:\text{sc}}$, $t_{1:\text{su}}$ and $t_{1:\text{sf}}$, respectively.
    \label{fig:fig_11}}
\end{figure}

\nomenclature{$t_{1:\text{su}}$}{Time taken for the spheroidization of uncapped rod}%
\nomenclature{$t_{1:\text{sf}}$}{Time taken for the spheroidization of faceted rod}%
In addition to influencing the initial stages of the spheroidization, owing to the unique curvature difference introduced by the sharp edges and corners, the transformation kinetics is significantly effected by the characteristic morphology of the uncapped and the faceted rods. 
Since the curvature along the orthogonal edges and corners of the uncapped and faceted rods are analytically ill-defined, the theoretical approach adopted for the predicting the transformation kinetics.
Moreover, an alternate numerical approach has been employed to the understand the spheroidization kinetics and the mechanism of structures which accommodate analytically ill-defined junctions~\cite{lee1989two}.
However, the consideration of surface diffusion as the only mode of diffusion, limits the applicability of this treatment.
Therefore, in the present analysis, the transformation kinetics is elucidated entirely based on the outcomes of the simulation.

Fig.~\ref{fig:fig_11} shows the time taken for the spheroidization of uncapped and faceted rods of different aspect ratio.
For comparison, the spheroidization rate exhibited by the capped rods are included.
Fig.~\ref{fig:fig_11} unravels that, irrespective of the aspect ratio, the time taken for the transformation of the uncapped rod is visibly lower than the corresponding capped and faceted precipitates.
Two factors contribute to the increased rate of the evolution in the uncapped structures.
One, for a given aspect ratio, the amount of required mass-transfer for spheroidization ($\delta V$) is minimal for the uncapped rod, when compared to the other rods.
In the capped precipitates, the mass associated with the hemispherical inclusions, which are augmented in the longitudinal ends, increases the amount of mass transfer required to spheroidise the structure.
Moreover, owing to the geometrical configuration of the faceted precipitate, the required mass-transfer is inherently higher in these rods.
These morphological factors are responsible for the relatively low amount of required mass-transfer in uncapped structures.
The other factor contributing to the enhanced spheroidization rate is the curvature difference at the longitudinal ends of the uncapped rod.
The removal of the hemispherical caps introduces sharp edges in the terminations of the uncapped precipitates.
The sharp edges relatively increases the driving force in the early stages of the spheroidization, thereby reducing the time taken for the transformation of the uncapped precipitate.

For a given aspect ratio, the volume of the faceted rod is significantly higher than the corresponding capped and uncapped rods due to its geometrical consideration, as shown in Fig.~\ref{fig:fig_9}.
Consequently, an increased amount of mass transfer is required to spheroidise the faceted precipitate.
However, unlike the uncapped rods wherein the sharp edges are confined to the terminations, the faceted structures includes orthogonal corners that extend along the entire length of the rods.
The sharp corners, along with the orthogonal edges, enhance the driving force for the spheroidization.
Interplay of these factors, increased mass transfer and enhanced driving-force, dictate the time taken for the spheroidization of the faceted rods.
The influence of the aspect ratio on the spheroidization rate of the faceted precipitate is included in Fig.~\ref{fig:fig_11}.
Similar to the other rods, the time taken for the transformation progressively increases with the aspect ratio of the faceted rod.

In Fig.~\ref{fig:fig_11}, it is observed that the time taken for the spheroidization of the capped rod of aspect ratio 8 is higher than its corresponding faceted rod.
Despite the low amount of required volume-transfer, the transformation rate is evidently lower in the capped rod.
Two factors are responsible for this unlikely dependency of the spheroidization kinetics.
One is the influence of the contra-diffusion.
As elucidated in Sec.~\ref{lab:contra}, during spheroidization, the capped rod of aspect ratio 8 experiences a significant amount of contra-diffusion, due to the inherent transformation mechanism.
The contra-diffusion prolongs the time taken for the spheroidization.
The other factor is the characteristic morphology of the faceted rod.
The curvature difference introduced by the orthogonal corners and edges enhances the spheroidization kinetics of the faceted rod, which proportionately reduces the time taken for spheroidization.
These factors collectively yield the spheroidization kinetics associated with the faceted and capped rod of aspect ratio 8.

\section{Ovulation-assisted spheroidization}

\begin{figure}
    \centering
      \begin{tabular}{@{}c@{}}
      \includegraphics[width=0.6\textwidth]{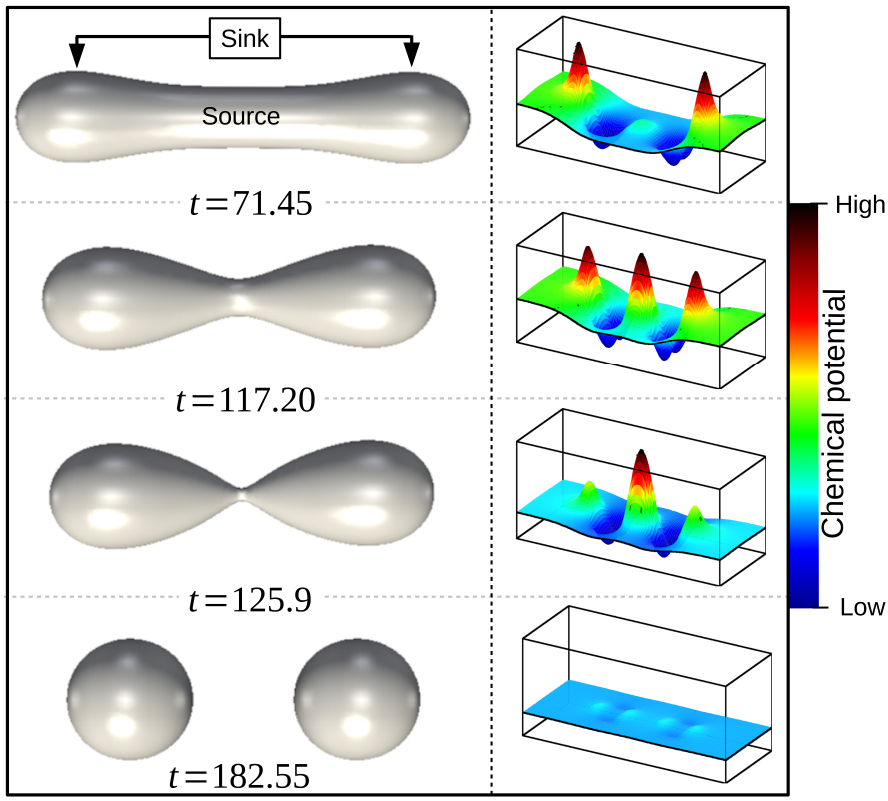}
    \end{tabular}
    \caption{ The temporal evolution of the capped rod of aspect ratio 9. The spheroidization mechanism is predominantly governed by the ovulation of the precipitate. 
    \label{fig:fig_12}}
\end{figure}

It has been shown that, unlike the two-dimensional structures, during the morphological transformation of the three-dimensional precipitate the curvature difference does not exhibit a smooth-monotonic decrease with time.
Since the curvature difference predominantly governs the morphological changes of the precipitate, the transformation mechanism is subsequently influenced by this characteristic change in the driving force.
During the spheroidization of the capped rods, it has been identified that the temporal decrease in the driving force experiences a definite period of the stagnation, which is replaced by the non-monotonic hump in larger rods.
The sluggish, or non-monotonic, evolution of the curvature difference indicates the necking of the rod through the mass transfer from the central region of the precipitate to the receding perturbations.
However, with time the contra-diffusion is reversed and no substantial change is observed in the transformation mechanism.
Despite the lack of any considerable change in the transformation mechanism, it has been realised that the magnitude of contra-diffusion increases with the aspect ratio of the rods.
In order to extend the current understanding on the role of contra-diffusion in the volume-diffusion governed spheroidization, the present numerical investigation is extended to larger rods.

\subsection{Mechanism}

Fig.~\ref{fig:fig_12} illustrates the morphological evolution of the capped rod of aspect ratio 9.
As opposed to the other smaller rods, a significant difference is evident in the spheroidization mechanism.
The morphological evolution encompasses breaking-up of single capped rod into two individual precipitates.
The fragmentation of a single structure into two (or many) sub-structures is referred to as \lq ovulation\rq \thinspace~\cite{nichols1976spheroidization,mclean1973kinetics}, and correspondingly, the transformation is categorised as ovulation-assisted spheroidization.

The capped rod of aspect ratio 9, owing to its geometrical configuration, begins to evolve like the smaller precipitate, governed by the high potential established around the longitudinal ends of the plate.  
The potential difference induces mass transfer from the terminations of the plate to the adjacent flat surface which result in the formation of the longitudinal perturbations.
Driven by the distribution of the chemical potential, the ridges grow at the expense of the remnant region of the precipitate.
During the growth of the perturbations, in three-dimensional rods, appropriate curvature-difference is established between the termination ridges and the remnant region, which increase the potential in the midriff of the rod, as shown in Fig.~\ref{fig:fig_12} at $t=71.45$.
Consequently, the central region becomes the source of the mass transfer, referred to as contra-diffusion, to the receding edges.
In the smaller rods, the sink, which is generally confined to the foot of the receding ridges, coalesce at the central region of the rod, primarily due to the size.
This coalescence of the low-potential region at the midriff of the rod, impedes the mass flow pertaining to the contra-diffusion, and transforms it to sink.
However, as shown in Fig.~\ref{fig:fig_12} at $t=117.20$, interplay of the particle size and the migration rate prevents the low-potential sinks from the reaching the central region of the rod of aspect ratio 9.
Consequently, the contra-diffusion continually transfers mass from the midriff to the adjacent sink, Fig.~\ref{fig:fig_12} at $t=125.9$, governed by the progressively increasing the difference in the chemical potential.
Ultimately, the unhindered contra-diffusion from the central region of the rod results in the fragmentation, or ovulation, of the precipitate.
The pear-shaped precipitate emerging from the ovulation, governed by the inherent difference in the curvature, transforms independently into spheroid, as shown in Fig.~\ref{fig:fig_12} at $t=182.55$.

The aspect ratio above which the ovulation occurs is referred to as critical aspect-ratio~\cite{nichols1976spheroidization}.
Experimental observations indicate that, for volume-diffusion governed spheroidization, the capped rod with aspect ratio greater than 8 fragments during the morphological evolution~\cite{mclean1973kinetics}.
Consistent with the experimental observation, the present numerical studies show that the spheroidization is governed by the termination migration below the critical aspect-ratio of 8, while ovulation is induced in the capped rod of aspect ratio 9.

\subsection{Kinetics}

\begin{figure}
    \centering
      \begin{tabular}{@{}c@{}}
      \includegraphics[width=0.8\textwidth]{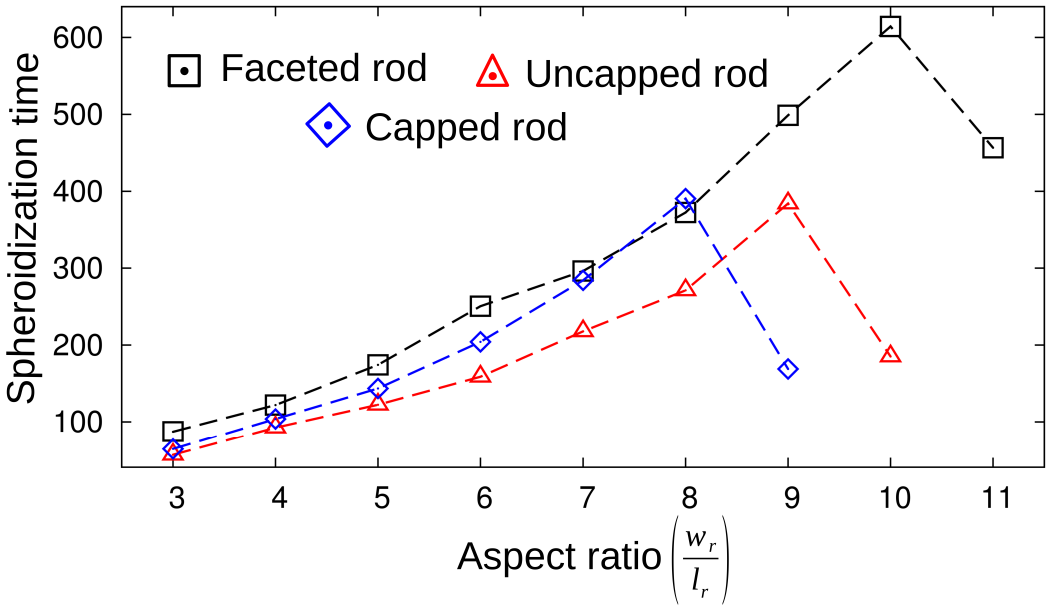}
    \end{tabular}
    \caption{ Drastic change in the spheroidization rate due to the shift in the transformation mechanism of the capped, uncapped and faceted rods. 
    \label{fig:fig_13}}
\end{figure}

To realise the impact of ovulation of the spheroidization kinetics, the plot in Fig.~\ref{fig:fig_11} is extended to encompass the rods which ovulate during the transformation.
For the comparative analysis, the transformation rates exhibited by the uncapped and faceted rods are included.
In the uncapped rods, as shown in Fig.~\ref{fig:fig_11}, the time taken for the spheroidization drastically decreases due to the fragmentation of the precipitate.
Moreover, similar behaviour is observed in all the rods.
However, the critical aspect-ratio are different for the capped, uncapped and faceted rods.
While the capped rod of aspect ratio 9 spheroidises through ovulation, the critical aspect-ratios of uncapped and faceted rods are 9 and 10, respectively.

As discussed in the previous section, two factors dictate the onset of ovulation; the size of the receding perturbation and the transformation rate.
The introduction appropriate curvature-difference, which induces contra-diffusion, primarily depends on the size of the migrating termination.
Moreover, the contra-diffusion is sustained by the transformation rate.
Although the contra-diffusion is induced, the receding terminations prevent the mass transfer by facilitating the coalescence of sink in the central region, when the transformation rate is higher.
The interplay of these factors are responsible for the non-conformity in the critical aspect-ratio between the difference rods in Fig.~\ref{fig:fig_11}.

For a given aspect ratio, due to the lack of hemispherical caps, the volume of the uncapped rod is comparatively low.
Therefore, in relation to the capped precipitate, the appropriate ridges which favour the contra-diffusion are formed in the uncapped rod of higher aspect ratio.
Consequently, a corresponding shift in the critical aspect-ratio is observed in Fig.~\ref{fig:fig_11}. 
In contrast to the uncapped rod, the volume of the faceted rod is greater than the other rods of similar aspect ratio.
Despite this size-advantage, the critical aspect-ratio of the faceted rod is higher (10) than the capped structure (8).
This shift in the critical aspect-ratio can be attributed to the increased rate of the transformation exhibited by the faceted rod.
Owing to the morphology of the faceted precipitate, particularly, the sharp corners that extend along the length of the rod, the spheroidization rate is enhanced in the faceted rod.
The enhanced transformation rate hinders the contra-diffusion which consequently prevents the ovulation.
Since the spheroidization rate decreases with size, ovulation is introduced in the faceted rod of aspect ratio 11, as indicated in Fig.~\ref{fig:fig_11}.

\subsection{Formation of \lq satellite\rq \thinspace particle}

\begin{figure}
    \centering
      \begin{tabular}{@{}c@{}}
      \includegraphics[width=0.55\textwidth]{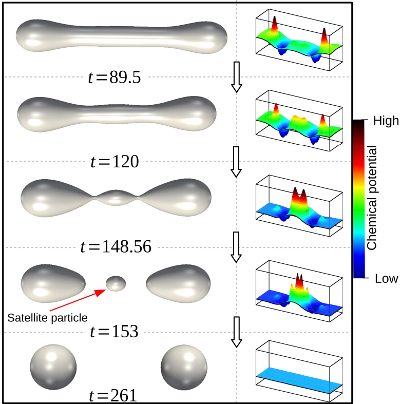}
    \end{tabular}
    \caption{ The spheroidization mechanism exhibited by the capped rod of aspect ratio 12, which yields a smaller \lq satellite\rq \thinspace particle during the morphological evolution.
    \label{fig:fig_14}}
\end{figure}

The aspect ratio, particularly the length, of the finite rod in a material depends predominantly on the process chain~\cite{ashbrook1977directionally}.
In the termination migration-assisted spheroidization, the rod size renders a direct influence on the kinetics by proportionately defining the amount of required mass-transfer for spheroidization.
The indirect effect on the transformation rate is due to the influence aspect ratio on the spheroidization mechanism.
A substantial change in the mechanism is observed in rods larger than the critical aspect-ratio, wherein spheroidization is characterised by the onset of ovulation.
To delineate the influence of the aspect ratio on the ovulation-assisted spheroidization, the numerical investigation is extended to the rods of higher aspect ratio ($\frac{w_{\text{r}}}{l_{\text{r}}}>9$).

Capped rods of aspect ratio upto 11 adhere to the mode of spheroidization exhibited by the rod of aspect ratio 9, Fig.~\ref{fig:fig_12}.
However, a substantial deviation in the transformation mechanism is observed during the evolution of the precipitate of aspect ratio 12.
The morphological transformation leading to the spheroidization of the capped rod of aspect ratio 12 is illustrated in Fig.~\ref{fig:fig_14}.
Although the mechanism is assisted by the fragmentation of the precipitate, the ovulation site is shifted from the midriff of the rod.
This change in the region of break-off yields three distinct entities at the end of the ovulation.
While two of the three entities are identical, the third one is considerably small, and is referred to as satellite particle.
This variation in the ovulation-assisted spheroidization of the capped rod of aspect ratio 12, Fig.~\ref{fig:fig_14}, is due to the size of the precipitate.

The volume of the precipitate invariably increases with aspect ratio of the rods.
Therefore, the receding termination ridges of the capped rod of aspect ratio 12 assume a size, which establishes an appropriate curvature difference with the remnant body of the precipitate, that favour contra-diffusion much earlier than the smaller rods.
In the capped rod of aspect ratio 9, the remnant body is restricted to the midriff of the precipitate, when the curvature difference favouring the contra-diffusion is established.
Subsequently, the  unhindered mass transfer, turns the source (midriff) into ovulation site.
However, in the rod of the aspect ratio 12, owing to the increased size, the source for the contra-diffusion is proportionately large and extends beyond the midriff of the precipitate, as shown in Fig.~\ref{fig:fig_14} at $t=120$.
Since the diffusion follows the least mean-free path, the specific regions of the extended source which are adjacent to the foot of the longitudinal ridges lose mass to these low-potential sink.
Correspondingly, as shown in Fig.~\ref{fig:fig_14} at $t=148.56$, the thickness of the regions close to the sinks begins to decrease.
The high potential established in these regions, due to the increasing difference in the curvature, progressively favours the mass flow to the longitudinal ridges.
Furthermore, as illustrated in Fig.~\ref{fig:fig_14}, governed by the distribution of the chemical potential,  during this stage of the transformation, the longitudinal migration of the ridges become dormant while the mass gets transferred predominantly by contra-diffusion.
This unhindered flux of mass from the source to the corresponding sink in the ridges, ultimately, results in the ovulation as shown in Fig.~\ref{fig:fig_14} at $t=153$.

Owing to the shift in the source of contra-diffusion from the midriff to the region adjacent to the foot of the ridges, the ovulation site correspondingly changes.
Furthermore, the extended source renders individual site for each longitudinal perturbation thereby increasing the number of ovulation events.
Although the number of ovulation events increases to two, when compared to the single event in spheroidization of the smaller rods,  the fragmentation occurs simultaneously. 
The simultaneous breaking-up of rods in two distinct sites yields three individual entities as shown in Fig.~\ref{fig:fig_14} at $t=261$.
While the identical sub-structures correspond to the longitudinal ridges, the smaller satellite particle pertains to the extended source.
The significant disparity in the size of the precipitate fragments induces coarsening through Ostwald ripening.
Consequently, the identical precipitates grow at the expense of the satellite particle, which ultimately disappears. 

The influence of size on the ovulation-assisted spheroidization can be understood by considering the corresponding change in the size of the source.
When the source is confined to the midriff, as in the rod of aspect ratio 9, the ovulation occurs in the central region which splits the precipitate into two identical pear-shaped structures.
However, with increase in the aspect ratio,  the remnant region correspondingly widens and a disparity in the chemical potential is established within the extended source.
Subsequently, the specific sites of the remnant body which are adjacent to the foot of the termination ridges become the source for the contra-diffusion and eventually, transforms into ovulation site.
The shift in the ovulation site from the midriff results in the formation of the satellite particle.
Since the size of the satellite particle, in this spheroidization mechanism, corresponds to the size of the remnant body, the size of the smaller entity proportionately increases with the aspect ratio of the rod. 

\subsubsection{Ostwald ripening}

\begin{figure}
    \centering
      \begin{tabular}{@{}c@{}}
      \includegraphics[width=0.8\textwidth]{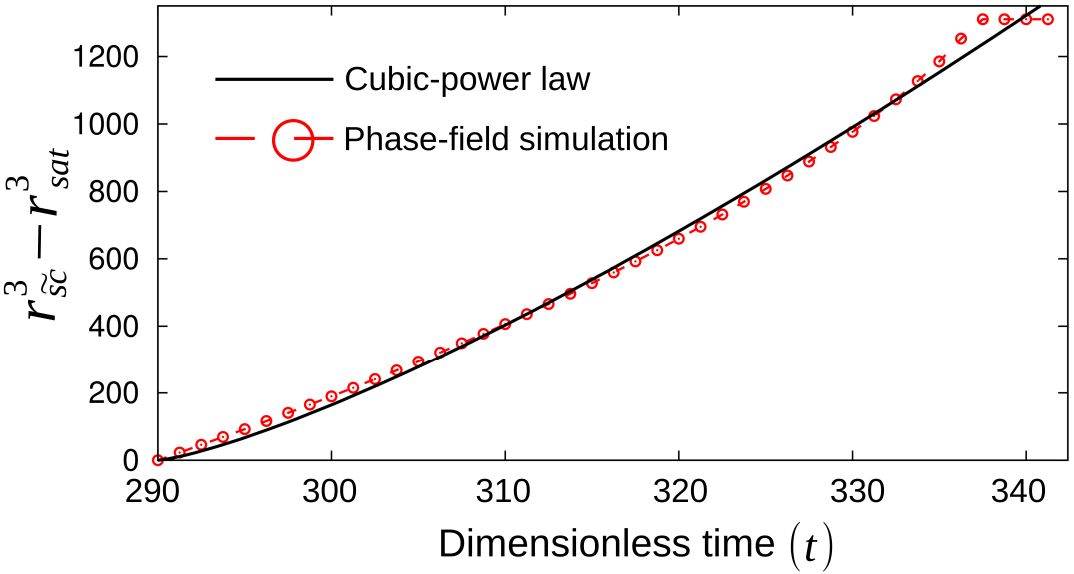}
    \end{tabular}
    \caption{ The coarsening rate of the identical entities at the expense of the satellite particle during the transformation of the capped of aspect ratio 16 is compared to the conventional power law.
    \label{fig:fig_15}}
\end{figure}

The simultaneous ovulation during the transformation of the rod of aspect ratio 12, owing to the dissimilarity in the sizes of the emerging particle, introduces an additional curvature-driven transformation.
Unlike spheroidization which involves the morphological evolution of the individual structure, the coarsening is introduced in a \lq multiphase\rq \thinspace system where the larger particles grow at the expense of the smaller ones.
The ovulation of the rod of aspect ratio 12, as shown in Fig.~\ref{fig:fig_14}, results in a distribution appropriate for the onset of Ostwald ripening.
Subsequently, owing to the significant disparity in the sizes, which introduces a difference in curvature, the identical sub-structures grow by consuming the satellite particle.
Since the size-difference between the satellite particle and the primary spheroids is large, the rate of coarsening is considerably high.
Therefore, the Ostwald ripening associated with the spheroidization of the rod of aspect ratio 12 is not considered for the present analysis.
However, since it has been realized that with increase in the initial size of the precipitate, the size of the satellite particle correspondingly increases, the investigation is extended to larger in order to understand the coarsening kinetics.

The capped rods of aspect ratio upto 20 adopt the transformation mechanism illustrated in Fig.~\ref{fig:fig_14}.
However, it is observed that with increase in the size of the rod, the satellite particle becomes larger and more stable to coarsening.
Particularly, the satellite particle which result during the spheroidization of the capped rod of aspect ratio 20 is almost identical to the primary entities.
Since the Ostwald ripening is prolonged with the increase in the size of the satellite particle, evolution of the rod of aspect ratio 16 is considered for understanding the coarsening.

Unlike the conventional coarsening, the Ostwald ripening which is considered in the present analysis occurs as an integral part of the another curvature-driven transformation, spheroidization.
Therefore, the Ostwald ripening is independently studied to identify any influence of the prior morphological evolution on the coarsening kinetics.
To that end, spheroidization of the capped rod of aspect ratio 16 is monitored.
The particles emerging from the simultaneous ovulation are allowed to spheroidise.
After the entities assume the spherical shape, the respective radii are measured, and the precipitates are allowed to evolve.
The conventional coarsening of the three-dimensional structures adhere to the cubic power law~\cite{voorhees1985theory}.
In order to verify the consistency of the present evolution, the temporal change in the difference of the cubic radius, $r_{\tilde{\text{sc}}}^3-r_{\text{sat}}^3$, where $r_{\tilde{\text{sc}}}$ and $r_{\text{sat}}$ are the respective radius of the primary and satellite spheroid, is analysed and plotted in Fig.~\ref{fig:fig_15}.
Evidently, the coarsening of the primary spheroids at the expense of the satellite particle adhere to the cubic power law.

\subsubsection{Interparticle distance}

\begin{figure}
    \centering
      \begin{tabular}{@{}c@{}}
      \includegraphics[width=0.8\textwidth]{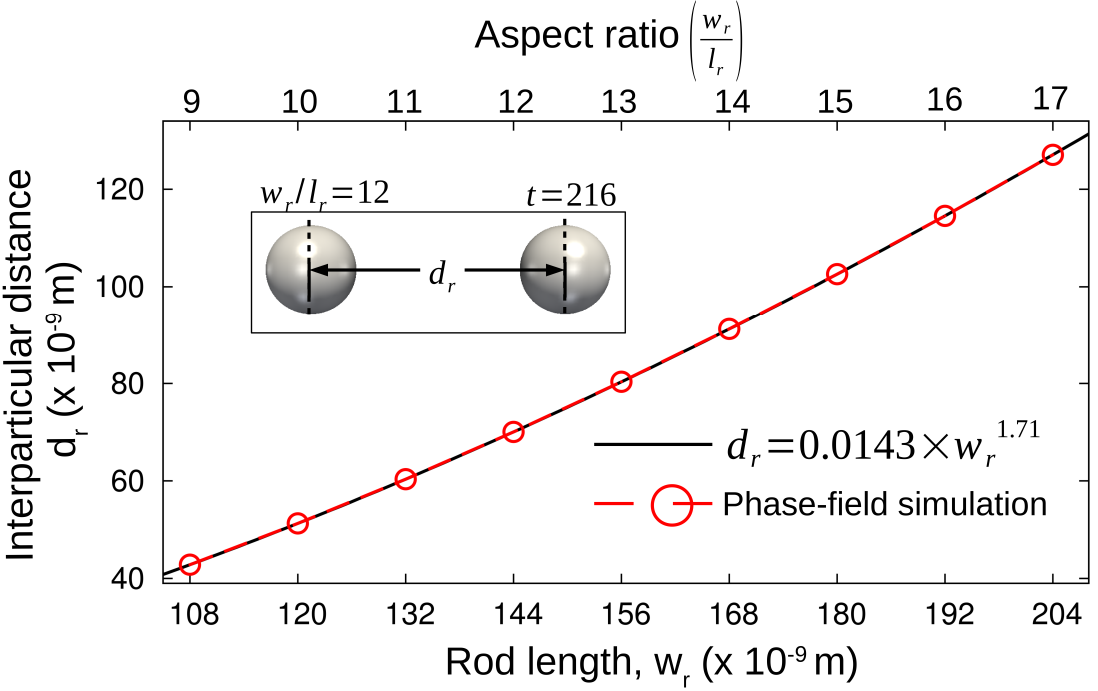}
    \end{tabular}
    \caption{ The distance between the primary spheroids which result from the spheroidization of the capped rods of different aspect ratio, involving the ovulation and subsequent coarsening.
    \label{fig:fig_16}}
\end{figure}

In addition to the morphology, the properties of a material is governed by the distribution of the precipitates~\cite{papazian1990tensile,knipling2008precipitation}.
One factor characterising the distribution of the precipitate, in a microstructure, is the distance separating them.
Since the onset of ovulation in the spheroidization mechanism yields more than one spheroid, the influence of the aspect ratio on the distance separating the precipitate is analysed in this section.

Contra-diffusion is induced by the appropriate curvature-difference established between the receding termination ridges and the remnant body of the rod.
With the onset of the contra-diffusion, the recession of longitudinal perturbation becomes dormant.
Therefore, the position of the ridges, which correspond to the position of the resulting primary precipitate, changes minimally following the introduction of the contra-diffusion.
In other words, the position of the source of contra-diffusion primarily governs the distance separating the primary spheroids.

In relatively smaller rods, the migrating termination ridges confine the remnant body of the precipitate to the midriff.
Therefore, the central region of the rod becomes the source of the contra-diffusion, which ultimately becomes the ovulation site.
However, with increase in the aspect ratio, the spheroidization mechanism shifts from the midriff-fragmentation to the simultaneous ovulation away from the central region.
This shift in the transformation mechanism is due to the widening of the remnant body with increase in the aspect ratio of the rod.
The widening of the remnant body moves the source of contra-diffusion from the midriff to the region adjacent to the foot of the termination ridges.
Correspondingly, larger the size of the remnant body, farther apart are the source of the contra-diffusion.
This increase in the size of the remnant body, and respective change in the position of the contra-diffusion source, influences the distribution of the primary particles.
Accordingly, the distance between the primary spheroids proportionately increases with increase in the aspect ratio of the rods.

In order to ascertain the influence of the aspect ratio on the interparticle distance, the position of the primary spheroids following the ovulation and subsequent, coarsening, is determined from the simulation and plotted in Fig.~\ref{fig:fig_16}.
The present analysis yields a relation
\begin{align}\label{eq:x_mid5}
 d_{\text{r}}=0.0143w_{\text{r}}^{1.71},
\end{align}
where $w_{\text{r}}$ and $d_{\text{r}}$ are the initial length of the rod and interparticle distance, respectively.
\nomenclature{$d_{\text{r}}$}{Interparticle distance}%
Although it is the contra-diffusion source which primarily dictates the position of the spheroids, it should be noted that these sources transform to ovulation sites during the spheroidization.
Therefore, the formation satellite particle prevents the any migration of the primary sub-structures, which eventually spheroidise.

\subsubsection{Ovulation criterion}

\begin{figure}
    \centering
      \begin{tabular}{@{}c@{}}
      \includegraphics[width=0.8\textwidth]{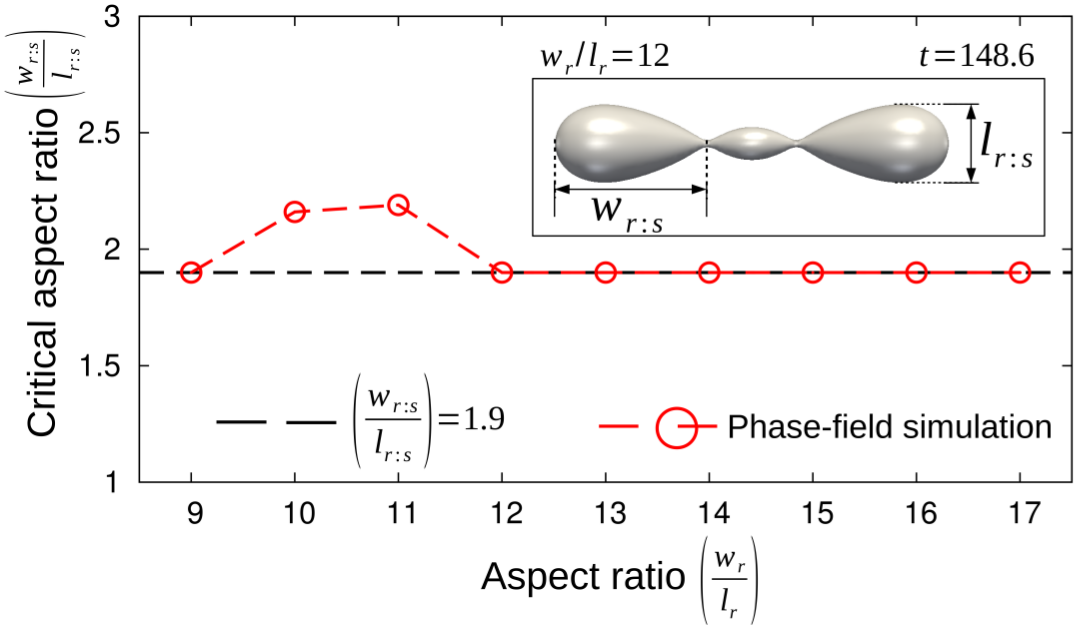}
    \end{tabular}
    \caption{ The aspect ratio of the pear-shaped sub-structure at the moment of ovulation during the spheroidization of larger capped rods of aspect ratio beyond 9.
    \label{fig:fig_17}}
\end{figure}

The proportionate increase in the size of the satellite particle with the aspect ratio of the capped rod indicates the shift in the ovulation site.
The shift in the ovulation in-turn signify that a definite criterion governs the fragmentation of the precipitate.
Although the ovulation criterion can be related to the curvature difference established between the receding edges and remnant which induces the contra-diffusion, the reversal of the mass transfer in the smaller rods disproves such direct comparisons.
Therefore, to identify the ovulation criterion in the volume-diffusion governed spheroidization of finite rods, the transformation mechanism of the larger rods ($\frac{w_{\text{r}}}{l_{\text{r}}}>9$) are
analysed.

In the morphological evolution of the infinite rods, the ovulation criterion is defined based on the parameters which characterise the introduced perturbations~\cite{nichols1965surface,qian1998non}.
Since, in the finite structure, the evolution is governed by the curvature difference which is inherent to the morphology of the precipitate, the ovulation criterion is determined by estimating the aspect ratio of the pear-shaped sub-structure at the point of fragmentation.
Fig.~\ref{fig:fig_17} shows the aspect ratio of the pear-shaped precipitates which result during the spheroidization of the capped rods of different aspect ratios.
For this illustration, the aspect ratio is measured at the ovulation point.

Fig.~\ref{fig:fig_17} unravels that, expect for the rods of aspect ratio 10 and 11, the aspect ratio of the pear-shaped precipitates are equal for the all the other rods.
Furthermore, it is evident that the aspect ratio of the precipitate at the point of fragmentation  is equal to, $\frac{w_\text{r:s}}{l_\text{r:s}} = 1.9$.
The independent nature of the $\frac{w_\text{r:s}}{l_\text{r:s}}$ substantiates the choice of this parameter for defining the ovulation criterion.
The deviations in Fig.~\ref{fig:fig_17} pertaining to the rods of aspect ratio 10 and 11 can be elucidated by their spheroidization mechanism.
The rods of the aspect ratio 10 and 11 are sandwiched between the structures which exhibit two different transformation mechanism.
While the rod of aspect ratio 9 experiences midriff fragmentation, due to the restricted remnant region, a shift in the ovulation site is observed the larger rod of aspect ratio 12, owing to the larger size remnant region.
The remnant region of the rods of aspect ratio 10 and 11 is relatively extended when compared to the corresponding region of smaller rod aspect ratio 9.
However, despite this extension, the remnant region in the rods of aspect ratio 10 and 11 are not large enough to shift the ovulation site away form the midriff.
The lack of adequate remnant region is responsible for the deviations in Fig.~\ref{fig:fig_17}.

\subsubsection{Ovulation time}

\begin{figure}
    \centering
      \begin{tabular}{@{}c@{}}
      \includegraphics[width=0.8\textwidth]{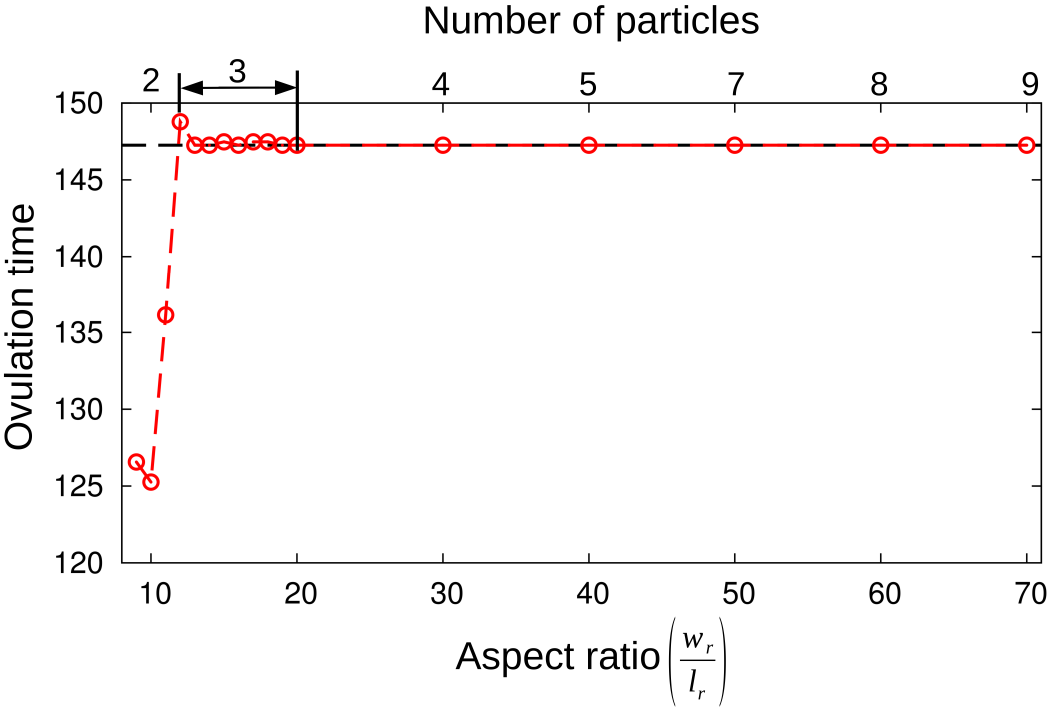}
    \end{tabular}
    \caption{ The time taken for the primary ovulation during the transformation of the rods with different aspect ratio.
    \label{fig:fig_18}}
\end{figure}

The present numerical study is systematically extended to seemingly infinite capped rods of aspect ratio upto 70, in order to identify any other variant of the ovulation-assisted spheroidization.
Outwardly, no significant deviation is observed in the spheroidization mechanism of the larger rods.
However, number of ovulation events, which are not simultaneous, increase with the size of the rods.
Furthermore, the non-simultaneous fragmentation correspondingly increased the resulting number of spheroids.
This intense ovulation-assisted spheroidization can be concisely described by considering the evolution of the rod of aspect ratio 30.
Similar to the spheroidization of the smaller rods, the evolution of the rod of aspect ratio 30 encapsulates an ovulation event wherein the primary pear-shaped sub-structures are formed leaving behind a precipitate fragment.
Since the fragment resulting from the initial ovulation is larger than the primary substructure, it cannot be referred to as satellite particle in this rod.
Owing to the size and morphology, the fragment precipitate continues to the transform and eventually ovulate which lead to the formation of two more pear shaped entities (secondary).
Therefore, the two non-simultaneous ovulation events in the spheroidization of the capped rod of aspect ratio 30, ultimately, yields four spheroids.
With increase in the size of the rods, the number of progressive ovulation increase which results in increased number of the spheroids.
The number of the precipitates emerging from the spheroidization of the rods of different aspect-ratio is shown in Fig.~\ref{fig:fig_18}.

Theoretical treatment of the surface-diffusion governed spheroidization of finite rods predicts that the first (primary) ovulation in the large rods occurs at the same time~\cite{nichols1965surface}.
This claim is yet to be verified for volume-diffusion governed spheroidization.
Therefore, the primary ovulation time of the rod considered in the present study is determined and plotted in Fig.~\ref{fig:fig_18}.
Evidently, above the aspect ratio 12, the no visible change is observed in the time taken for the primary ovulation.
Furthermore, Fig.~\ref{fig:fig_18}, confirms that although the specific criticality vary with the dominant mode of mass transfer, volume or surface diffusion, the overall scheme of evolution remains unchanged.

\section{Conclusion}

Unlike surface-diffusion governed transformations, the theoretical investigation of the volume-diffusion governed evolution demands the consideration of entire system.
Therefore, the studies reporting on the volume-diffusion governed transformations are limited.
The theoretical treatment get furthermore complicated when the evolution of three-dimensional structures are analysed.
Since the phase-field approach, while encompassing the entire domain, elegantly handles the curvature difference, the spheroidization of the finite three-dimensional rods are numerically investigated in this chapter.

The present analysis unravels that, the hitherto assumed smooth-monotonic temporal evolution of the curvature difference~\cite{courtney1989shape,park2012prediction,semiatin2005prediction,park2012mechanisms}, is not an appropriate description of the transformation of three-dimensional structures.
During the spheroidization of the three-dimensional capped rods, depending on the aspect ratio, the smooth-monotonic decrease in the driving force is disrupted by a definite period of stagnation or non-monotonic hump.
The deviation from the expected monotonic decrease in driving force is due to the introduction of an appropriate curvature-difference between the receding perturbation and remnant body which induces contra-diffusion.
Although in the smaller rods the contra-diffusion is reversed, in rods larger than the critical aspect-ratio, the progressive contra-diffusion leads to the fragmentation of the precipitate.
This substantial shift in the transformation mechanism is observed in the capped rod of aspect ratio 9.
The change in the spheroidization mechanism from termination migration-assisted to ovulation-assisted substantially influence the kinetics.

Increase in the size of the capped rod yields a different variant of the ovulation-assisted spheroidization.
In this transformation mechanism, owing to the increased size of the rod, the size of the remnant body involved in the contra-diffusion proportionately increases.
The widening of the region between the termination ridges shifts the source from the midriff to the foot of the perturbation.
Therefore, the initial ovulation leads to the formation of primary sub-structure and precipitate fragment.
Depending on the initial aspect ratio, the fragments (satellite particle) either disappears through Ostwald ripening or spheroidises independently.

In the present study, since a phase-field model which convincingly handles analytically ill-defined curvatures is employed, morphological evolution of the uncapped and faceted rods are analysed.
Furthermore, by exploiting the ability of the adopted approach to render consistent evolution of the singularity events like fragmentation, a ovulation-criterion for volume-diffusion governed spheroidization is identified.

\newpage\null\thispagestyle{empty}\newpage

\afterpage{\blankpage}

\newpage
\thispagestyle{empty}
\vspace*{8cm}
\begin{center}
 \Huge \textbf{Part IV} \\
 \Huge \textbf{Stability of three-dimensional plates}
\end{center}

\chapter{Globularisation of unidirectionally-equiaxed \lq pancake\rq \thinspace structure}\label{chap:pancake}

The compositional make-up of a system critically governs the morphology of the accommodating phases in the microstructure.
Correspondingly, in certain alloy systems, the lamellar arrangement of the phases comprises of shapes which are significantly different from the conventional ribbon- or rod-like structures~\cite{zhao2008influence,day1968microstructure}.
The morphology of the precipitate, when compared between two different materials, is predominantly governed by the specific crystallographic relation between the alternating phases~\cite{buchanan2012crystallography,makhlouf2001aluminum}.
However, it has been observed that, within a alloy system, the shape of the phases noticeably vary with the change in the chemical composition~\cite{bedolla2005effect,shin2003effect,hu2002effect}.
The influence of the alloying elements on the morphology of the precipitate is attributed to the mismatch in the lattice parameter which contributes to the coherency strain~\cite{cline1970effect,nathal1985influence}.
Furthermore, the parameters involved in the process chain, particularly the thermal cycle, additionally effect the precipitate shape in a microstructure~\cite{sohn2013effect,culpan1978microstructural}.
Considering the spectrum of possible structures in a lamellar arrangement of the phases, the investigation of the shape-instability is extended to unconventional morphologies.

Analysing the morphological evolution of three-dimensional structure, which are exclusively governed by surface diffusion, is relatively straightforward, since the mass transfer is restricted to the two-dimensional surface~\cite{nichols1965morphological}.
But in volume-diffusion governed transformation, the atomic fluxes which drift through the matrix and precipitate dictate the evolution~\cite{nichols1965surface}.
Therefore, the migration of atoms in the entire domain is examined to understand the microstructural changes governed by the volume-diffusion.
Owing to this numerical intricacies, the theoretical treatment of volume-diffusion governed evolution are limited~\cite{ho1974coarsening,nakagawa1972stability,tian1987kinetics}.
Moreover, the existing works are primarily confined to two-dimension or conventional structures like rods~\cite{nichols1965surface}.
Despite the lack of complete understanding, the volume-diffusion governed shape-instability exhibited by the unconventional structures is employed to enhance the mechanical properties of highly applicable materials~\cite{atasoy1989pearlite,ogris2002silicon,rack2006titanium}.

A prime example of a material wherein the precipitate hardly assumes a regular rod-like structure is the two-phase titanium alloy which includes alloying elements like aluminium and vanadium~\cite{christoph2003titanium}.
Like any another alloy systems, the morphology of the phases ($\alpha$ and $\beta$) in the two-phase titanium alloy is governed by the manufacturing process.
However, the general processing-route which involves a series of hot working and heat treatment yields a lamellar microstructure of alternating $\alpha$ and $\beta$ phases~\cite{lutjering1998influence}.
The mechanical properties of the two-phase titanium alloy is further improved by an isothermal annealing treatment, referred to as static globularisation~\cite{filip2003effect,stefansson2002kinetics}.
The thermal cycle of the static globularisation is primarily devised to avert any phase-transformation, while facilitating an accelerated morphological evolution.
In the early stages of the annealing, owing to the prior hot working, the $\alpha$-colonies of the two-phase lamellar arrangement disintegrate~\cite{fan2012mechanism,roy2013microstructure}.
The breaking-up of the $\alpha$-precipitate through the splitting of the sub-boundaries disrupts lamellar arrangement and the results in a microstructure which comprises unconventionally-shaped $\alpha$-precipitate distributed in the $\beta$-matrix.
The morphological evolution of the unconventional structures is the dominant transformation which accompanies the static globularisation.
The shape of the $\alpha$-precipitate and the dominant mode of diffusion, volume diffusion, have hitherto restricted the theoretical investigation of the static globularisation.
Since it has already been shown that the present numerical approach elegantly handles the curvature-driven transformation and yields thermodynamically-consistent outcomes, this technique is adopted to analyse the morphological evolution of the refined three-dimensional structures.

\section{Domain setup}

\begin{figure}
    \centering
      \begin{tabular}{@{}c@{}}
      \includegraphics[width=0.8\textwidth]{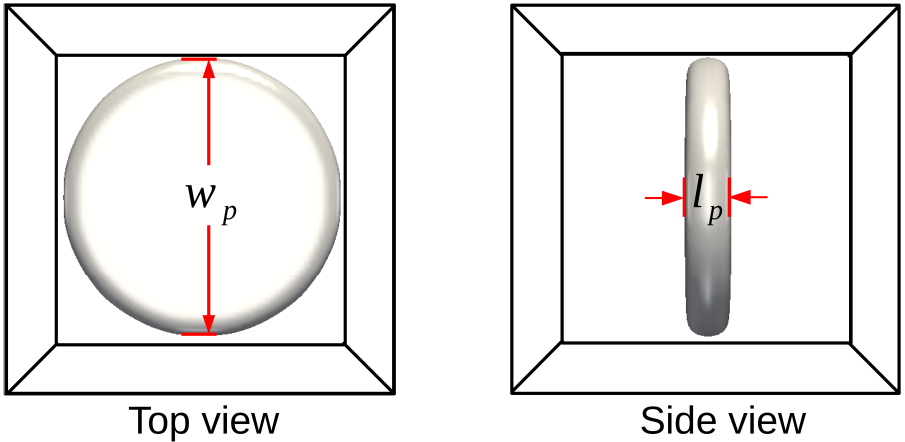}
    \end{tabular}
    \caption{ Top- and side-view of the unidirectionally-equiaxed \lq pancake\rq \thinspace structure of diameter $w_\text{p}$ and thickness $l_\text{p}$. 
    \label{fig:setup}}
\end{figure}

The splitting of the continuous precipitates at the sub-boundaries, which are introduced during the process chain, yields a wide range of individual structures.
Experimental observation have identified that the considerable fraction of $\alpha$-precipitate, subsequently after the initial fragmentation, appears equiaxed along a particular direction~\cite{lutjering1999property,semiatin1994microstructure}.
Therefore, the theoretical treatment which attempts to predict the kinetics of the globularisation assumes a \lq pancake\rq \thinspace structure which seemingly corroborates the observed microstructure~\cite{semiatin2005prediction}.
Despite considering a simplified variant of the precipitate shape, in the existing work, hemispherical caps are augmented for defining the curvature difference at the termination.
Given that the pancake structure seemingly resembles the morphology of the observed precipitate, the hemispherical inclusion hampers this corroboration. 
Therefore, by formulating an appropriate geometrical configurations, the edges of the precipitate are smoothened without any inclusions.

The precipitate morphology adopted for the present analysis is shown in Fig.~\ref{fig:setup}.
The diameter of the pancake shape is represented by $w_\text{p}$ while $l_\text{p}$ is the thickness of the structure.
\nomenclature{$w_\text{p}$}{Diameter of the pancake structure}%
\nomenclature{$l_\text{p}$}{Thickness of the pancake structure}%
The aspect ratio is defined as the ratio of the diameter and thickness ($\frac{w_\text{p}}{l_\text{p}}$).
The thickness of the precipitate is fixed at $l_\text{p}=0.01 \times 10^{-6}$m, and the diameter is varied in relation to the required aspect ratio.
Appropriate concentration is assigned to the phases to establish chemical equilibrium, and the domain is sufficiently large to avoid the influence of the boundary conditions.

\section{Transformation kinetics}

Generally, the pancake morphology is considered as a three-dimensional cylinder of very low height, with the hemispherical cap encapsulating the otherwise sharp edges of the precipitate~\cite{semiatin2005prediction}. 
In order to obviate the geometrical inclusions, which are virtually non-physical and distort the resemblance of the structure to the microscopically observed precipitate, the unidirectionally equiaxed pancake shape is configured differently.
For the present theoretical approach, the pancake morphology is assumed to be a segment, with height $l_\text{p}$, of a larger solid sphere of diameter $w_\text{p}$.
The pancake shape, defined as a part of a larger spheroid, retains smooth termination which consequently averts the need for hemispherical caps.

\subsection{Geometrical treatment}

\begin{figure}
    \centering
      \begin{tabular}{@{}c@{}}
      \includegraphics[width=0.4\textwidth]{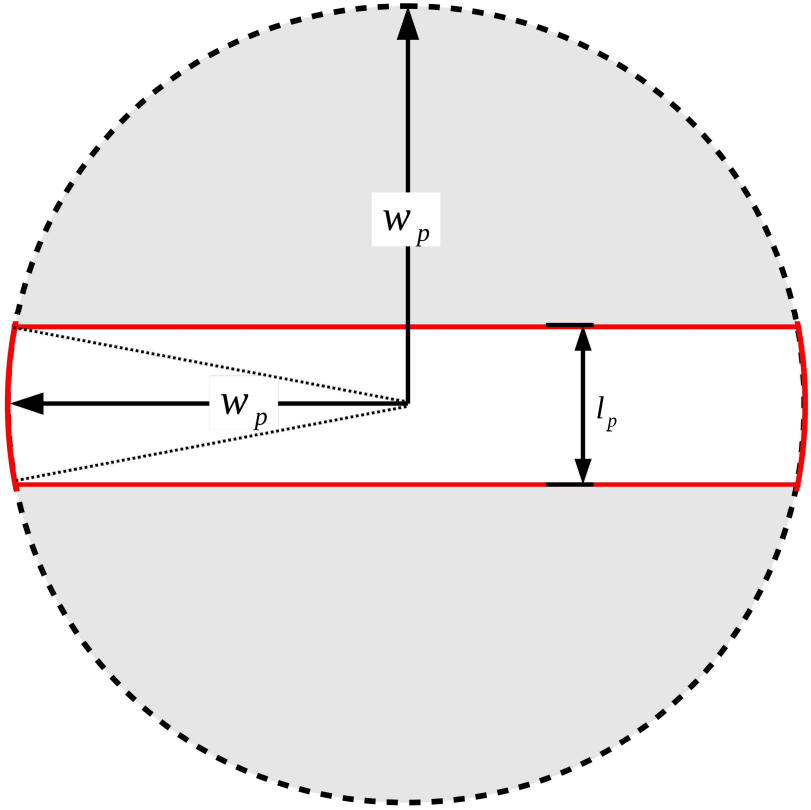}
    \end{tabular}
    \caption{ A cross-sectional view of the geometrical consideration adopted to define the precipitate structure. 
    \label{fig:fig_2}}
\end{figure}

A schematic representation of the precipitate cross-section, which depicts its overall morphological configuration, is shown in Fig.~\ref{fig:fig_2}.
Owing to the unique description of the pancake structure, in the present analysis, the existing geometrical treatment, Ref.~\cite{semiatin2005prediction}, cannot be directly adopted.
Therefore, in this section, the parameters required for estimating the globularisation kinetics is ascertained for the considered morphology.

Fig.~\ref{fig:fig_2} indicates that the precipitate, which is considered for this investigation, is a central segment of a solid sphere.
Correspondingly, as opposed to the shapes hitherto analysed, the pancake morphology comprises of flat surfaces which are circular in the respective normal direction.
When normal to the flat surfaces is along the $y$-axis, the volume of the precipitate is calculated by
\begin{align}\label{eq:ppt_volume1}
 V_{\text{p}}=\int_{\frac{w_\text{p}}{2}-\frac{l_\text{p}}{2}}^{\frac{w_\text{p}}{2}+\frac{l_\text{p}}{2}}\pi \tilde{r}^2 \diff y,
\end{align}
where $\tilde{r}$ is the radius of the equiaxed flat surface associated with the pancake morphology.
Owing to the geometrical nature of the precipitate, the radius of the circular surface is related to the diameter and thickness as $\tilde{r}^2=\left(\frac{w_\text{p}}{2}\right)^2-\left(\frac{l_\text{p}}{2}\right)^2$.
Substituting this relation in Eqn.~\ref{eq:ppt_volume1}, and subsequently integrating it, yields
\begin{align}\label{eq:ppt_volume2}
 V_{\text{p}}=\pi\left[\left(\frac{w_\text{p}}{2}\right)^2-\left(\frac{l_\text{p}}{2}\right)^2\right][y]^{\frac{w_\text{p}}{2}+\frac{l_\text{p}}{2}}_{\frac{w_\text{p}}{2}-\frac{l_\text{p}}{2}}.
\end{align}
From the above Eqn.~\ref{eq:ppt_volume2}, the entire volume of the precipitate reads
\begin{align}\label{eq:ppt_volume3}
 V_{\text{p}}=\pi l_\text{p}\left[\left(\frac{w_\text{p}}{2}\right)^2-\left(\frac{l_\text{p}}{2}\right)^2\right].
\end{align}

A critical parameter which dictates the transformation kinetics is the amount of mass transfer required to globularise the precipitate, $\delta V_{\text{gp}}$.
The required mass-transfer is calculated by the eliminating the volume shared by the initial and final morphology of the precipitate.
Since the region shared by the initial structure and globularised precipitate is the central segment of the ultimate spheroid, the geometrical approach used for calculating the precipitate volume can be adopted.
Correspondingly, the volume of the overlapping region can be expressed as
\begin{align}\label{eq:sh_volume}
 V_{\text{p:sh}}&=\int_{r_{\text{gp}}-\frac{l_\text{p}}{2}}^{r_{\text{gp}}+\frac{l_\text{p}}{2}}\pi\left[ r_{\text{gp}}^2-\left(\frac{l_\text{p}}{2}\right)^2 \right]\diff y\\ \nonumber
 &=\pi l_\text{p} \left[ r_{\text{gp}}^2-\left(\frac{l_\text{p}}{2}\right)^2 \right],
\end{align}
where $r_{\text{gp}}$ is the radius of the globularised precipitate.

From Eqns.~\ref{eq:ppt_volume3} and ~\ref{eq:sh_volume}, the amount of mass transfer required for the globularisation of the pancake structure can be written as
\begin{align}\label{eq:req_volume}
 \delta V_{\text{gp}}=V_{\text{p}}-V_{\text{p:sh}}=\pi l_\text{p}\left[\left(\frac{w_\text{p}}{2}\right)^2-r_{\text{gp}}^2\right].
\end{align}
\nomenclature{$\delta V_{\text{gp}}$}{Required mass-transfer for globularisation of the pancake structure}%
Owing to the chemical equilibrium established between the phases, the volume of the precipitate is conserved all through the evolution.
Therefore, the radius of the globularised structure, can be related to the geometrical parameters pertaining to the initial morphology by
\begin{align}\label{eq:radius_s}
 r_{\text{gp}}=\left\{\frac{3}{4}l_\text{p}\left[\left(\frac{w_\text{p}}{2}\right)^2-\left(\frac{l_\text{p}}{2}\right)^2\right]\right\}^{1/3}.
\end{align}

In the present analysis, two different approaches are employed to ascertain the globularisation kinetics of the pancake structures.
Since both these techniques share the existing theoretical framework~\cite{courtney1989shape,park2012prediction,semiatin2005prediction,park2012mechanisms}, the analytical treatment begins by identifying three distinct stages of the transformation which are referred to initial ($t_{0:\text{gp}}$), midpoint ($t_{1/2:\text{gp}}$) and final ($t_{1:\text{gp}}$).
First of the two approaches, referred to as cylinderization approach, directly extends the conventional analytical treatment\cite{courtney1989shape} to the re-defined pancake structure, while the second involves the in-situ data from the phase-field simulations.
Despite the differences, the delineation of the driving force in the initial stage of the globularisation is identical in both the formulations.

The driving force at the beginning of the evolution ($t_{0:\text{gp}}$) is ascertained by
\begin{align}\label{eq:dr_initial1}
 \Gamma_{0:\text{gp}}\propto A_{0:\text{gp}}\left( \frac{\delta c}{\delta x} \right)_{0:\text{gp}},
\end{align}
where $A_{0:\text{gp}}$ is the area available for the migration of the atomic flux and $\left( \frac{\delta c}{\delta x} \right)_{0:\text{gp}}$ is the concentration gradient introduced by the inherent  curvature-difference which is primarily dictated by the precipitate shape.
In the initial stages of the transformation, the diffusion area $A_{0:\text{gp}}$ corresponds to the surface area of the curved termination associated with the pancake structure.
The area of the curved termination can be determined by treating it as a surface of revolution (\textit{zone}) along the y-axis.
Accordingly, the area available for diffusion is expressed as
\begin{align}\label{eq:surface_0}
A_{0:\text{gp}} =2\pi\int_{\frac{w_\text{p}}{2}-\frac{l_\text{p}}{2}}^{\frac{w_\text{p}}{2}+\frac{l_\text{p}}{2}} f(x) \sqrt{1+|f'(x)|^2}\diff y,
\end{align}
where $f(x)$ is the function involving the rotating curve.
\nomenclature{$A_{i:\text{gp}}$}{Active diffusion-area during pancake-precipitate globularisation}%
Since, in the present case, the rotating curve is a sphere, the respective function reads
\begin{align}\label{eq:surface_func1}
 f(x)=\sqrt{\left(\frac{w_\text{p}}{2}\right)^2-y^2},
\end{align}
and correspondingly,
\begin{align}\label{eq:surface_func2}
 |f'(x)|^2=\frac{y^2}{\left(\frac{w_\text{p}}{2}\right)^2-y^2}.
\end{align}
By substituting Eqns.~\ref{eq:surface_func1} and ~\ref{eq:surface_func2} in Eqn.~\ref{eq:surface_0}, the diffusion area is written as
\begin{align}\label{eq:surface_01}
 A_{0:\text{gp}}=2\pi\int_{\frac{w_\text{p}}{2}-\frac{l_\text{p}}{2}}^{\frac{w_\text{p}}{2}+\frac{l_\text{p}}{2}}\left\{ \left[ \left(\frac{w_\text{p}}{2}\right)^2-y^2\right]\left[ 1+ \frac{y^2}{\left(\frac{w_\text{p}}{2}\right)^2-y^2}\right] \right\}^{1/2} \diff y.
\end{align}
Solving the above Eqn.~\ref{eq:surface_01} yields
\begin{align}\label{eq:surface_02}
 A_{0:\text{gp}}=\pi w_\text{p}l_\text{p}.
\end{align}

The concentration gradient is analytically solved by separately treating $\delta c_{0:\text{gp}}$, the difference from the equilibrium concentration introduced by the curvature, and the distance associated with the resulting mass transfer, $\delta x_{0:\text{gp}}$.
Before describing the components of the concentration gradient, the length of the curved termination associated with the cross-section of the precipitate, shown in Fig.~\ref{fig:fig_2}, is determined.
The arc-length in the curved edges of the precipitate is calculated by $S_{\text{tip}}=\frac{w_\text{p}}{2}\theta$, where $\theta=2\sin^{-1}\left(\frac{l_\text{p}}{w_\text{p}}\right)$.
Correspondingly, the length of the curved termination is written as
\begin{align}\label{eq:arc_length}
 S_{\text{tip}}=w_\text{p}\sin^{-1}\left(\frac{l_\text{p}}{w_\text{p}}\right).
\end{align}

The difference in the equilibrium concentration which induced due to the morphology of the precipitate is related to the curvature by
\begin{align}\label{eq:conc0}
 \delta c_{0:\text{gp}}\propto \left[\Big(\frac{1}{R_1}+\frac{1}{R_2}\Big)_{\text{sources}}-\Big(\frac{1}{R_3}+\frac{1}{R_4}\Big)_{\text{sinks}}\right],
\end{align}
where $R_1$ and $R_2$ are the principal radii of curvature at the source, while the respective radii at sink is $R_3$ and $R_4$.
\nomenclature{$\delta c_{i:\text{gp}}$}{Disparity in equilibrium induced by pancake shape}%
Since the sinks are flat surfaces of the pancake structure, the respective terms in Eqn.~\ref{eq:conc0} become zero.
Subsequently, by appropriately formulating the principal radii of curvature at the smooth edges of the precipitate, the influence of curvature on the equilibrium concentration can written as
\begin{align}\label{eq:conc01}
 \delta c_{0:\text{gp}}\propto \left[ \frac{2}{w_\text{p}}+\frac{\pi}{w_\text{p}\sin^{-1}\left(\frac{l_\text{p}}{w_\text{p}}\right)} \right].
\end{align}
Furthermore, the diffusion length, which the flux covers to compensate for the concentration gradient, in the initial stage is expressed as
\begin{align}\label{eq:x0}
 \delta x_{0:\text{gp}}\propto \left[ \frac{w_\text{p}}{2}+\frac{\pi}{2w_\text{p}\sin^{-1}\left(\frac{l_\text{p}}{w_\text{p}}\right)}-r_{\text{gp}} \right].
\end{align}
\nomenclature{$\delta x_{0:\text{gp}}$}{Diffusion length associated with pancake-shape globularisation}%
The driving force at the beginning of the globularisation can be estimated by substituting Eqns.~\ref{eq:surface_02},~\ref {eq:conc01} and ~\ref{eq:x0} in Eqn.~\ref{eq:dr_initial1}.
Since the description of the driving force at the initial stage remains unaltered irrespective of the theoretical approach, this delineation will be adopted to ascertain the transformation kinetics in the forthcoming treatments.

\subsection{Cylinderization approach}\label{lab:cy1}

\begin{figure}
    \centering
      \begin{tabular}{@{}c@{}}
      \includegraphics[width=0.7\textwidth]{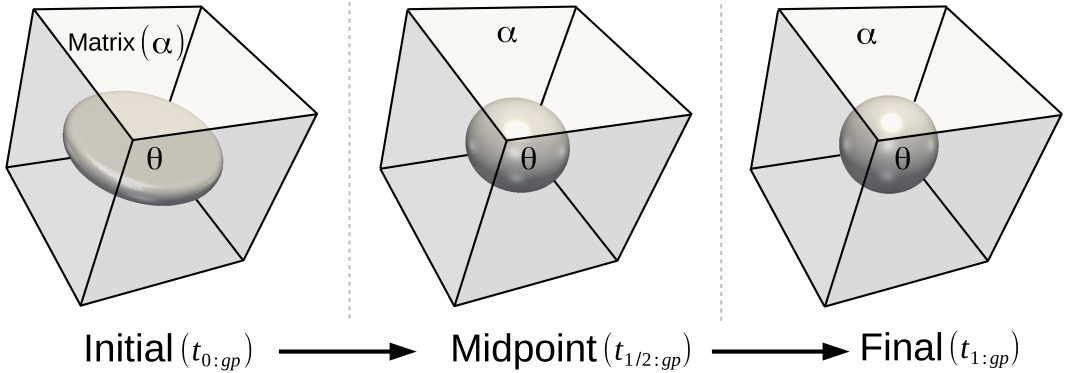}
    \end{tabular}
    \caption{ Morphology of the precipitate at the initial ($t_{0:\text{gp}}$), midpoint ($t_{1/2:\text{gp}}$) and final stage ($t_{1:\text{gp}}$) of the globularisation. 
    \label{fig:fig_1}}
\end{figure}

Owing to the lack of in-situ report on the globularisation of the pancake structure, two assumptions are made to formulate the driving force at the midpoint of the transformation.
The first assumption pertains to the morphology of the precipitate at the midpoint.
In the conventional treatment, Ref.~\cite{semiatin2005prediction}, the precipitate is treated as an ellipsoid to estimate the midpoint driving force.
The shapes of the precipitate at the three distinct stages of the globularisation, simulated using the phase-field approach, are shown in Fig.~\ref{fig:fig_1}.
This illustration apparently supports the theoretical consideration that precipitate exhibits ellipsoidal shape at the midpoint of the evolution.
Secondly, since the geometrical nature of the midpoint ellipsoid is unknown, the parameters, the diffusion area and the concentration gradient, which dictate the driving force at the midpoint of the globularisation is approximated as
\begin{align}\label{eq:midpoint_approx1}
 A_{\frac{1}{2}:\text{gp}}=\frac{A_{0:\text{gp}}+A_{1:\text{gp}}}{2}
\end{align}
and
\begin{align}\label{eq:midpoint_approx1}
 \left(\frac{\delta c}{\delta x}\right)_{\frac{1}{2}:\text{gp}}=\frac{\left(\frac{\delta c}{\delta x}\right)_{0:\text{gp}}+\left(\frac{\delta c}{\delta x}\right)_{1:\text{gp}}}{2},
\end{align}
respectively, where $A_{1:\text{gp}}$ and $\left(\frac{\delta c}{\delta x}\right)_{1:\text{gp}}$ correspond to the diffusion area and concentration gradient at the end of the transformation.

Earlier, it has been elucidated that treating the entire surface area as the diffusion area in the final state ($A_{1:\text{gp}}$) entails an unphysical condition wherein an appropriate sink cannot be accommodated.
Therefore, like the previous analysis, surface area pertaining to a segment of the spheroid is considered as final diffusion area.
Consequently, the diffusion area at the midpoint, which is assumed to be the mean of the respective area in the initial and final state is expressed as
\begin{align}\label{eq:surface_12}
 A_{\frac{1}{2}:\text{gp}}=\frac{\pi}{2}(w_\text{p}l_\text{p}+r_{\text{gp}}^2).
\end{align}
Furthermore, the influence of the curvature on the equilibrium concentration is considered as the half of the summation of its effect at the beginning and the ends of the globularisation.
Correspondingly, the principal radii of the curvature are averaged between the initial and final state.
Based on the resulting curvature difference, the deviation introduced in  the equilibrium concentration is quantified as
\begin{align}\label{eq:conc_12}
 \delta c_{\frac{1}{2}:\text{gp}} \propto \left[ \left(\frac{4}{w_\text{p}+2r_{\text{gp}}} \right)+\left( \frac{2\pi}{w_\text{p}\sin^{-1}\left(\frac{l_\text{p}}{w_\text{p}}\right)+\pi r_{\text{gp}}} \right) \right].
\end{align}
In the similar way, the distance associated with the concentration gradient is formulated, and the driving force at the midpoint of the globularisation is written as
\begin{align}\label{eq:gamma_12}
 \Gamma_{\frac{1}{2}:\text{gp}} \propto A_{\frac{1}{2}:\text{gp}} \delta c_{\frac{1}{2}:\text{gp}}\underbrace{\left\{\frac{1}{4}\left[ w_\text{p} + r_{\text{gp}}(\pi-2) + \frac{\pi}{2w_\text{p}\sin^{-1}\left(\frac{l_\text{p}}{w_\text{p}}\right)}\right]\right\}^{-1}}_{:=\left(\delta x_{\frac{1}{2}:\text{gp}}\right)^{-1}}.
\end{align}

The morphological evolution halts with the precipitate assuming a shape of negligible curvature difference.
Accordingly, the driving force at the end of the transformation is infinitesimal and almost inoperative.
By relating the transitory driving forces with the required mass-transfer, the time taken for the globularisation of the pancake precipitate is determined by
\begin{align}\label{eq:time_cy}
 t_{1:\text{gp}} \propto 3\frac{\delta V_{\text{gp}}}{\Gamma_{0:\text{gp}}+\Gamma_{\frac{1}{2}:\text{gp}}}.
\end{align}
The pre-factor 3 in the above Eqn.~\ref{eq:time_cy} indicates the consideration that the overall driving force of the transformation is the average of the instantaneous driving-forces. 
\nomenclature{$t_{1:\text{gp}}$}{Time taken for the globularisation of pancake-shape}%
\nomenclature{$\bar{\Gamma}_{\text{gp}}$}{Overall driving-force governing the globularisation of pancake structure}%

\subsection{Semi-analytical treatment}\label{lab:sm1}

\begin{figure}
    \centering
      \begin{tabular}{@{}c@{}}
      \includegraphics[width=0.7\textwidth]{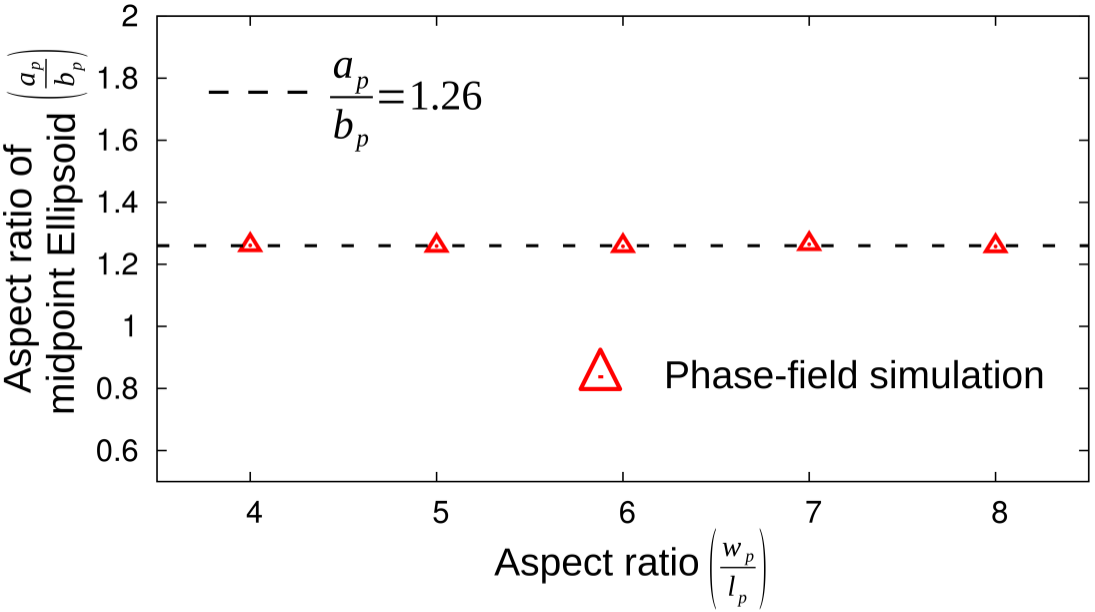}
    \end{tabular}
    \caption{ Aspect ratio of the ellipsoid, or specifically \lq oblate spheroid\rq \thinspace, which are formed at the midpoint of the globularisation of pancake precipitates with various aspect ratio. 
    \label{fig:fig_3}}
\end{figure}

The revisited semi-analytical treatment, which is elucidated in this section, follows the similar framework of the cylinderization approach.
Therefore, three specific stages of the transformation are identified, the driving force at these states are estimated based on the instantaneous curvature-difference and the kinetics is determined by relating the average of the transitory driving-forces with required mass-transfer (Eqn.~\ref{eq:time_cy}).
However, in contrast to the cylinderization approach wherein the dimensions of the midpoint ellipsoids are approximated from the initial and final configuration, in this analysis, the respective geometrical parameters are directly ascertained  by tracking the morphological evolution.

The phase-field simulations unravel that the precipitate assumes an ellipsoidal structure at the midpoint of the transformation.
Moreover, as opposed to the conventional ellipsoid, which is characterized by $a \neq b \neq c$, the precipitate exhibits an unique morphology called oblate spheroid ($a_{\text{p}}=c_{\text{p}}>b_{\text{p}}$) at the midpoint, irrespective of the initial aspect ratio of the pancake structure.
The parameters $a_{\text{p}}$, $b_{\text{p}}$ and $c_{\text{p}}$ are length of the ellipsoid along the major- and minor-axes.
The aspect ratio of the midpoint ellipsoid, which are formed during the globularisation of the precipitates of different initial sizes, are plotted in Fig.~\ref{fig:fig_3}.
This illustration shows that, the aspect ratio of the midpoint ellipsoid is independent of the initial size of the pancake structure, and more importantly, the corresponding geometric parameters are related by $a_{\text{p}}=c_{\text{p}}=1.26b_{\text{p}}$.
The geometric relation between the dimensions of the midpoint structure is employed in this analytical approach to estimate the driving force. 
Therefore, the resulting formulation of the globularisation kinetics encompasses the in-situ information of the morphological transformation.

Since the chemical equilibrium between the constituent phases preserves the volume of the precipitate during the evolution, the geometrical parameters of the midpoint ellipsoid can be related to the radius of the globularised precipitate as $(abc)_{\text{p}}=r_{\text{gp}}\p3$.
By substituting the observed geometric relation in Fig.~\ref{fig:fig_3}, the length along the minor axis is calculated by
\begin{align}\label{eq:ep_r}
 b_{\text{p}}=\left(\frac{r_{\text{gp}}\p3}{1.6}\right)\p{\frac{1}{3}}.
\end{align}
Therefore, from Eqns.~\ref{eq:radius_s} and ~\ref{eq:ep_r}, through the observed relation $a_{\text{p}}=c_{\text{p}}=1.26b_{\text{p}}$,  the dimensions of the midpoint ellipsoid can be determined from initial aspect-ratio of the pancake

In this analysis, the driving force at the midpoint of the morphological evolution is expressed as
\begin{align}\label{eq:dr_12}
 \Gamma_{\frac{1}{2}:{\text{gp}}}(a_{\text{p}},b_{\text{p}}) \propto A_{\frac{1}{2}:\text{gp}}(a_{\text{p}},b_{\text{p}}) \frac{\delta c_{\frac{1}{2}:\text{gp}}(a_{\text{p}},b_{\text{p}})}{\delta x_{\frac{1}{2}:\text{gp}}(a_{\text{p}},b_{\text{p}})}.
\end{align}
In order to calculate the area associated with the atomic fluxes at the midpoint of the evolution $A_{\frac{1}{2}:\text{gp}}(a_{\text{p}},b_{\text{p}})$, it is assumed that the sources are confined to the major axes of the oblate spheroid.
Accordingly, the midpoint diffusion-area is calculated by eliminating the surface area of the segment along the minor axis from the overall surface area of the oblate spheroid.
Using \textit{Knud-Thomsen} approximation, the surface area of the oblate spheroid is expressed as
\begin{align}\label{eq:SA_12}
 S_{\text{obs}}=4\pi\left[  \frac{2(a_{\text{p}}b_{\text{p}})\p{\mathcal{P}}+a_{\text{p}}\p{2\mathcal{P}}}{3} \right]\p{\frac{1}{\mathcal{P}}},
\end{align}
where the constant $\mathcal{P}=1.6$.
The area of the segment along the minor axis is determined by $S_{\text{seg}}=4\pi r_{\text{gp}} b_{\text{p}}$.
Therefore, the area involved in the mass transfer at the midpoint of the globularisation is estimated as
\begin{align}\label{eq:A_12}
 A_{\frac{1}{2}:\text{gp}}(a_{\text{p}},b_{\text{p}}) = S_{\text{obs}}-S_{\text{seg}}.
\end{align}

The principal radii of curvature of the midpoint precipitate is determined by delineating the respective fundamental forms of the surface.
The co-coefficients of the first fundamental form, for a regular three-dimensional ellipsoid, is derived in the appendix.
Since the pancake precipitate assumes a oblate spheroid structure in the midpoint of the globularisation, substituting the characteristic condition ($a_{\text{p}}=c_{\text{p}}>b_{\text{p}}$), the co-efficients of the first fundamental form is expressed as
\begin{align}\label{eq:coeff_first}
 E_{\text{p}}&=b_{\text{p}}\p2+\left(a_{\text{p}}\p2-b_{\text{p}}\p2\right)\left(\sin\p2\theta+\sin\p2\Theta\cos\p2\theta\right) \\ \nonumber
 F_{\text{p}}&=\left(a_{\text{p}}\p2-b_{\text{p}}\p2\right)\sin\theta\cos\theta\sin\Theta\cos\Theta\\ \nonumber
 G_{\text{p}}&=b_{\text{p}}\p2\sin\p2\theta+\left(a_{\text{p}}\p2-b_{\text{p}}\p2\right)\sin\p2\theta\cos\p2\Theta.
\end{align}
Similarly, the second fundamental-form co-efficients reads
\begin{align}\label{eq:coeff_second}
\tilde{E}_{\text{p}}&=a_{\text{p}}\p2b\left[ (a_{\text{p}}b_{\text{p}})\p2+a_{\text{p}}\p2 \left(a_{\text{p}}\p2-b_{\text{p}}\p2\right) \sin\p2\theta\cos\p2\Theta \right]\p{-\frac{1}{2}} \\ \nonumber
\tilde{F}_{\text{p}}&=0 \\ \nonumber
\tilde{G}_{\text{p}}&=a_{\text{p}}\p2b_{\text{p}}\sin\p2\theta\left[ (a_{\text{p}}b_{\text{p}})\p2+a_{\text{p}}\p2 \left(a_{\text{p}}\p2-b_{\text{p}}\p2\right) \sin\p2\theta\cos\p2\Theta \right]\p{-\frac{1}{2}}.
\end{align}
It has been shown that the influence of curvature difference on the equilibrium concentration can be expressed as
\begin{align}\label{eq:c_12}
\delta c_{\frac{1}{2}:\text{gp}}\propto\left[ H_{\text{sink}}-H_{\text{source}}\right],
\end{align}
where $H$ is the mean curvature.
Based on the co-efficients of the fundamental form, the mean curvature is calculated by
\begin{align}\label{eq:mean_coeff}
 H &=\frac{1}{2}\left[\frac{\tilde{E}_{\text{p}}G_{\text{p}}+\tilde{F}_{\text{p}}E_{\text{p}}-2\tilde{G}_{\text{p}}F_{\text{p}}}{E_{\text{p}}G_{\text{p}}-F_{\text{p}}^2}\right]. 
\end{align}
By adopting the appropriate angular variables, $\theta$ and $\Theta$, the change in the equilibrium concentration induced by the inherent curvature-difference, which is associated with the oblate spheroid, is written as
\begin{align}\label{eq:c_12a}
\delta c_{\frac{1}{2}:\text{gp}}\propto\left(\frac{b_{\text{p}}}{2a_{\text{p}}\p2}\right).
\end{align}
The distance over which the concentration gradient extends is estimated by considering the cross-section of the midpoint ellipsoid.
Since the diffusion length can be related to the perimeter of the cross-sectional ellipse, $\delta x_{\frac{1}{2}:\text{gp}}(b_{\text{p}},r_{\text{gp}})$ is expressed as
\begin{align}\label{eq:x_12a}
 \delta x_{\frac{1}{2}:\text{gp}} \propto \frac{\pi}{2}\left(1.14b_{\text{p}}-r_{\text{gp}}\right).
\end{align}
Substituting Eqns.~\ref{eq:A_12}, ~\ref{eq:c_12a} and ~\ref{eq:x_12a} in Eqn.~\ref{eq:dr_12}, the instantaneous driving-force at the point of the globularisation can be approximated.
Subsequently, the transformation kinetics is estimated by relating the amount of required mass-transfer with the initial and midpoint driving force.

\subsection{Comparative analysis}

\begin{figure}
    \centering
      \begin{tabular}{@{}c@{}}
      \includegraphics[width=0.7\textwidth]{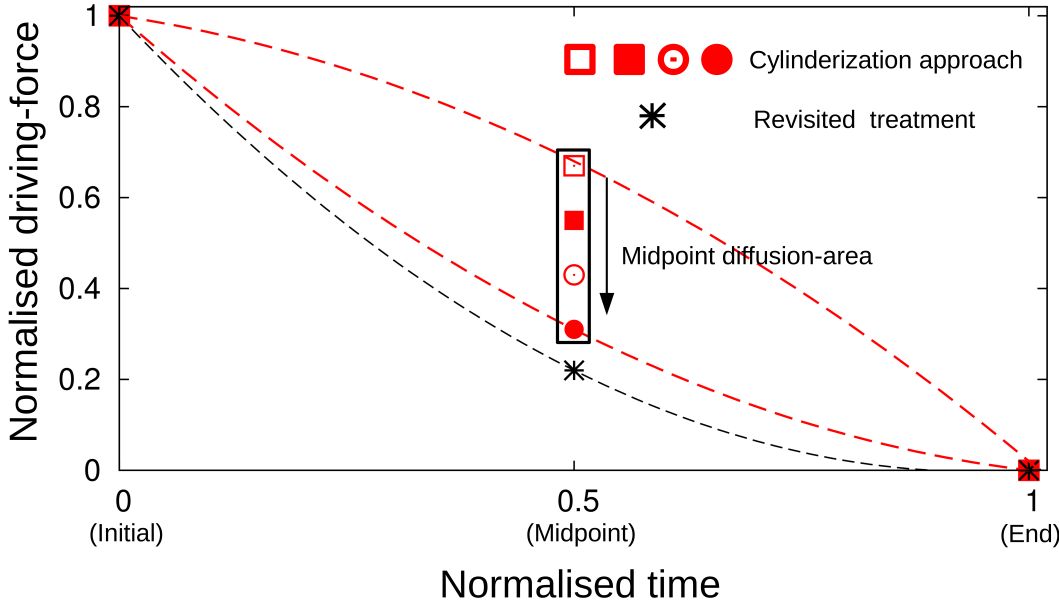}
    \end{tabular}
    \caption{ The transitory driving-force at the initial, midpoint and final stage of the globularisation calculated using the cylinderization and semi-analytical approach. The disparity in the cylinderization-based midpoint driving-forces is introduced by the different final diffusion-area considered.  
    \label{fig:fig_4}}
\end{figure}

The driving force at the specific states of the transformation estimated using the cylinderization approach, wherein the governing parameters at the midpoint are approximated as the average of the initial and final stages, is graphically compared with the outcomes of the semi-analytical treatment in Fig.~\ref{fig:fig_4}.
The driving forces are normalised for the purpose of comparison and correspondingly, the curvature difference at the end of the transformation is assumed to zero.
As shown in Fig.~\ref{fig:fig_4}, for the conventional treatment in Sec.~\ref{lab:cy1}, several variants of the midpoint driving-force can be defined by appropriately considering the diffusion area at the end of the transformation. 
Assuming that the a quadrant of the final spheroidal area participates in the diffusion yields a midpoint driving force which is greater than the average of the initial and final potential difference.
However, as shown in Fig.~\ref{fig:fig_4}, by progressively reducing the final diffusion area, the driving force at the midpoint is correspondingly decreased.
Although the area associated with the final stage of the diffusion is varied, since the midpoint diffusion-area is assumed to the mean of the respective initial and final state, ultimately, the midpoint parameters are influenced.

The driving forces, as shown in Fig.~\ref{fig:fig_4}, increasing become closer to the semi-analytical prediction with decrease in the midpoint diffusion-area.
In this illustration, it is important to note that the instantaneous driving-forces are numerically fitted, assuming that the curvature difference exhibits a smooth monotonic decrease during the evolution.
In other words, both the cylinderization and semi-analytical treatments yields driving forces exclusively at the specific stages of the evolution.
Therefore, in order to ascertain the transformation kinetics, it is inherently assumed that the driving force decreases smoothly and monotonically during the globularisation.
This assumption, although implicit, undergirds the entire analytical treatment.
Consequently, as observed during the morphological transformation of the three-dimensional rods, any deviation in the evolution scheme of the driving force, owing to the transformation mechanism, would result in a disparity between the analytical and simulation results.  

\section{Transformation mechanism}

\begin{figure}
    \centering
      \begin{tabular}{@{}c@{}}
      \includegraphics[width=0.7\textwidth]{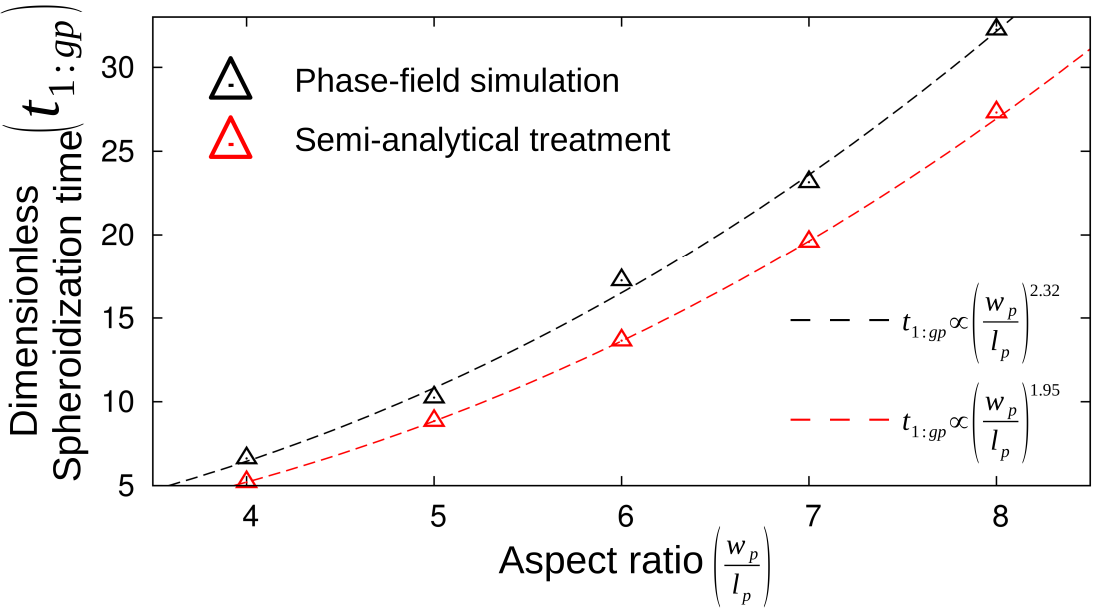}
    \end{tabular}
    \caption{ The time taken for the globularisation of the pancake precipitates of different aspect ratio in the phase-field simulations is compared to the predictions of the semi-analytical. 
    \label{fig:fig_5}}
\end{figure}

Framework of the theoretical approaches, irrespective of the inclusion of the in-situ information, entails the assumption that the morphological evolution of the pancake structure is analogous to the cylinderization of the ribbon-like precipitate.
Accordingly, the transitory driving-forces at the initial, midpoint and final stages of the evolution are fitted by smooth monotonic curve.
However, it has been shown that the temporal change in the curvature difference is characteristically governed by the transformation mechanism and varies with the initial morphology of the precipitate.
Therefore, to realise the influence of the transformation mechanism on the globularisation kinetics, the predictions of the semi-analytical approach for the pancake structures of various aspect ratio is plotted along with the outcomes of the phase-field simulation in Fig.~\ref{fig:fig_5}.

Both the numerical simulation and the analytical treatment, in Fig.~\ref{fig:fig_5}, show that with increase in the aspect ratio of the pancake structure the time taken for the globularisation proportionately increases.
Although such trend is expected in the analytical treatment, wherein the evolution of the driving force is assumed, the analogous outcomes of the phase-field simulation which unravels a monotonic decrease in the transformation rate with increase in aspect ratio indicates that the transformation mechanism remains unaltered for the all plates, irrespective of its size.
Despite predicting similar influence of size on the globularisation kinetics, definite non-conformity between the analytical and simulation results is visible Fig.~\ref{fig:fig_5}.
Furthermore, similar to the spheroidization kinetics of the rods, the disparity increases with the initial size of the precipitate.

The analytical approach, which includes the in-situ information about the geometry of the precipitate at the midpoint of the evolution, relates the aspect ratio of the pancake structure with the time taken for the transformation as
\begin{align}\label{eq:ana_predict}
 t_{1:\text{gp}}\propto \left( \frac{w_\text{p}}{l_\text{p}} \right)^{1.95}.
\end{align}
Whereas, the influence of the precipitate size on the globularisation kinetics, based on the phase-field simulations, can be expressed as
\begin{align}\label{eq:ana_predict}
 t_{1:\text{gp}}\propto \left( \frac{w_\text{p}}{l_\text{p}} \right)^{2.32}.
\end{align}
The gradual increase in the globularisation time with the aspect ratio of the precipitate is primarily due to the proportional increase in the amount of required mass-transfer.
However, the consistent under-estimation of the transformation rate by the analytical treatment indicates that the characteristic mechanism governing the evolution of the pancake structure prolongs to the time taken for the globularisation of the precipitate.
In order to identify the transformation mechanism and its influence on the kinetics, the morphological evolution of the pancake structure is comprehensively analysed.

\subsection{Modified perturbation theory}\label{lab:mp1}

\begin{figure}
    \centering
      \begin{tabular}{@{}c@{}}
      \includegraphics[width=0.7\textwidth]{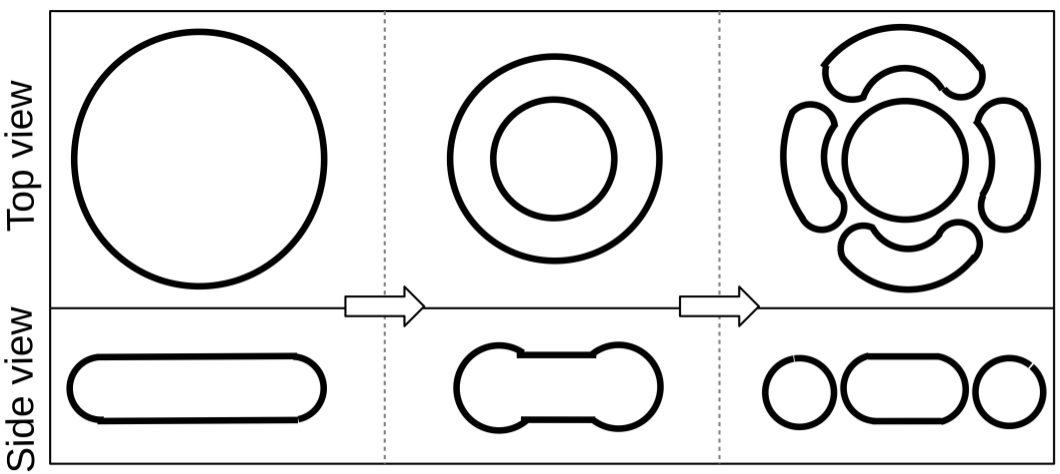}
    \end{tabular}
    \caption{ A schematic representation of the temporal change in the shape of the pancake structure as proposed by the \lq modified perturbation theory\rq \thinspace. 
    \label{fig:fig_7}}
\end{figure}

In comparison to the conventional rod-like morphology, one characteristic feature which distinguishes the pancake structure is the significant amount of flat surfaces.
Owing to the flat surfaces, the inherent curvature-difference becomes ill-suited for the analytical treatment which aims to delineate the transformation mechanism.
Therefore, based on the theoretical investigation of the stability of the infinitely long rod to the imposed perturbations, a globularisation mechanism for the evolution of the pancake structure has been proposed~\cite{nichols1965surface}. 
The morphological transformation of the pancake shape which accompanies the globularisation, according to the \lq modified perturbation theory\rq \thinspace, is schematically represented in Fig.~\ref{fig:fig_7}.
Since this theory considers the pancake morphology as the series of rod-like structures of varying aspect ratio welded together, it elucidates the temporal change in the shape correspondingly.
Therefore, as shown in Fig.~\ref{fig:fig_7}, the modified perturbation theory suggests that the globularisation of the pancake precipitate begins the stable growth of termination ridges, similar to the three-dimensional rods.
The mass transfer from the curve edges to the adjacent flat surface, owing to the inherent difference in the curvature, governs the growth of the perturbation.
With the progressive growth of the ridges, it is postulated that an appropriate curvature-difference is established between the receding perturbation and the pancake flat-surfaces which induces contra-diffusion and ultimately, leads to the fragmentation of the precipitate.
Despite the seemingly reasonable view of the morphological evolution of the pancake shape, this theory is rarely accepted and often argued for its implicit inconsistencies.

The modified perturbation theory, as shown in Fig.~\ref{fig:fig_7}, postulates that the morphological evolution of the pancake structure involves fragmentation of the 
precipitate into central island and numerous surrounding bean-shaped entities.
It is conceivable that, with the stable growth of the termination ridges, a fragmentation analogous to the ovulation in rods can be induced, which results in the formation central island and a ring-like network.
However, the driving force for the breaking-off of the ring-like network into individual bean-like precipitates is unclear.
Additionally, it is well-established that the stability of the flat surface is unaltered by any definite perturbation~\cite{mullins1959flattening}.
In other words, when a perturbation, either external or induced by the mass transferred from the edges, is introduced onto a flat surface, owing to the extensive disparity in the curvature, the perturbation is expected to decay.
Moreover, recognising this stability to the flat surface to the perturbation, theoretical investigations analyse the stability of the thin films by introducing discontinuities or holes, which are assumed as a form of the large perturbations~\cite{srolovitz1986capillary,srolovitz1986capillary2}.
Therefore, the entire morphological evolution of the pancake structure governed by the stable growth of the termination ridges, particularly the growth of the perturbations at the expense of the flat surface is viewed skeptically.
In spite of these plausible inconsistencies in the modified perturbation theory, an alternate theory describing the morphological evolution of the pancake structure has not been reported.

\subsection{Flattening ridge theory}\label{lab:fr1}

\begin{figure}
    \centering
      \begin{tabular}{@{}c@{}}
      \includegraphics[width=0.7\textwidth]{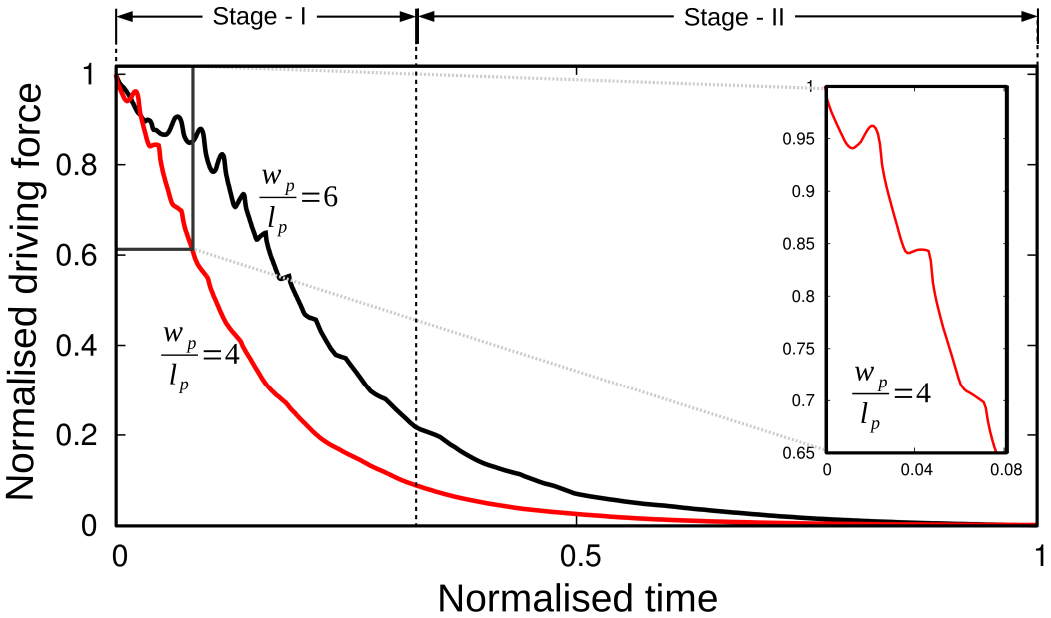}
    \end{tabular}
    \caption{ The change in the potential difference, induced by the inherent disparity in the curvature, with time during the morphological evolution of the pancake precipitate of aspect ratio 4 and 6. 
    \label{fig:fig_6}}
\end{figure}

Previous investigations on the morphological evolution of the three-dimensional rods have shown that the temporal change in the instantaneous chemical-potential difference,   $\Delta \mu(x,t) = \mu(x)|^{+}_{t} - \mu(x)|^{-}_{t}$ where $\mu(x)|^{+}_{t}$ and $\mu(x)|^{-}_{t}$ are highest and lowest potential at time $t$, is a reasonable indicator of the influence of transformation mechanism on the kinetics.
Therefore, the progressive decrease in the driving force ($\Delta \mu(x,t)$) which accompanies the curvature-governed transformation of the pancake shape of aspect ratio 4 and 6 are monitored and plotted in Fig.~\ref{fig:fig_6}.
As opposed to the analytical assumption, the temporal evolution of the potential difference is characterized by a series of non-monotonic sharp peaks.
Although these peaks are significant in the larger precipitate of aspect ratio 6, it is \textit{not} completely absent in the smaller structure.

Fig.~\ref{fig:fig_6} includes a resolved sub-plot of the change in the curvature difference at the initial stages of the globularisation of the precipitate of aspect ratio 4.
This illustration unravels that the non-monotonic peaks, although for a shorter duration, are induced during the morphological transformation of the smaller structure ($\frac{w_\text{p}}{l_\text{p}}=4$). 
Therefore, in contradiction to the spheroidization of the three-dimensional rods wherein the deviation from the smooth-monotonic decrease in the driving force is evident only in the larger structures of aspect ratio greater than 5, the series of sharp peaks are noticeable in small pancake structure as well.
In addition to disrupting the monotonic decrease in the curvature difference, the series of peaks which characterise in the temporal evolution pancake shape, prolong the time taken for the globularisation.
Correspondingly, the disparity between the analytical and the simulation results, which were confined to the larger rods, are visible in smaller pancake precipitates, as shown in Fig.~\ref{fig:fig_5}.

For all the pancake structures, the series of peaks which are associated with the progressive change in the driving force are confined to the initial stage of the transformation, as shown in Fig.~\ref{fig:fig_6}.
With time, the intensity of these peaks continually decreases and the driving force begins to evolve smoothly.
To elucidate the transformation mechanism, this characteristic behaviour of the driving force is exploited and the entire evolution of the pancake structure is distinguished as \textit{stage-I} and \textit{stage-II}.
In the \textit{stage-I} of the globularisation the morphological changes in the pancake structures are governed by the non-monotonic peaks and the corresponding evolution of the driving force, whereas the transformation in \textit{stage-II} is dictated by the smooth-monotonic decrease in the curvature difference.
A schematic distinction of the stages based on the temporal evolution of the driving force is shown in Fig.~\ref{fig:fig_6}.

\subsubsection{Stage - I}\label{lab:fr1}

\begin{figure}
    \centering
      \begin{tabular}{@{}c@{}}
      \includegraphics[width=0.8\textwidth]{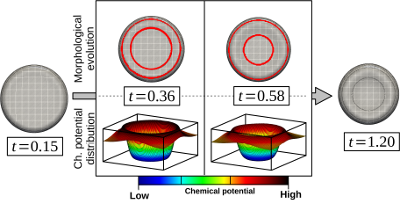}
    \end{tabular}
    \caption{ The morphological transformation of the pancake structure of aspect ratio 6 during the \textit{stage-I} of the globularisation. 
    \label{fig:fig_8}}
\end{figure}

The temporal evolution of the driving force in the \textit{stage-I} of the globularisation of the pancake precipitates are characterised by a series of non-monotonic peaks with progressively decreasing intensity.
Since the morphological changes in the globularisation are governed entirely by the potential difference, a corresponding characteristic evolution is exhibited by the pancake shape.
Fig.~\ref{fig:fig_8} shows the change in the morphology of the precipitate during the \textit{stage-I} of the globularisation.
It is important to note that this illustration does not encompass the entire \textit{stage-I} but shows the morphological evolution associated with a non-monotonic peak.
Owing to the restricted focus on a single peak in the evolution of the potential difference, the duration considered in Fig.~\ref{fig:fig_8} is relatively small.

Irrespective of the morphology, for a finite structure, high potential is established along the termination of the precipitate.
Accordingly, for the pancake structure, the potential in the curved edges of the precipitate are substantially higher than the central flat surfaces.
The disparity in the potential difference, induced by the inherent difference in the curvature, eventuates a mass transfer from the curved edges to the adjacent flat surfaces.
This mass transfer from the termination to the immediate flat surface introduces a perturbation as shown in Fig.~\ref{fig:fig_8} at $t=0.36$.
The transformation mechanism which results in the perturbation along the curved edges of the pancake precipitate is similar to the formation of the longitudinal ridges in initial stages of the spheroidization of rods.
Therefore, the introduction of the circular perturbation in the flat surface corresponds to the smooth-monotonic decrease in the potential difference.
The morphological change associated with the raise in the driving force which disrupts the monotonic decrease and initiates the characteristic peak is shown in Fig.~\ref{fig:fig_8} at $t=0.58$.

Unlike the cylindrical structures wherein the longitudinal perturbation are allowed to grow at the expense of the remnant body, the stable growth of the perturbations are not favoured in the flat surfaces of the pancake structures, owing to the extensive difference in the curvature.
Therefore, instead of a stable thickening (growth) of the termination ridges, the perturbation decays by spreading over the surface, as shown in Fig.~\ref{fig:fig_8} at $t=0.58$.
The driving force for the lateral expansion of the ridges is rendered by the increase in the potential difference.
The three-dimensional representation of the chemical-potential distribution in Fig.~\ref{fig:fig_8} shows that, the increase in the potential difference, which introduces the peak, is established by deepening of the restricted low-potential region at $t=0.58$.
The increased potential-difference in the central region of the precipitate facilitates the growth of the perturbation along flat surface, as opposed to its thickening (stable growth).
With time, the laterally expanding ridges coalesce, and the flat surface of the pancake precipitate is resumed.
Although the overall morphology of the precipitate is preserved by the coalescence of the perturbation, the area of the flat surface after the coalescence of the ridges is significantly lower than its initial state.

In other words, the morphological evolution of the pancake structure during the \textit{stage-I} of the globularisation can be described by considering the evolution of the termination ridges.
The perturbation introduced in the flat surface, owing to the mass transfer from the termination, instead of exhibiting a stable growth by thickening, decays by expanding along the surface of the precipitate.
The mass transfer from the termination, which facilitates the lateral expansion of the ridges, is governed by a unique change in the low potential distribution which initiates the characteristic non-monotonic peak seen in Fig.~\ref{fig:fig_6}.
As the perturbation expands and coalesces, the flat surface and overall morphology of the precipitate is resumed, while the aspect ratio is considerably reduced.
The formation of the subsequent perturbation is governed by the mass transfer from the termination to the neighbouring flat surface which, through the consequent decrease in the curvature difference, completes the peak.
Therefore, while the lateral expansion of the existing ridge is driven the ascending section of the peak, the formation of the subsequent perturbation is governed by the descending segment.

The gradual decrease in the aspect ratio of the pancake structure, while the overall shape is preserved through the unique temporal evolution of the driving force, characterises the \textit{stage-I} of the globularisation.
When compared to the spheroidization of the conventional rod-like structures, two pivotal differences are observed during the \textit{stage-I} evolution of the pancake structure.
One, instead of the thickening or the stable growth of the ridges along the direction normal to the remnant body, the perturbation expands over the flat surface of the pancake precipitate.
Two, owing to the lateral growth of the termination ridges the overall shape of the precipitate is preserved, whereas in the rods, the longitudinal perturbation disrupts the morphology.

\subsubsection{Stage - II}\label{lab:fr1}

\begin{figure}
    \centering
      \begin{tabular}{@{}c@{}}
      \includegraphics[width=0.8\textwidth]{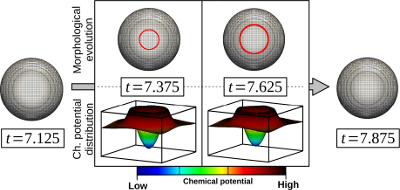}
    \end{tabular}
    \caption{ The subsequent transformation following the \textit{stage-I} evolution which results in the globularisation of the pancake structure .
    \label{fig:fig_9}}
\end{figure}

At the end of the \textit{stage-I} globularisation, which is characterised by the progressive disappearance of the non-monotonic peaks, the precipitate assumes a button-like morphology, a pancake structure with low aspect ratio.
Correspondingly, as shown in Fig.~\ref{fig:fig_9} at $t=7.125$, the precipitate accommodates flat surfaces along with thick curved edges.
However, owing to the continual formation and coalescence of the termination ridges in \textit{stage-I}, the amount of flat surface is considerably reduced.
The morphological transformation which globularises the button-like precipitate is shown in Fig.~\ref{fig:fig_9}.

The reduction in the flat surface of the precipitate, owing to the characteristic peaks in the \textit{stage-I}, confines the low-potential region.
As shown in the Fig.~\ref{fig:fig_6}, the curvature difference in the \textit{stage-II} of the globularisation exhibits a smooth and monotonic decrease.
Therefore, the evolution following the \textit{stage-I}, is governed by the mass transfer from the termination to the central low-potential region of the structure.
The mass from the thick curved-edges of the button-like precipitate which gets deposited in the confined central-region of the precipitate, leads to the formation a circular perturbation, as shown in Fig.~\ref{fig:fig_9} at $t=7.375$.
As opposed to the termination ridges in the \textit{stage-I} which grow inwardly, the central perturbation in \textit{stage-II} expands outwardly along the flat surface, Fig.~\ref{fig:fig_9} at $t=7.625$.
The central perturbations in \textit{stage-II} transforms the button-like precipitate to ellipsoidal structure.
Therefore, at the midpoint the globularisation the precipitate assumes ellipsoidal, more specifically oblate spheroidal shape as shown in Fig.~\ref{fig:fig_1}.
The subsequent morphological changes which transform the precipitate are conventional and governed by smooth-monotonic decrease in the driving force.

The present theory, referred to the \lq flattening ridge theory\rq \thinspace renders an alternate understanding to the modified perturbation theory on the globularisation of the pancake structures.
Accordingly to this theory, the uniqueness in the transformation mechanism which is introduced by the flat surfaces is progressively reduced in the \textit{stage-I} of the globularisation by the characteristic peaks in the temporal evolution of the driving force.
Consistent with the well-established understanding of the stability of the flat surface to the induced perturbation, it is shown that the termination ridges expand laterally as opposed to the stable growth.
The progressive expansion and coalescence of the perturbation substantially decreases the amount of flat surface and leads to the \textit{stage-II} of the globularisation shown in Fig.~\ref{fig:fig_9}.
In \textit{stage-II}, the mass transfer from the curved edges to the central region of the reduced flat-surface transforms the button-like precipitate into ellipsoid structure, which subsequently evolves into a globular shape governed by the edge-recession.
According to the flattening ridge theory, since the considerable amount flat surface in the initial stage of the transformation prevents the stable growth of perturbation, no significant change in the transformation mechanism is expected with increase in the aspect ratio of the pancake structure.

\section{Conclusion}

One of the non-conventional shape that is prevalent in the microstructure of the highly applicable alloys, and is often introduced during the process chain, is analysed to understand its morphological stability.
The unidirectionally-equiaxed pancake structure, owing to its inherent curvature difference, transforms to a globular shape under appropriate thermodynamical conditions.
Since such morphological evolution, which are governed entirely by the curvature difference, enhance the mechanical properties of the materials like two-phase titanium alloys, heat treatment techniques are adopted to facilitate the globularisation.
The globularisation kinetics of the pancake precipitate have hitherto been analysed by considering a morphology which includes well-defined hemispherical caps.
These inclusions, though facilitate an appropriate description of the curvature difference in the initial stages of the evolution, distorts the resemblance of the morphology to the physically observed structures.
Therefore, in this analysis, a different geometrical approach is adopted to define the pancake structure which, while inherently introducing curved edges, obviate the need for the hemispherical caps.

For the uniquely defined pancake structure, the kinetics of the globularisation is investigated by extending the existing approach, where the parameters which govern the driving force at the midpoint of the evolution is assumed to the average of the corresponding initial and final condition.
Furthermore, this treatment is revisited and semi-analytical approach which includes in-situ information from the simulation is derived.
The outcomes of the two analytical treatments are compared and it is shown that the existing delineation can be made close to the revisited treatment by appropriately considering the parameters.
Despite the introduction of the in-situ data, it is observed that the analytical and the simulation results, on the globularisation kinetics, are not entirely compatible.
Recognising the assumptions in the analytical treatments, which pertain to the temporal evolution of the driving force, the change in the curvature difference accompanying the morphological transformation is analysed from the numerical simulations.

Interestingly, as opposed to the assumed smooth-monotonic decrease in the driving force, it is identified the decrease in the potential difference is interrupted by series non-monotonic peaks, during the evolution of pancake structures of all aspect ratio.
Since the temporal change in the potential difference reflects the mechanism of the evolution, the transformation of the pancake structure leading to the globularisation is extensively analysed.

The investigation of the morphological transformation of the pancake structures unravels a unique evolution mechanism much different from the one postulated by the modified perturbation theory.
According to the present flattening ridge theory, the globularisation mechanism is elucidated by distinguishing the transformation into two distinct stages.
In \textit{stage-I} of the evolution, the flat surfaces associated with the pancake morphology are progressively reduced by successive lateral growth and coalescence of a series of perturbations.
The formation, expansion and coalescence of the termination ridges correspond to the peaks observed in the temporal evolution of the driving force.
The decay of the perturbations by expanding along the flat surface is consistent with the well-established understanding on the stability of the flat surface to the induced perturbations.
\textit{stage-I} which transforms the pancake precipitate into button-like structure of analogous morphology, leads to the 
\textit{stage-II} transformation wherein the button-like shape transforms to an ellipsoid by the formation of central perturbation through edge recession.
Morphological evolution which transforms the ellipsoid, or more specifically the oblate spheroid, is similar to the evolution of the conventional structures, and is characterised by smooth-monotonic decrease in the driving force.
In other words, the postulated flattening ridge theory renders an alternate transformation mechanism for the globularisation of the pancake structure, much different from the modified perturbation theory which is based on the inconsistent stable-growth of ridges.
Since the stable growth of perturbations are not favoured in the flat surfaces, a significant change in the mechanism is not expected with increase in the aspect ratio of the pancake structure.

\chapter{Morphological stability of finite elliptical-plate}\label{chap:epplate}

The manufacturing techniques involved in the production of highly applicable materials include mechanical processing, in addition to the heat treatment~\cite{groover2007fundamentals}.
The mechanical treatments adopted in the process chain, although depend on the material and its application, conventionally encompass hot or cold rolling and forging.
These processing techniques, through enormous loads, considerably deform the external shape of the material.
Apart from the deformation, significant amount of stress is introduced into the material during the mechanical treatment which consequently alters the microstructure.
One of the prominent changes in the polycrystalline material is the decrease in the average grain size~\cite{wert1981grain,valiev2006principles}.
Along with the grain refinement, sub-boundaries are introduced are within the constituent phases during the mechanical treatment~\cite{lindroos1967structure}.
These sub-boundaries, which are induced in the precipitate of the lamellar microstructure, influence the morphological evolution associated with the annealing technique~\cite{song2005mechanical,shin2003spheroidization}.

The sub-boundaries in the precipitate breaks-up the seemingly infinite structure, which extends across the grain, into finite shape~\cite{sandim2004annealing,allen2004microstructural}.
The morphological evolution which achieves the fragmentation of the semi-infinite precipitate is illustrated in Fig.~\ref{fig:fig_1}.
At higher temperature, the enhanced diffusivity, accelerates the migration of atoms.
Therefore, depending on the interfacial energy between the region separating the phases and the sub-boundary, a groove is introduced at the triple junction, in accordance with the Young's law~\cite{young1805iii}.
The thermal grooving along the sub-boundary of the precipitate is accompanied by the mass transfer to the adjacent flat surfaces.
The accumulation of the mass in the neighbouring flat surfaces renders a characteristic profile to the triple junction~\cite{mullins1957theory}.

It is well-established that any disturbance to the flat surface, in the form of external or induced perturbation, eventually decays owing to the substantial disparity in the curvature~\cite{mullins1959flattening}.
Accordingly, the characteristic profile, which is induced in the triple junction due to the thermal grooving, gets disrupted.
Particularly, the mass accumulated on the regions adjacent to the grooving gets dispersed, governed by the curvature difference, and turns flat with time.
In order to the re-establish the characteristic morphology at the triple junction, the mass is transferred from the sinks around the groove to the neighbouring regions.
The mass transfer, while creating an appropriate profile, increases the depth of the groove.

A considerable fraction of the sub-boundaries, which are induced during the mechanical treatment, extend all-through the thickness of the precipitate~\cite{lee2002role}.
The mass transfer that facilitates the penetration of the groove occurs on both sides of the semi-infinite structure along the through-thickness sub-boundaries, as shown in Fig.~\ref{fig:fig_1}.
The continual flattening of the source and subsequent, mass transfer to establish the characteristic profile at the triple junction, progressively increases the depth of the groove.
With time, the incrementally deepening grooves lead to the fragmentation of the seemingly infinite precipitate.
The morphological evolution wherein the breaking-up of the continuous precipitate into finite structures is aided by the through-thickness sub-boundaries is referred to as boundary-splitting~\cite{kampe1989shape}.
In other words, the thermal grooving of the through-thickness sub-boundaries leads to the fragmentation of the precipitates by boundary-splitting.

\begin{figure}
    \centering
      \begin{tabular}{@{}c@{}}
      \includegraphics[width=0.5\textwidth]{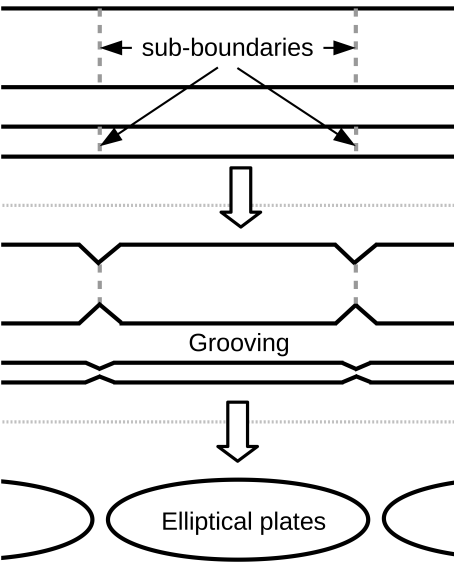}
    \end{tabular}
    \caption{ The boundary-splitting which, through the fragmentation of the continuous precipitate along the sub-boundaries, results in the formation of finite elliptical plates.
    \label{fig:fig_1}}
\end{figure}

The morphology of the finite structure which result from boundary-splitting depend on the initial configuration of the continuous precipitate and the distance separating the through-thickness sub-boundaries.
During the processing of two-phase titanium alloys, a significant fraction of the precipitate assume a uni-directionally equi-axed pancake morphology subsequently after the boundary-splitting~\cite{geetha2004effect,guo2016microstructure}.
The morphological evolution of the pancake structure which accompanies the static annealing treatment has been analysed in the previous chapter.
However, recent experimental observations reveal that, in addition to the pancake shape, the fragmentation of the continuous structure through boundary-splitting yields finite precipitate of elliptical plate morphology~\cite{stefansson2003mechanisms,park2012mechanisms}, as shown in Fig.~\ref{fig:fig_1}.
Therefore, in this chapter, the transformation kinetics and mechanism of the globularisation of the elliptical plate is investigated.

\section{Domain setup}

A chemically stable three-dimensional domain consisting of two distinct phases, $\alpha$ and $\theta$, is configured for the present investigation.
The composition of the constituent phases corroborate the CALPHAD-informed equilibrium concentration, which consequently establishes chemical equilibrium between the phases.
Furthermore, the phases are distributed in such a way that the matrix-$\alpha$ completely encapsulates the precipitate-$\theta$.

The initial morphology of the precipitate-$\theta$, which is analysed in upcoming sections, is illustrated in Fig.~\ref{fig:fig_2}.
From the top, the geometrical configuration of the plate is defined by parameters $\frac{w_\text{e}}{2}$ and $b_\text{e}$, which represent the respective length along the semi-major and -minor axes of the ellipse.
The parameter $l_\text{e}$ is the thickness of the precipitate.
\nomenclature{$\frac{w_\text{e}}{2}$}{Semi-major axis length of elliptical plate}%
\nomenclature{$b_\text{e}$}{Semi-minor axis length of elliptical plate}%
\nomenclature{$l_\text{p}$}{Thickness of elliptical plate}%
The aspect ratio of the elliptical structure is described as the ratio of the major-axes length and the thickness of the plate, $\frac{w_\text{e}}{l_\text{e}}$.
Correspondingly, the thickness of the plate is affixed at $l_\text{e}=0.001\times 10^{-6}$m, while $w_\text{e}$ is varied in relation to the required aspect ratio.
The ratio between the major- and minor-axes length of the elliptical plate is set as a constant, independent of the aspect ratio, $\frac{w_\text{e}}{b_\text{e}}=4$.

A sufficiently large simulation domain is considered to avoid any influence of the boundary on the morphological transformation.
Moreover, the domain size is proportionately varied in relation to the initial aspect-ratio of the elliptical plate.
 
\begin{figure}
    \centering
      \begin{tabular}{@{}c@{}}
      \includegraphics[width=0.7\textwidth]{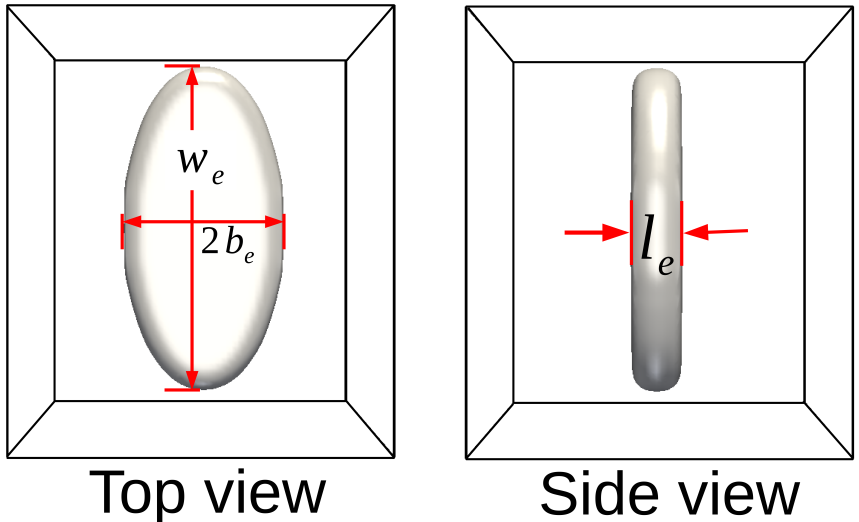}
    \end{tabular}
    \caption{ The morphological configuration of a three-dimensional elliptical plate of the thickness $l_\text{e}=0.001\times 10^{-6}$m and aspect ratio ($\frac{w_\text{e}}{l_\text{e}}=$) 6.
    \label{fig:fig_2}}
\end{figure}

\section{Transformation kinetics}

During static annealing, immediately following the boundary-splitting and fragmentation of the continuous precipitate, the finite structure evolves morphologically.
The morphological transformation is governed by the inherent curvature-difference in the shape assumed by the fragmented finite structure.
Analytical treatments which attempt to predict the kinetics of the curvature-driven shape-change adhere to a relatively straightforward framework, wherein a smooth temporal evolution of the driving force is assumed~\cite{courtney1989shape,park2012prediction,semiatin2005prediction,park2012mechanisms}.
For a shape-instability induced transformation, apart from the thermodynamical constants, the driving force is primarily dictated by the inherent difference in the curvature.
Therefore, the theoretical approaches consider suitable shapes which facilitate an elegant description of the curvature difference.
Often, such considerations result in the introduction of the termination caps which are hemispherical in shape and the diameter identical to the thickness of the plate.
Although the hemispherical inclusions mitigate the complexity in the formulation of the curvature difference, the resemblance of the finite precipitate to the microscopically observed structure is noticeably compromised.
Moreover, these termination caps additionally influence the predictions of the existing treatments.

For investigating the globularisation kinetics of the pancake structure, in the previous chapter, a different approach is adopted to geometrically define the morphology which obviates the need for the spherical inclusions and retains a well-defined termination.
Similar consideration is extended to delineate the elliptical plate.

\subsection{Geometrical treatment}\label{lab:geo_ep}

A prolate spheroid, which is a three-dimensional ellipsoid characterised by the relation $a_\text{E}>b_\text{E}=c_\text{E}$, where $a_\text{E}$, $b_\text{E}$ and $c_\text{E}$ correspond to the lengths along major and minor axes, is treated as the \textit{geometrical-parent} of the elliptical precipitate.
The cross-section of the parent ellipsoid is shown in Fig.~\ref{fig:fig_3}.
For the present geometrical treatment, the length along the major axis of the prolate spheroid is considered as $a_\text{E}=\frac{w_\text{e}}{2}$ and the equivalent lengths along the minor axes are $b_\text{E}=c_\text{E}=b_\text{e}$.
A central segment of the parent structure with thickness $l_\text{e}$, as shown in Fig.~\ref{fig:fig_3}, yields an elliptical plate with morphology relatable to the experimental observations.
Furthermore, the terminations of the emerging elliptical structure are inherently curved, which precludes any need for the additional inclusions.
Therefore, the precipitates of different aspect ratios, that are analysed in this study, are the central segment of prolate spheroid with various major-axis length.

\begin{figure}
    \centering
      \begin{tabular}{@{}c@{}}
      \includegraphics[width=0.7\textwidth]{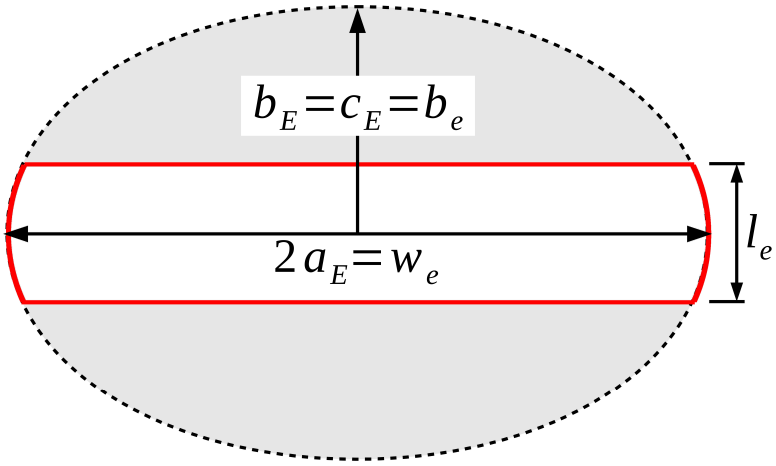}
    \end{tabular}
    \caption{ The cross-section of the parent three-dimensional ellipsoid of major-axes length $a_\text{E}=\frac{w_\text{e}}{2}$ and identical minor axis length, $b_\text{E}=c_\text{E}=b_\text{e}$. The central segment of the prolate spheroid which is treated as the elliptical plate of thickness $l_\text{e}$ is distinguished.
    \label{fig:fig_3}}
\end{figure}

As elucidated in the previous chapters, the two theoretical approaches which are employed to investigate the transformation kinetics identify three specific stages of the globularisation.
The driving forces at these three particular stages, initial ($t_{0:\text{ge}}$), midpoint ($t_{1/2:\text{ge}}$) and final ($t_{1:\text{ge}}$), are estimated from the curvature difference established by the instantaneous morphology of the plate.
The shape of the precipitate at these three distinct stages of the globularisation are illustrated in Fig.~\ref{fig:fig_4}.
Since the transformation ends with the precipitate assuming a shape which is characterised by the negligible difference in the curvature, in both analytical treatments, the driving force at the final stage of the evolution is considered as infinitesimal.

For a given morphology of the precipitate, irrespective of the approach, the formulation of the intial driving-force ($\Gamma_{0:\text{ge}}$) is identical.
The disparity between the treatments is introduced in the estimation of the midpoint driving-force ($\Gamma_{1/2:\text{ge}}$), wherein one delineation includes in-situ information from the simulation.
Since the thermodynamical description at the beginning of the evolution is independent of the treatment, the initial driving-force is ascertained separately based on the inherent curvature-difference.

In addition to the driving force, the time taken for the globularisation of the precipitate is governed by another factor which referred to as the required mass-transfer ($\delta V_\text{ge}$).
The morphological evolution of the precipitate is established by the migration of the atoms.
The amount of the mass-diffusion which is required to the complete the transformation is the required mass-transfer ($\delta V_\text{ge}$).
\nomenclature{$\delta V_\text{ge}$}{Required mass-transfer for globularisation of elliptical plate}%
For the elliptical plate, the required mass-transfer is determined by
\begin{align}\label{eq:req_volume}
 \delta V_\text{ge}=V_{\text{e:pt}}-V_{\text{e:sh}},
\end{align}
where $V_{\text{e:pt}}$ is volume of the precipitate and $V_{\text{e:sh}}$ is the volume shared by the initial and final structure.
Since $V_{\text{e:sh}}$ does not necessitate any migration during the globularisation, in Eqn.~\ref{eq:req_volume}, it is eliminated from the overall volume to determine the required mass transfer $\delta V_\text{ge}$.

For the present geometrical consideration, wherein the precipitate is the segment of a larger prolate spheroid, the volume of the precipitate is calculated by $V_{\text{e:pt}}=V_{\text{E}}-2V_{\text{E:cap}}$, where $V_{\text{E}}$ and $V_{\text{E:cap}}$ are the volume of the parent ellipsoid and the cap above (and below) the central segment, as shown in Fig.~\ref{fig:fig_3}.
Although the volume of the precipitate is estimated directly during the analysis of the pancake structure, the inherent geometrical complexities associated with the ellipsoidal structure necessitates an indirect approach~\cite{tee2005surface,harris2006curvature}.

\begin{figure}
    \centering
      \begin{tabular}{@{}c@{}}
      \includegraphics[width=0.8\textwidth]{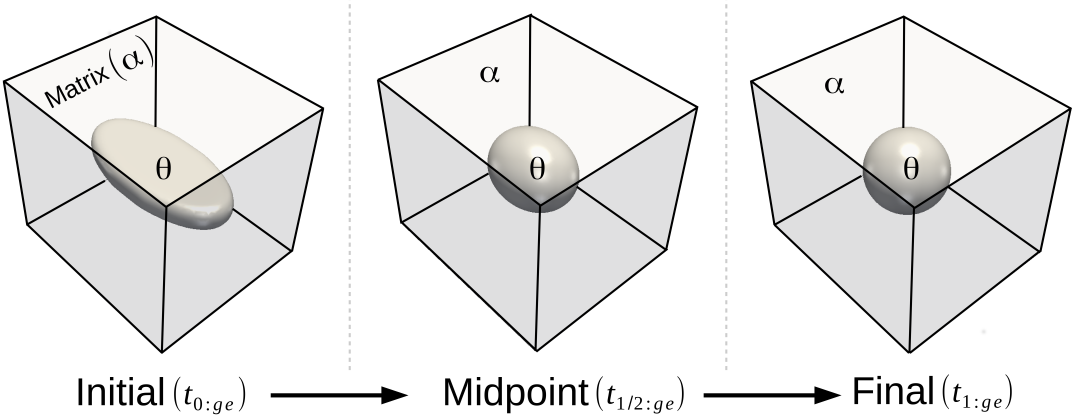}
    \end{tabular}
    \caption{ The temporal change in the shape of the elliptical plate of aspect ratio 6 at the three specific stages, initial ($t_{0:\text{ge}}$), midpoint ($t_{1/2:\text{ge}}$) and final ($t_{1:\text{ge}}$), of the globularisation.
    \label{fig:fig_4}}
\end{figure}

From Fig.~\ref{fig:fig_3}, assuming that the flat surface of the elliptical plate extends along the $xz$-plane, the equation of the parent ellipsoid can be expressed as
\begin{align}\label{eq:parent_epE}
 \frac{x^2}{\left(w_{\text{e}}/2 \right)^2}+\frac{y^2}{b_{\text{e}}^2}+\frac{z^2}{b_{\text{e}}^2}=1.
\end{align}
The volume of this prolate spheroid is calculated by $V_{\text{E}}=\frac{2}{3}\pi w_{\text{e}} b_{\text{e}}^2$.
The ellipsoidal cap in Fig.~\ref{fig:fig_3} extends from flat surface of the precipitate to the end of the parent structure along a minor axis.
Given that the minor-axis, which is normal to the flat surface, is along the $y$-axis, the volume of the cap can be estimated by
\begin{align}\label{eq:parent_V}
 V_{\text{E:cap}}=\int_{b_{\text{e}}-\frac{l_{\text{e}}}{2}}^{b_{\text{e}}}\pi (xz) \diff y.
\end{align}
Based on Eqn.~\ref{eq:parent_epE}, the equation of the larger ellipsoid along the x- and z- can be written as 
\begin{align}\label{eq:x_eqn}
 x=\frac{w_{\text{e}}}{2}\left[1-\left(\frac{y}{b_{\text{e}}}\right)^2 \right]^{\frac{1}{2}} && \text{and} && z=b_{\text{e}}\left[1-\left(\frac{y}{b_{\text{e}}}\right)^2 \right]^{\frac{1}{2}},
\end{align}
respectively.
By substituting Eqn.~\ref{eq:x_eqn} in Eqn.~\ref{eq:parent_V}, the volume of cap is expressed as
\begin{align}\label{eq:Vparent_cap}
 V_{\text{E:cap}}=\int_{b_{\text{e}}-\frac{l_{\text{e}}}{2}}^{b_{\text{e}}}\pi \frac{w_{\text{e}}}{2} b_{\text{e}} \left(1-\frac{y^2}{b_{\text{e}}^2} \right) \diff y.
\end{align}
Solving the above Eqn.~\ref{eq:Vparent_cap} yields the relation
\begin{align}\label{eq:Vparent_cap2}
 V_{\text{E:cap}}=\frac{\pi}{2}w_{\text{e}} b_{\text{e}}\left[ \frac{2b_{\text{e}}}{3}-\frac{l_{\text{e}}}{2}+\frac{\left(l_{\text{e}}/2\right)^3}{3b_{\text{e}}^2} \right],
\end{align}
which can be used to calculate the volume of the ellipsoidal cap from the known geometrical parameters, $w_\text{e}$, $l_\text{e}$ and $b_\text{e}$.
By eliminating the volume of the caps, which are above and below the central segment, from the overall volume of the parent ellipsoid, the precipitate volume can be determined.
Correspondingly, the volume of the elliptical precipitate reads
\begin{align}\label{eq:Vppt}
 V_{\text{e:pt}}=\pi w_{\text{e}} b_{\text{e}}\left[\frac{l_{\text{e}}}{2}-\frac{\left(l_{\text{e}}/2\right)^3}{3b_{\text{e}}^2} \right].
\end{align}

In order to determine the required mass-transfer ($\delta V_\text{ge}$), as given the Eqn.~\ref{eq:req_volume}, the volume shared by the initial and final morphology of the precipitate should be estimated.
While the initial shape of the precipitate is defined by the parameters $w_\text{e}$, $l_\text{e}$ and $b_\text{e}$, the final morphology is characterised by the radius of the ultimate spheroid, $r_\text{ge}$.
Therefore, the radius of the final spheroid should be known for calculating the shared volume, $V_{\text{e:sh}}$.
Since the volume of the precipitate is conserved during the morphological evolution, owing to the chemical equilibrium established between the phases, the relation between the geometrical parameters can be ascertained by equating the initial and final volume of the precipitate.
Correspondingly, the radius of the final spheroid is calculated from the initial geometrical parameters by
\begin{align}\label{eq:r_sWe}
 r_\text{ge}=\left\{ \frac{3}{4}w_\text{e}b_\text{e}\left[\frac{l_\text{e}}{2}-\frac{\left(l_\text{e}/2\right)^3}{3b_\text{e}^2} \right] \right\}^{\frac{1}{3}}.
\end{align}
The region, which is shared by the initial and final shape of the precipitate, and is hardly involved in the morphological transformation is a central segment of the final spheroid of thickness $l_\text{e}$.
Accordingly, the shared volume which do not contribute to the required mass-transfer is expressed as
\begin{align}\label{eq:sharedV}
 V_{\text{e:sh}}=\pi l_\text{e} \left[ r_\text{ge}^2 - \left(\frac{l_\text{e}}{2}\right)^2\right].
\end{align}
By substituting Eqns.~\ref{eq:Vppt} and ~\ref{eq:sharedV} in Eqn.~\ref{eq:req_volume}, the mass transfer required to complete the globularisation ($\delta V_\text{ge}$) can be determined from the initial geometrical configuration of the elliptical plate.

The driving force at the initial stages of the evolution is dictated by two factors.
One is the concentration gradient introduced by the inherent curvature-difference in the morphology of the precipitate, while the other is area available for the migration of the atoms.
In the presence of the hemispherical caps, the area available for the diffusion is determined from the geometry of the inclusions.
Since in the present study the inclusions are averted by an appropriate geometrical description of the precipitate, a different approach is employed to ascertain the diffusion area $A_{0:\text{ge}}$.

A characteristic feature which differentiates the elliptical plate, considered in the present study, from a regular elliptical cylinder, of thickness $l_\text{e}$, is the curved termination.
Owing to the finite nature of the precipitate, these curved terminations act as the source of the mass transfer at the initial stages of the globularisation.
Therefore, the surface area of the curved edges is the area available for diffusion at the beginning of the evolution ($A_{0:\text{ge}}$).
The area along the termination is determined by quantifying the volume of the curved region ($V_{\text{cur}}$).
The volume of the curved edges can be approximated as
\begin{align}\label{eq:curV}
 V_{\text{cur}}=V_{\text{e:pt}}-V_{\text{e:cy}},
\end{align}
where $V_{\text{e:cy}}$ is the volume of a hypothetical cylinder with its radial cross-section identical to the surface of the precipitate and thickness $l_\text{e}$.
Considering that the thickness is much smaller than the major- and minor-axis lengths of the precipitate, the volume of the elliptical cylinder is written as
\begin{align}\label{eq:cy_epV}
 V_{\text{e:cy}}=\pi l_\text{e} \left[ \left(\frac{w_\text{e}}{2}\right)^2-\left(\frac{l_\text{e}}{2}\right)^2 \right]^{\frac{1}{2}}\left[ \left( b_\text{e}\right)^2-\left(\frac{l_\text{e}}{2} \right)^2 \right]^{\frac{1}{2}}.
\end{align}
Substituting Eqns.~\ref{eq:Vppt} and ~\ref{eq:cy_epV} in Eqn.~\ref{eq:curV}, the volume of the curved edges of the precipitate is quantified separately.
Based on the geometric configuration of the elliptical plate, shown in Fig.~\ref{fig:fig_3}, the smooth termination can be treated as a longitudinal segment of a cylinder of radius $r_{\text{cur}}$.
Correspondingly, from Eqns.~\ref{eq:curV} and ~\ref{eq:cy_epV}, the termination radius is ascertained by
\begin{align}\label{eq:r_cur}
 r_{\text{cur}}=\frac{V_{\text{cur}}^{\frac{1}{2}}}{\pi}\left\{ \frac{1}{2} \left[ \left(\frac{w_\text{e}}{2}\right)^2 + b_\text{e}^2 + \frac{l_\text{e}^2}{2} \right] \right\}^{-\frac{1}{4}}
\end{align}
The geometrical nature of the curved-termination can now be described by the radius $r_{\text{cur}}$, which is encompassed in the major-axis length $w_\text{e}$ of the elliptical precipitate.

For the hypothetical elliptical-structure of thickness $l_\text{e}$, which considered in Eqn.~\ref{eq:cy_epV}, the diffusion area at the initial stages of the transformation is its surface area.
Therefore, for this sharp-edged precipitate, the initial diffusion-area is expressed as
\begin{align}\label{eq:A0_sharpEp}
 S_{\text{e:cy}}=\underbrace{2\pi\left \{ \frac{1}{2}\left[ \left(\frac{w_\text{e}}{2}\right)^2 + b_\text{e}^2 + \frac{l_\text{e}^2}{2}  \right] \right \}^{\frac{1}{2}}}_{:=\mathcal{P}_\text{el}}l_\text{e},
\end{align}
where $\mathcal{P}_\text{el}$ is the perimeter of the elliptical cross-section of the hypothetical sharp-cylinder.
Since the terminations of the precipitate are curved, the initial diffusion-area is relatively larger than the surface area of the sharp elliptical-cylinder, $S_{\text{el:cy}}$.

By projecting the curved-edges of the precipitate along the thickness of the elliptical structure, and based on Eqn.~\ref{eq:r_cur}, the diffusion area $A_{0:\text{ge}}$ is approximated as
\begin{align}\label{eq:A0_plEp}
A_{0:\text{ge}}=4\pi\left\{\left[ \left(\frac{l_\text{e}}{2}\right)^2+r_{\text{cur}}^2 \right]\left[{\frac{1}{2}} \left( \frac{w_\text{e}^2}{4}+b_\text{e}^2-\frac{l_\text{e}^2}{2} \right)\right] \right\}^{\frac{1}{2}} .
\end{align}
\nomenclature{$A_{i:\text{ge}}$}{Diffusion area contributing to globularisation of elliptical plate}%
The diffusion area determined from the above Eqn.~\ref{eq:A0_plEp} is greater than $S_{\text{el:cy}}$ and more importantly, includes the geometrical nature of the curved terminations of the precipitate.

In the analytical approach~\cite{courtney1989shape}, the gradient in the concentration which is introduced by the inherent curvature-difference in the shape of the precipitate, is solved by separating the concentration and associated spatial term.
In other words, at a given stage of the evolution, the curvature-induced difference in the equilibrium concentration ($\delta c_{i:\text{ge}}$) and distance across which the disparity is established ($\delta x_{i:\text{ge}}$) are treated independently.

The difference in the equilibrium concentration, at the beginning of the globularisation, is dictated by the curvature difference associated with the morphological configuration of the elliptical plate.
The disparity in the equilibrium concentration actuates mass transfer from the source to sink.
The source, which is curved terminations of the precipitate, is an integral part of the parent ellipsoid illustrated in Fig.~\ref{fig:fig_3}, whereas the sinks are the flat surfaces of the elliptical structure.
The influence of the curvature on the equilibrium concentration is formulated based on the principal radii of curvature at the source and sink.
Correspondingly, the curvature-induced disparity in the equilibrium concentration is expressed as
\begin{align}\label{eq:c0_plEp}
\delta c_{i:\text{ge}} \propto \frac{w_\text{e}}{2b_\text{e}^2}+\frac{{w_\text{e}}^2 + 4b_\text{e}^2}{2w_\text{e}^2 b_\text{e}}.
\end{align}
\nomenclature{$\delta c_{0:\text{ge}}$}{Influence of elliptical plate on equilibrium concentration}%
Since the sink of the mass transfer is the flat surface of the elliptical plate, it does not influence the equilibrium concentration.
Furthermore, the principal radii of the source, which are adopted in Eqn.~\ref{eq:c0_plEp}, are determined by considering the large prolate spheroid with parameters $\frac{w_\text{e}}{2}$ and $b_\text{e}$.

The spatial term associated with the concentration gradient is determined by identifying the primary diffusion-path(s) and estimating the effective migration distance~\cite{courtney1989shape,park2012mechanisms}.
In the pancake structure of circular cross-section, the realisation of the diffusion path, and the calculation of the corresponding distance, is straightforward, whereas in the elliptical plate, the  migration distance vary centro-symmetrically.
Therefore, by considering the two extreme migration paths, which correspond to the major- and minor-axes of the elliptical plate, the diffusion distance at the initial stages of the globularisation is expressed as
\begin{align}\label{eq:x0_plEp}
\delta x_{0:\text{ge}} \propto \frac{w_\text{e}}{2}+b_\text{e}+V_{\text{cur}}^{\frac{1}{2}}\left\{ \frac{1}{2} \left[ \left(\frac{w_\text{e}}{2}\right)^2 + b_\text{e}^2 + \frac{l_\text{e}^2}{2} \right] \right\}^{-\frac{1}{4}} -2r_\text{ge}.
\end{align}
\nomenclature{$\delta x_{i:\text{ge}}$}{Diffusion length associated with globularisation of elliptical plate}%
The initial diffusion-distance, formulated in the above Eqn.~\ref{eq:x0_plEp}, includes the geometrical parameters of the curved termination, in addition to the elliptical sections of the precipitate.
From Eqns.~\ref{eq:A0_plEp}, ~\ref{eq:c0_plEp} and ~\ref{eq:x0_plEp}, the curvature-induced driving force at the beginning of the morphological evolution can be determined by $\Gamma_{0:\text{ge}} \propto A_{0:\text{ge}}\left(\frac{\delta c}{\delta x} \right)_{0:\text{ge}}$.

\subsection{Cylinderization approach}\label{lab:cy1}

In the analytical treatment of the transformation kinetics, referred to as the cylinderization approach~\cite{courtney1989shape}, the temporal change in the morphology of the precipitate during the globularisation is assumed.
The lack of in-situ information, experimental or theoretical, on the shape-changes accompanying the volume-diffusion governed curvature-induced transformation is responsible for such consideration.
Correspondingly, the geometrical parameters which dictate morphological configuration of the precipitate at the midpoint of the evolution is treated as the average of the respective parameters at the initial and final stages.

The area available for the diffusion in the beginning of the globularisation is expressed in Eqn.~\ref{eq:A0_plEp}.
The corresponding terms which define the midpoint diffusion-area ($A_{\frac{1}{2}:\text{ge}}$) is ascertained from the surface area of the ultimate globular shape of the precipitate.
In any given stage of the transformation, the diffusion area is predominantly associated with the source of the mass transfer.
Therefore, assuming the entire surface area of the final spheroid as the diffusion area entails an unphysical configuration which lacks sink.
Considering that only a fraction of the precipitate acts as the source of mass transfer, the final diffusion-area can be formulated exclusively based on the segment of the globular morphology, rather than the entire surface area.
In the early chapters, it has been shown that the driving force at the midpoint can be convincingly calculated by assuming the quadrant of the ultimate spheroid as the source.
Accordingly, for the globularisation of the elliptical plate, the midpoint diffusion-area is expressed as
\begin{align}\label{eq:A12c_plEp}
 A_{\frac{1}{2}:\text{ge}}= \frac{1}{2}\left[ A_{0:\text{ge}}+\frac{\pi}{4}\left\{ \frac{3}{4}w_\text{e}b_\text{e}\left[\frac{l_\text{e}}{2}-\frac{\left(l_\text{e}/2\right)^3}{3b_\text{e}^2} \right] \right\}^{\frac{2}{3}} \right].
\end{align}

Similar to the diffusion area, the influence of the curvature on the equilibrium concentration at the midpoint of the globularisation is formulated  by considering the mean of the appropriate parameters at the initial and final stages.
These parameters include the principal radii at the source and sink.
The morphological changes in the precipitate is achieved by the progressive mass transfer from the source to sink.
Therefore, it is reasonable to assume that moments prior to the complete globularisation of the elliptical plate, the relatively flat-regions are present in the structure which act as sinks.
Correspondingly, the change in equilibrium concentration, owing to the inherent curvature-difference in the midpoint-morphology of the precipitate, is written as
\begin{align}\label{eq:c12c_plEp}
 \delta c_{\frac{1}{2}:\text{ge}} \propto \frac{1}{2}\left[ \frac{w_\text{e}^3+w_\text{e}^2b_\text{e}+4b_\text{e}^3}{2w_\text{e}^2b_\text{e}^2} + 2\left\{ \frac{3}{4}w_\text{e}b_\text{e}\left[\frac{l_\text{e}}{2}-\frac{\left(l_\text{e}/2\right)^3}{3b_\text{e}^2} \right] \right\}^{-\frac{1}{3}} \right].
\end{align}
While the above Eqn.~\ref{eq:c12c_plEp} quantifies the disparity introduced in the equilibrium concentration, the spatial term associated with the gradient is ascertained separately.
The mean of the diffusion length at the initial and final stages of the shape-change yields
\begin{align}\label{eq:x12c_plEp}
 \delta x_{\frac{1}{2}:\text{ge}} \propto \frac{1}{2}\left[ \frac{w_\text{e}}{2}+b_\text{e}+V_{\text{cur}}^{\frac{1}{2}}\left\{ \frac{1}{2} \left[ \left(\frac{w_\text{e}}{2}\right)^2 + b_\text{e}^2 + \frac{l_\text{e}^2}{2} \right] \right\}^{-\frac{1}{4}}+r_\text{ge}(\pi-2) \right].
\end{align}
Based on the aforementioned analytical delineations and through Eqn.~\ref{eq:A12c_plEp}, ~\ref{eq:c12c_plEp} and ~\ref{eq:x12c_plEp}, the instantaneous driving-force at the midpoint of the globularisation is determined.

\subsection{Semi-analytical treatment}\label{lab:sm1}

\begin{figure}
    \centering
      \begin{tabular}{@{}c@{}}
      \includegraphics[width=0.8\textwidth]{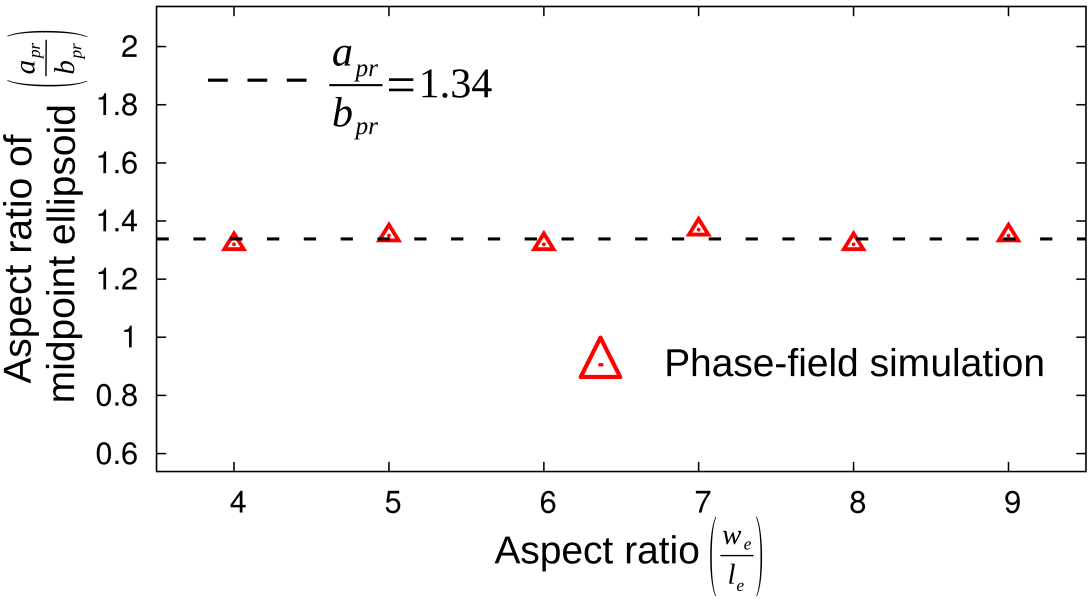}
    \end{tabular}
    \caption{ The aspect ratio of the prolate ellipsoid formed at the midpoint of the transformation during the globularisation of elliptical plates of different aspect ratios.
    \label{fig:fig_5}}
\end{figure}

As opposed to the cylinderization approach, the semi-analytical treatment considers the transitory morphology of the precipitate by encompassing the in-situ data.
The geometric configuration of the precipitate, particularly at the midpoint, is ascertained by tracking the phase-field simulation.
As shown in Fig.~\ref{fig:fig_4}, the morphology of the precipitate at the midpoint is considerably closer to the final structure than the initial elliptical plate.
Accordingly, the cylinderization approach, wherein the morphological parameters at the midpoint are estimated from the mean of the respective terms at the initial and final stages, appears to deviate from the observed shape-change.
Therefore, the geometrical configuration of the precipitate at the midpoint is analysed and employed in the semi-analytical formulation.

It is identified that, similar to the other finite structures, the precipitate assumes ellipsoidal shape at the midpoint of the globularisation.
However, in contrast to the pancake structure, the midpoint ellipsoid exhibits a different geometric relation which is associated with the prolate spheroid, $a_\text{pr}>b_\text{pr}=c_\text{pr}$, where $a_\text{pr}$, $b_\text{pr}$ and $c_\text{pr}$ are semi-major and minor axes lengths, respectively.
In other words, irrespective of the initial aspect ratio, the pancake structure transforms into a oblate spheroid ($a_\text{pr}<b_\text{pr}=c_\text{pr}$), while the elliptical plate assumes prolate spheroid morphology at the midpoint.
The aspect ratio of the midpoint prolate-spheroid ($\frac{a_\text{pr}}{b_\text{pr}}$), which is formed during the transformation of the elliptical plates of various sizes, is ascertained and plotted in Fig.~\ref{fig:fig_5}.
It is evident that, irrespective of the initial aspect-ratio of the plate, a definite relation is established between the semi-major and -minor lengths of the midpoint ellipsoid, $a_\text{pr}=1.34b_\text{pr}$.
In the present approach, the midpoint driving-force is formulated based this characteristic aspect-ratio ($\frac{a_\text{pr}}{b_\text{pr}}=1.34$) exhibited by the prolate spheroid.

Since the midpoint ellipsoid which is formed during the globularisation of the elliptical plate exhibits a characteristic relation $a_\text{pr}>b_\text{pr}=c_\text{pr}$, it is assumed that the source is confined to the major-axis.
Therefore, the diffusion area at the midpoint is determined by eliminating the area along the minor axes from the entire surface area of the prolate spheroid.
Correspondingly, the midpoint diffusion-area reads
\begin{align}\label{eq:A12s_plEp}
 A_{\frac{1}{2}:\text{ge}}=4\pi\left\{\left[\frac{2(a_\text{pr}b_\text{pr})^{\mathcal{P}}+b_\text{pr}^{2\mathcal{P}}}{3}\right]^{\frac{1}{\mathcal{P}}}-(a_\text{pr}b_\text{pr}^5)^{\frac{1}{3}} \right\},
\end{align}
where $\mathcal{P}$ is a constant and is equal to 1.6.
In Eqn.~\ref{eq:A12s_plEp}, the surface area of the prolate spheroid is calculated based on \textit{Knud Thomsen} approximation.
Furthermore, in the formulation of the midpoint diffusion-area, the relation between the radius of the ultimate globular-precipitate and prolate ellipsoid parameters, $r_\text{ge}=(a_\text{pr}b_\text{pr}^2)^{\frac{1}{3}}$ which is derived by equating the volume, is implicitly employed.

\begin{figure}
    \centering
      \begin{tabular}{@{}c@{}}
      \includegraphics[width=0.8\textwidth]{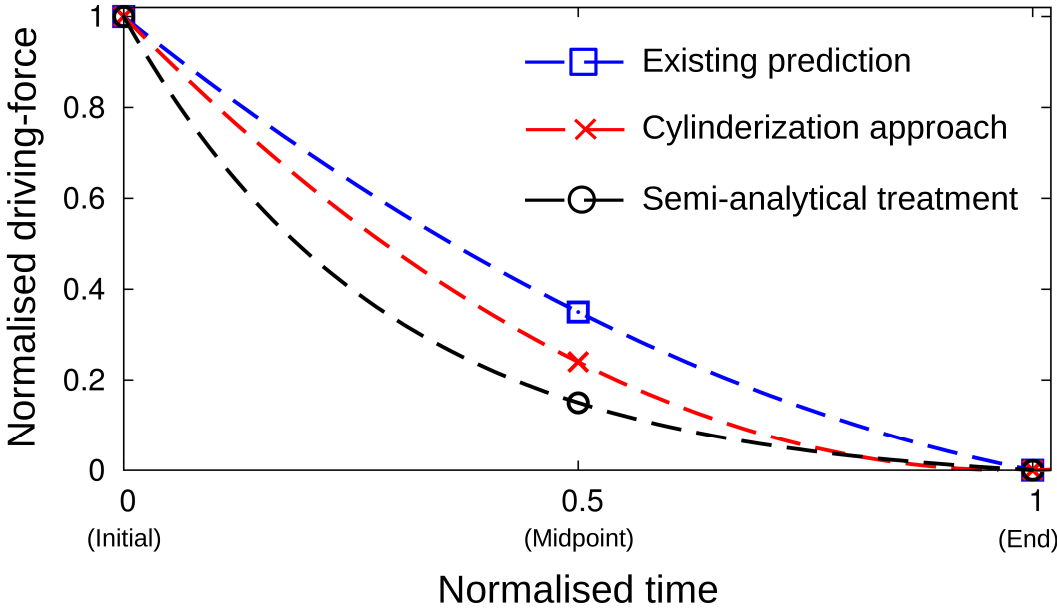}
    \end{tabular}
    \caption{ The instantaneous driving-force at the initial, midpoint and final stage of the globularisation of the elliptical plate of aspect ratio $6$. The predictions of the existing approach is included for comparison.
    \label{fig:fig_6}}
\end{figure}

Unlike the two-dimensional structures, the influence of the inherent curvature-difference of a three-dimensional shape on the equilibrium concentration is elegantly described by the mean curvature.
Therefore, the mean curvature of the midpoint ellipsoid formed during the globularisation of the elliptical precipitate is defined appropriately.
For the delineation of the mean curvature of the prolate spheroid, the fundamental forms of the respective surfaces, as elucidated in the preceding chapters, are realised.
The co-efficients of the first fundamental form of the ellipsoidal surface, which is characterised by geometric relation $a_\text{pr}>b_\text{pr}=c_\text{pr}$, is expressed as 
\begin{align}\label{eq:first_fund}
 E_\text{e}&=a_\text{pr}^2\sin^2\theta+b_\text{pr}^2\cos^2\theta \\ \nonumber
 F_\text{e}&=0\\ \nonumber
 G_\text{e}&=b_\text{pr}^2\sin^2\theta.
\end{align}
Furthermore, the second fundamental form of the midpoint prolate-ellipsoid is defined by the co-efficients
\begin{align}\label{eq:second_fund}
 \tilde{E}_\text{e}&=\frac{a_\text{pr}b_\text{pr}^2}{(b_\text{pr}^4\cos^2\theta+a_\text{pr}^2b_\text{pr}^2\sin^2\theta)^\frac{1}{2}}\\ \nonumber
 \tilde{F}_\text{e}&=0\\ \nonumber
 \tilde{G}_\text{e}&=\frac{a_\text{pr}b_\text{pr}^2\sin^2\theta}{(b_\text{pr}^4\cos^2\theta+a_\text{pr}^2b_\text{pr}^2\sin^2\theta)^\frac{1}{2}}.
\end{align}
In Eqns.~\ref{eq:first_fund} and ~\ref{eq:second_fund}, the angular variable which defines the curvature in a distinct orthogonal co-ordinate system is represented by $\theta$.
Based on the co-efficients of the first and second fundamental forms, the mean radius at a given point on the surface of the prolate ellipsoid reads
\begin{align}\label{eq:mean_curv}
 H=\frac{a_\text{pr}(b_\text{pr}^2+a_\text{pr}^2\sin^2\theta+b_\text{pr}^2\cos^2\theta)}{2(a_\text{pr}^2\sin^2\theta+b_\text{pr}^2\cos^2\theta)(b_\text{pr}^4\cos^2\theta+a_\text{pr}^2b^2\sin^2\theta)^\frac{1}{2}}.
\end{align}
Upon comparing the mean-curvature formulation for the midpoint ellipsoid of pancake and elliptical plate, it is evident that $H$ is governed by both angular variables in the former, while in the latter, it is exclusively dictated by $\theta$.
With $H_{\text{sink}}$ and $H_{\text{source}}$ representing the mean curvature at the sink and source of the prolate spheroid, its influence on the equilibrium concentration is written as
\begin{align}\label{eq:c12s_plEp}
\delta c_{\frac{1}{2}:\text{ge}}\propto \left[ H_{\text{sink}}-H_{\text{source}}\right] = \left( \frac{a_\text{pr}}{b_\text{pr}^2}-\frac{a_\text{pr}^2+b_\text{pr}^2}{2a_\text{pr}^2b_\text{pr}} \right).
\end{align}
Moreover, the spatial component of the concentration gradient which is induced at the midpoint of the morphological evolution is described based on the geometric variable of the prolate spheroid.
Correspondingly, the midpoint diffusion-length is expressed as
\begin{align}\label{eq:x12s_plEp}
 \delta x_{\frac{1}{2}:\text{ge}}=\frac{\pi}{4}\left[ 4(2a_\text{pr}^2+2b_\text{pr}^2)^{\frac{1}{2}}-(a_\text{pr}b_\text{pr}^2)^{\frac{1}{3}} \right].
\end{align}
By combining the diffusion area and the concentration gradient at the midpoint, the respective driving force is estimated by $\Gamma_{\frac{1}{2}:\text{ge}} \propto A_{\frac{1}{2}:\text{ge}}\left(\frac{\delta c}{\delta x}\right)_{\frac{1}{2}:\text{ge}}$.

\subsection{Comparative analysis} 

By adopting the formulation of the cylinderization approach and the semi-analytical treatment, the driving force at three specific stages, initial ($t_{0:\text{ge}}$), midpoint ($t_{1/2:\text{ge}}$) and final ($t_{1:\text{ge}}$), of the globularisation is calculated.
The respective driving forces are normalised and plotted in Fig.~\ref{fig:fig_6}.
The outcomes of the recent analytical treatment~\cite{park2012mechanisms}, wherein the precipitate morphology is assumed to include termination caps, is also presented in Fig.~\ref{fig:fig_6}.
Despite the deviations introduced by the inclusion of the hemispherical caps in the formulation of the driving force, the results are normalised proportionately to render a comparative discussion.

\begin{figure}
    \centering
      \begin{tabular}{@{}c@{}}
      \includegraphics[width=0.8\textwidth]{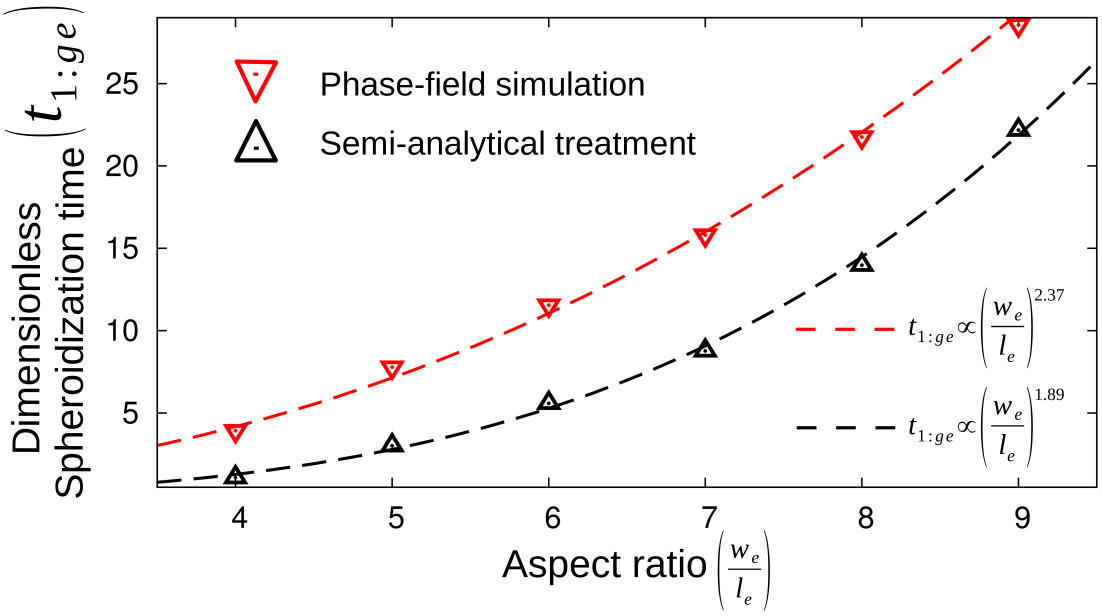}
    \end{tabular}
    \caption{ The time taken for the globularisation of the elliptical plates of different aspect ratio in the phase-field simulation is compared with the prediction of the semi-analytical treatment which is formulated in Sec.~\ref{lab:sm1}.
    \label{fig:fig_7}}
\end{figure}

Fig.~\ref{fig:fig_6} shows that, when compared to the estimations of the present analysis, the recent report on the globularisation kinetics predicts a higher driving-force at the midpoint, approximately $40\%$ of the initial driving-force $\Gamma_{0:\text{ge}}$.
Since the morphology of the precipitate at the midpoint is closer to its final shape, the corresponding curvature-difference (driving force) is expected to be noticeably lower than the prediction of Ref.~\cite{park2012mechanisms}.
The over-estimation of the driving force by this approach can be attributed to two factors.
One is the inclusion of the hemispherical termination-caps.
As opposed to the edges which are the integral part of the precipitate morphology, the caps are augmented inclusions which preserve the initial curvature-difference independent of the plate size (aspect ratio).
The other factor contributing to the excess driving-force is the unphysical consideration that the entire surface of the globularised precipitate is actively involved in the diffusion at the final stages of the transformation.
A reasonable treatment of the final diffusion-area ($A_{1:\text{ge}}$) yields a convincing driving-force through the present cylinderization approach, as shown in Fig.~\ref{fig:fig_6}.

The cylinderization approach, which is elucidated in Sec.~\ref{lab:cy1}, ascertains the midpoint driving-force similar to Ref.~\cite{park2012mechanisms}, wherein the governing parameters at the midpoint is treated as the average of the respective initial and final variables.
However, by the restricting the final diffusion-area to a quadrant of the globularised precipitate, the relatively lower driving-force is predicted.
Furthermore, the unique geometrical consideration of the initial morphology, which renders implicit curved edges without any inclusions (Sec.~\ref{lab:geo_ep}), additionally enhances the cylinderization approach.

A driving force apparently consistent with the midpoint morphology of the precipitate is predicted by the semi-analytical treatment, as shown in Fig.~\ref{fig:fig_6}.
The relation between the geometric parameters of the midpoint structure, which is ascertained from the phase-field simulation, provides the in-situ information hitherto absent in the reported studies~\cite{courtney1989shape,park2012prediction,semiatin2005prediction,park2012mechanisms}.
In the semi-analytical treatment discussed in Sec.~\ref{lab:sm1}, the geometric relation of the prolate spheroid is incorporated in the formulation, which ultimately yields a convincing midpoint driving-force as illustrated in Fig.~\ref{fig:fig_6}.

The time taken for the globularisation of elliptical plate is calculated by
\begin{align}\label{eq:rate}
t_{1:\text{ge}}=\frac{3\delta V_\text{ge}}{\Gamma_{0:\text{ge}}+\Gamma_{\frac{1}{2}:\text{ge}}},
\end{align}
where $\Gamma_{0:\text{ge}}$ and $\Gamma_{\frac{1}{2}:\text{ge}}$ are the driving force at the initial and midpoint of the evolution which is determined by the semi-analytical approach, and $\delta V_\text{ge}$ is the size-dependent required mass-transfer for the globularisation.
\nomenclature{$t_{1:\text{ge}}$}{Time taken for the globularisation of elliptical plate}%
\nomenclature{$\bar{\Gamma}_{\text{ge}}$}{Overall driving-force dictating globularisation of elliptical plate}%
Owing to the geometrically coherent and in-situ informed delineation of the transitory driving-force, of the different formulations, the prediction of the semi-analytical treatments is compared to the outcomes of the phase-field simulation in Fig.~\ref{fig:fig_7}.
Both the simulation and analytical study show a monotonic increase in the time taken for the globularisation with increase in the aspect ratio.
The decrease in the overall transformation rate with increase in the precipitate size is primarily due to the proportionate increase in the required mass-transfer ($\delta V_\text{ge}$).
However, while the semi-analytical treatment predicts the influence of the aspect ratio on the globularisation time as
\begin{align}\label{eq:sma_el}
 t_{1:\text{ge}} \propto \left( \frac{w_\text{e}}{l_\text{e}}\right)^{1.89},
\end{align}
the phase-field simulation of the elliptical plate with the geometric relation $\frac{w_\text{e}}{b_\text{e}}=4$ yields the relation
\begin{align}\label{eq:pfs_el}
 t_{1:\text{ge}} \propto \left( \frac{w_\text{e}}{l_\text{e}}\right)^{2.4}.
\end{align}
Despite being a simplified representation of the effect of precipitate size on the transformation kinetics, Eqns.~\ref{eq:sma_el} and ~\ref{eq:pfs_el} capture the disparity between these theoretical analysis.

Non-conformity between the semi-analytical predictions and the phase-field simulation is expected, since the analytical treatment assumes an \textit{ideally} decreasing driving force.
However, in relation to the comparative studies extended in the previous chapters, difference between the simulation results and the analytical predictions is relatively larger in the elliptical plates.
The primary factor contributing to the noticeable disparity between the semi-analytical and the simulation studies is the intricacy in the morphology of the precipitate.
Unlike the conventional shapes like rods, the delineation of the geometrical parameters which govern the transformation kinetics of three-dimensional elliptical and ellipsoidal structure is inherently complex~\cite{tee2005surface,harris2006curvature}.
Although ellipsoidal shapes are formed during the morphological evolution of the rod and pancake structures, they are predominantly confined to the midpoint of the globularisation.
However, in the present study, the intricate geometrical formulation associated with the elliptical and ellipsoidal structure influences both initial and midpoint driving-force.
The deviations introduced in the driving forces by the inherent geometrical intricacies result in the visible difference between the theoretical studies in Fig.~\ref{fig:fig_7}.

\section{Transformation mechanism}

\begin{figure}
    \centering
      \begin{tabular}{@{}c@{}}
      \includegraphics[width=0.8\textwidth]{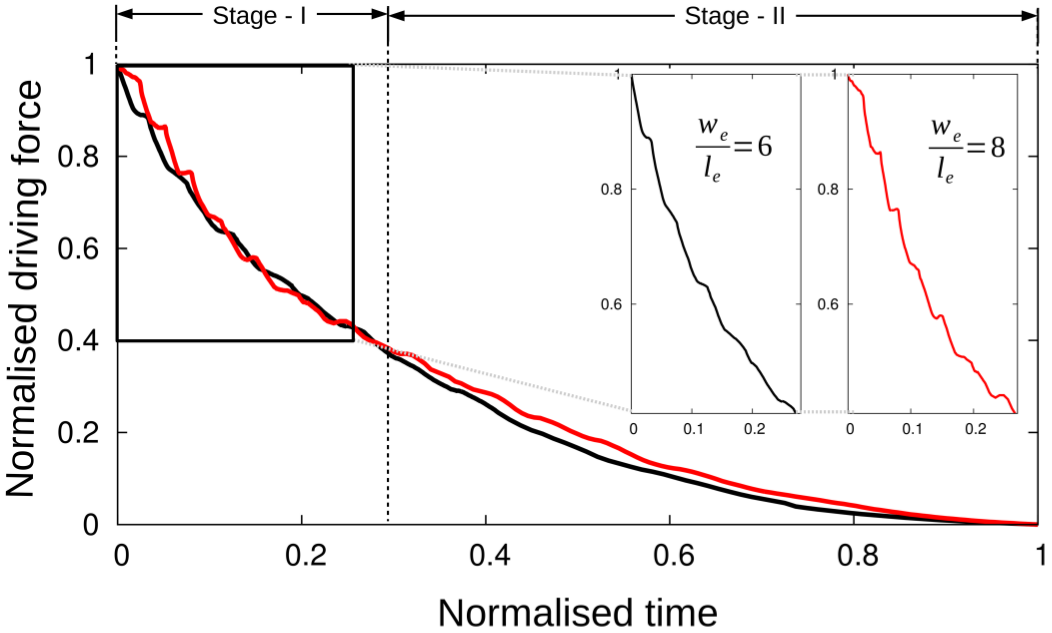}
    \end{tabular}
    \caption{ The temporal evolution of the potential difference which govern the shape-changes exhibited by the elliptical plates of aspect ratios 6 and 8.
    \label{fig:fig_8}}
\end{figure}

To investigate the progressive increase in the disparity between the simulation and analytical results, as shown in Fig.~\ref{fig:fig_7}, the transformation mechanism, which is largely assumed in the analytical treatment, is examined.
Since the morphological evolutions like globularisation are governed by the inherent curvature-difference, that consequently introduce a inhomogeneity in the chemical potential, the transformation mechanism is primarily guided by the incremental change in the potential difference.
Therefore, similar to the previous studies, the temporal decrease in the driving force is analysed by considering the change in the curvature induced potential-difference with time. 
Fig.~\ref{fig:fig_8} illustrates the gradual change in the driving force which undergird the globularisation of the elliptical plates of aspect ratio 6 and 8.

A geometric feature which relates the pancake shape, studied in the previous chapter, to the elliptical plate is the presence of a significant amount of flat surfaces.
Correspondingly, as observed in the pancake structure, Fig.~\ref{fig:fig_8} unravels that the globularisation of the elliptical plate is not established by the smooth monotonic decrease in the curvature difference.
However, while the progressive decrease in the driving force is interrupted by a series of non-monotonic peaks in the globularisation of the pancake shapes, a unique \textit{step-wise} evolution is exhibited by the elliptical plates.
Despite these difference in the trend, the characteristic temporal evolution of the potential difference in both pancake and elliptical plates is confined to the early stages of the transformation.
In other words, analogous to the pancake-shaped precipitate, the morphological evolution of the elliptical plate can be distinguished into two stages based on the change in the driving force with time.
The \textit{stage-I} of the shape-change is dictated by the characteristic temporal evolution of the potential difference, while the driving force decreases smoothly and monotonically in the second stage.

The framework of the analytical treatments, including the revisited formulation wherein the in-situ information are adopted to describe the driving force, do not comprehensively encompass the temporal change in the driving force.
Therefore, the characteristic evolution of the potential difference which is substantially different from the monotonic decrease, in the early stages of the globularisation, introduces the disparity between the analytical and the simulation results.
Furthermore, with increase in the aspect ratio of the elliptical structure, the step-wise evolution of the driving force gets more resolved as shown in Fig.~\ref{fig:fig_8}.
Consequently, the outcomes of the simulation progressively deviate from the analytical prediction.

\subsection{Stage - I}\label{lab:fr1}

\begin{figure}
    \centering
      \begin{tabular}{@{}c@{}}
      \includegraphics[width=0.8\textwidth]{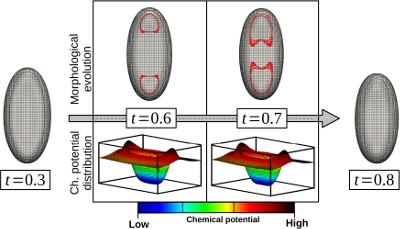}
    \end{tabular}
    \caption{ The shape change exhibited by the elliptical plate of aspect ratio 6 in the \textit{stage-I} of globularisation.
    \label{fig:fig_9}}
\end{figure}

Analogous to other structures, the characteristic evolution of the curvature (potential) difference results in a unique transformation mechanism which is pertinent to both geometrical configuration and aspect ratio of the precipitate.
In Fig.~\ref{fig:fig_9}, the \textit{stage-I} of the morphological changes, which are governed by the step-wise decrease in the driving force, accompanying the globularisation of the elliptical structure of aspect ratio 6 is illustrated.    
Moreover, a three dimensional representation of the distribution of the chemical potential that reflect the inherent curvature-difference is included.

The curvature-driven shape change in the elliptical plate, similar to all the finite structures, begins with the onset of high potential along the terminations of the precipitate.
Correspondingly, the remnant region which, in the elliptical plate, predominantly comprises of flat surfaces assume low potential.
However, owing to the geometrical nature of the precipitate, and in contrast to the pancake structure, the high potential established along the curved edges is not uniform.
Particularly, in the edges along the major axis, the potential is noticeably higher than the minor-axis terminations.
This disparity within the distribution of high potential along the termination, consequently yields a different degree of mass transfer.
Accordingly, in the initial stages of the transformation, the mass transfer from the termination to the adjacent flat surface is primarily confined to the major axis.
The predominant migration of the mass along the major axis results in unique termination ridges which, unlike the pancake perturbation, are restricted to certain regions as shown in Fig.~\ref{fig:fig_9} at $t=0.6$.

In the conventional structures like rods, the formation of termination ridges is followed by its stable growth.
However, in the elliptical plate, the morphological evolution succeeding the formation of major-axis perturbation is similar to the transformation of the pancake structure.
The substantial curvature-difference between the termination ridges and flat surface, prevents the stable growth along the direction normal to the elliptical surface.
Therefore, the perturbation expands across the flat region governed by the characteristic change in the potential distribution as shown in Fig.~\ref{fig:fig_9} at $t=0.7$.
The lateral growth of the ridges, as opposed to the thickening, is consistent with the stability of the flat surface to any morphological disturbances~\cite{mullins1959flattening}. 
With time, $t=0.7$ in Fig.~\ref{fig:fig_9}, the expanding major-axis perturbations coalesce and re-establish the flat surface.
However, as observed during the globularisation of the pancake precipitate, the flat surface which result from the coalescence of the perturbation a significantly smaller that its initial distribution.

A single \textit{step}, in the characteristic step-wise evolution of the potential difference, comprises of a drastic change in the driving force followed by a relatively unchanged curvature-difference.
Analysing the morphological transformation, in relation to the temporal evolution of the potential difference, unravels that the drastic change dictates the formation of the major-axis ridges while the constant driving force governs its expansion along the flat surface.
Therefore, the \textit{stage-I} of the globularisation, which is characterised by the series of \textit{steps}, corresponds to the successive formation, growth and coalescence of numerous major-axis ridges that ultimately reduce the flat region, and alter the morphology of the precipitate.  

\subsection{Stage - II}\label{lab:fr1}

\begin{figure}
    \centering
      \begin{tabular}{@{}c@{}}
      \includegraphics[width=0.8\textwidth]{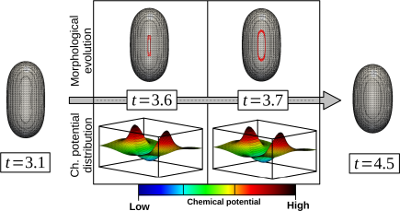}
    \end{tabular}
    \caption{ The transformation mechanism following the \textit{stage-I} globularisation of the elliptical plate of aspect ratio 6.
    \label{fig:fig_10}}
\end{figure}

With the considerable decrease in the amount of flat surfaces, through the repetitive formation and the lateral expansion of numerous perturbations, the elliptical plate transforms to a capsule-like structure at the end of \textit{stage-I} of the globularisation. 
As shown in Fig.~\ref{fig:fig_8}, in the subsequent stage of the evolution, the driving force decreases ideally.
The transformation mechanism corresponding this smooth-monotonic evolution of the curvature difference is illustrated in Fig.~\ref{fig:fig_10}.

During the spheroidization of the smaller three-dimensional rods, the undisturbed monotonic decrease in the driving force results in a transformation mechanism governed by the stable growth of the perturbations, which are introduced in the longitudinal ends.
However, since the \textit{stage-I} of the globularisation of the elliptical plate yields capsule-like structure, the low potential is confined to the midriff of the precipitate.
Therefore, in contrast to the rods, in \textit{stage-II}, the perturbation is induced in the central region of the precipitate directed by the inherent curvature-difference, as shown in Fig.~\ref{fig:fig_10} at $t=3.6$.
The central ridge grows stably through the mass transfer from the high-potential termination to the widening low-potential region.
The outward growth of the perturbation transforms the capsule-like precipitate to ellipsoid structure, and more specifically to prolate spheroid, at $t=3.7$ in Fig.~\ref{fig:fig_10}.
The continued mass transfer, dictated by the potential distribution, and the consequent growth of the perturbations lead to the formation of the globular precipitate.

\section{Conclusion}

Generally, the manufacturing techniques which are devised to achieve required properties, encompass both mechanical and thermal treatment.
One noticeable influence of the mechanical processing, which often involves deformation of the material through excessive stress, on the microstructure is the introduction of the sub-boundaries.
During annealing, the sub-boundaries contribute to the evolution of the phases.
Particularly, the through-thickness sub-boundaries leads to the fragmentation of the seemingly continuous structure, through an unique phenomenon known as boundary-splitting (Fig.~\ref{fig:fig_1}).
The geometrical configuration of the phase (precipitate) emerging from the boundary-splitting is finite and often, unconventional.
In the present chapter, the shape-instability of one such unconventional three-dimensional structure is analysed.

Experimental observations indicate that, in the two-phase titanium alloy, the precipitates assume elliptical plate shape, owing to the previous mechanical treatment, in addition to the other shapes.
The kinetics of the morphological evolution exhibited by the elliptical plates during static annealing is theoretically investigated.
The existing analytical approach delineates the driving force which governs this curvature-driven shape-change by assuming hemispherical caps at the terminations~\cite{courtney1989shape,park2012prediction,semiatin2005prediction,park2012mechanisms}.
Although the hemispherical inclusions facilitate an elegant description of the curvature difference, these caps, despite being unphysical, influence the outcomes of the analytical treatment.
Furthermore, owing to the lack of a comprehensive understanding on the morphological evolution of the plates, the existing study assumes the governing parameters at the midpoint of the globularisation is the average of the its initial and final condition~\cite{park2012mechanisms}.

In order to address the aforementioned limitations, in this analysis, the theoretical treatment is revisited.
By adopting a different geometrical consideration, the need of the inclusions is obviated while retaining curved terminations.
Additionally, the in-situ information on the geometrical parameters of the precipitate which dictate the driving force at midpoint is recovered from the phase-field simulation and incorporated in the formulation.
When compared to the existing work, the current formulations yield relatively consistent midpoint driving-force.

Despite the incorporation of the in-situ data in the delineation, it is identified that the noticeable disparity exists between the analytical predictions and the simulation.
Since the influence of the transformation mechanism is not convincingly included in the analytical treatment, the morphological evolution accompanying the globularisation of the elliptical plate is examined.
In contrast to the current understanding of the curvature-driven shape-change being governed by smooth-monotonic decrease in the driving force, it is realised that the globularisation of the elliptical plate is associated with step-wise decrease in the potential difference.
Correspondingly, as opposed to the stable growth of perturbations, the elliptical structure evolves by successive formation, expansion and coalescence of numerous major-axis ridges.
The characteristic morphological evolution of the elliptical plate, which is consistent well-established insight on the stability of the flat surface~\cite{mullins1959flattening}, while accounting for the disparity between the outcomes of simulation and analytical treatment, unravels a unique transformation mechanism.

\chapter{Pearlite spheroidization: Stability of three-dimensional faceted plates}\label{chap:pearlite}

Owing to the significant influence of the microstructure  on the properties of the material, the shape-instabilities are induced to achieve desired properties.
In addition to the titanium alloys, the practice of invoking morphological changes to enhance the properties is employed to the most-widely used alloy, steel.
A stable phase transformation in steel, which is generally defined as an alloy of iron and carbon, invariably yields a combination of ferrite and cementite, with decrease in temperature.
Often these phases form a interpenetrating bicrystals called pearlite~\cite{hillert1962decomposition}.
Pearlite, which results from the eutectoid decomposition of austenite, is an integral part of the plain carbon steel, and microscopically, appears as an alternating layer of ferrite and cementite.
The characteristic morphology of the self-accommodating phases in pearlite renders high strength and wear resistance to the material~\cite{taleff2002microstructure}.
Therefore, steels with extensive lamellar microstructure, referred to as the pearlitic steels, are preferred for engineering components which include cords and tee rails.
However, the microstructure which is responsible for these mechanical properties, also limits its applicability.

Despite the strength and wear resistance, the workability of steels with lamellar microstructure is inadequate.
Correspondingly, apart from the elementary components, which do not demand forming or other pervasive mechanical treatment during manufacturing, intricate shapes are rarely produced from these steels.
However, a balance in the mechanical properties, which improves the workability is achieved by disrupting the microstructure of the pearlitic steel~\cite{rastegari2015warm,hwang2016influence}.
Specifically, it has been identified that breaking down the lamellar arrangement of the phases improves the toughness, ductility and fatigue-life of steel~\cite{kim2012role,lu2016effect}.
Moreover, the change in the morphology of the constituent phases expands the applicability to pearlitic steel to bearings, bolts and nuts.

Several thermal and thermo-mechanical techniques have been adopted, in large-scale, to disrupt the lamellar microstructure and achieve the required properties.
The most commonly employed heat-treatment techniques include sub-critical, inter-critical and cyclic annealing.
While the morphological changes in the constituent phases are established through phase transformation in inter-critical and cyclic annealing~\cite{hernandez1992spheroidization,lu2016effect}, the lamellar arrangement deteriorates without any transformation in sub-critical annealing~\cite{o2002spheroidization}.
The applicability of the inter-critical and cyclic annealing is restricted, since phase transformation aides the microstructural evolution in these techniques.
Moreover, the inter-critical and cyclic annealing are adopted exclusively to soften the hyper-eutectoid steels.
On the other hand, the microstructural changes during sub-critical annealing occurs in a chemical equilibrium.
Therefore, irrespective of the chemical composition of the material, this technique operates effectively.

Despite its advantages, one factor which hampers the applicability of the sub-critical annealing is the lack of  comprehensive understanding on the microstructural transformations ensuing the treatment.  
The absence of a definite knowledge on the kinetics and the mechanism of the morphological changes, often leads to an inaccurate formulation of the heat treatment cycle, which consequently portrays sub-critical annealing as relatively uneconomical treatment.
Understanding the microstructural transformations, henceforth referred to as spheroidization, through experimental observations is an arduous task, as it requires both three-dimensional representation of the precipitate (cementite) structure and an in-situ observation of its temporal evolution. 
Therefore, theoretical treatments are cooperatively adopted to offer substantial insights on the kinetics and the mechanism of the spheroidization.

The morphological evolution of the cementite, during sub-critical annealing, is governed by the stability of its shape.
Furthermore, by estimating the activation energy, analytical investigations indicate that the transformation of cementite, i.e pearlite spheroidization, is governed by volume diffusion~\cite{chattopadhyay1982kinetics,tian1987kinetics,tian1987mechanisms}.
As elucidated earlier, the theoretical treatment of the shape-changes governed by this specific mode of mass transfer demands three-dimensional consideration.
Therefore, only few attempts have hitherto been made.
To that end, the phase-field approach which has already been employed, in the preceding chapters, to investigate the volume-diffusion governed curvature-driven transformations in three-dimensional rods and other unconventional structures is extended to analyse the pearlite spheroidization.

\section{Domain set-up}

\begin{figure}
    \centering
      \begin{tabular}{@{}c@{}}
      \includegraphics[width=0.8\textwidth]{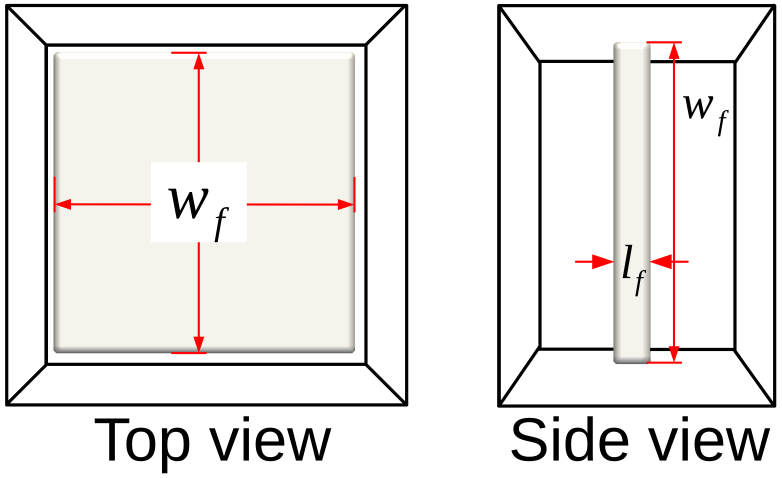}
    \end{tabular}
    \caption{ The morphological configuration of the cementite plate embedded in the ferrite matrix, with length (or width) and length represented by $w_\text{f}$ and $l_\text{f}$, respectively.
    \label{fig:fig_1}}
\end{figure}

Three-dimensional microscopic observations of pearlite reveal that the cementite plates are inherently associated with discontinuities~\cite{tian1987mechanisms,wang2010quantitative}.
The size and distribution of the discontinuities are varied and often, treated as an anomaly in the co-operative growth of ferrite and cementite.
In the absence of these anomalies, the cementite phase in the pearlite resemble a solid square-plate with faceted terminations that extends across the grain.
Therefore, in the present analysis, the domain shown in Fig.~\ref{fig:fig_1} is considered, wherein the faceted three-dimensional cementite structure is encapsulated by ferrite.
Analogous to pearlite, these phases ferrite and cementite are in chemical equilibrium.

The geometrical parameters which dictate the dimensions of the precipitate are included in Fig.~\ref{fig:fig_1}.
The length (or width) of the structure is represented by $w_\text{f}$, while $l_\text{f}$ denotes its thickness.
\nomenclature{$w_\text{f}$}{Length (or width) of cementite plate}%
\nomenclature{$l_\text{f}$}{Thickness (or width) of cementite plate}%
The aspect ratio of the plate, which characterises the size of the structure, is the ratio of its length and thickness ($\frac{w_\text{f}}{l_\text{f}}$).
Since the aspect ratio of the cementite in the microstructure varies with the composition and heat treatment, the evolution of plates with different aspect ratio is investigated.
To achieve the desired aspect ratio, the length of the structure is varied while its thickness remains unchanged.
Moreover, consistent with the experimental reports on fine pearlite structures~\cite{arruabarrena2014influence}, the thickness of the plate is set at $l_\text{f}=0.01$\text{$\mu$}m.
The size of the simulation domain is changed proportionately with the aspect ratio of the plate. 
Furthermore, for a given size, the influence of the boundary on the transformation is averted by considering an optimum domain size.

\section{Termination-migration assisted spheroidization}

Experimental studies on sub-critical annealing indicate that the distortion of the lamellar arrangement begins with the breaking-down of continuous structure into finite isolate precipitates~\cite{chattopadhyay1982kinetics}. 
This break-down of the seemingly infinite structures is achieved either by ovulation near the grain boundaries~\cite{arruabarrena2014influence} or thermal grooving~\cite{mullins1957theory}.
Although the spheroidization initiates with the disintegration of the continuous structures, the morphological transformation of the isolated plate predominantly dictates the kinetics and distribution of the cementite particles. 

\subsection{ Transformation kinetics}

Owing to two factors, theoretical analysis on the pearlite spheroidization have hitherto been limited.
One, the faceted morphology of the precipitate, which makes the curvature difference analytically ill-defined.
Two, the dominant mode of mass-transfer that demands three dimensional consideration.
However, since the phase-field approach elegantly handles these limiting factors, the present study renders a critical insight on the morphological evolution of cementite during pearlite spheroidization. 

The temporal change in the shape, which accompany the spheroidization of the cementite plate of aspect ratio $15$ is shown in Fig.~\ref{fig:fig_2}. 
In view of the existing framework of the analytical treatment, the morphology assumed by the precipitate at the initial ($t_{0:\text{sf}}$), midpoint ($t_{1/2:\text{sf}}$) and final stages ($t_{1:\text{sf}}$) of the spheroidization is chosen to represent the morphological evolution in Fig.~\ref{fig:fig_2}.
It is evident that, similar to the pancake morphology, the faceted plate transforms to a button-like structure at the midpoint.
Furthermore, Fig.~\ref{fig:fig_2} unravels that, although the cementite plate assumes an the ellipsoidal shape at the midpoint of the spheroidization, it holds a unique geometrical relation associated with the oblate spheroid, $a_\text{f}=b_\text{f}>c_\text{f}$, where $a_\text{f}$, $b_\text{f}$ and $c_\text{f}$ are semi-major and -minor axes lengths, respectively.

\begin{figure}
    \centering
      \begin{tabular}{@{}c@{}}
      \includegraphics[width=0.9\textwidth]{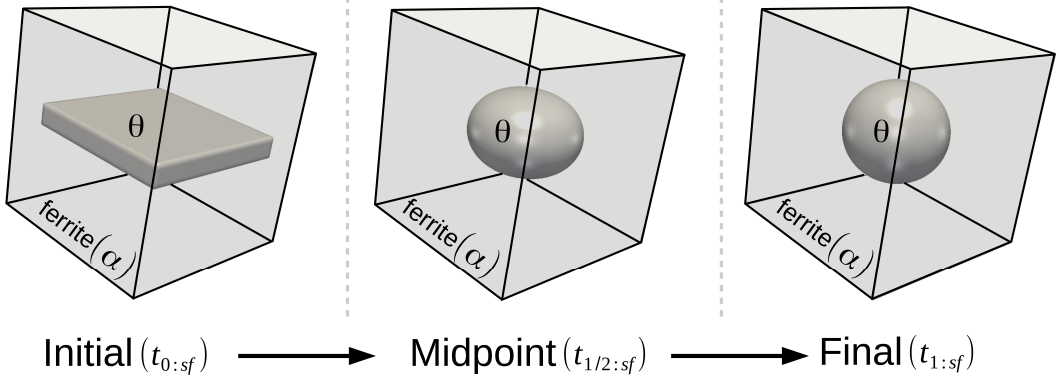}
    \end{tabular}
    \caption{ The shape of the faceted precipitate of aspect ratio $15$ at the initial ($t_{0:\text{sf}}$), midpoint ($t_{1/2:\text{sf}}$) and final stage ($t_{1:\text{sf}}$) of the spheroidization.
    \label{fig:fig_2}}
\end{figure}

Previous theoretical analyses of curvature-driven transformations in two- and three-dimensional shapes suggest that, for a given structure, the aspect ratio of the precipitate at the midpoint of the evolution is independent of its initial aspect ratio. 
Often, this characteristic behaviour of the shape-instability is exploited to analytically ascertain the kinetics of evolution.
However, faceted morphology of the cementite in pearlite prevents such analytical delineation of the kinetics.
Although hemispherical caps can be augmented along the edges and corners of the plates~\cite{courtney1989shape,ho1974coarsening,park2012prediction}, these inclusions disrupt the physical resemblance of the structure and influence the analytical treatment.
Therefore, the transformation kinetics governing the volume-diffusion governed pearlite spheroidization is ascertained exclusively through the phase-field simulation.

Despite the preclusion of the analytical treatment, in order realise the geometrical configuration of the precipitate at the midpoint of spheroidization, the aspect ratio of the midpoint oblate-spheroid is ascertained from the phase-field simulations.
The influence of the initial size of the cementite structure on the aspect ratio of the midpoint ellipsoid is illustrated in Fig.~\ref{fig:fig_3}.
Analogous to the previous observations, the curvature-driven transformation yields a midpoint morphology which is independent of its initial aspect ratio.
Furthermore, it is identified that the aspect ratio of the midpoint oblate-spheroids is $\frac{a_\text{f}}{c_\text{f}}\approx1.3$. 

A cementite precipitate is considered spheroidised when its aspect ratio is equal to 1.
Fig.~\ref{fig:fig_3} indicates that, at the midpoint of the transformation, the precipitate is much closer to its final stage than its initial configuration.
Therefore, by estimating the influence of oblate spheroids on the mechanical properties, which is intuitively expected to be similar to the spheroidal structures, the thermal cycle for the sub-critical annealing can be formulated correspondingly. 
Furthermore, the relation $\frac{a_\text{f}}{c_\text{f}}\approx1.3$ can be employed to the evaluate the degree of spheroidization.

\begin{figure}
    \centering
      \begin{tabular}{@{}c@{}}
      \includegraphics[width=0.9\textwidth]{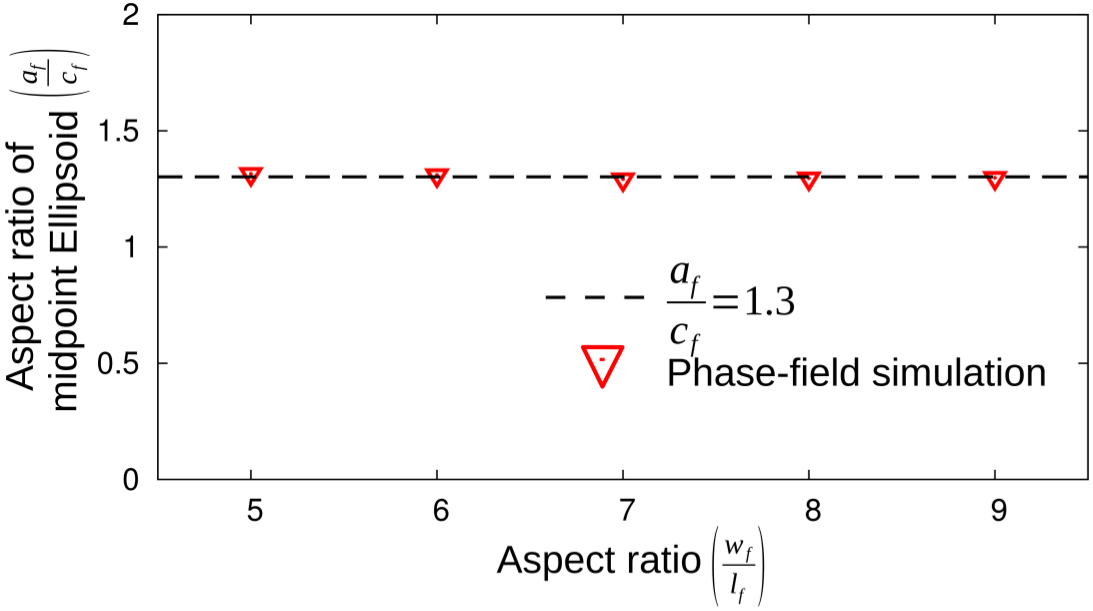}
    \end{tabular}
    \caption{ The aspect ratio of the prolate spheroid, which is formed at the midpoint of the spheroidization of different faceted plates.
    \label{fig:fig_3}}
\end{figure}

The other factor which facilitate the calculation of the spheroidization rate is the amount of atomic migration required to the completely spheroidise the structure, referred to as the required mass transfer ($\delta V_{\text{sf}}$)~\cite{courtney1989shape}. 
Analytically, the required mass-transfer is approximated by eliminating the volume shared by initial and final configuration of the precipitate from the overall volume of the structure.
The characteristic feature of the curvature-driven transformation, wherein the volume of the phases are preserved owing to the chemical equilibrium, facilitates the calculation of the required mass-transfer.
However, with increase in the aspect ratio of the plate the cementite volume increases, and correspondingly, the amount of mass transfer which is required to establish the spherical shape  increases~\cite{courtney1989shape}. 
Therefore, the transformation kinetics is determined by monitoring the time taken for the spheroidization of plates with different initial aspect-ratio.

Based on the outcomes of the phase-field simulations, the time taken for the spheroidization of cementite plates of various aspect-ratio is plotted in Fig.~\ref{fig:fig_4}.
Owing to the increase in the required mass-transfer with the aspect ratio, the transformation time monotonically increases with size.
By fitting the data points, the influence of size on the spheroidization rate can be expressed as
\begin{align}\label{eq:plate_kinetics}
  t_{1:\text{sf}}\propto \left(\frac{w_\text{f}}{l_\text{f}}\right)^{2.39},
\end{align}
where $t_{1:\text{sf}}$ is the dimensionless-time taken for the spheroidization of the cementite and $\frac{w_\text{f}}{l_\text{f}}$ is the initial aspect ratio of the cementite plate.
\nomenclature{$t_{1:\text{sf}}$}{Time taken for cementite spheroidization}%
It is vital to note that the relation in Eqn.~\ref{eq:plate_kinetics} encompasses the effect of the transformation mechanism, unlike the analytical treatment wherein the temporal evolution of the curvature difference is assumed.
Furthermore, decrease in the spheroidization rate with precipitate size, as shown in Fig.~\ref{fig:fig_4}, indicates that the mechanism governing evolution is identical in the plates considered.
However, the substantial understanding on the spheroidization mechanism cannot be gained from the kinetics plot alone.
Therefore, the morphological evolution of the cementite plate is extensively studied to recognize the transformation mechanism.

\begin{figure}
    \centering
      \begin{tabular}{@{}c@{}}
      \includegraphics[width=0.9\textwidth]{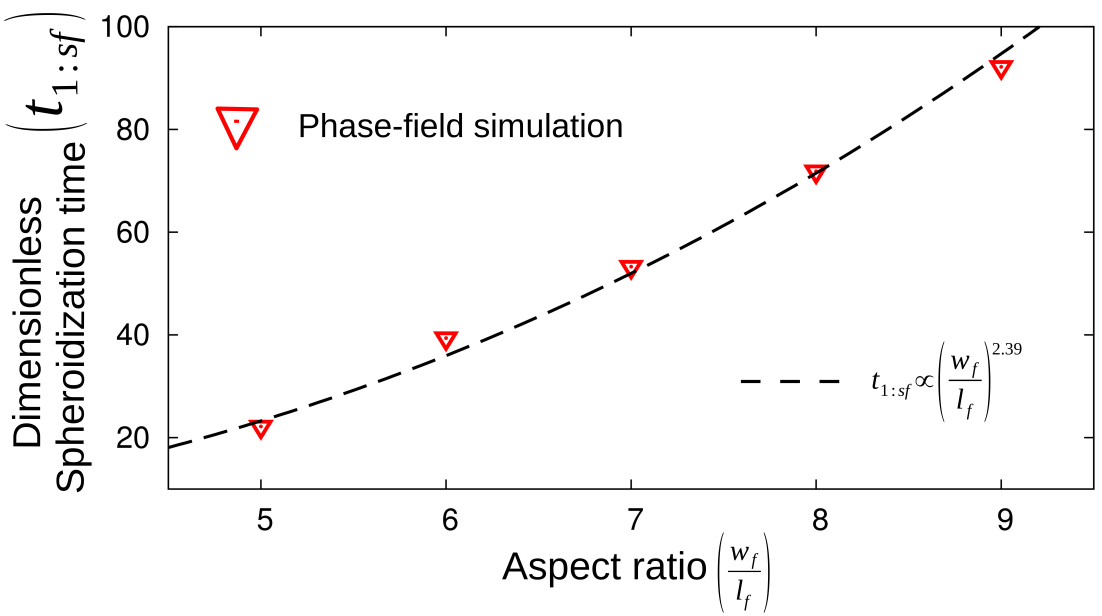}
    \end{tabular}
    \caption{ The monotonic increase in the time taken for the spheroidization of the cementite plate with increasing aspect ratio.
    \label{fig:fig_4}}
\end{figure}

\subsection{Spheroidization mechanism}

Fig.~\ref{fig:fig_5} illustrates the progressive change in the shape of the precipitate of aspect ratio 15, with time, during spheroidization.
A similarity in morphology of the faceted plate, when compared to the pancake and elliptical structure, is the predominant presence amount of flat surfaces. 
However, despite the presence of the significant amount of flat surfaces, the transformation mechanism of the cementite plate is substantially different from other unconventional three-dimension structures studied in the preceding chapters.
While the pancake and elliptical plates evolved through the flattening of the numerous termination ridges, Fig.~\ref{fig:fig_5} indicates that the faceted plates transforms through the stable growth of the perturbation.
The morphological aspect which facilitates the stable growth of the ridges in the faceted plates is the presence of the sharp corners.

\begin{figure}
    \centering
      \begin{tabular}{@{}c@{}}
      \includegraphics[width=0.7\textwidth]{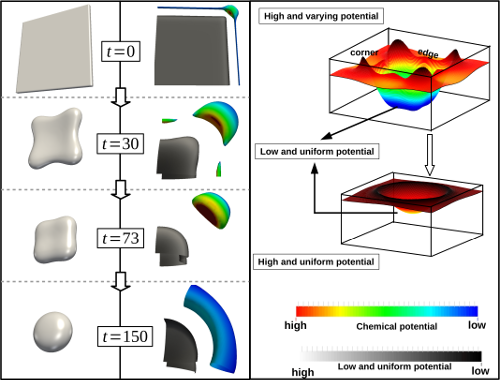}
    \end{tabular}
    \caption{ The termination-migration assisted spheroidization of cementite plate of aspect ratio $15$. The potential distribution in the internal flat-region of the plate and the faceted termination is shown by considering a quadrant. The distinct changes in the distribution is included.
    \label{fig:fig_5}}
\end{figure}

For an exhaustive elucidation of the spheroidization mechanism, the distribution of the chemical potential, which directs the morphological changes, is included Fig.~\ref{fig:fig_5}.
From the outset, owing to the curvature difference, the flat surfaces assume low potential while the high potential is established along the termination of the plate.
The termination which assumes relatively high-potential includes faceted edges and sharp corners, wherein the edges abut.
Since the morphological configuration of the edges are significantly different from the sharp corners, a non-uniformity is established in the distribution of the high potential.
This inhomogeneity in the high-potential distribution, which reflects the disparity of the curvature within the termination, is graphical represented in Fig.~\ref{fig:fig_5} at $t=0$.
Irrespective of the difference in the high-potential region, the overall potential-gradient induces mass transfer from the termination to the near-by flat faces.
This mass transfer results in the recession of the edges and corners, which consequently leads to the formations of thick ridges around the flat surfaces of the plate.
The unique morphology of the precipitate at $t=30$ in Fig.~\ref{fig:fig_5}, particularly the the difference in the size of the edge and corner perturbations vindicates the non-uniformity in the high-potential distribution. 

With time, as the transformation proceeds, this disparity in the terminations continues to decrease through the migration of the termination towards the center of the plate. 
At $t=73$,  the variation in the perturbation size along the termination is visibly subsided.
Any minimal difference between the edges and corners completely vanishes as the precipitate assumes an ellipsoidal shape at $t=150$. 
In other words, the high and non-uniform potential established initially around the plate ($t=0$) becomes uniform as the precipitate transforms to an ellipsoidal shape ($t=150$). 
Despite becoming uniform, the potential surrounding the precipitate remains high when compared to the central region of the plate.
This difference between the uniform potentials, ultimately spheroidises the cementite structure.
Since the morphological evolution primarily results from the recession of the corners and edges, the transformation mechanism is referred to as termination-migration assisted spheroidization. 

The spheroidization of the faceted plates is dictated by the stable growth of the termination ridges.
Therefore, the transformation cannot be distinguished into \textit{stages} based on the morphological evolution, as in the globularisation of pancake and elliptical structures.
However, a similar distinction can be made based on the temporal change in the potential distribution.

The entire mechanism of termination-migration assisted spheroidization can be summarised in two steps as illustrated by the three-dimensional representation of potential distribution in Fig.~\ref{fig:fig_5}. 
In the early stages of the evolution, the potential around the plate is high and varying (non-uniform) due to the inhomogeneous curvature difference between the edges and corners ($t=30$). 
Although the variation in the potential introduces a considerable difference in the mass transfer, as the precipitate evolves, the disparity  is progressively reduced.
Accordingly, the first step of the spheroidization is characterised by the complete neutralisation of the curvature difference within the termination.
At the end of this step, as the high and varying potential becomes high and uniform potential, the precipitate transforms to an ellipsoidal shape ($t=150$).
The subsequent step is driven by the difference in the uniform high- and low-potential established along the termination and central region of the precipitate, respectively.
As the curvature difference completely vanishes, the precipitate becomes spheroidal.

\begin{figure}
    \centering
      \begin{tabular}{@{}c@{}}
      \includegraphics[width=0.6\textwidth]{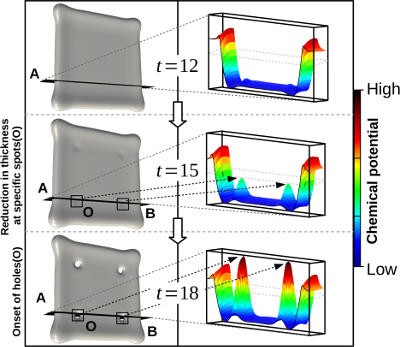}
    \end{tabular}
    \caption{ The transformation of a relatively-large faceted cementite plate of aspect ratio $35$ at the beginning of the spheroidization. The increase in potential at the specific spots (O) of the flat surface leads to the formation circular discontinuities called \lq holes\rq \thinspace.
    \label{fig:fig_6}}
\end{figure}

\section{Discontinuities assisted spheroidization}

Experimental observations indicate the a wide range of isolated plate with different aspect ratios are formed in the initial stages of the static (or sub-critical) annealing~\cite{chattopadhyay1977quantitative}. 
Therefore, the simulation study is extended to the plates with relatively large aspect-ratio to examine a plausible change in the transformation mechanism.

\subsection{Onset of the holes}

\begin{figure}
    \centering
      \begin{tabular}{@{}c@{}}
      \includegraphics[width=0.56\textwidth]{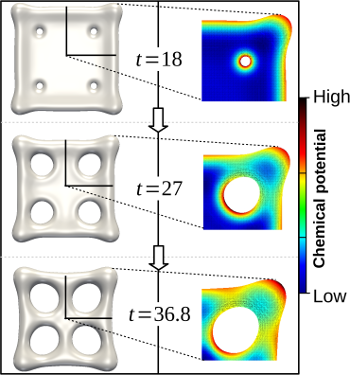}
    \end{tabular}
    \caption{ The temporal evolution of the circular discontinuities accompanying the morphological transformation of cementite plate of aspect ratio $35$. The potential distribution is overlayed on the plate (quadrant) to unravel the diffusion associated with the growth of the holes.
    \label{fig:fig_7}}
\end{figure}

Fig.~\ref{fig:fig_6} shows the initial stages of the spheroidization of the plate with aspect ratio $35$.
As discussed in the previous section, the morphological evolution begins with the termination migration, which consequently leads to the formation of ridges around the flat faces of cementite plate ($t=12$). 
Owing to the disparity in the curvature within the termination, thicker perturbations are introduced at the corners, analogous to the smaller plates.
However, in contrast to the smaller precipitates, wherein the non-uniformity in the termination ridges vanishes with time, in the plate of aspect ratio $35$, the thicker corner-perturbation continues remain stable, as shown Fig.~\ref{fig:fig_6} at $t=15$.
Moreover, directed by the curvature difference established between the thick corner-ridges and adjoining regions, the potential at the 
specific symmetric-spots of the flat surfaces,  which invariably assume low potential in the smaller plates, increases.
The three-dimensional representation of the potential distribution along the appropriate section of the plate, which is included in Fig.~\ref{fig:fig_6}, illustrates the raise in chemical potential.

The increase in the potential at specific spots adjacent to the corner perturbations, induces mass transfer identical to the \lq contra-diffusion\rq \thinspace observed during the curvature-driven transformation of the finite rods~\cite{nichols1976spheroidization}.
This contra-diffusion in cementite facilitates the mass from the flat-surface spots to the receding corner perturbations.
Consequently, as shown in Fig.~\ref{fig:fig_6} at $t=15$, the thickness of the plate at these spots are visibly decreases.
The unhindered mass transfer from the spots, governed by the contra-diffusion, ultimately results in the formation of circular discontinuities called \lq holes\rq \thinspace, at $t=18$ in Fig.~\ref{fig:fig_6}.
Although several postulations have been made for the formation of discontinuities in cementite during the pearlite growth,~\cite{bramfitt1973transmission,frank1956cementite}, a clear understanding on the onset of the holes during spheroidization, though experimentally observed, has not been convincingly reported. 
Based on the phase-field simulations, this study renders the first theoretical description for the formation of the holes during morphological evolution associated with the sub-critical annealing. 

\subsection{Growth of the holes}

The shape-change exhibited by the cementite plate, subsequently following the onset of discontinuities, is presented in Fig.~\ref{fig:fig_7}.
Additionally, the distribution of the chemical potential is overlayed on a quadrant of the precipitate, and included in Fig.~\ref{fig:fig_7}, to explicate the diffusion paths that aide the growth of the discontinuities.
It is evident from the illustration that, with time, the holes begin to grow by transferring mass from the inner rim of the discontinuities, high-potential region, to the adjacent flat surface, which act as the low-potential sink.
Despite the growth, the morphology of the discontinuities remains circular.

The deposition of mass, which is transferred from the inner rim of the discontinuities, to the adjacent flat region results in the formation of thick \lq bank\rq \thinspace around the circular holes.
Although the banks introduce a curvature difference in the flat region which favours its flattening, since the rate of the mass transferred from the discontinuities is greater, the banks surrounding the holes remain stable.
These ridges around the holes grow proportionately with the expansion of the discontinuities and become noticeably thick at $t=27$ in Fig.~\ref{fig:fig_7}.
As the transformation proceeds, with the growth of discontinuities, the banks converge at the region separating the holes.
Consequently, as shown in Fig.~\ref{fig:fig_7} at $t=36.8$, the precipitate sandwiched between the holes assume an unique morphology.
  
\subsection{Coalescence of the discontinuities}

\begin{figure}
    \centering
      \begin{tabular}{@{}c@{}}
      \includegraphics[width=0.35\textwidth]{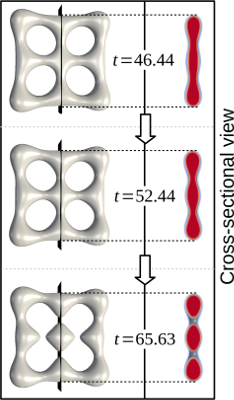}
    \end{tabular}
    \caption{ The fragmentation of the region separating the discontinuities in the cementite plate of aspect ratio $35$, resulting in the coalescence of holes.
    \label{fig:fig_8}}
\end{figure}

The morphology of the region separating the discontinuities is shown in Fig.~\ref{fig:fig_8} by considering its cross-section.
At $t=46.44$, the corresponding section of the plate which is between the holes, appears to have been disrupted by definite perturbations.
As elucidated in the previous section, this characteristic morphology is due to the convergence of the discontinuity-banks at the regions separating the holes.
Although in the subsequent transformation, the size of the holes appears to remain unchanged, the precipitate extending across the holes exhibit shape-change.

As shown in Fig. at $t=52.44$, the apparent \textit{perturbations} in the cross-section of the plate begin grow, not governed by the expansion of the discontinuities, but similar to the Rayleigh-instabilities.
The growth of the ridges in the bridging-region, which separates the holes, subsequently changes the morphology of the discontinuities.
Analogous to the Rayleigh-instabilities, wherein the continuous rod fragments in response to the externally introduced perturbation, the bridging-section of the precipitate breaks-off governed by the perturbation induced by the converging discontinuity-banks.
The disintegration of the region separating the holes, as shown in Fig.~\ref{fig:fig_8} at $t=65.63$, results in the coalescence of the discontinuities.

The coalescence of the holes splits a single precipitate into two distinct entities, at $t=65.63$, a cementite island and surrounding network.
The subsequent shape-change in the spheroidization is governed by the combinatory curvature-driven transformation of the island and network.
Furthermore, it is evident from Figs.~\ref{fig:fig_6}, ~\ref{fig:fig_7} and ~\ref{fig:fig_8} that during the onset, growth and coalescence of the holes, the size of the precipitate does not vary significantly.
In other words, although the initial stages leading-up to the onset of holes is governed by the termination migration, the subsequent morphological evolution of the precipitate is predominantly dictated by the discontinuities.
Therefore, this mode of transformation is referred to as discontinuities-assisted spheroidization.

\subsection{Thermodynamical consistency of the growth and coalescence of the holes}

Conventionally, since the three-dimensional plates are stable to any perturbations, its morphological stability is ascertained by introducing discontinuities.
In order to investigate the evolution of the pre-existing circular discontinuities in an infinitely large plate of thickness $l$, two geometrical parameters $k$ and $n$ are defined~\cite{werner1991growth,wang2010quantitative}.
\begin{figure}
    \centering
      \begin{tabular}{@{}c@{}}
      \includegraphics[width=0.6\textwidth]{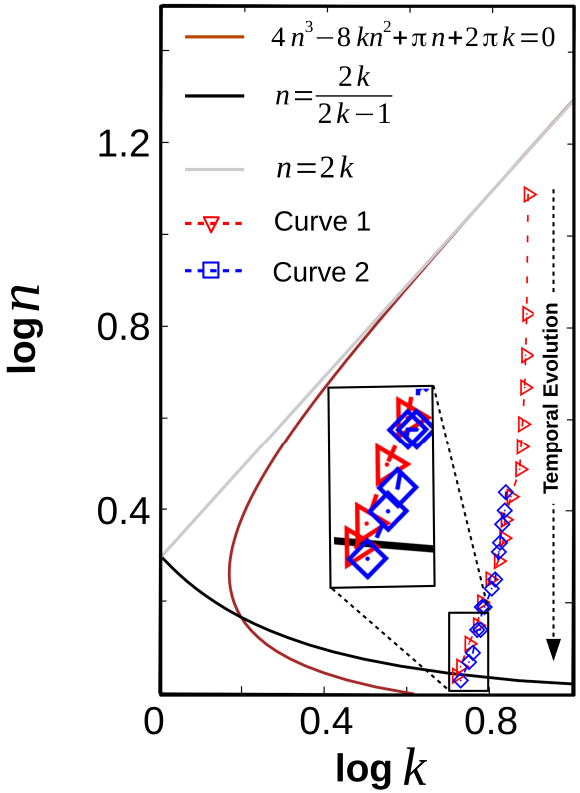}
    \end{tabular}
    \caption{ The temporal change in the geometric parameter $k$ and $n$ which monitor the growth of the circular discontinuities in the cementite plates of aspect ratio 35 (curve 1) and 40 (curve 2) is plotted along with  thermodynamic criterion which ensures the decrease in the interfacial energy of the system~\cite{werner1991growth,wang2010quantitative}.
    \label{fig:fig_9}}
\end{figure}
For a given morphological configuration, the parameters $k$ and $n$ are determined by
\begin{align}\label{eq:k}
  k=\frac{\hat{a}}{l}
\end{align}
and 
\begin{align}\label{eq:k}
  n=\frac{\hat{a}}{\hat{R}},
\end{align}
where $\hat{R}$ and 2$\hat{a}$ correspond to the radius of the circular discontinuity and distance between them from the center.
\nomenclature{$\hat{R}$}{Radius of circular discontinuity}%
\nomenclature{2$\hat{a}$}{Distance between discontinuities}%
In the absence of the phase transformations $i.e,$ when the volume of the constituent phases are preserved, it has analytically been shown that the discontinuities grow only under a specific geometric-condition, which is expressed as
\begin{equation}\label{hole_eqn}
4n^3-8kn^2+\pi n+2\pi k \leq 0.
\end{equation}
The criterion in the above Eqn.~\ref{hole_eqn} ensures that the growth of the discontinuities is consistent with the reduction of the overall interfacial area per unit volume.

As opposed to the holes which are introduced apriori in the analytical treatments~\cite{werner1991growth,wang2010quantitative}, the discontinuities in the cementite plates are inherently formed during the morphological evolution.
Despite this difference, since the evolution of the holes in both these cases is expected to be governed by the same thermodynamical driving-force, which decreases interfacial energy of the system, the evolution of the discontinuities accompanying the spheroidization of the cementite plate is analysed in relation to the criterion in Eqn.~\ref{hole_eqn}. 

In Fig.~\ref{fig:fig_9}, the geometrical criterion for the growth of the discontinuities is graphically represented by solid-continuous brown line.
The parameters $k$ and $n$ are measured for the circular discontinuities induced during the spheroidization of plates of aspect ratio $35$ and $40$.
The temporal changes in these parameters $k$ and $n$ are plotted in Fig.~\ref{fig:fig_9} as curve $1$ and $2$, representing the plates of aspect ratio $35$ and $40$, respectively. 
The visible difference in the initial position of the curves in Fig.~\ref{fig:fig_9} is due to the difference in the transformation rate between the plates.
As shown Fig.~\ref{fig:fig_9}, all-through the evolution, the parameters $k$ and $n$ pertaining to both the plates are confined to the the region defined by the criterion in Eqn.~\ref{hole_eqn}.
This characteristic evolution of the parameters $k$ and $n$ indicates that the growth of the holes, accompanying spheroidization, is thermodynamically consistent and is driven towards the decrease in the interfacial energy of the system.

In Fig.~\ref{fig:fig_9}, the tail-end of the temporal change in the geometric parameters $k$ and $n$ is zoomed-in and included as a subset.
This graphical representation unravels that towards the end of the discontinuities-growth, the parameters $k$ and $n$ cross over a curve defined as $n=\frac{2k}{2k-1}$.
The onset of Rayleigh-like instabilities is analytically described by this curve.
Therefore, the migration of the geometric parameters across the Rayleigh curve (solid-continuous black line) suggest that the coalescence of the holes occurs by the growth of perturbations.
Although, in the previous section, the disintegration of the region separating the discontinuities is elucidated based on the shape-change exhibited by its cross-section, Fig.~\ref{fig:fig_9} reveals that the coalescence is established by the Rayleigh-instabilities.

\subsection{Fragmentation of the cementite network}

\begin{figure}
    \centering
      \begin{tabular}{@{}c@{}}
      \includegraphics[width=0.9\textwidth]{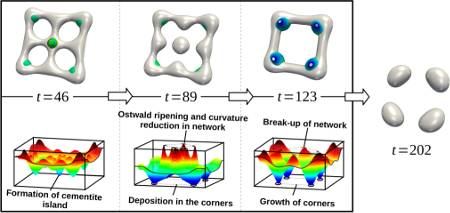}
    \end{tabular}
    \caption{ The formation of the separate cementite entities through the breaking-up of the cementite network. A three-dimensional representation of the chemical-potential distribution which direct the mass transfer is included.
    \label{fig:fig_10}}
\end{figure}

The coalescence of the discontinuities results in the formation of individual island and surrounding cementite.
The morphological changes ensuing the merging of the holes is shown in Fig.~\ref{fig:fig_10}.
Furthermore, three-dimensional representation of the transitory potential-distribution which underpin the transformation in included in Fig.~\ref{fig:fig_10}.
The regions of low potential are highlighted to show the mass-transfer sinks in the precipitate.

Moments before the coalescence of the holes, the perturbations formed by the coalescence of banks introduces potential gradient in the bridging section of the precipitate.
Correspondingly, while specific sections of the region separating the holes assume high-potential, as shown in Fig.~\ref{fig:fig_10} at $t=46$, the central region becomes a dominant sink to the mass transfer.
Moreover, it is evident from the potential-distribution plot that, in addition to the central region, low potential is established in the inner corners of the precipitate.
Governed by the potential difference, the mass gets transferred towards the low-potential sinks which ultimately disintegrates the precipitate into two distinct entities, Fig.~\ref{fig:fig_10} at $t=89$.
The formation of the island and surrounding network has been observed in the theoretical studies involving pre-existing discontinuities~\cite{srolovitz1986capillary,srolovitz1986capillary2}.

As shown in Fig.~\ref{fig:fig_10} at $t=89$, the residual section of the precipitate which bridged the holes introduces inhomogeneity in the inner surface of the cementite network.
This inhomogeneity in the inner section of the network, consequently induces potential gradient wherein the protrusions become the source of mass transfer.
Moreover, since the volume of the island is substantially lower than the network, the potential in the correspondingly region increases in relation to its surroundings.
Dictated by the potential-distribution, and the resulting mass transfer to the adjacent sink, the protrusion in the inner section of the cementite network abates, .
Similarly, due to Ostwald ripening, the island begins to shrink with the mass predominantly transferred to the inner corners of the precipitate
At $t=123$, as shown in Fig.~\ref{fig:fig_10}, the precipitate island and the inhomogeneity in the network disappears due to the progressive mass transfer to the sink.

The accumulation of the mass in the inner corners of the network, renders an appropriate morphology, which reduces the chemical-potential in that the region.
This decrease in potential at the corners induces a potential distribution, as shown in Fig.~\ref{fig:fig_10} at $t=123$, which favours mass transfer from the edges of the network to the sink.
The continued migration of the mass to the low-potential corner, intensifies the distribution and raises the potential at the midriff of the edges.
Consequently, as shown in Fig. at $t=202$, the precipitate network fragments and ultimately, yields four independent entities each pertaining to a corner of the network.
These entities, governed by the inherent difference in the curvature, transform independently into individual spheroids.

In the termination-migration assisted spheroidization, the sharp corners of the faceted plates enables the stable growth of the perturbation.
Additionally, in the discontinuities assisted spheroidization, these corners introduce an appropriate potential-distribution which facilitates the fragmentation of the cementite structure.
While the pinching-off of the network in the pancake structure is not consistently elucidated in the modified perturbation theory~\cite{nichols1965surface}, the orthogonal corners in the cementite structure govern the network break-off during the spheroidization.

\subsection{Variations in discontinuities assisted spheroidization}

Experimental observations suggest that the aspect ratio of the cementite plate involved in the pearlite spheroidization extend upto $70$~\cite{chattopadhyay1977quantitative}.
Therefore, the present numerical analysis is extended to plates of aspect ratio greater than $40$ to explicate any deviation in the previously discussed transformation mechanism.
The shape-change associated with the spheroidization of the faceted plate of aspect ratio $50$ is shown in Fig.~\ref{fig:fig_11}.

Irrespective of the size, owing to the finitude of the precipitate, the morphological transformation begins introduction of high potential along the edges and corners, which consequently leads to the formation of termination ridges.
However, with increase in the aspect ratio, the size of the ridges proportionately increases.
Correspondingly, the larger perturbations induced in the plate of aspect ratio 50 introduce an appropriate curvature-difference, which initiates contra-diffusion, along both edges and corners of precipitate.
In smaller plates, as shown in Fig.~\ref{fig:fig_7}, such curvature-differences are confined to the corners.
As a result, the potential in the regions adjacent to the edges and corners increases, actuating a reverse mass-transfer towards the termination.
The continued contra-diffusion, as shown in Fig.~\ref{fig:fig_11} at $t=24$, nucleates discontinuities that extend along the edges of the plate, in addition to the holes that are close to the corners.
In other words, with the increase in the aspect ratio, the morphology and the size of the discontinuities which are induced by the contra-diffusion consistently vary.

\begin{figure}
    \centering
      \begin{tabular}{@{}c@{}}
      \includegraphics[width=1\textwidth]{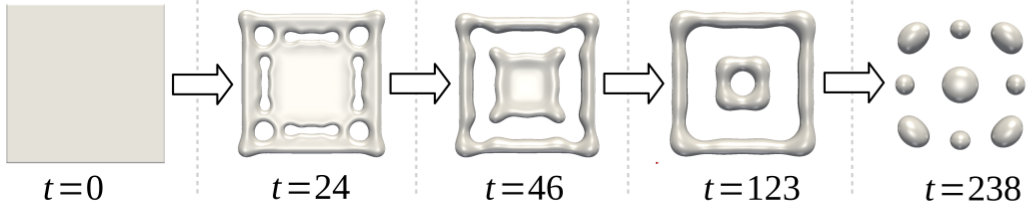}
    \end{tabular}
    \caption{ The morphological evolution associated with the spheroidization of the cementite plate of aspect ratio $50$.
    \label{fig:fig_11}}
\end{figure}

In addition to the corners, since the discontinuities in the larger plates extend along the edges, the coalescence occurs at a relatively-shorter duration.
Fig.~\ref{fig:fig_11} shows the formation of cementite island and network by the merging of the discontinuities at $t=46$.
The evolution of the island and cementite network are analogous to its evolution in the smaller plate (Fig.~\ref{fig:fig_10}).
The inhomogeneity introduced in the network by the coalescence of the discontinuities abates through the mass transfer to the sinks.
However, owing to the increased size, the island independently spheroidises, instead of shrinking.
At $t=123$, while the inner morphology of the network becomes visibly uniform, whereas a central hole in induced in the island.

The network, similar to the smaller plates, governed by the accumulation of the mass in the sinks, and the resulting potential distribution, disintegrates into individual entities as shown in Fig. at $t=238$.
However, much different from the plate of aspect ratio 35, the fragmentation of network yields satellite particles along with the independent corner entities in the larger structures.
Furthermore, the island which vanished due to the coarsening in the smaller precipitate, remains stable throughout the morphological evolution.

It is evident in Fig.~\ref{fig:fig_11} that there exists a significant difference in the size of the individual precipitates at $t=238$.
Particularly, the satellite particles are much smaller than the neighbouring corner entities and central island.
Therefore, owing to this disparity in size, the Ostwald ripening would set-in with time, which ultimately reduces the number of independent spheroids emerging from the morphological transformation.

\section{Cut-off aspect ratio}

\begin{figure}
    \centering
      \begin{tabular}{@{}c@{}}
      \includegraphics[width=1\textwidth]{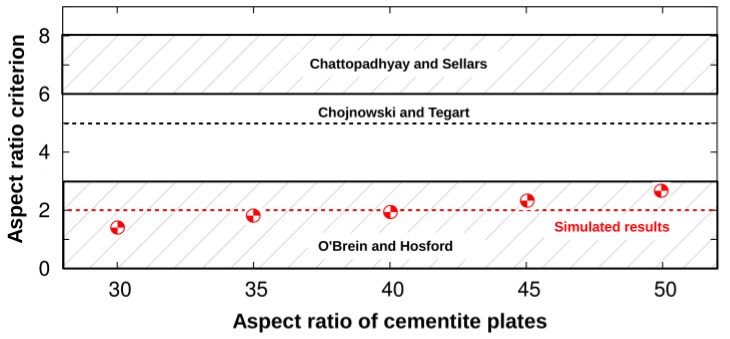}
    \end{tabular}
    \caption{ The cut-off criterion, which is adopted in the industrial heat-treatment techniques to distinguish spheroidised precipitate in a given microstructure, is ascertained from the evolution of the cementite and plotted along with the existing criteria.
    \label{fig:fig_12}}
\end{figure}

In large-scale sub-critical annealing, the precipitates are rarely allow to transform till the aspect ratio becomes $1$.
However, in order to distinguish the spheroidised precipitate from the rest, a cut-off criterion is defined, based on the size of the individual cementite structure.
Conventionally, the cut-off is the maximum aspect-ratio, which renders a single particle without exhibiting any more fragmentations. 
Therefore, the cut-off is always greater than $1$, and when the size of the evolving structure gets smaller than this criterion, it is deemed as a spheroid.

Owing to the lack of in-situ observation, a complete consent in the cut-off criterion has not been achieved yet.
The existing criteria include the early suggestion of Chojnowski and Tegart of $5:1$~\cite{chojnowski1968accelerated}, which was increased to $8:1$ (or $6:1$) by Chattopadhyay and Sellars~\cite{chattopadhyay1977quantitative}, and recently, reduced again to $3:1$ by O'Brein and Hosford~\cite{o2002spheroidization}.
The present theoretical investigation of the morphological evolution of the faceted plates unravels that the disintegration of the network is the final fragmentation of the spheroidization.
Therefore, the aspect ratio of the structure immediately following the pinch-off of the network is plotted along with the existing criteria in Fig.~\ref{fig:fig_12}.
This illustration suggests that the cut-off aspect ratio of $3:1$ is the most appropriate criterion for defining the spheroidization in large-scale industrial heat treatments.

\section{Conclusion}

By employing phase-field approach, which is recovers the sharp-interface laws, the morphological evolution associated with the pearlite spheroidization is analysed.
It is identified that the spheroidization of the plates of aspect ratio $27$ and below, is entirely governed by the recession of the edges. 
This mechanism, wherein the plate appears to be shrinking towards its center by the migration of the boundaries, is referred to as termination-migration assisted spheroidization. 
Since the given plate transforms into a single spheroid, the position of the precipitate is affixed to its center. 

In case of the larger plates, aspect ratio $28$ and above, the transformation mechanism is significantly altered by the introduction of discontinuities. 
Although the spheroidization is initiated by the termination-migration, in these plate, the transformation is predominantly governed by the growth and coalescence of the discontinuities.
Therefore, these morphological evolutions are called as discontinuities assisted spheroidization.
In contrast to the termination-migration assisted transformation, in this mechanism a single plate transforms into multiple spheroids. 
The size and distribution of the spheroids, as shown in  Figs.~\ref{fig:fig_10} and ~\ref{fig:fig_11}, depend on the morphology of the discontinuities which is dictated by the initial aspect-ratio of the plate.

\newpage\null\thispagestyle{empty}\newpage

\afterpage{\blankpage}

\newpage
\thispagestyle{empty}
\vspace*{8cm}
\begin{center}
 \Huge \textbf{Part V} \\
 \Huge \textbf{Conclusion}
\end{center}

\chapter{Conclusion}

\section{Summary}

Understanding shape-instability induced morphological transformations which are governed by the volume-diffusion requires consideration of the entire system.
In conventional approaches, encapsulating the temporal changes in the dynamic variables across the entire domain is  strenuous task. 
However, in the present work, the volume-diffusion governed curvature-driven evolutions are efficiently analysed by adopting phase-field technique. 
The morphological changes in the sharp-interface treatment are monitored by tracking the interface. 
For complex two- and three-dimensional evolutions the interface tracking becomes numerically convoluted. 
This intricacy is implicitly averted by the introduction of the scale variable, phase field. 
Furthermore, in the curvature-induced transformation, the volume fraction of the phases remains unchanged, despite the temporal change in the shape. 
Often, in the phase-field framework, the volume fraction is conserved by solving the Cahn-Hilliard type equation which is relatively more expensive than the time-dependent Ginzburg-Landau equation. 
As discussed in Chapter~\ref{chap:quant} , this thermodynamical condition is quantitatively achieved by establishing chemical equilibrium between the phases through the incorporation of the CALPHAD data.

The analysis of the shape-instability, through the present numerical approach, significantly deepens the current understanding, while being consistent with the existing studies. 
The investigation on the temporal evolution of the lamellar arrangement, by unravelling the diminishing influence of the neighbours with the spacing, renders a convincing support for analysing the stability of the isolated structures. 
Subsequent analyses on the morphological transformation of the isolated finite-structures, in two and three dimension,  offer critical insights on the mechanism and the kinetics of the evolution in relation to the initial configuration of the precipitate.
While the transformation of the two-dimensional rod-like structure is independent of the initial size, a significant change in the mechanism is observed in the three-dimensional rods, beyond a critical aspect-ratio, through the introduction of ovulation.
Furthermore, a unique evolution scheme, which is consistent with the well-established analytical prediction, is exhibited  during the transformation of the circular and elliptical plates.
While the mechanism is independent of the size in pancake and elliptical structures, the morphological evolution of the larger faceted plate are directed by the onset and growth of discontinuities. 
The effect of these mechanism on the kinetics and the distribution of the resulting structures have been extensively analysed in the present work.

\section{Outlook}

Despite the extensive study on the volume-diffusion governed shape-instability in the isolated structures, several other microstructural transformations which can adequately be examined through the present approach are not included in this work.
Therefore, the established applicability of the phase-field approach, elucidated in Chapters~\ref{sec:grand_potentialM} and ~\ref{chap:chm_elast}, is concisely discussed in this section to explicate its potential.

\subsection{Influence of neighbouring structures}

As described in Chapter~\ref{chap:fault}, the morphological stability of the precipitate is considerably influenced by the surrounding structures.
Since this influence depends on the spacing between the phases, the subsequent studies assumed \textit{coarse-microstructural} condition wherein the role of the neighbours are minimal.
However, preliminary work has shown that the present phase-field approach can be employed to analyse the influence of the surroundings on the curvature-driven evolutions~\cite{mittnacht2018understanding}.
Particularly, the effect of interlamellar spacing on the stability of the lamellar microstructure.
It has been identified that the neighbouring precipitates influence both the kinetics and the mechanism of the morphological evolution.
Therefore, the phase-field technique can be adopted to extensively examine the role of neighbouring structures, of identical and (or) different shape, on the transformation.
Such investigations would significantly expand the current understanding on the stability of the microstructure, and possibly, the realisation of an \lq acceleration field\rq \thinspace~\cite{jones1975thermal}.

\subsection{Volume and surface diffusion}

\begin{figure}
    \centering
      \begin{tabular}{@{}c@{}}
      \includegraphics[width=0.65\textwidth]{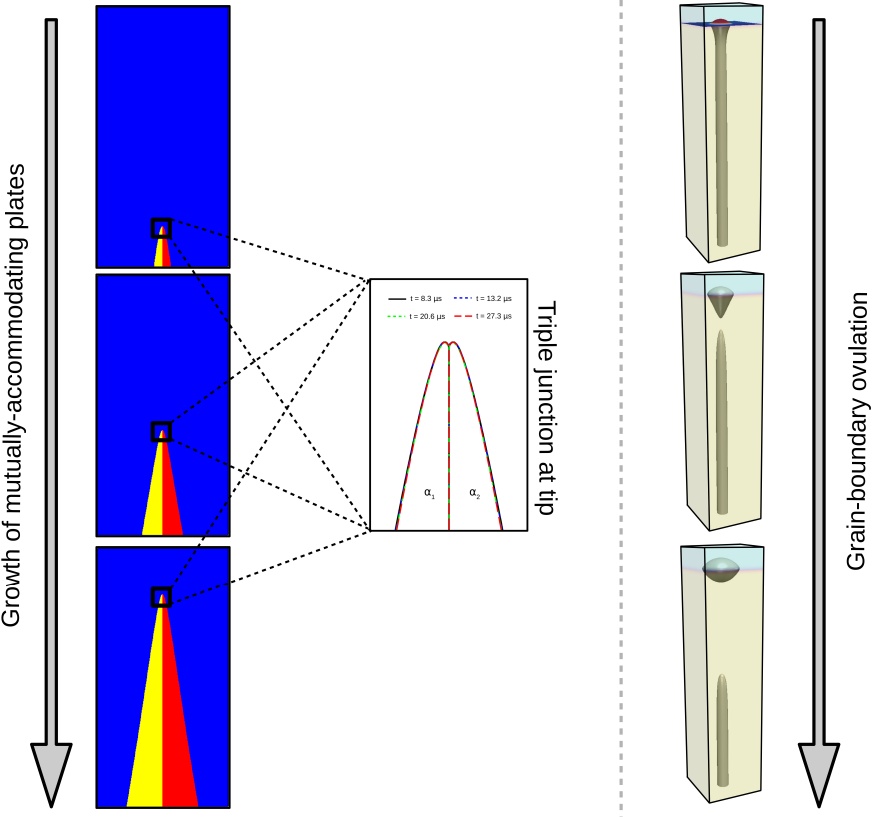}
    \end{tabular}
    \caption{ The outcomes of the preliminary analysis on the co-operative growth of the mutually-accommodating plates and ovulation at the grain boundary.
    \label{fig:coop_growth}}
\end{figure}

Theoretical analyses of the shape-instability are largely confined to one mode of mass transfer, either volume or surface diffusion~\cite{amos2018phase1,amos2018phase2,nichols1976spheroidization,nichols1965morphological}.
Both these modes are rarely considered in a single framework.
Physically, microstructural transformations, though governed by one of these modes of mass transfer, include the other mode as well.
Therefore, a quantitative analysis should accommodate both volume and surface diffusion.
In the phase-field framework, the different modes of mass transfer have been incorporated and employed to study the curvature-driven transformations~\cite{tian2014phase,amos2018globularization,amos2018mechanisms,amos2018volume}.
However, an exhaustive study encapsulating the prevalent duality in the mass transfer and its influence on the morphological evolution is yet to be reported.
To that end, the present numerical approach can be extended to consistently analyse the definitive role of the surface and volume diffusion in a given transformation.

\subsection{Crystallographic orientation}

Depending on the chemical make-up of the system, a specific crystallographic relation exists between the phases in a microstructure.
The crystallographic relation introduces anisotropy to the interfacial energy, which consequently dictates the resulting morphology of the phases during the transformation.
In the phase-field modelling, the anisotropy is introduced by appropriately formulating the contribution of the interfacial energy in the functional~\cite{tschukin2017concepts}.
Moreover, it has recently been reported that the model described in Chapter~\ref{chap:chm_elast} can be adopted to incorporate the anisotropy through inelastic strains~\cite{amos2018chemo}.
As shown in Fig.~\ref{fig:coop_growth}, this multiphase-field model simulates the growth of the mutually-accommodating plate while recovering the sharp-interface law at the triple junctions.
Therefore, the present work, which assumes isotropic interfacial energy, can be extended to included the effect of the crystallographic orientation on the shape-instability induced transformation.

\subsection{Polycrystalline system}

On a mesoscopic scale, conventionally, the microstructure includes numerous grains.
The evolution of the grains, during the static annealing, is entirely governed by the curvature difference.
Since the phase-field approach inherently recovers the Gibbs-Thomson relation, the grain growth has extensively been analysed in resent framework~\cite{perumal2017phase,perumal2018phase}.
As introduced in Chapter~\ref{chap:shape}, one substantial role of the grain boundary in the morphological transformation of the lamellar structure is the fragmentation of the seemingly continuous structure.
Preliminary analysis of this grain-boundary ovulation unravels that the present model renders a physically-consistent morphological transformation, as shown in Fig.~\ref{fig:coop_growth}.
Therefore, the postulated multiphase-approach can be adopted to polycrystalline setup to understand the stability of the microstructure on a mesoscopic scale.

\afterpage{\blankpage}

\newpage
\thispagestyle{empty}
\vspace*{8cm}
\begin{center}
 \Huge \textbf{Part VI} \\
 \Huge \textbf{Appendices}
\end{center}
\appendix

\chapter{Curvilinear co-ordinate system}\label{sec:app1}

\begin{figure}
    \centering
      \begin{tabular}{@{}c@{}}
      \includegraphics[width=0.5\textwidth]{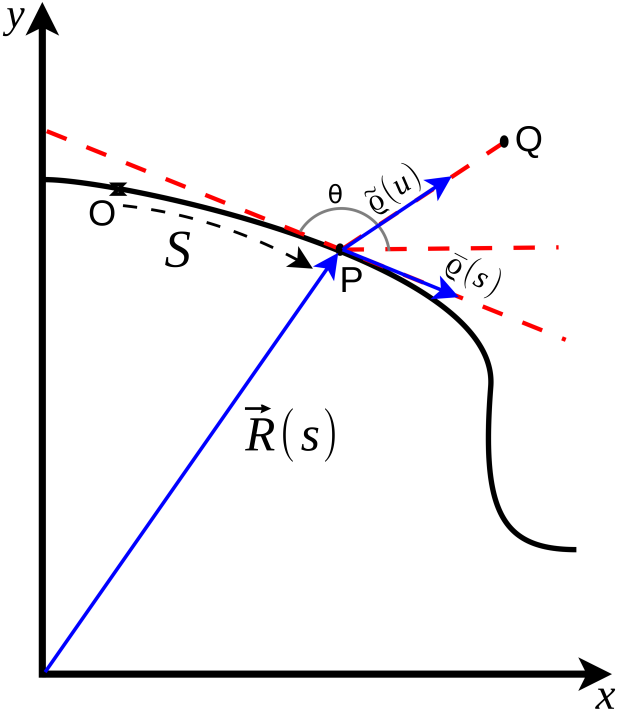}
    \end{tabular}
    \caption{ Transformation of the 2-dimensional Cartesian co-ordinate system to the curvilinear system.
    \label{fig:curvilinear}}
\end{figure}

Cartesian co-ordinate system is conventionally adopted for the theoretical treatment of the problem.
Often, these axes are transformed to other co-ordinate systems like spherical for the ease of handling the variables.
Since the curvature, which plays a critical role in the Gibbs-Thomson relation, is elegantly defined in a curvilinear co-ordinate system, this system is adopted for the present analysis.
Furthermore, the following asymptotic treatment is confined to two-dimension, as the role of the curvature is non-trivial in both two- and three-dimensional setup.

To briefly understand the transformation of the co-ordinate systems, a curved interface (solid black line) and a point Q$(x,y)$, as illustrated in Fig.~\ref{fig:curvilinear}, is considered in a Cartesian system.
From a reference point O on the interface, a point P can be ascertained which is at a distance of arclength $S$.
Point Q lies along the normal to the interface at point P ($u$).
Correspondingly, a local curvilinear co-ordinate system $(u,s)$ can be established by including $s$, which is the tangent to the interface at P.
The unit normal and tangential vector are represented by $\tilde{\bm{\varrho}}$ and $\bar{\bm{\varrho}}$, respectively.
The location of P from the origin is given by the position vector $\vec{R}$.
Furthermore, the angle subscribed between the tangent and the axis parallel to the x-axis is defined as $\theta$.
It is evident from Fig.~\ref{fig:curvilinear} that all the fundamental parameters $\tilde{\bm{\varrho}}$, $\bar{\bm{\varrho}}$, $\vec{R}$ and $\theta$ are influenced by arclength $S$.

The transformation from the Cartesian to the curvilinear co-ordinate system can be made by
\begin{align}\label{j_matrix}
 \begin{bmatrix}
  \frac{\partial \phi}{\partial s} \\\\
  \frac{\partial \phi}{\partial u}
 \end{bmatrix}
=
\bm{J}^{-1}
 \begin{bmatrix}
  \frac{\partial \phi}{\partial x} \\\\
  \frac{\partial \phi}{\partial y}
 \end{bmatrix}
\end{align}
where $\bm{J}^{-1}$ inverse of the Jacobian matrix $\bm{J}$ which can be expressed as
\[
 \bm{J}=
  \begin{bmatrix}
    \frac{\partial s}{\partial x} 		& \frac{\partial u}{\partial x}\\\\
  \frac{\partial s}{\partial y}		& \frac{\partial u}{\partial x}
  \end{bmatrix}.
\]
In order to ascertain the matrix entities, a relation between the Cartesian and curvilinear system is established.
The Cartesian axes $x$ and $y$ are related to the unit base vectors $\tilde{\bm{\varrho}}$ and $\bar{\bm{\varrho}}$ through
\begin{align}\label{eq:baseVx_aysm}
 x= \tilde{\bm{\varrho}}_{x}u+\bar{\bm{\varrho}}_{x}s
\end{align}
and
\begin{align}\label{eq:baseVy_aysm}
 y= \tilde{\bm{\varrho}}_{y}u+\bar{\bm{\varrho}}_{y}s,
\end{align}
respectively.
These base vectors can be written in terms angle $\theta$ as $\tilde{\bm{\varrho}}\to (\text{sin}\theta, -\text{cos}\theta)$ and $\bar{\bm{\varrho}}\to (-\text{cos}\theta, \text{sin}\theta)$, since $\bar{\bm{\varrho}} \sim -\frac{\partial \tilde{\bm{\varrho}}}{\partial \theta}$.
Substituting the $\theta$ based description of the unit vectors in Eqn.~\ref{eq:baseVx_aysm} and ~\ref{eq:baseVy_aysm}, the matrix entries pertaining to the X-axis are expressed as
\begin{align}\label{eq:partialx_u}
 \frac{\partial x}{\partial u}=\text{sin}\theta
\end{align}
and
\begin{align}\label{eq:partialx_s}
 \frac{\partial x}{\partial s}=\frac{\text{sin}(\theta(s))}{\partial s}+\bar{\bm{\varrho}}_{x}=\bar{\bm{\varrho}}_{x}-uk\text{cos}\theta.
\end{align}
It is vital to note that, in Eqn.~\ref{eq:partialx_s} the term $k$ which represents the curvature is introduced because $\frac{d\theta}{ds}=-k$. 
Extending similar treatment to the terms pertaining the Y-axis, the corresponding matrix entities read
\begin{align}\label{eq:partialy_su}
 \frac{\partial y}{\partial u}=-\text{cos}\theta 	&&& \frac{\partial y}{\partial s}=\bar{\bm{\varrho}}_{y}-uk\text{sin}\theta.
\end{align}
By substituting Eqns.~\ref{eq:partialx_u},~\ref{eq:partialx_s} and~\ref{eq:partialy_su} in Eqn.~\ref{j_matrix} the co-ordinate system can be transformed.
From the transformed Eqn.~\ref{j_matrix}, the corresponding expression for the gradient and the laplacian in the curvilinear co-ordinate is deduced and written as
\begin{align}\label{eq:curvi_grad}
 \nabla  \to \tilde{\bm{\varrho}}\frac{\partial}{\partial u}+\Big(\frac{1}{1+uk}\Big)\bar{\bm{\varrho}}\frac{\partial}{\partial s}
\end{align}
and 
\begin{align}\label{eq:curvi_lap}
 \nabla \p{2}  \to \frac{\partial \p{2}}{\partial u\p{2}}+\Bigg(\frac{1}{1+uk}\Bigg)k\frac{\partial}{\partial u}+\Bigg(\frac{1}{1+uk}\Bigg)\p{2}\frac{\partial\p{2}}{\partial s\p{2}}-\Bigg(\frac{1}{1+uk}\Bigg)\p{3}u\frac{\partial k}{\partial s}\frac{\partial}{\partial s},
\end{align}
respectively~\cite{provatas2011phase}.

\chapter{Geometrical treatment of three-dimensional ellipsoid}\label{sec:app2}

Extending the analytical investigation to three-dimension correspondingly increases the complexity of formulating the curvatures.
Particularly, at the midpoint of the spheroidization wherein the precipitate assumes an ellipsoidal structure, an appropriate description of the curvature at the sources and sinks is required for the formulation of the free-energy.
The geometrical treatment employed to describe the curvature of the three-dimensional ellipsoid is presented here.

\section{Generalised approach}

Principal curvatures at the point on a surface can be derived by defining a position vector $\vR$.
In the framework of an orthogonal coordinate system, the point can be expressed as
\begin{align}\label{eq:x1}
 \vR = (x,y,z)
\end{align}
Considering that the each coordinate is function of two additional parameters, the position vector reads 
\begin{align}\label{eq:x2}
 \vR = (x(u,v),y(u,v),z(u,v)).
\end{align}
Based on these variables, a surface can be expressed as $\vR = \vR(u,v)$.
Following the existing geometrical treatment~\cite{harris2006curvature}, the curvature is derived by formulating the \lq fundamental forms of the surface\rq \thinspace which involves derivatives of the position vector with respect to the fundamental variable $u$ and $v$.

A curve on the surface, using the position vector $\vR = \vR(u,v)$, can be defined by fixing the parameter $v$ at $v_0$.
Accordingly, the curve which is similar to latitude in sphere is written as
\begin{align}\label{eq:x3}
 \vR = \vR(u,v_0).
\end{align}
A curve referred to as $v$-curve,  can be defined similar to the $u$-curve in Eqn.~\ref{eq:x3}, by fixing $u$ at $u_0$ which is expressed as $\vR = \vR(u_0,v)$.
The point wherein the $u$-and $v$-curves intersect is represented by the position vector as $\vR(u_0,v_0)$.
The tangent vectors to the $u$-and $v$-curves at the point of intersection is written as
\begin{align}\label{eq:x4}
 \frac{\partial \vR}{\partial u}=\frac{\partial \vR}{\partial u}(u_0,v_0) &&& \text{and} && \frac{\partial \vR}{\partial v}=\frac{\partial \vR}{\partial v}(u_0,v_0),
\end{align}
respectively.
The second derivatives are correspondingly expressed at the intersection point, P.

In order to describe the fundamental forms of the surface, a point Q is chosen of the surface which is adjacent to the intersection point of the $u$-and $v$-curves.
Therefore, the position vector of the point Q reads $\vR(u_0+du,v_0+dv)$, where $du$ and $dv$ are close to zero but not equal to $0$.
The differential vector between the points is written as
\begin{align}\label{eq:x5}
d\vR=\vR(u_0+du,v_0+dv)-\vR(u_0,v_0).
\end{align}
Through linear expansion, the differential vector between points P and Q can be simplified as
\begin{align}\label{eq:x6}
 d\vR=\frac{\partial \vR}{\partial u}du+\frac{\partial \vR}{\partial v}dv.
\end{align}

The \lq first fundamental form\rq \thinspace of the surface at the point of intersection between the $u$-and $v$-curves (P) is described by the dot product of the differential vector, $d\vR$.
Correspondingly, from Eqn.~\ref{eq:x6}, the first fundamental form at point P is expressed as
\begin{align}\label{eq:x7}
 d\vR \bm{\cdot} d\vR =\underbrace{\left[\frac{\partial \vR}{\partial u} \bm\cdot \frac{\partial \vR}{\partial u} \right]}_{\mathclap{:=E}} du^2 + 2\underbrace{\left[\frac{\partial \vR}{\partial u} \frac{\partial \vR}{\partial v}  \right]}_{\mathclap{:=F}}dudv  + \underbrace{\left[\frac{\partial \vR}{\partial v}\bm{\cdot}\frac{\partial \vR}{\partial v} \right]}_{\mathclap{:=G}} dv^2,
\end{align}
where, $E$, $F$ and $G$ first fundamental co-efficients.
By adopting these co-efficients, the first fundamental form is simplified as $Edu^2+2Fdudv+Gdv^2$.

For the formulation of the second fundamental form, the cross product of the tangent vectors in Eqn.~\ref{eq:x4} is considered.
This normal vector at point P is written as
\begin{align}\label{eq:x8}
 \bm{n}=\frac{\partial \vR}{\partial u}\times\frac{\partial \vR}{\partial v}.
\end{align}
The unit normal at point P on the surface, based on Eqn.~\ref{eq:x8}, is ascertained by 
\begin{align}\label{eq:x9}
 \hat{\bm{n}}=\frac{\bm{n}}{n},
\end{align}
where the scalar length $n$ is expressed as
\begin{align}\label{eq:x10}
 n=\left|\frac{\partial \vR}{\partial u}\times\frac{\partial \vR}{\partial v}\right|.
\end{align}
By using unit normal $\hat{\bm{n}}$, the \lq second fundamental form\rq \thinspace of the surface is written as
\begin{align}\label{eq:x11}
 \underbrace{\left[\frac{\partial^2 \vR}{\partial u^2}\bm{\cdot}\hat{\bm{n}}\right]}_{\mathclap{:=\tilde{E}}}du^2 + 2\underbrace{\left[\frac{\partial^2 \vR}{\partial u \partial v} \bm{\cdot}\hat{\bm{n}}\right]}_{\mathclap{:=\tilde{F}}}dudv + \underbrace{\left[\frac{\partial^2 \vR}{\partial v^2}\bm{\cdot}\hat{\bm{n}}\right]}_{\mathclap{:=\tilde{G}}}dv^2.
\end{align}
The second fundamental form of the surface can be simplified using the co-efficients as $\tilde{E}du^2+2\tilde{F}dudv+\tilde{G}dv^2$.

According to Ref.~\cite{poelaert2011surface}, from Eqns.~\ref{eq:x7} and ~\ref{eq:x11}, the curvature at the point P can be described as
\begin{align}\label{eq:x12}
 k = \frac{\tilde{E}du^2+2\tilde{F}dudv+\tilde{G}dv^2}{Edu^2+2Fdudv+Gdv^2}.
\end{align}
The above Eqn.~\ref{eq:x12}, expressed as the ratio of the second and first fundamental form, describes the curvature at P along the direction PQ.
Therefore, by appropriate consideration of the orthogonal system and the fundamental variables ($u$ and $v$), the principal curvatures at point for a given three-dimension structure can be estimated.

\section{Principal curvatures of ellipsoidal structures}\label{sec:app3}

The equation of a three-dimensional ellipsoid reads
\begin{align}\label{eq:x13}
 \frac{x^2}{a^2}+\frac{y^2}{b^2}+\frac{z^2}{c^2}=1.
\end{align}
When the angular variables, like eccentric anomalies, are considered as the fundamental parameters ($u$ and $v$), the variables pertaining to the orthogonal coordinate system are expressed as
\begin{align}\label{eq:x14}
  x(\theta) &= a\cos\theta, \\ \nonumber
  y(\theta,\Theta) &= b\sin\theta\cos\Theta, \\ \nonumber
  z(\theta,\Theta) &= c\sin\theta\sin\Theta.
\end{align}
Correspondingly, based on the angular parameters $\theta$ and $\Theta$, the position vector $d\vR = d\vR(\theta,\Theta)$ describes a point on the ellipsoidal surface.

By adopting the aforementioned generalised geometrical treatment to the present ellipsoidal system, the co-efficients of the first fundamental form of the surface is written as
\begin{align}\label{eq:x15}
 E &= a^2\sin^2\theta+\cos^2\theta(b^2\cos^2\Theta+c^2\sin^2\Theta),\\ \nonumber
 F &= (c^2-b^2)\sin\theta\cos\theta\sin\Theta\cos\Theta, \\ \nonumber
 G &= \sin^2\theta(b^2\sin^2\Theta+c^2\cos^2\Theta). 
\end{align}
Furthermore, the second fundamental co-efficients for the present ellipsoidal surface reads
\begin{align}\label{eq:x16}
 \tilde{E} &= abc \left( b^2c^2\cos^2\theta+a^2c^2\sin^2\theta\cos^2\Theta+a^2b^2\sin^2\theta\sin^2\Theta\right)^{-\frac{1}{2}}, \\ \nonumber
 \tilde{F} &= 0, \\ \nonumber
 \tilde{G} &= abc\sin^2\theta \left( b^2c^2\cos^2\theta+a^2c^2\sin^2\theta\cos^2\Theta+a^2b^2\sin^2\theta\sin^2\Theta\right)^{-\frac{1}{2}}.
\end{align}

Although by involving Eqns.~\ref{eq:x15} and ~\ref{eq:x16}, the curvature at a point on the ellipsoidal surface can be determined from the ratio of the first and second fundamental forms (Eqn.~\ref{eq:x12}), a simpler approach is employed to directly calculate the radius of curvature~\cite{poelaert2011surface}.
The principal radii of the curvature, which is the reciprocal of the principal curvature, is expressed as
\begin{align}\label{eq:x17}
 R_1 &=\frac{1}{\mathcal{H}-\sqrt{\mathcal{H}^2-\mathcal{K}}} \\ \nonumber
 R_2 &=\frac{1}{\mathcal{H}+\sqrt{\mathcal{H}^2-\mathcal{K}}},
\end{align}
where $\mathcal{H}$ and $\mathcal{K}$ represent the mean and Gaussian curvature.
The mean and Gaussian curvature, based on the first and second fundamental co-efficients of the ellipsoidal, are ascertained by
\begin{align}\label{eq:x18}
 \mathcal{H} &=1/2\left[\frac{\tilde{E}G+\tilde{F}E-2\tilde{G}F}{EG-F^2}\right] \\ \nonumber
 \mathcal{K} &=\frac{\tilde{E}\tilde{F}-\tilde{G}^2}{EG-F^2},
\end{align}
respectively.

\newpage\null\thispagestyle{empty}\newpage

%
%



\pagestyle{fancy}
\fancyhf{}
\lhead[\thepage]{Bibliography}      
\rhead[\thesection]{\thepage}
\bibliographystyle{unsrt}
\addcontentsline{toc}{chapter}{Bibliography}
\bibliography{references.bib}
\clearpage

\printnomenclature[2cm] 
\cleardoublepage

%

\end{document}